\documentclass[10pt,a4paper,twoside,openright]{book}

\usepackage{PhDThesis}

\numberwithin{equation}{section}

\makeindex

\begin{document}

\fancyhead[LE]{ }
\fancyhead[RO]{ }
\fancyhead[LO,RE]{ }

%-------------------------------------------------------------------------------
%------------------------  TITLE PAGE (ENGLISH)  ---------------------
%-------------------------------------------------------------------------------

\thispagestyle{empty}

\begin{center}
\begin{figure}
\begin{tabular}{ccc}
\includegraphics[scale=0.2]{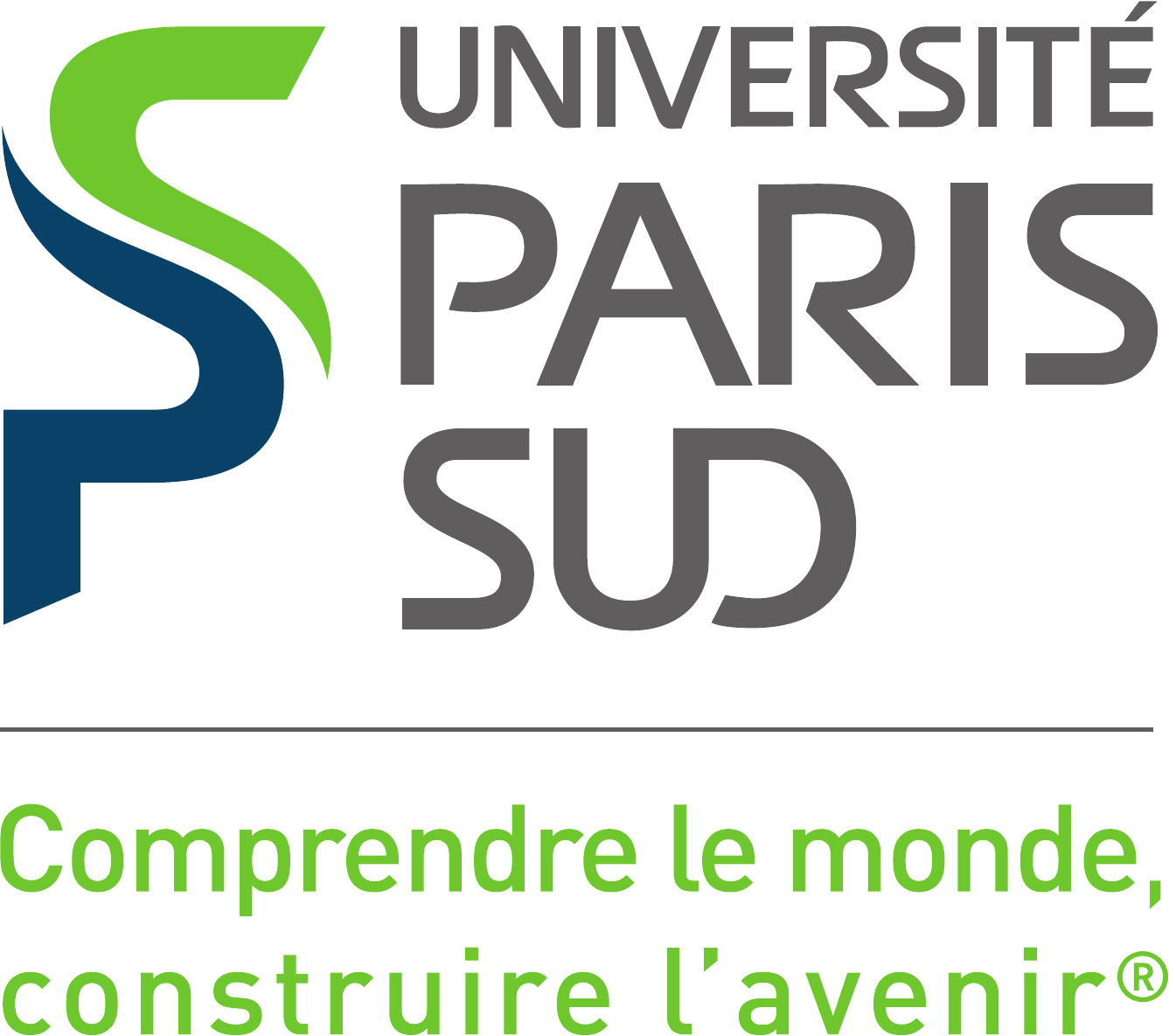}$\qquad\qquad\qquad\qquad$ &
\includegraphics[scale=0.2]{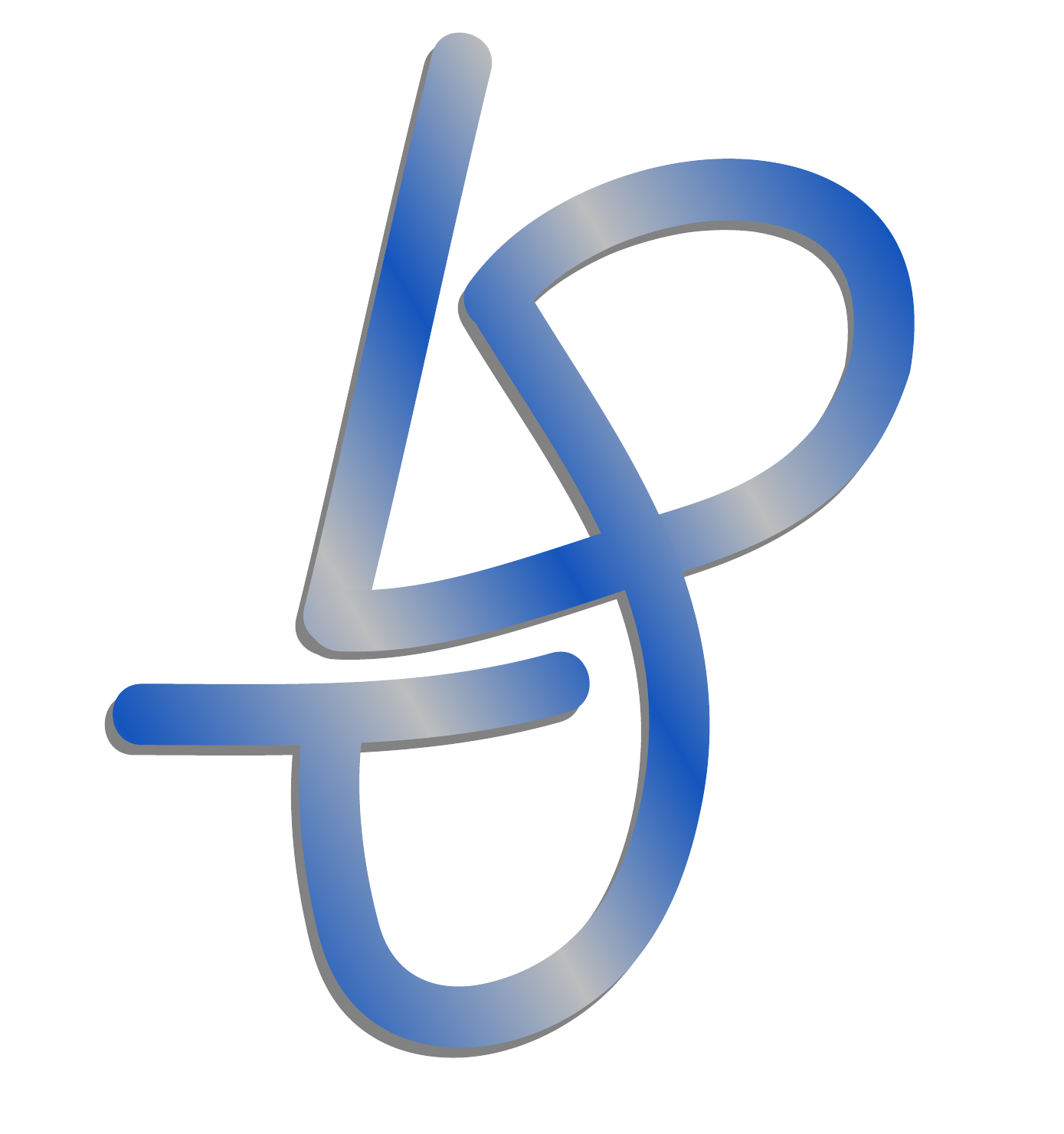}$\qquad\qquad\qquad\qquad$ &
\includegraphics[scale=0.4]{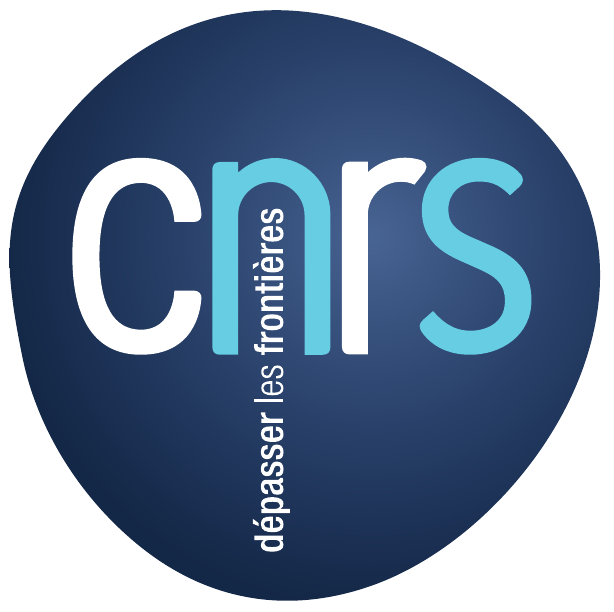}\\
\vspace{-0.5cm}\\
\end{tabular}
\end{figure}
 \begin{flushright}
 LPT-Orsay-13-94 \\ 
 \end{flushright}
{\Large {\bf{THÈSE}}}\\
\vspace{0.3cm}
Présentée pour obtenir\\
\vspace{0.3cm}
\Large {\bf{
LE GRADE DE DOCTEUR EN SCIENCES\\
DE L'UNIVERSITÉ PARIS-SUD XI}}\\
\vspace{0.3cm}
ÉCOLE DOCTORALE\\ de Physique de la Région Parisienne - ED 107\\
\vspace{0.3cm}
Laboratoire de Physique Théorique - LPT
\\
\vspace{0.3cm}
DISCIPLINE: Cosmologie\\
\vspace{0.5cm}
\Large {\bf{Second-order cosmological perturbations in  two-field inflation and predictions for non-Gaussianity}} \\
\vspace{0.5cm}
par\\
\vspace{0.5cm}
Eleftheria TZAVARA\\
\vspace{0.5cm}
Soutenue le Lundi 30 septembre 2013 devant la Commission d'examen:\\
\vspace{1cm}
{\normalsize
\begin{tabular}{crll}
Pr. & Ana & ACH\'UCARRO  & (Rapporteur)\\ 
Pr. & Ulrich & ELLWANGER & (Examinateur)\\
Pr. & David & LANGLOIS   & (Examinateur)\\
Dr. & Karim & MALIK      & (Examinateur)\\
Pr. & Sabino & MATARRESE  & (Examinateur)\\
Pr. & Renaud & PARENTANI & (Directeur de thèse)\\
Pr. & Paul & SHELLARD    & (Rapporteur)\\
Dr. & Bartjan & VAN TENT & (Directeur de thèse)\\
\end{tabular}
}
\end{center}

%%PhD Advisor
%\begin{flushleft}
%{\bf {Ph.D. Advisor:}}\\
%{\bf{ -- }}
%\end{flushleft}
%%Date
%\begin{flushright}
%{\bf{--~Date~--}}
%\end{flushright}

\newpage

~
\vspace{17cm}
\begin{flushleft}
\hspace{4cm}
Thèse préparée dans le cadre de l'École Doctorale 107 au
\\ 
\hspace{4cm}
{\bf Département de Physique d'Orsay}\\ 
\hspace{4cm}
Laboratoire de Physique Théorique (UMR 8627), Bât. 210
\\ 
\hspace{4cm}
Université Paris-Sud 11
\\ 
\hspace{4cm} 
91 405 Orsay CEDEX
\end{flushleft}

\thispagestyle{empty}

%-------------------------------------------------------------------------------

\frontmatter

\fancyhead[LE]{\thepage~~~~Abstract} 
\fancyhead[RO]{Abstract~~~~~\thepage} 

\begin{center}
\textbf{Abstract}
\end{center}

Inflationary predictions for the power spectrum of the curvature perturbation have been verified to an excellent degree, leaving many models compatible with observations. In this thesis we studied third-order correlations, that might allow one to further distinguish between inflationary models. From all the possible extensions of the standard inflationary model, we chose to study two-field models with canonical kinetic terms and flat field space. The new feature is the presence of the so-called isocurvature perturbation. Its interplay with the adiabatic perturbation outside the horizon gives birth to non-linearities characteristic of multiple-field models. In this context, we established the second-order gauge-invariant form of the adiabatic and isocurvature  perturbation and found the third-order action that describes their interactions. Furthermore, we built on and elaborated the long-wavelength formalism in order to acquire an expression for the parameter of non-Gaussianity fNL as a function of the potential of the fields. We next used this formula to study analytically, within the slow-roll hypothesis, general classes of potentials and verified our results numerically for the exact theory. From this study, we deduced general conclusions about the properties of fNL, its magnitude depending on the characteristics of the field trajectory and the isocurvature component, as well as its dependence on the magnitude and relative size of the three momenta of which the three-point correlator is a function.

\chapter*{Acknowledgements}
\fancyhead[LE]{\thepage~~~~Acknowledgements} 
\fancyhead[RO]{Acknowledgements~~~~~\thepage} 
%\thispagestyle{empty}

%Set spanish as the language for the "Acknowledgements" in spanish
%\selectlanguage{spanish}

%Restore English as language
%\selectlanguage{english}

%Set spanish as the language for the "Acknowledgements" in spanish
%\selectlanguage{spanish}

%Restore English as language
%\selectlanguage{english}

%%------------------------------------------------------------------------------------------

No matter how much I try, this section cannot really accommodate the gratitude and warm feelings for all the people that supported me and stood by me during the years of my PhD. To start with, 
I would never have been able to accomplish my thesis without the valuable contribution of my advisor, Bartjan van Tent. Bartjan welcomed me, guided me, encouraged me and trusted me all through these years, offering me an ideal environment for doing research.  I hope he will also remember these years, spent working together, as a fruitful and enjoyable period during his research life.   
  
I would like to thank colleague and friend, Christos Charmousis, who introduced me to LPT and helped me in my first steps, as well as all through these years, with precious advice. I am also very thankful to Renaud Parentani who accepted to be my official advisor and thus offered me the chance to work in LPT. LPT has proven to me to be a very warm and inspiring research background.  
Furthermore, I would like to express my gratitude to all the members in the jury for my defense for the time they took and especially Ana Achucarro and  Paul Shellard for accepting to review my manuscript. 

Finally, I would like to express my gratitude and love to all my close ones who bore with me during all these years with affection,  patience and understanding. I know it hasn't always been easy. Thank you.

\tableofcontents
\fancyhead[LE]{\thepage~~~~Contents}
\fancyhead[RO]{Contents~~~~~\thepage}
\fancyhead[LO,RE]{}
\newpage

\mainmatter

%------------------------------ Chapter 1 --------------------------------------

{\Chapter{Introduction} \label{Ch1}}
%------------------------------------------------------------------------------------------

In this thesis we study the observational predictions of a period in the early universe called inflation. In this introductory chapter, we present the standard cosmological model and explain why this needs to be supplemented by inflation. Next we introduce inflation and connect it with  observations. This introduction is based on \cite{Weinberg:2008zzc,Kolb:1990vq,Lyth:1998xn}.  Finally, we give an outline of the further contents of the thesis.

\Section{The standard cosmological model}

%"What is that which always is and has no becoming; and what is that which is always becoming and never is?", this is how Tima

Observing the night sky has always intrigued the human mind and raised questions about the nature of the luminous  objects seen, as well as about the medium that separates us from them. Today, the answer to these questions is the goal of astronomy, that studies the physics of stars and galaxies, and cosmology, that studies the evolution of the universe as a whole.  
Looking at the sky one can observe different types of objects at different scales. Roughly speaking, at  small scales one can see stars, massive luminous spherical objects of plasma held together by gravity,  like our own sun. There exist different kinds of stars with respect to their radius, luminosity and composition depending on their stage of life and the material available in their surrounding region (the interstellar region) at the time of their creation.    

\begin{figure}
\begin{center}
\begin{tabular}{cc}
\includegraphics[scale=0.4]{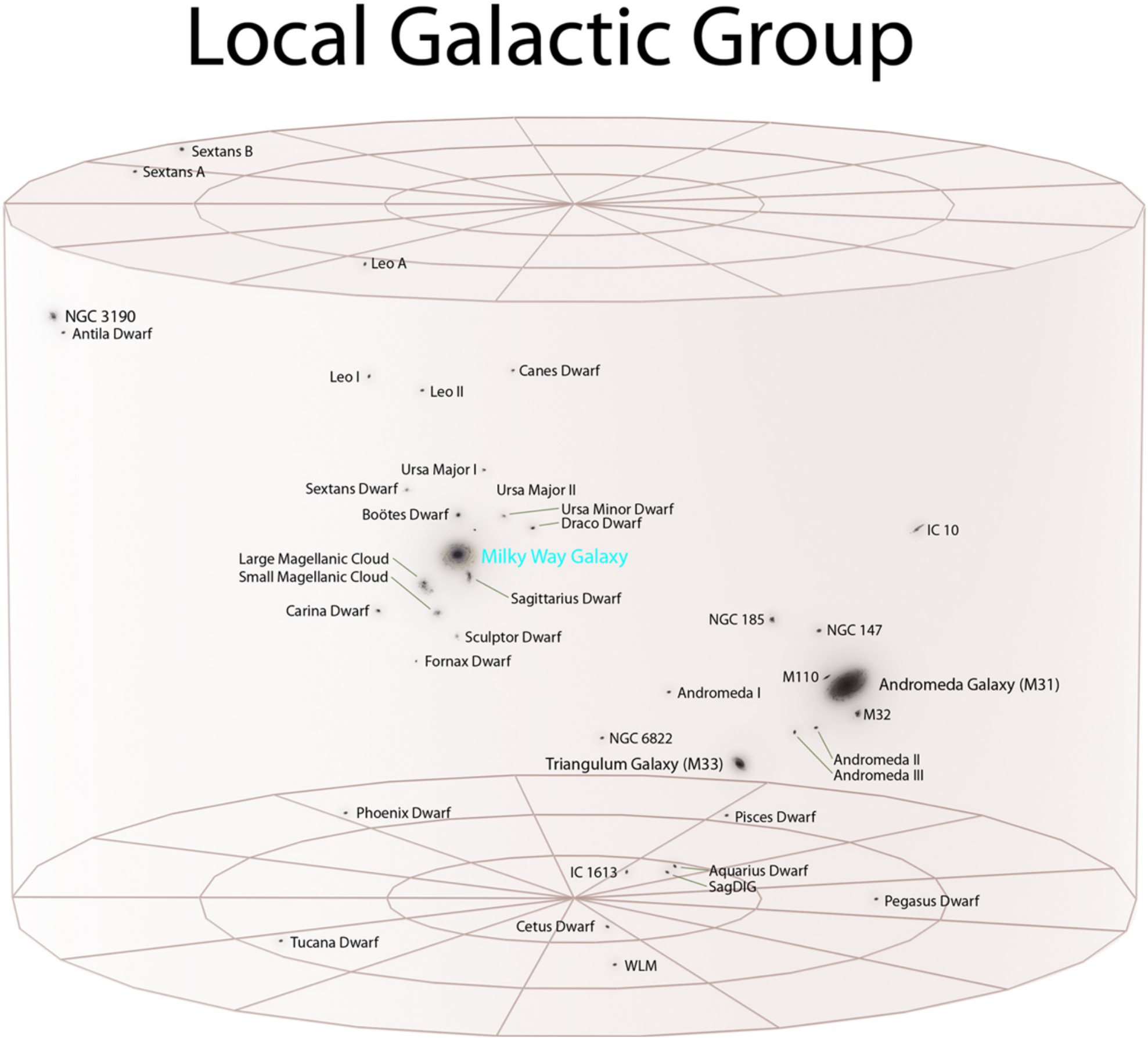}&
\includegraphics[scale=0.4]{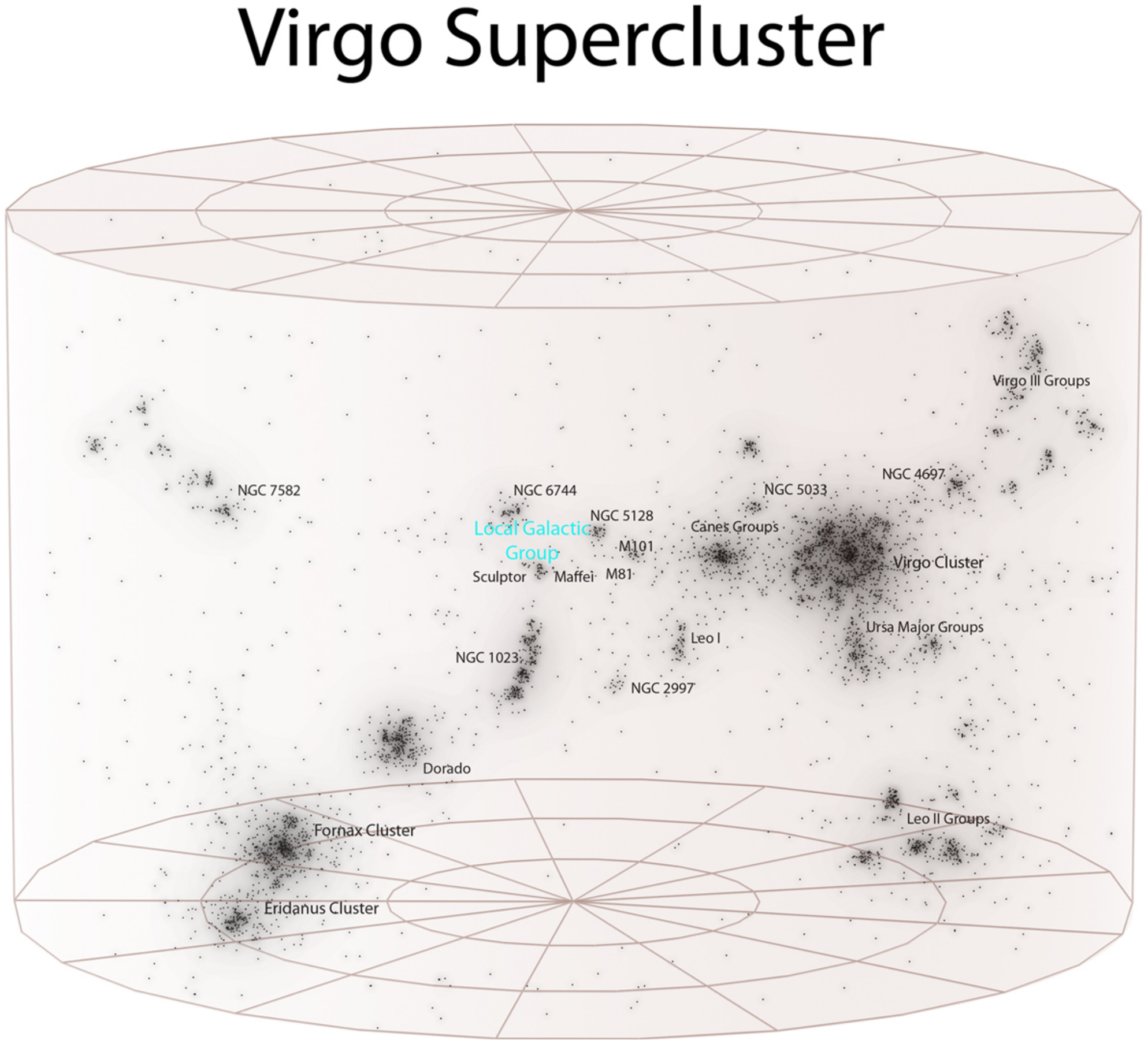}\\
\includegraphics[scale=0.4]{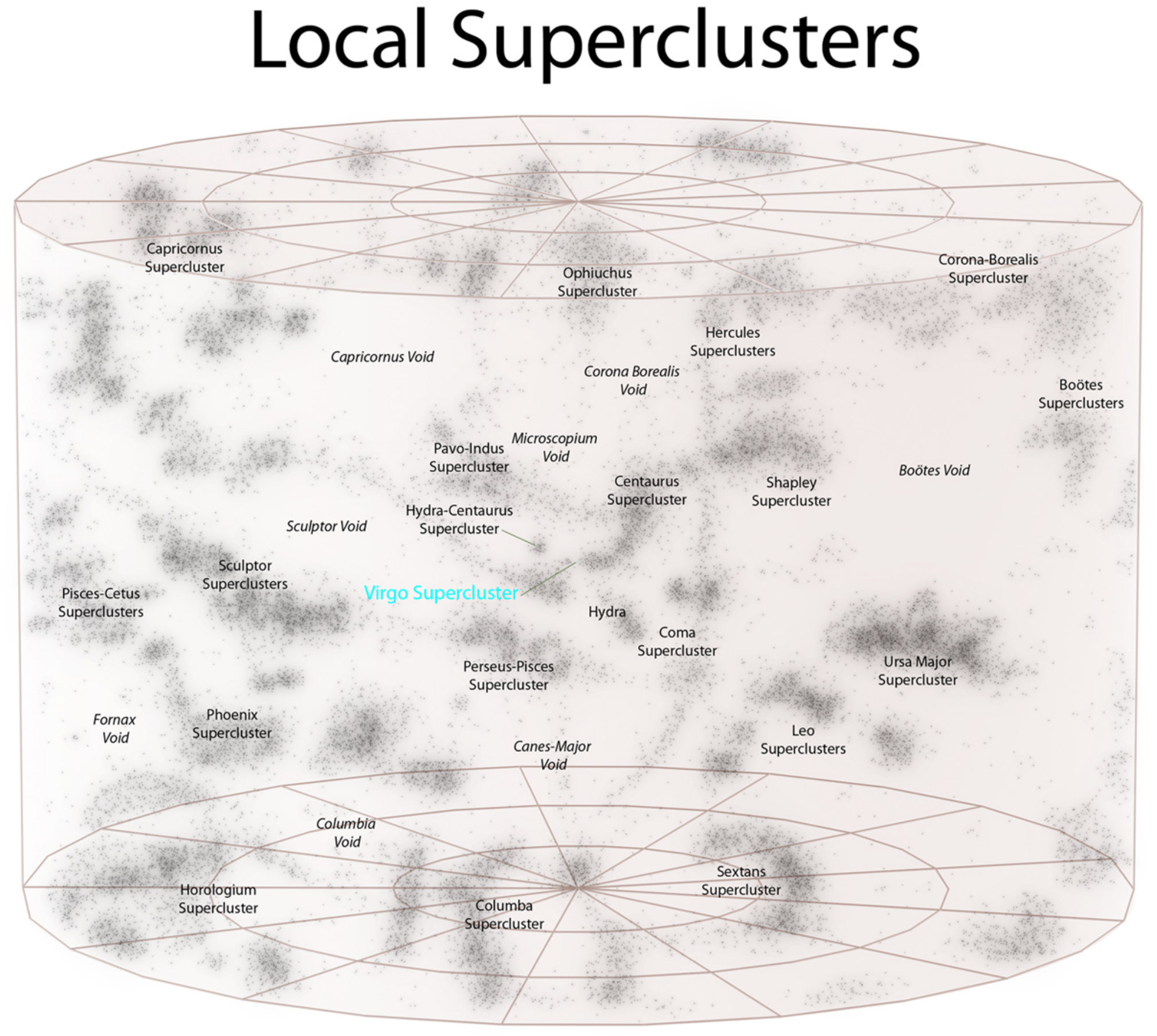}&
\includegraphics[scale=0.4]{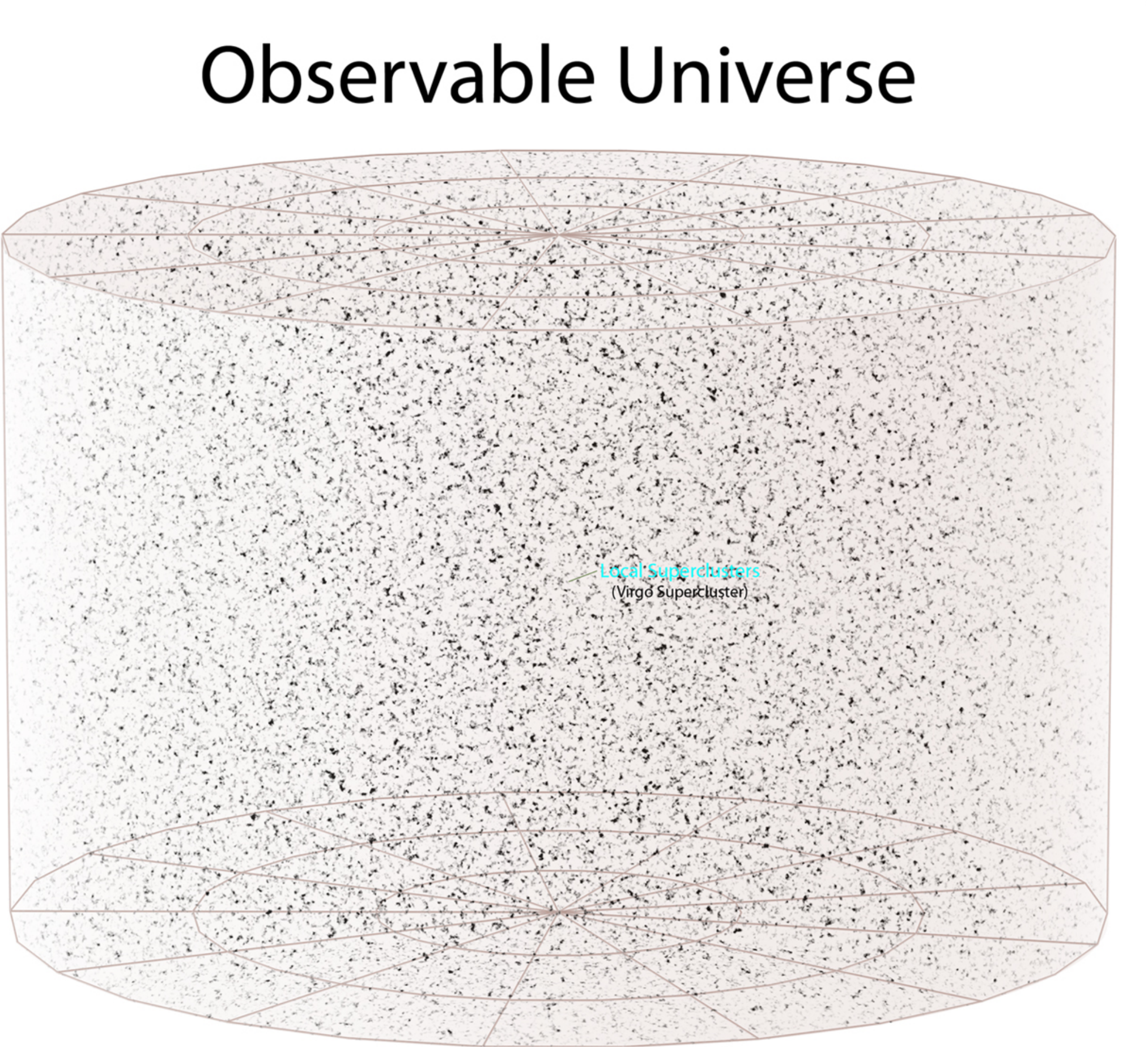}
\end{tabular}
\caption{Structure at different scales as depicted in  \cite{photolarge}.}
\label{fig100}
\end{center}
\end{figure}

Stars group together to form star systems and star clusters, while at larger scales they form galaxies, 
gravitationally bound systems that can very in radius from $1\ \mathrm{kpc}$ to $100\ \mathrm{kpc}$. The unit of distance $1\ \mathrm{pc}=30.9\times 10^{15}\ \mathrm{m}$ can be thought of as the mean distance from the sun to its neighbour stars. The sun belongs to our local galaxy, the Milky Way, that consists of about $10^{11}$ stars. 
Although there are some galaxies known to be isolated, most of them are part of larger structures such as groups, clusters and superclusters of galaxies. The Milky Way for example is part of the Local Group of galaxies, which in turn belongs to the Virgo supercluster, a supercluster that contains at least $100$ galaxy groups within its diameter of $33\  \mathrm{Mpc}$. The different structure at different scales is depicted in figure \ref{fig100}.

At even larger scales the clusters of galaxies form   structures resembling walls, that are called filaments. These galaxy sheets have a typical length of $50\ \mathrm{Mpc}$ to $100\ \mathrm{Mpc}$ and form the boundaries between large voids, i.e.\ almost empty space, in the universe. The universe hereafter is defined as our space-time. Beyond these scales the universe appears to have on average the same matter distribution in all directions, i.e.\ it is isotropic. All these observations are made from our own galaxy. One can make the further assumption that observations would be the same from any other galaxy in the universe, that is assume that the universe is also homogeneous. This is called the Cosmological Principle. Cosmological models that are spatially homogeneous and isotropic are called Friedmann Lema\^itre Robertson Walker models and are studied in section \ref{tcb}.     

The picture of the universe described above corresponds to a photograph of the universe as we see it today. However, one is also interested in demystifying the evolution of the universe through time. The first observational evidence for the time evolution of the universe was the discovery made by E. Hubble in 1929 that all galaxies observed today move away from our own galaxy  with a velocity $v$ proportional to their distance $d_l$, $v=Hd_l$. This is commonly known as Hubble's law. The constant of proportionality is called the Hubble parameter $H$ and its value today is approximately $H_0=67.3\  \mathrm{km\ s^{-1} Mpc^{-1}}$ (see table \ref{table1}). The usual convention is to parametrise  the Hubble constant at the present day as $H_0=100h\  \mathrm{km\ s^{-1} Mpc^{-1}}$. Given the Cosmological Principle this discovery leads to the idea that the universe is  expanding with time. One can think of a universe filled with galaxies and clusters of galaxies that do not expand themselves because of the gravitational attraction, but which recede from one another because of the expansion of the universe. 

Extrapolating back in time, one discovers a universe that was smaller and more dense. So, the distance between two galaxies changes with time as $d_l(t)\propto a(t)$, where $a(t)$ is called the scale factor. Indeed the Hubble constant is just the ratio of the rate of change of the scale factor to the scale factor itself,  
%There is a simple relation between the Hubble constant, which is not really a constant but is time dependent, and the scale factor 
as explained in subsection \ref{FRW}. Extrapolating even further back in time, one reaches a singularity where all the matter in the universe was concentrated at a point at time $t=0$. This singularity is called the Big Bang. The standard Big Bang theory assumes that the universe today emerged from such a singularity, like an explosion that fired its expansion and gave it an initial expanding velocity. From there on, the evolution of the universe is governed by the laws of physics at high energies. Since we do not have a quantum theory of gravity for energies higher than $m_{pl}=10^{19}\ \mathrm{GeV}$,  corresponding with times  before the Planck time $10^{-43}\ \mathrm{s}$, we cannot really study the universe at its very beginning and therefore we do not have a clue about the nature of such a singularity. Although the initial singularity itself and the emergence of a universe out of such a singularity raise mathematical, physical and above all philosophical questions, we shall not further comment on it and we will only follow the evolution of the universe after the Planck time.     

None of the above observations and findings would be possible unless the celestial objects were emitting light. Because of the finite speed of light  $c=2.99792458\times10^8\ \mathrm{m\ s^{-1}}$, when one observes a light signal from a light source at the sky, one looks back in time. Light emitted from a far away galaxy needs more time to reach us than light from a galaxy in our Local Group, and as a result the far away galaxy is observed in a much earlier stage of its life than the nearby one. Furthermore, the evolution of space-time itself is imprinted in the light signals. Because of the expansion the wavelength of the observed light signal $\lambda_{obs}$ is redshifted, i.e.\ is higher than the wavelength of the emitted light signal $\lambda_{em}$ as $1+z\equiv\lambda_{obs}/\lambda_{em}=a_{obs}/a_{em}$. Hence, a certain redshift $z$ corresponds to a certain  distance and time.

\begin{table}[t]
\begin{center}
\begin{tabular}{|l|l|l|}
\hline
Cosmological quantity & Symbol & Value \\[0.5ex]
\hline   & & \\[-0.5ex]
Hubble constant & $H_0$ & $67.3\pm 1.2\mathrm{km\ s^{-1}\ Mpc^{-1}}$ $(68\%\ \mathrm{c.l.})$\\
Temperature of CMB & $T_0$ & $2.7255\pm 0.0006\ \mathrm{K}$ $(68\%\ \mathrm{c.l.})$\\
Age of universe & $t_0$ & $13.817\pm 0.048\ \mathrm{Gyr}$ $(68\%\ \mathrm{c.l.})$\\
%Radiation density parameter & & \\
Baryonic matter density parameter & $\Omega_bh^2$ & $0.02205\pm0.00028$ $(68\%\ \mathrm{c.l.})$\\
Cold dark matter density parameter & $\Omega_{c}h^2$ & $0.1199\pm0.0027$ $(68\%\ \mathrm{c.l.})$\\
Total matter density parameter & $\Omega_m$ & $0.315^{+0.016}_{-0.018}$ $(68\%\ \mathrm{c.l.})$\\
Dark  energy density parameter & $\Omega_{\Lambda}$ & $0.685^{+0.018}_{-0.016}$ $(68\%\ \mathrm{c.l.})$\\
Curvature & $\Omega_K$ & $-0.0010^{+0.0062}_{-0.0065}$ $(95\%\ \mathrm{c.l.})$\\
Redshift of matter-radiation equality & $z_{eq}$ & $3391\pm 60$ $(68\%\ \mathrm{c.l.})$\\
Redshift of recombination 
& $z_{rec}$ & $1090.43\pm 0.54$
\\
Curvature power spectrum at $k_0=0.05\ \mathrm{Mpc^{-1}}$ &  
$A_\zeta$ &
$\lh 2.196^{+0.051}_{-0.060}\rh\times 10^{-9}$  $(68\%\ \mathrm{c.l.})$
\\
Scalar spectral index & $n_\zeta$ & $0.9603\pm 0.0073$ $(68\%\ \mathrm{c.l.})$\\
\hline
\end{tabular}
%\\[1ex]
\caption{Present values of some cosmological parameters according to Planck (Planck data combined with WMAP polarization data at low multipoles (Planck$+$WP)) \cite{Ade:2013zuv} and their confidence limits.}
\label{table1}
\end{center}
\end{table}

Today, combining supernovae and CMB observations we have managed to understand the content and evolution of our universe to a very good degree. Supernovae are stellar explosions that can occur depending on the mass of a star at the end of its life. The type SNIa supernovae can be used as standard candles to generate diagrams of the Hubble  parameter as a function of redshift (Hubble diagrams) and therefore as a function of time. 
CMB is an abbreviation for the Cosmic Microwave Background, the relic photons created when the universe was hot and dense. These photons survive today with a much lower temperature $T_0$ since their wavelength is increased because of the expansion of the universe. The CMB will be further discussed below. 

The evolution of the universe depends on its matter and energy content. In table \ref{table1} we give the  values of the density parameters $\Omega_i$ of the components of the universe today. The density parameter is the ratio of the energy density of a species $\rho_i$ to the critical energy density of the universe $\rho_c$, i.e.\ the energy density that the universe would have if it was spatially flat, so that $\Omega_{tot}=1$ (equivalently, the curvature density today $\Omega_K=\Omega_{tot}-1$ would be zero). As   can be seen in table \ref{table1} the universe is dominated by matter and an exotic form of energy, called the dark energy.  
Radiation, comprised of photons and neutrinos, contributes only a tiny fraction to the total energy density of the universe since the wavelengths of these particles are redshifted to very low values due to the expansion of the universe.   

\begin{figure}[h]
\begin{center}
\includegraphics[scale=0.9]{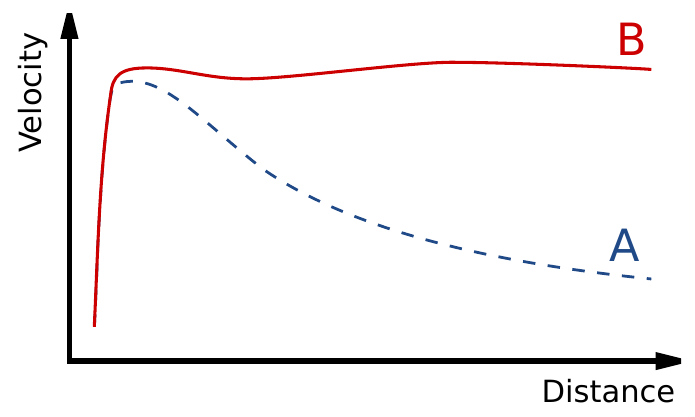}
\caption{Rotation curve of a typical spiral galaxy: predicted (A) and observed (B) \cite{Begeman:1991iy}.}
\label{fig101}
\end{center}
\end{figure}

As for the matter component this is comprised of  baryonic matter and non-baryonic dark matter. Baryonic matter is only a small fraction of the total matter content of the universe. The first evidence for dark matter was based on its gravitational influence on galaxies. Indeed if the only matter present in galaxies was the baryonic matter then the rotational velocity of stars around the center of the galaxy should decrease at large distances following Kepler's laws. In figure \ref{fig101} we see that this is not true. The discrepancy between the observed and predicted curves can be attributed to dark matter. In fact it turns out that the total matter in the universe is much larger than the luminous matter we see directly in stars and even larger than the total baryonic matter predicted by nucleosynthesis (see the following for details). Baryonic dark matter consists of non-luminous gas and condensed objects such as black holes, planets and brown dwarfs. On the other hand, non-baryonic dark matter is supposed to consist of particles beyond the Standard Model of particle physics, such as supersymmetric partner particles.
%, whose existence is inferred only from its gravitational effect. 
Dark matter can also be  divided into cold dark matter (CDM) consisting of massive particles and hot dark matter (HDM) consisting of light particles with relativistic speeds.

As can be seen in table \ref{table1}, dark energy contributes about $70\%$ to the total energy density of the universe. A possible candidate for the dark energy is a cosmological constant $\Lambda$ (and hence the name $\Lambda$CDM of the standard cosmological model). However, this does not mean that such a dark energy dominated the universe through all its time evolution. Indeed the density of each species depends on time. For  relativistic matter and radiation the energy density scales like $a^{-4}$. This can be understood since the energy density is the product of the energy per particle times the number density of particles. The number density decreases with the expansion of the universe as the inverse of its volume, $a^{-3}$, while the energy per relativistic particle decreases as $a^{-1}$  due to the redshift of the particle's wavelength. Matter density decreases as $a^{-3}$, since each particle's energy does not depend on the expansion. Curvature density scales like $a^{-2}$ (for more details see section \ref{tcb}). The energy density of a cosmological constant remains constant. Hence, it is possible for a cosmological constant to  dominate the universe only on late times, when the rest of the components have diluted away.

\Section{Thermal history of the universe}

Here we will briefly discuss the thermal history of the universe. 
%, extrapolating back from today
This history depends on the interactions of the various species of particles present at each era. Any set of species interacting among themselves at a high enough rate $\Gamma_A\mg H$, so that their mean free path is smaller than the Hubble scale $1/H$, 
%horizon
will share the same temperature. 
%Furthermore, the total entropy in a comoving volume $sa^3$ of a system in thermal equilibrium is conserved \cite{}, since there is no heat flow. 
When the Hubble parameter becomes larger than the interaction rate, the particles are said to have decoupled from the rest of the thermal bath and from there on they are travelling along geodesics. Hence, the universe can be populated by different sets of  species each with its own temperature. 

Since we do not have a quantum theory of gravity, we do not know what happened in the universe before the time $10^{-43}\ \mathrm{s}$ (or equivalently for mass scales larger than $10^{19}\ \mathrm{GeV}$). Presumably, before this time the universe was highly distorted since quantum fluctuations were very large compared to the size of the universe itself. 
%As a matter of fact, 
%we do not have a concrete physical theory 
% there is not much we can say
Much later, at the time $10^{-10}\ \mathrm{s}$ the electroweak interaction split into the electromagnetic and weak interactions and the universe became as it is known today. In between we do not have many observational data, but presumably at about $t\sim 10^{-37}\ \mathrm{s}$, the strong interaction  decoupled from the electroweak at the GUT scale $10^{16}\ \mathrm{GeV}$. At some energy below the GUT scale we assume that inflation took place. Inflation is a period of rapid expansion of the universe.   The reasons for introducing inflation will be explained in the next section. After inflation the universe reheated, while the field driving inflation decayed into particles and filled the universe with Standard Model particles, starting the radiation dominated era of the universe. 

At $10^{-10}\ \mathrm{s}$ the universe was filled with all kinds of elementary particles. At about $10^{-6}\ \mathrm{s}$, when the energy of the universe was $1\ \mathrm{GeV}$, quarks and gluons were bound in hadrons, i.e.\ baryons and mesons. Due to the low temperature even the lightest hadrons and mesons  annihilated (as did the heavier leptons) and by the time $10^{-4}\ \mathrm{s}$, when the universe's mass scale was of the order of $100\ \mathrm{MeV}$, the lepton era began. During that period, the universe was filled primarily with photons, neutrinos, electrons and positrons. In addition there were some protons and even less neutrons. At $0.01\ \mathrm{s}$ neutrinos decoupled from electrons and from there on they travelled freely. 
% on geodesics. 
 This era ended when electrons and positrons could not be created any more and most of them annihilated by the time $10\ \mathrm{s}$ at a scale $30\ \mathrm{keV}$.

Nucleosynthesis started shortly after. When the universe was $3\ \mathrm{min}$ old, the photon energy had dropped enough to allow for the formation of deuterium and then consequently helium could be produced. Except for a little lithium and beryllium no heavier elements could be produced. Indeed these could be formed only later inside stars, where the necessary   conditions for high density and temperature are fulfilled. The homogeneous abundances of helium and deuterium observed today in the universe, can have but a cosmological origin and are attributed to the nucleosynthesis period. Nucleosynthesis ended when all free neutrons disappeared, and it is considered as  one of the most important successes of the standard Big Bang theory. 

Nothing much happened for the next  $10^5\ \mathrm{years}$. The universe continued to expand under the domination of radiation. However, although the number of photons and neutrinos was much higher than the number of non-relativistic particles (matter), their energy density dropped more rapidly due to the expansion and at about $10^4\ \mathrm{years}$ (at a mass scale of $2\ \mathrm{eV}$) their energy density became equal to that of matter. Thereafter began the matter domination era of the universe. 

At the beginning of matter domination, the universe was filled with photons and ionized matter in thermal equilibrium. Soon electrons and protons combined into neutral hydrogen. This process is called recombination. The largest amount of recombination did not however happen at $13.6\ \mathrm{eV}$ (the ionization energy of hydrogen), but rather later, at an energy scale of $0.35\ \mathrm{eV}$, due to the low baryon to photon number. At this time, $380,000\ \mathrm{years}$ after the Big Bang, the photons decoupled and the universe became transparent. Ever since, photons travel along geodesics and are continuously cooled down by the expansion of the universe.
\begin{figure}[h]
\begin{center}
\includegraphics[scale=0.4]{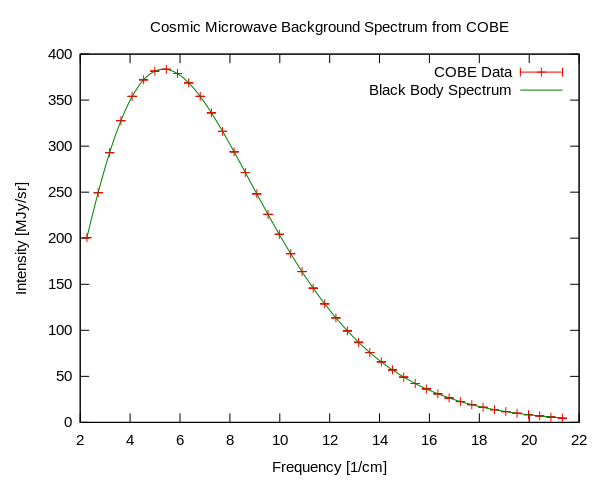}
\caption{The CMB spectrum from COBE. No spectral distortions from a black body have been discovered to date \cite{Hu:2008hd}}
\label{fig102}
\end{center}
\end{figure}
%At some during the history of the universe the energy density of radiation became equal to the matter energy density and from there on matter started dominating the evolution of the universe (remember that radiation energy density scales like $a^{-4}$, while matter energy density scales like $a^{-3}$ and hence after some time the energy of radiation will be less than that of matter). Using the black body spectrum of $T_0=2.7255\mathrm{K}$ of the CMB photons today and the matter density parameter in table \ref{table1} one can find the ratio of the two energy densities today. The evolution of this ratio backwards in time is just $\rho_r/\rho_m\propto a_0/a=1+z$ and hence one can calculate the redshift of matter and radiation equality. The Planck satellite's prediction is $z_{eq}=3391\pm 60$ corresponding to $T_{eq}\sim 0.75eV$. After this redshift the universe was matter dominated until dark energy became important at low redshift, $\sim 1$.
%
%Next we consider the photon decoupling. The present number density of photons can be found by integrating $n\d\nu$ to be $n_{\gamma 0}=410\ \mathrm{photons/cm^3}$. On the other hand, the nucleons density today is $n_{B0}=\Omega_B\rho_c/m_N=1.1\times10^{-5}\Omega_Bh^2
%\mathrm{nucleons/cm^3}$. Since both densities scale like $a^{-3}$, their ratio has been the same ever since their decoupling $t_{dec}$, $n_B/n_\gamma\sim10^{-9}$. The nucleons, which were mostly protons, were very tightly coupled to the free electrons by Thomson scattering.
% 
Today the energy of these relic photons is in the microwave regime and hence they are called the Cosmic Microwave Background (CMB). They follow at an excellent accuracy the spectrum of a black body of  temperature $T_0=2.7255\ \mathrm{K}$ (see figure \ref{fig102}) with a number density of photons with frequency between $\nu$ and $\nu+\d\nu$:
\be
n_T\d\nu=
\frac{8\pi\nu^2\d\nu}{\mathrm{exp}\lh h\nu/k_BT\rh -1
},
\ee
where $k_B=1.3806503\times10^{-23}\ \mathrm{J\ K^{-1}}$ is Boltzmann's constant and $h=6.62606957\times10^{-34}\ \mathrm{J\ s}$ is the Planck constant. \footnote{Note that in this thesis  we will work with natural units, $h/(2\pi)=c=k_B=1$.} 
This spectrum was created and maintained during the photons'  thermal equilibrium, before their decoupling. 
%, while it was frozen on later times with temperature that scales like $a^{-1}$. 
%Such a spectrum cannot but have been created under conditions of thermal equilibrium between radiation and hot dense matter, implying that in the past the universe was not transparent as it is today, i.e.\ photons were not flying free, but were rather tightly coupled with charged particles. During this period,  one can find the energy density of radiation by  multiplying the above equation by the energy per photon $h\nu$ and integrating, to find that $\rho\propto T^4$. Hence, during thermal equilibrium the temperature of the radiation scales like $T\propto a^{-1}$. 
%As explained above the photon temperature scales like $a^{-1}$.  
%As time passed, the matter became cooler and less dense and eventually the radiation began a free expansion after $t_{dec}$. 
Afterwards, the number of photons with frequency $\nu$ remained constant. These photons had frequency $\nu_{d}=\nu a/a_{d}$ at decoupling and hence their distribution function $n$ afterwards obeys
%\be 
%na^{-3}\d\nu =n_{T_{dec}}(\nu_{dec})a_{dec}^{-3}\d\nu_{dec}
%=n_{T(t)}a^{-3}\d\nu,
%\ee 
%\be
%\d N\propto f(\vc{x},\vc{p},t)\d^3\vc{x}\d^3\vc{p}= f_T(\vc{x}_d,\vc{p}_d,t_d)\d^3\vc{x}_d\d^3\vc{p}_d= f_T(\vc{x}_d,\vc{p}\ a/a_d,t_d)\d^3\vc{x}\d^3\vc{p},
%\ee
\be
\d N\propto n(\nu,T)\d^3\vc{x}\d\nu=
%n_T(\nu_d,T_d)\d^3\vc{x}_d\d\nu_d= 
\frac{8\pi\nu_d^2\d\nu_d\d^3\vc{x_d}}{\mathrm{exp}\lh h\nu_d/(k_BT_d)\rh -1
}=
\frac{8\pi\nu^2\d\nu\d^3\vc{x}}{\mathrm{exp}\lh h\nu a/(k_BTa_d)\rh -1
},
%n_T(\nu a/a_d,T a_d/a)\d^3\vc{x}\d\nu,
\ee
where $\d N$ is the number of photons in a physical volume $\d^3\vc{x}$ with a frequency between $\nu$ and $\nu+\d\nu$. 
Therefore, the photon density 
follows the black body form even after the photons went out of equilibrium with matter, but with a redshifted temperature $T\propto a^{-1}$. 

The main process taking place after recombination and until now, when the universe is $13.81\times 10^9\ \mathrm{years}$ old (see table \ref{table1}) is the gravitational collapse of matter, forming the large-scale structure observed   in the universe. Only recently the universe started to be dominated by some kind of cosmological constant or perhaps a scalar field, leading once more during its history to an exponential expansion. We still do not know the nature of this dark energy and hence we cannot make any predictions about the future of our universe, but if it is really a cosmological constant the universe will continue eternally to expand exponentially towards an increasingly cold, empty state.

\Section{Inflation}

The standard Big Bang model is very successful. It  incorporates and explains nucleosynthesis, the CMB and the growth of the large scale structure (although not its origin). Furthermore, assuming  the standard Big Bang model, different estimates by very different observational methods concerning the age of the universe, the Hubble parameter today and so on are in very good agreement. 
However there are still some issues that remain unresolved. The most important are

\begin{description}
\item[The flatness problem] According to observations, the total density parameter today is very close to $1$ and the curvature of the universe is practically zero. Hence the universe is almost flat. Since the curvature density scales like $a^{-2}$, it decreases more rapidly than matter or radiation and hence in the past the curvature density parameter would have been even smaller and the universe even more close to flat.  Indeed, comparing to radiation that scales like $a^{-4}$ we find
\be
\frac{\rho_K}{\rho_r}\sim 10^2\lh\frac{a}{a_0}\rh^2 
=10^2\lh\frac{\rho_{r,0}}{\rho_r}\rh^{1/2},\nn
\ee
where we used that $\Omega_K\sim 10^{-3}$ today (see table \ref{table1}) and $\Omega_r\sim 10^{-5}$ \footnote{One can calculate the radiation density parameter $\Omega_r=\rho_r/\rho_c$, using the expressions $\rho_c=3H_0^2m_{pl}^2/(8\pi)$ for the critical density (see subsection \ref{FRW}) and $\rho_r=a_BT_0^4$ for the radiation energy density assuming this is only comprised by photons ($a_B=7.56577\times10^{-16}\mathrm{J\ m^{-3}\ K^{-4}}$ is the radiation energy constant). 
}
.  
Today the radiation energy density is of order $(10^{-4}\mathrm{eV})^4$. 
Then at Planck time, when the energy density of the universe was $(10^{19}\mathrm{GeV})^4$, this ratio was $10^{-62}$. 
Given the random initial conditions at the end of a quantum gravity stage one would expect that all contributions to the energy density of the universe, namely radiation and curvature, would be of the same order of magnitude. The flatness problem can therefore be formulated as: why should we start from initial conditions in the very early universe such that the curvature density parameter should be fine-tuned to $10^{-62}$ of the total energy density, when it could have taken any value.

\item[The horizon problem] The (particle) horizon encompasses that part of space within which causal contact is possible. Since no information can travel faster than light, an  event taking place at $(t_1,\vc{x_1})$ cannot be perceived by an observer at $(t_2,\vc{x_2})$ if   his/her physical distance is larger than the distance light can travel in this time interval. For a  matter or radiation dominated universe, the horizon grows faster than space itself \footnote{For matter domination the particle horizon grows like $a^{3/2}$, while for radiation domination like $a^2$ and  hence in both cases faster than the scale factor $a$ (subluminal expansion).}  or in other words, going back in time the horizon shrinks faster than space. This means that today we can perceive events that in the past were not in causal contact with us. Hence, the early universe consisted of different regions not exchanging information with each other. Nevertheless, the observable universe today is highly isotropic, apparently nucleosynthesis took place the same way everywhere and the CMB temperature is uniform.  But the present observable universe was split into about $10^6$ disconnected patches at the time of recombination, when the CMB was formed, and into  about $10^{24}$ patches during nucleosynthesis. So the horizon problem is how such disconnected parts of the universe could have evolved in such a similar way.    

\item[The monopole problem] Phase transitions in the early universe are expected to create topological defects. Among these, monopoles are considered as ``dangerous relics'', since their very large density would dominate the total density of the universe. These relics are typically non-relativistic, with en energy density decaying like $a^{-3}$, so they would never allow for radiation or matter domination to take place. 

\item[The large-scale homogeneity problem] At large scales the universe is highly homogeneous and isotropic. Already at recombination the departure from homogeneity of the background radiation was only  $10^{-5}$. 
This raises questions concerning the initial conditions that produced such a homogeneous universe. Indeed at the Planck time these conditions are expected to be chaotic and it is extremely unlikely  that they could evolve into a universe so isotropic and homogeneous.

\item[The small-scale inhomogeneity] At the same time this highly uniform universe exhibits inhomogeneities at small scales, like galaxies and clusters of galaxies. It is believed that the inhomogeneities of order $10^{-5}$ of the CMB sourced the formation of this large-scale structure through gravitational collapse. So the question is, how were these small fluctuations created in an otherwise homogeneous universe.  
\end{description} 

These problems can be solved by introducing a period in the early universe when the expansion was superluminal. Such a period is called (cosmological) inflation. We do not know exactly when this took place, however the upper limit of the energy scale of inflation according to Planck is $1.9\times 10^{16}\ \mathrm{GeV}$.  
%and is thought to have taken place somewhere  below the GUT scale, that is between $10^{-43}\mathrm{s}$ and $10^{-35}\mathrm{s}$ after the Big Bang. 
It is believed that during at most $10^{-30}\mathrm{s}$ the universe inflated at least by a factor $10^{26}$. Such an expansion can be achieved when the universe is filled by a scalar field, the inflaton, usually denoted by $\phi$,  whose slowly decreasing potential energy dominates the total energy density. During this rapid expansion, the universe cooled down while the rest of its components diluted away. Hence, a period of reheating is needed afterwards, to recover the temperature of the universe to its value needed to preserve all the successful features of the standard Big Bang theory. During reheating the energy of the scalar field decays into particles and fills the universe with Standard Model particles. The first stages of reheating can occur in a regime of parametric resonance called preheating, where the inflaton decays extremely efficiently to another scalar field, which then decays into the Standard Model particles (see for example \cite{Kofman:1994rk}).

The possibility of an early exponential expansion was first noticed by Starobinsky in 1979-1980 \cite{Starobinsky:1980}, but at first it attracted little attention. It was Guth in 1981 \cite{Guth:1980zm} that noted that an inflationary period could solve the flatness and horizon problems.  In the model proposed by Guth,  known as old inflation,  a scalar field is trapped at the origin in a local minimum of its potential and hence the universe is dominated by the field's false vacuum energy. Inflation ends when the field tunnels through the barrier and descends quickly to the minimum of the potential. However, this model could not provide sufficient reheating and it was soon abandoned. 

In 1982, Linde \cite{Linde:1981mu} and Albrecht and Steinhardt \cite{Albrecht:1982wi} proposed the new inflation model. Assuming a phase transition, the inflaton is initially situated on a maximum of its  potential at the origin. The field starts slowly rolling down the rather flat potential. Inflation ends when it reaches its minimum and starts oscillating around it, reheating the universe. Although this type of potential was abandoned due to observational constraints, new inflation first introduced the concept of slow-roll inflation (further discussed in chapter \ref{Ch2}). 

Later on, Linde \cite{Linde:1983gd} proposed chaotic inflation, during which the field rolls towards the  origin in a $\phi^2$ or $\phi^4$ potential. Its name derives from the chaotic initial conditions that are used to explain the needed large initial value of the field. During observable inflation the field's magnitude is of the order of $m_{pl}$, and hence it is not easy to make connection with particle physics theories. However, because of their simplicity,  monomial models became the favoured paradigms of inflation.

Many models have been built since then. Among these we distinguish hybrid inflation, where two scalar fields were first introduced \cite{Linde:1993cn}. One field is responsible for the main part of inflation, while the second achieves a graceful exit. Originally the model used was a combination of chaotic inflation and a second-order phase transition. In hybrid inflation models the magnitude of the inflaton is typically much less than $m_{pl}$ and hence connection with particle theory becomes a realistic possibility. In this thesis we discuss the observational consequences of two-field models. The  basics of such models are discussed in chapter \ref{Ch3}.
    
It is the rapid expansion during inflation that resolves all the problems described in the beginning of this section. Indeed, by the end of inflation, all the components of the universe are diluted away due to the exponential growth of the volume. Among these the curvature and any catastrophic topological defects or inhomogeneities. After inflation the universe is effectively flat and uniform, without requiring that it started with negligible curvature after gravity decoupled from the rest of the forces. 

The horizon problem is solved, since the (particle) horizon becomes much larger than it would be in the standard Big Bang theory. Hence, all the observable universe today originated inside the inflationary horizon and causal contact between all its points was achieved already at the GUT scale.      
  
Finally, inflation offers a natural way to produce the initial fluctuations responsible for the small-scale inhomogeneities.
Quantum fluctuations of the inflaton are produced during inflation and because of their interplay with gravity, they give rise to curvature perturbations, which one can loosely think of as a gravitational potential. These perturbations  become classical once their wavelengths are  stretched beyond the Hubble length, the fundamental length scale of the universe,  which coincides with the event horizon. The event horizon encompasses the space where causal contact will become possible in the future. In absence of an inflationary period, the event horizon is infinite. During inflation however, the event horizon remains almost constant. As a result any physical scale, that increases like $a$, will eventually exit the event horizon and become causally disconnected with the universe inside. After inflation, the Hubble length (that now coincides with the particle horizon)  grows faster than any other scale and hence the curvature perturbations eventually re-enter (more details about horizons can be found in section \ref{tcb}). Single-field models of inflation predict the creation of almost Gaussian, adiabatic perturbations, which once outside the horizon preserve their amplitude  and are characterized by an almost scale-invariant profile, i.e.\ their amplitude does not depend on their scale (see chapter \ref{Ch2}). However, the subject of inflationary perturbations within two-field inflation is the main subject of this thesis and it will be extensively  studied in the next chapters. There is a qualitative difference compared to the single-field case, in the sense that they introduce a new type of perturbation, called the isocurvature perturbation. The adiabatic and isocurvature perturbations interact and can produce new features in the statistics of the perturbations, most notably non-linearities (non-Gaussianities) created outside the horizon, that could in principle distinguish them from other models.

\Section{Connection to observations}

\begin{figure}[h]
\begin{center}
\includegraphics[scale=0.4]{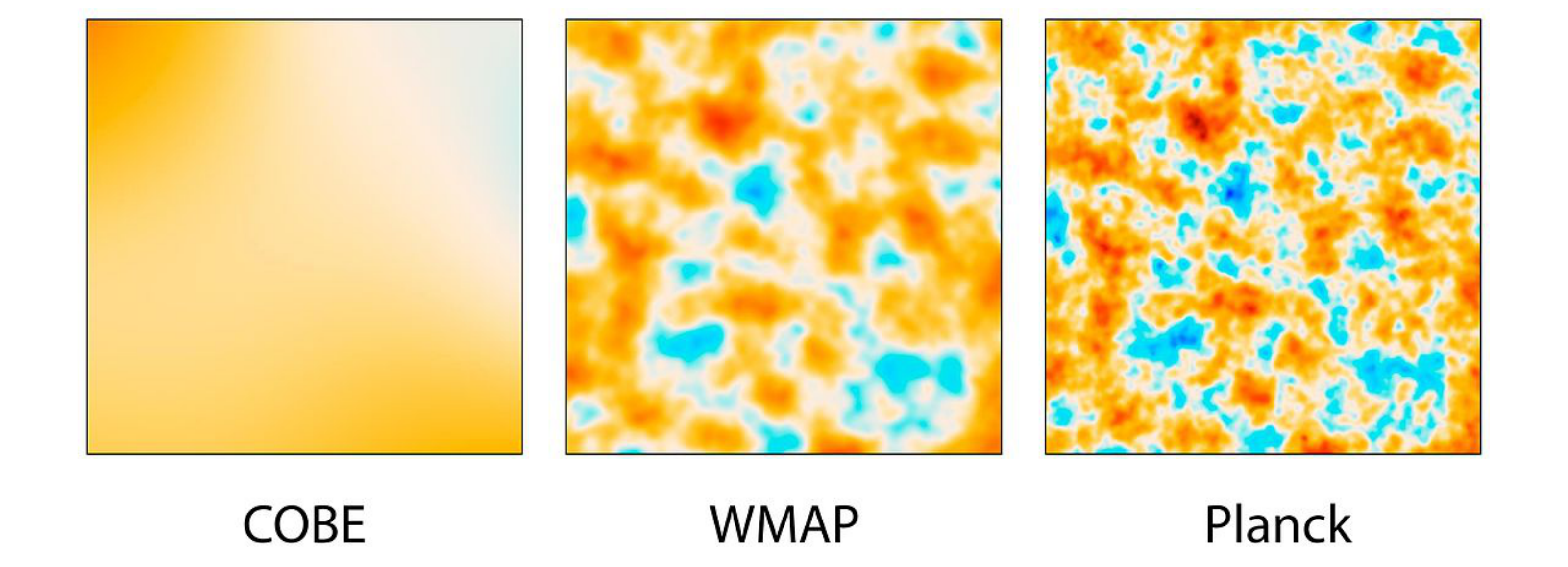}
\caption{Comparison of $10$-square-degree patches of all-sky maps created by COBE, WMAP and Planck  \cite{photo}}
\label{fig103}
\end{center}
\end{figure}

%The existence of inflation has not been proved.  
%Furthermore, its
%The origins of inflation remain a problem, since 
We do not know what are the initial conditions that can trigger an inflationary period and we still lack the connection  with realistic high energy models.  
% still remains to be made. 
However, 
%inflation is widely accepted today due to its 
%success in providing the primordial perturbations and solving at the same time  important problems of the standard Big Bang theory. 
%
%
%Even though we do not know how inflation started,
its success in solving the problems discussed in the previous section as well as in providing the seeds for the CMB fluctuations that gave rise to all structure in the universe today, has established it as part of the  standard cosmological model. 
Observationally, inflation can be tested by extrapolating the evolution of the primordial perturbations up to today and comparing to data. The richest sets of data we have today are related to the CMB and hence in this section we will discuss the information we can gain from the observed CMB power spectrum.   

Since the 60's when Penzias and Wilson first confirmed the existence of the CMB, much progress has been made and the fluctuations of the order of $10^{-5}$ of the CMB temperature can be measured today to an excellent precision. 
The COsmic Background Explorer (COBE) satellite launched in 1989 \cite{cobe}, first revealed the perfect fit between the relic photon black-body curve predicted by the Big Bang theory and that observed  in the CMB, see figure \ref{fig102}. Furthermore, after four years of measurements it provided the first full-sky maps of the anisotropy of the CMB, by subtracting galactic emissions and dipole contributions at various frequencies. 

In 2001 NASA launched a follow-up mission to COBE, the Wilkinson Microwave Anisotropy Probe (WMAP) in order to clarify and expand COBE's accomplishments. WMAP observed in five frequencies allowing for the measurement and subtraction of foreground contamination from the Milky Way and extragalactic sources. During nine years of measurements, WMAP  determined several cosmological parameters and revealed the geometry, the content and evolution of the universe, while it confirmed the present domination of the universe by dark energy and verified the nearly scale-invariant spectrum predicted by inflation  \cite{Hinshaw:2012aka}.

\begin{figure}[t]
\begin{tabular}{cc}
\includegraphics[scale=0.3]{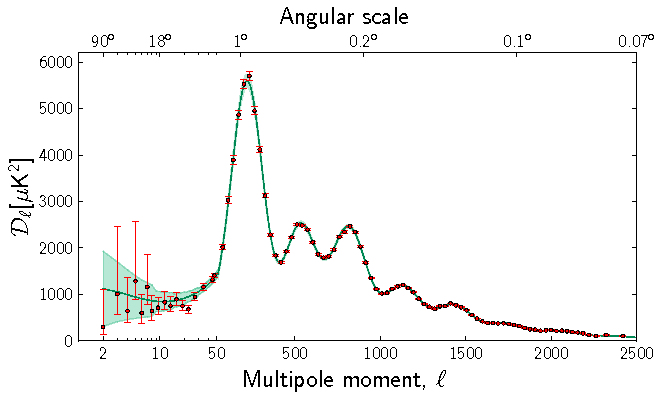}&
\includegraphics[scale=0.3]{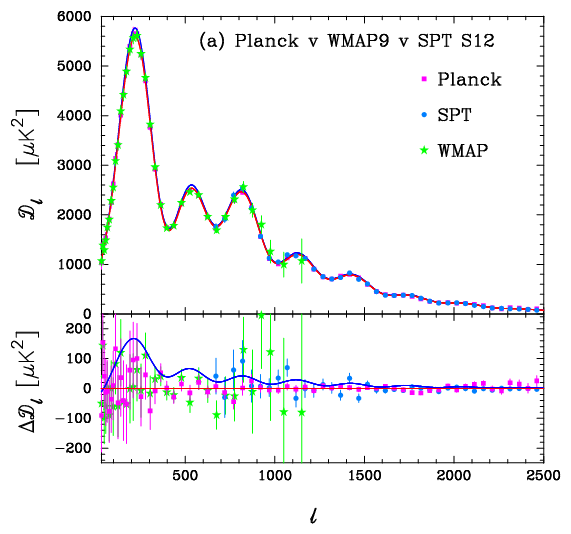}
\end{tabular}
\caption{Left: The observed power spectrum from Planck, along with a $\Lambda$CDM theoretical model.  The parameters of the theoretical curve are calculated combining Planck and low-multipole WMAP polarization data. Right: Comparison of the Planck best-fit power spectrum (pink curve) and the combined South Pole Telescope and $9$-year WMAP power spectrum (blue line). The green points show the $9$-year WMAP power spectrum. The lower panel shows the residuals. Both figures are taken from \cite{Ade:2013zuv}.}
\label{fig104}
\end{figure}

Following WMAP, the ESA's probe, Planck, has continued to increase the resolution at which the CMB is mapped, measuring the anisotropies up to smaller angular scales and with higher sensitivity than ever before (see figure \ref{fig103}, where we compare the resolution of COBE, WMAP and Planck for a $10$-square degree patch of the sky). In 2013, Planck gave its first full-sky maps and estimates of the cosmological parameters  \cite{Ade:2013ktc,Planck:2013kta}. Planck ruled out exact scale invariance of the primordial spectrum, measuring the spectral index to be $0.9603\pm 0.0073$ (see table \ref{table1}) and constrained the allowed standard inflationary models, showing that exponential, simple hybrid and monomial models of degree $n\geq 2$ do not provide a good fit to the data \cite{Ade:2013uln}. 
 
The main observable related to the primordial  perturbations is the power spectrum of the CMB fluctuations. Roughly speaking, the power spectrum is the amplitude of the perturbations squared as a function of their scale. 
When the almost scale-invariant primordial curvature perturbations re-enter inside the horizon, 
they begin to evolve. 
Due to the Einstein equations this is translated into the evolution of 
%give rise to 
the matter-photon plasma perturbations.    
After recombination, when photons decouple from baryons, baryonic matter fluctuations grow and eventually gravitationally collapse, giving rise to the large-scale structure we observe today. On the other hand, photons mainly cool down due to the expansion of the universe, carrying the profile of their temperature fluctuations at the time of recombination (in reality, their profile also  evolves with time due to  sources of secondary anisotropies, like the ISW effect discussed later in this section). 
Their angular power spectrum  today, as measured by Planck, is shown in figure \ref{fig104}. 
%Tracing back all the effects on its evolution, one can  extrapolate back to the primordial profile and gain valuable information  on the inflationary parameters as well as on the cosmological model. 

The fluctuations of the CMB are statistically isotropic and it turns out that a harmonic description is more efficient than a real space description. The appropriate harmonics for sky observations are the spherical harmonics $Y_{lm}$, so we express the deviation of the temperature from the mean temperature as
\be
\Theta(\vc{n})\equiv\frac{T(\vc{n})-\bT}{\bT}=\sum_{lm}\Theta_{lm}Y_{lm}(\vc{n}),\nn 
\ee
where $\vc{n}$ is the unit vector in the direction of observation. Then the ensemble average of the temperature fluctuations is described by the angular power spectrum
\be 
\langle\Theta^*_{lm}\Theta_{l'm'}\rangle=\delta_{ll'}
\delta_{mm'}C_l.\nn
\ee 
The amplitude of the angular power spectrum $C_l$ depends only on the angle under which we observe a certain scale on the sky and not on its orientation. If the fluctuations are Gaussian, the power spectrum contains all of the statistical information. 
%In figure \ref{fig104} we show the angular power spectrum as measured after nine years of WMAP and as measured by Planck. 
%On the other hand, the spatial power spectrum at recombination can be likewise described by its harmonic modes, which for a flat geometry are Fourier  modes
%\be
%\Theta(\vc{x})=\int\frac{\d^3k}{(2\pi)^{3/2}}\Theta(\vc{k})\mathrm{exp}^{\mathrm{i}\vc{k}
%\cdot\vc{x}}. \nn
%\ee
%The two-point function is described by the (spatial) power spectrum
%\be
%\langle\Theta(\vc{k})^*\Theta(\vc{k'})\rangle
%\equiv\delta^3(\vc{k}
%+\vc{k'})\frac{2\pi^2}{k^3}\cP(k).\nn 
%\ee
%It can be shown that for a nearly scale invariant spatial power spectrum, $\cP$ can be related to $C_l$ 
%as 
%\be
%C_l=
%\frac{2\pi}{l(l+1)}\cP(l/D_*),\nn 
%\ee   
%where $D_*$ is the angular diameter distance to recombination. Hence, looking at figure \ref{fig104} one can think of the multipole moment $l$ as the inverse of the angle under which a scale of wavelength $\lambda$ at recombination is seen, so that the larger $l$ is, the smaller the scales we observe. 
The different features of the power spectrum in figure \ref{fig104} at different scales convey information about the primordial perturbations themselves, as well as about the processes that created these features, and hence the components of the universe  (remember that the primordial perturbations had an almost flat power spectrum).

Next we will describe briefly the evolution of the primordial curvature perturbations up to today and relate it with the various features of the angular spectrum.
%The CMB photons that we observe today are a cooled down picture of the universe at recombination. Thereafter, the photons are decoupled and fly freely preserving the information that carried at the time. Within single-field inflationary models, the  primordial perturbations remain frozen while they are outside the (event) horizon, i.e.\ their amplitude is constant. Furthermore, the amplitude of the perturbations is almost scale-independent. 
%After inflation, the horizon starts growing faster than the wavelength of the perturbations and these re-enter and become again causally connected with an observer inside the horizon. 
Let us first discuss scales that entered the horizon before recombination but in a matter dominated universe.  During that time photons are strongly coupled to electrons through Thomson scattering, and electrons on their side are strongly coupled with baryons by Coulomb interactions. The photon pressure resisted the gravitational compression of the baryons, leading to acoustic oscillations in the baryon-photon plasma \cite{Hu:2001bc,Hu:2008hd}. At recombination this coupling came to an end and photons travelled thereafter freely on perturbed geodesics. These oscillations can be seen in the CMB power spectrum for multipoles $l\gtrsim 100$. The modes that were in a minimum or a maximum at the time correspond to a peak in the CMB spectrum (since fluctuations are squared in the power spectrum). 

Since recombination is not an instantaneous process, there was a transition period where the photon coupling to baryons was not so effective and the photons had a certain short mean free path. When this mean free path became comparable to the wavelength of the perturbation, the photons diffused and smoothed out the inhomogeneities. This is called Silk damping and it is related to the damped oscillations for multipoles $l\gtrsim 1000$. 

The first region of the power spectrum $l\lesssim 100$, corresponds to scales that were still outside the horizon at the time of recombination. Hence, we should be able to see the flat primordial spectrum at these scales and therefore from the point of view of inflation this would be expected to be the most important part of the spectrum. However, this is not true due to the cosmic variance, i.e.\ the large statistical errors due to the fact that we only have one sky to observe. 
Furthermore, this region of the power spectrum is tilted due to the  late-time ISW effect \cite{Hu:2001bc}, namely the fact that the recent accelerated expansion, due to dark energy, causes changes in the gravitational potential during the time needed by photons to cross it. This effect is important for scales corresponding with the horizon scale after the transition from matter domination to dark energy domination and hence for the smallest values of $l$.
 
The features of the power spectrum depend on the exact   cosmological model, and hence we can gain valuable information about it. For example, the comparison between the predicted acoustic peak scale and its angular extent provides a measurement of the angular diameter distance to recombination. This in turn depends on the spatial curvature and the  expansion history of the universe. Furthermore, the baryon-photon ratio controls the even-odd (minimum-maximum) modulation of peak heights and hence the second peak, that represents a rarefaction of the acoustic wave in a gravitational potential, is suppressed in amplitude by the baryon inertia. So the CMB places limits on the baryon density. The peaks depend also on the existence of isocurvature perturbations, which if present change the position of the peaks. However, no significant contribution of isocurvature perturbations is detected. 

But the Planck satellite provides yet another tool to distinguish between different models of inflation and this is related to non-Gaussianity. Slow-roll single-field models of inflation with standard kinetic terms  predict primordial perturbations that are almost Gaussian. This is not true however for other types of models. For example in multiple-field models non-Gaussianity created outside the horizon depends on the potential of the fields and is not a priori negligible. If present, non-Gaussianity could manifest itself through a non-zero three-point function, which is related to what is known as $\f$, the parameter of non-Gaussianity. 
Planck has provided us with unprecedented constraints on primordial non-Gaussianity \cite{Ade:2013ydc} and in order to exploit them we need a good understanding of the mechanisms creating primordial non-Gaussianity. 
The aim of this thesis is to study how non-Gaussianity is produced in multiple-field models and examine how this depends on the model's properties and the scales of the three perturbations involved in the correlation function.  
%and finds no evidence for significant deviations from Gaussianity.

\Section{Outline of the thesis}

In this thesis we will study inflation driven by two scalar fields and the related non-Gaussianity. Given the discussion in the previous sections, such models can appear naturally in the early universe. We will consider the consequences of such models for the background cosmological model, but we will be mainly concerned about perturbations around this background. Two-field models are characterized by the existence of two types of perturbations, the adiabatic perturbation and the isocurvature perturbation. These perturbations interact while they are outside the horizon, so they evolve with time as opposed to the single field case. Furthermore, this interaction induces non-linearities that can manifest themselves as a  departure from Gaussianity of the perturbations. We will study the properties of this non-Gaussianity, i.e.\ its evolution, the criteria for it to be large by the end of inflation or not  and its scale dependence. The contents of this thesis are based on papers \cite{Tzavara:2010ge}, \cite{Tzavara:2011hn} and \cite{Tzavara:2012qq}. A fourth paper generalising the findings of \cite{Tzavara:2011hn} for inflationary models with general kinetic terms is in its final stages of preparation, but in this thesis we will restrict ourselves to scalar fields with standard kinetic terms.

In more detail, the following chapter \ref{Ch2}, is an introductory chapter where we start by describing the FLRW cosmological models and next we study the single-field paradigm of inflation. We introduce mathematically the concept of slow-roll and study the inflationary background. In section \ref{s21} we study linear perturbations around this background, introduce the concept of gauge invariance and the Hamiltonian (ADM) formalism of gravity which we will use to construct  the action that governs the  perturbations.  We also study the evolution  of the perturbations and introduce their power spectrum. In section \ref{S22} we review the second-order perturbation theory results and the related non-Gaussianity. Based on the single-field limit of our paper \cite{Tzavara:2011hn}, we extend the already known relation between gauge invariance and the action of the perturbations.     

In chapter \ref{Ch3} we present the basic tools for treating two-field inflation. We begin by studying the background (section \ref{S31}) and generalizing the slow-roll concept. In the following sections we present two different ways of studying perturbations within the context of multiple-field models: studying first and second order perturbations (sections \ref{S32} and \ref{S33}) as well as studying the fully non-linear quantities outside the horizon (section \ref{long}). In particular, in sections \ref{S32} and \ref{S33} we examine gauge invariance and construct the action that governs the perturbations at linear and second order respectively, based on the findings of \cite{Tzavara:2011hn}. In section \ref{long} we introduce the long-wavelength formalism, a powerful tool to study perturbations outside the horizon. 

In chapter \ref{Ch4} we start by presenting the procedure for solving for the evolution of the perturbations and calculating the parameter of non-Gaussianity within the long-wavelength formalism (section \ref{S41}). In section \ref{general} we construct the general solution for the parameter of non-Gaussianity, assuming slow-roll only at horizon crossing, while in section \ref{secSlowRoll} we use the  slow-roll approximation during the whole inflationary period, in order to compute a semi-analytical expression for the parameter of non-Gaussianity. This chapter as well the next one are based on paper \cite{Tzavara:2010ge}.

In chapter \ref{Ch5} we study large classes of inflationary models and examine whether non-Gaussianity can be formed and sustained by the end of inflation. We use the semi-analytical slow-roll expression for the parameter of non-Gaussianity found in chapter \ref{Ch4} for our estimates. In section \ref{numerical} we study numerically the quadratic model in order to compare our findings with other calculations in the literature and construct a model that can produce non-Gaussianity of order $\cO(1)$ by the end of inflation.

In chapter \ref{Ch6} we present the findings of our paper \cite{Tzavara:2012qq} concerning the scale dependence of non-Gaussianity. In section \ref{sour} we discuss the sources of this scale dependence, while in sections \ref{conf} and \ref{shape} we parametrize the scale dependence by spectral indices and study non-Gaussianity for different triangle configurations.

Finally, in chapter \ref{Ch7} we conclude and comment on the future of studying non-Gaussianity within two-field models. Finally there are four appendices. Appendix \ref{AppA} contains some intermediate steps of  calculations at first and second order in perturbation theory, while in appendix  \ref{AppB} we present the main steps of the calculation for the action of the perturbations. In appendix \ref{AppC} we present calculations related to the long-wavelength formalism and finally in appendix  \ref{AppD} we give analytical expressions for the spectral indices discussed in chapter \ref{Ch6}.

%------------------------------ Chapter---------------------- -------------------------------------------------------------

{\Chapter{Standard Inflationary Predictions}\label{Ch2}}
%------------------------------------------------------------

The paradigm of inflation is essentially related to a quasi-de Sitter universe, i.e.\ a homogeneous and isotropic universe that expands almost exponentially fast, with an almost constant event horizon. Historically the first and simplest realisation of inflation is achieved by a single scalar field  dominating the universe. 
In this chapter the basics of inflationary physics are introduced, starting from the description of the cosmological background and continuing with studying  perturbations on this background.

\Section{The cosmological background}\label{tcb}

\subsection{The FLRW metric}\label{FRW}

We will assume that the universe is dominated by a background  homogeneous matter field with a diagonal stress-energy tensor
\be
T^\mu_\nu=\mathrm{diag}[-\rho(t),p(t),p(t),p(t)],\label{diagt} 
\ee
where $\rho(t)$ is the energy density and $p(t)$ the pressure density of the matter field. The energy momentum tensor $T^\mu_{\nu}$ carries information about the energy and pressure density and energy and momentum flux. 
For an homogeneous and isotropic matter field the latter are zero.

Such a matter field is the source of an homogeneous and isotropic space background. The latter is described by the Friedmann Lema\^itre Robertson Walker (FLRW) metric
\be
\d s^2=-N(t)^2\d t^2+a(t)^2\lh\frac{\d r^2}{1-K r^2} + r^2\d\Omega^2\rh,\label{frw} 
\ee
where $K$ is the spatial curvature of the Universe, taking values $K=0$ for a flat, $K=1$ for a closed universe and $K=-1$ for an open universe, while $a(t)$ is the scale factor showing the rate of expansion of the proper distance between two comoving observers. $N(t)$ is merely a redefinition of the time coordinate, for instance $N(t)=1$ corresponds to the cosmological time $t$ and $N(t)=a(t)$ corresponds to the conformal time $\eta$.
%and $N(t)=1/H(t)$ corresponds to a time coordinate $N$ equal to the number of e-folds. 

The action for any system of gravity and matter is
\be
S=\int\d^4x\lh\kappa^{-2}\sqrt{-g}\lh\frac{R}{2}-\Lambda\rh+\mathcal{L}_m\rh , 
\ee\
where $g$ is the determinant of the metric, $R$ is the Ricci scalar, $\Lambda$ a cosmological constant and $\mathcal{L}_m$ the matter Lagrangian density. $\kappa$ is the inverse reduced Planck mass defined by 
\be 
\kappa^2\equiv 8\pi G=8\pi/m_{pl}^2.
\ee
Applying the action principle to the above action one gets the Einstein equations 
\be 
G^\mu_\nu+\Lambda\delta^\mu_\nu=\kappa^2T^\mu_\nu,\quad\mathrm{where}\quad
T^\mu_\nu=-2\frac{g^{\mu\kappa}}{\sqrt{-g}}\frac{\delta \mathcal{L}_m}{\delta g^{\kappa\nu}},\label{stress}
\ee
governing the evolution of the system. 
The Einstein tensor $G^\mu_\nu$ carries the symmetries of the energy momentum tensor and hence for an homogeneous and isotropic universe it is diagonal. Within the Einstein equations there is also hidden the energy-momentum conservation equation 
\be
T^\mu_{\nu;\mu}=0,\label{encons}
\ee
since by virtue of the Bianchi identity $G^\mu_{\nu;\mu}=0$. Here we introduced the notation $;$ or $\cD$ for the covariant derivative relative to the space-time metric $g_{\mu\nu}$
\be 
\cD_\mu A^\nu=A^\nu_{;\mu}=\partial_\mu A^\nu+\Gamma^\nu_{\mu\kappa}A^\kappa,
\ee
ensuring the parallel transport of the vector under differentiation. $\Gamma^\nu_{\mu\kappa}$ is the affine connection (for definitions of the affine connection and the Riemannian tensors, see \cite{weinberg1972gravitation,Misner:1974qy}). 

For the energy-stress tensor (\ref{diagt}) and the FLRW metric (\ref{frw}) the Einstein equations become
\bea 
&&H^2\equiv\frac{\dot{a}^2}{N^2a^2}=\frac{\kappa^2}{3}\rho+\frac{\Lambda}{3}-\frac{K}{a^2}\nn\\
&&\frac{1}{a}\frac{\d}{\d t}\lh\frac{\dot{a}}{N}\rh=-
\frac{\kappa^2}{6}\lh \rho+3p\rh +\frac{\Lambda}{3} \qquad\mathrm{or}\qquad 
\frac{\dot{H}}{N}=-\frac{\kappa^2}{2}\lh p+\rho\rh+\frac{K}{a^2},\label{ee}
\eea 
where we have defined the background Hubble parameter $H$ as the rate of change of the scale factor. The first of these equations is called the Friedmann equation. Inspecting the definition of the Hubble parameter we see that choosing $N=1/H$ the scale factor expands exponentially. In that case $N$ corresponds to a time variable $t$ that coincides with the number of e-folds.

On the other hand, (\ref{encons}) becomes
\be
\dot{\rho}+3HN(\rho+p)=0.\label{enecon}
\ee   
Dividing the Friedmann equation by $H^2$ and rearranging we find
\bea
&&1+\Omega_K=\Omega+\Omega_\Lambda\equiv\Omega_{tot}\nn\\
&&\Omega\equiv\frac{\rho}{\rho_c},\quad
\Omega_\Lambda\equiv\frac{\rho_\Lambda}{\rho_c},\quad
\Omega_K\equiv\frac{\rho_K}{\rho_c},\quad
\rho_\Lambda\equiv\frac{\Lambda}{\kappa^2},\quad 
\rho_K\equiv\frac{3K}{\kappa^2a^2},\quad 
\rho_c\equiv\frac{3H^2}{\kappa^2}, 
\eea
where we define the density parameters $\Omega_i$. The critical density $\rho_c$ is the total energy density necessary for a flat universe $\Omega_K=0$. The energy densities $\rho,\rho_c,\rho_K$ are all time dependent, because of their dependence on the scale factor. The general matter field energy density $\rho$ has a time dependency that can be found if its equation of state is known. For a perfect fluid the equation of state has the form $p=w\rho$ and hence using (\ref{enecon}) it is straightforward to find that 
\be
\rho\propto a^{-3(1+w)}
\ee
and then solving for the Friedmann equation one can find the time dependence of the scale factor. 
In the next subsection we will study separately the case of a scalar field. The most common examples of perfect fluids are non-relativistic matter with $p_m=0$ and radiation with $p_r=\rho_r/3$. Assuming that such a perfect fluid dominates the total energy density of the universe, one finds
\be 
\rho_m\propto a^{-3},\quad a\propto t^{2/3}\quad\mathrm{and}\quad
\rho_r\propto a^{-4},\quad a\propto t^{1/2},
\label{radmat}
\ee
where just for the sake of the example we have set $N=1$, i.e.\ $t$ is the cosmic time. The conclusion is that the energy density of matter or radiation reduces with time, due to the expansion of the universe. During inflation, which is defined by its quasi-exponential expansion, any such components are quickly decaying and hence we will not take them into account.

The energy density of the cosmological constant is by definition constant. Its pressure is $p_\Lambda=-\rho_\Lambda$ and hence we find
\be
\rho_\Lambda=\frac{\Lambda}{\kappa^2},\quad
a\propto\mathrm{e}^t,
\label{rhol} 
\ee
again for $N=1$. 
If this cosmological constant is the dark energy dominating the universe today, its constant energy density would have been much smaller than any other component in the past. Therefore, from now on we will ignore its  contribution and explicitly set $\Lambda=0$, since this is negligible during the era of inflation. 

The energy density of the curvature $K$ scales as $\rho_K\propto a^{-2}$. Observations indicate that the universe today is almost flat. This is also one of the reasons to introduce inflation. During inflation, the quasi-exponential expansion drives the curvature density quickly to zero and hence we will explicitly set $K=0$ from now on.

\subsection{Scalar-field cosmology}

The matter Lagrangian of a scalar field $\varphi(x^\mu)$ with canonical kinetic terms, obeying a potential $W(\varphi)$ is
\be
\cL_m\equiv\sqrt{-g}P_m=\sqrt{-g}\lh-\frac{1}{2}g^{\mu\nu}\partial_\mu\varphi\partial_\nu\varphi-W(\varphi)\rh. 
\ee
$P_m$ is introduced for later convenience. 
From this one can derive the field equations  
\be 
%\frac{1}{\sqrt{-g}}\partial_\mu\lh\sqrt{-g}\ g^{\mu\nu}\partial_\nu\varphi\rh-\frac{\partial W}{\partial\varphi}=
g^{\mu\nu}\cD_\mu\partial_\nu
\varphi-W_{,\varphi}=0,\label{fe}
%,\qquad\mathrm{where}\qquad
%\partial^\mu\varphi=g^{\mu\nu}\partial_\nu\varphi,
\ee
where $W_{,\varphi}\equiv\partial W/\partial\varphi$. 
%where we have defined the covariant derivative
%\be 
%\cD_\mu T^\kappa_\lambda=\partial_\mu T^\kappa_\lambda+\Gamma^\kappa_{\mu\rho}T^\rho_\lambda-\Gamma^\rho_{\mu\lambda} T^\kappa_\rho\qquad\mathrm{with}\qquad\cD g=0.
%\ee
The stress-energy tensor (\ref{stress}) becomes  
\be
%G^\mu_\nu=\kappa^2T^\mu_\nu,\qquad\mathrm{with}\qquad 
T^\mu_\nu
%=-\frac{2g^{\mu\kappa}}{\sqrt{-g}}\frac{\delta\cL_m}{\delta g^{\kappa\nu}}
=\partial^\mu\varphi\partial_\nu\varphi-\delta^\mu_\nu\lh \frac{1}{2}g^{\kappa\lambda}\partial_\kappa\varphi\partial_\lambda
\varphi+W\rh.\label{eee}
\ee
%
%\be
%G^0_0
%%=\kappa^2T_{00}
%=-\kappa^2\rho\qquad\mathrm{and}\qquad 
%G^i_i
%%=\kappa^2T_{ii}
%=\kappa^2 p\delta^i_i\label{eeback}
%\ee

%In order to solve these equations one has to choose an ansatz for the metric, i.e.\ choose the physical system to study. We will assume for the moment that the universe is dominated by a homogeneous field $\phi(t)$, leading to a homogeneous and isotropic space background. The latter is described by the Friedmann Lema\^itre Robertson Walker (FRLW) metric
%\be
%\d s^2=-N^2(t)\d t^2+a^2(t)\lh\frac{\d r^2}{1-K r^2} + r^2\d\Omega^2\rh,\label{frw} 
%\ee
%where $K$ is the spatial curvature of the universe, taking values $K=0$ for a flat, $K=1$ for a closed universe and $K=-1$ for an open universe, while $a(t)$ is the scale factor showing the rate of expansion of the proper distance between two comoving observers. $N(t)$ is merely a redefinition of the time coordinate: $N(t)=1$ corresponds to the cosmological time $t$, $N(t)=a(t)$ corresponds to the conformal time $\eta$ and $N(t)=1/H(t)$ corresponds to a time coordinate $N$ equal to the number of e-folds. Then (\ref{eeback}) become 
%\be
%H^2\equiv\frac{\dot{a}^2}{N^2a^2}=\frac{\kappa^2}{3}\rho-\frac{K}{a^2}\qquad\mathrm{and}\qquad \frac{\dot{H}}{N}=-\frac{\kappa^2}{2}\lh p+\rho\rh+\frac{K}{a^2},\label{ee}
%\ee 
In the background the scalar field depends only on time. We will denote the background field as $\phi(t)$ to distinguish it from the fully non-linear field $\varphi(t,\vc{x})$ that we will study in the next sections. The background energy and pressure density turn out to be
\be
\rho=\frac{1}{2}\Pi^2+W\qquad\mathrm{and}\qquad p=\frac{1}{2}\Pi^2-W,\label{rp}
\ee
where we defined the canonical momentum of the field $\phi$
\be
\Pi\equiv N\frac{\partial P_m}{\partial \dot{\phi}}=\frac{\dphi}{N}.  
\ee 
%These equations can be combined to give
%\be
%\dot{\rho}+3H(\rho+p)=0,
%\ee
%coinciding with the energy-momentum conservation condition $\cD_\mu T^\mu_0=0$. 
The Einstein equations (\ref{ee}) reduce to
\be 
H^2=\frac{\kappa^2}{3}\rho=\frac{\kappa^2}{3}\lh \frac{1}{2}\Pi^2+W\rh\qquad\mathrm{and}\qquad 
\frac{\dot{H}}{N}=-\frac{\kappa^2}{2}\Pi^2\label{ee1},
\ee 
while the background field equation (\ref{fe}) can be rewritten in terms of the momentum as
\be
\dot{\Pi}+3NH\Pi+NW_{,\phi}=0.\label{fe1} 
\ee

There exists a simple solution for this set of equations corresponding to a de-Sitter universe: the vacuum solution, i.e.\ when the field $\phi$ has reached a local or global minimum of its potential and hence $\Pi=0$, while $W=\mathrm{constant}$. It is then easy to check from (\ref{rp}) that $\rho=-p$ and the scalar field behaves like a cosmological constant. Therefore, the scale factor expands exponentially fast, while the Hubble parameter remains constant. However, this means that the domination of the vacuum would never end since all other components in the universe would dilute away and hence a domination of the vacuum can not serve as an inflationary period.

\begin{figure}
\begin{center}
\includegraphics[scale=0.75]{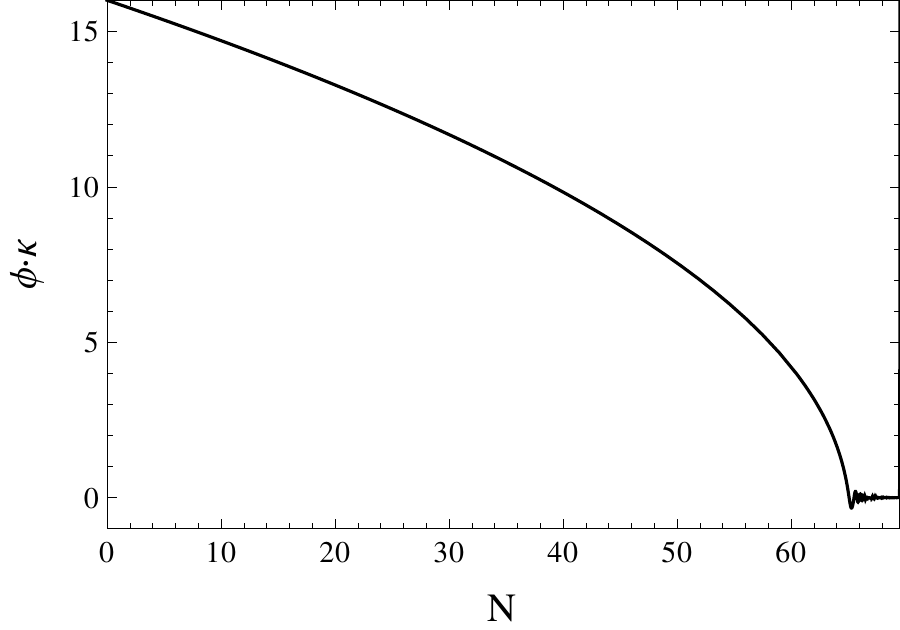}
\caption{The time evolution of the scalar field $\phi$ for the model (\ref{modelsingle}) with initial condition 
$\phi_0=16\kappa^{-1}$ and mass $m_{\phi}=12\times 10^{-6}\kappa^{-1}$.}
\label{fig21}
\end{center}
\end{figure}

%Finally, before studying the matter field that drives inflation, i.e.\ a scalar field, 
Let us introduce here the notion of the event horizon closely related to an inflationary period in the history of the universe. An event horizon defines how far into the future light emitted from an observer can reach. In other words, the event horizon bounds that region of space with which causal contact will be possible in the future.  
% This is defined as the maximum distance from which particles could have travelled to an observer during the age of the universe, or else the maximum distance from the past to an observer today at $t$ from which he/she can obtain information (causal contact has been possible). Hence in absence of inflation it represents the boundary of the observable universe\footnote{When an inflationary era is assumed during the expansion of the universe, the cosmological horizon is larger than the observable universe. This is how inflation solves the horizon problem.}. 
Taking the null geodesic, we find from (\ref{frw})
\be
%\int_0^t\frac{N(t')}{a(t')}\d t'=\int_0^{r_0}\frac{\d r}{\sqrt{1-Kr^2}}. 
\int_t^{t_\mathrm{max}}\frac{1}{a(t')}\d t'=\int_{r_0}^{r_\mathrm{max}}\frac{\d r}{\sqrt{1-Kr^2}},  
\ee
where for this and the next paragraph we set $N=1$, i.e.\ we work with cosmic time. 
The geometrical distance between two points at the same time $t$ separated by a comoving distance $r_0$ is 
\be
d(t)=a(t)\int_{r_0}^{r_\mathrm{max}}\frac{\d r}{\sqrt{1-Kr^2}} 
\ee
and hence we find for the cosmological horizon
\be 
d_E(t)=a(t)\int_t^{t_\mathrm{max}}\frac{1}{a(t')}\d t'.\label{eventhor}
\ee
For radiation or matter one can find using  (\ref{radmat}) and taking $t_\mathrm{max}=\infty$ that $d_E=\infty$. In that sense the old standard CDM cosmology does not have an event horizon. For a cosmological constant or vacuum domination we find using (\ref{rhol}) that $d_E=1/H$ and hence the exponential expansion produces an event horizon at a constant proper distance which is equivalent to a shrinking comoving horizon. 

On the other hand, the particle horizon encompasses that region in space with which causal contact has  been possible in the past. Analogously to equation (\ref{eventhor}) we find 
\be
d_P(t)=a(t)\int_0^t\frac{1}{a(t')}\d t'.\label{parthor} 
\ee
For a radiation ($a\propto t^{1/2}$) or matter ($a\propto t^{2/3}$) dominated universe, the particle horizon is $d_P=1/H$ and $d_P=2/H $ respectively. For vacuum domination one obtains $d_P=\lh \mathrm{exp}(Ht)-1\rh/H$.

In order to visualise how inflation is realised by a scalar field, we will consider the example of a free massive field that obeys the quadratic potential 
\be 
W=\frac{1}{2}m_\phi^2\phi^2. \label{modelsingle}
\ee
The only free parameter is the mass of the field, which we take to be $m_\phi=12\times10^{-6}\kappa^{-1}$. For our numerical calculations we use the initial condition  $\phi_0=16\kappa^{-1}$ at $t=0$. In figure \ref{fig21} we plot the evolution of the field as a function of the number of e-folds. The field $\phi$ dominates inflation while rolling down its potential and about $60$ e-folds after the initial time $t=0$ it starts oscillating around the minimum of its potential and inflation ends.   
\begin{figure}
\begin{center}
\includegraphics[scale=0.75]{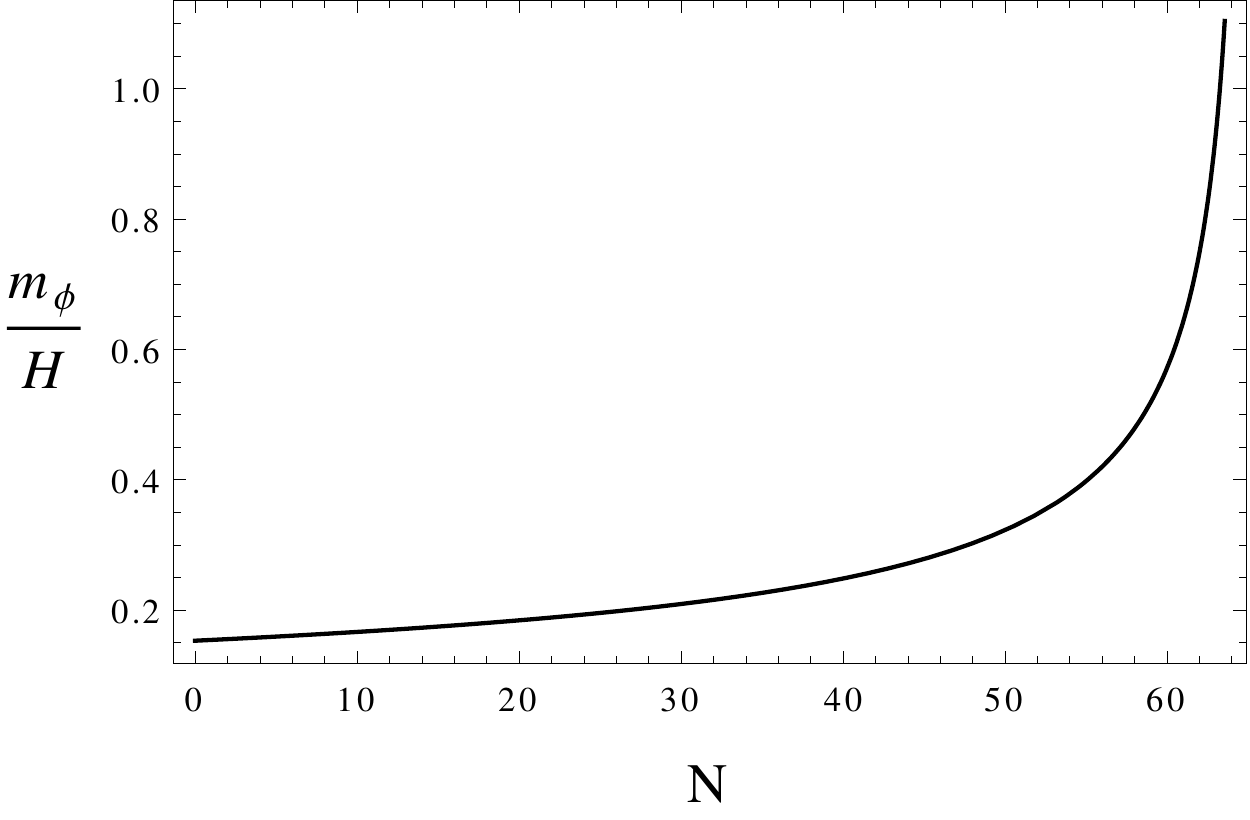}
\caption{The time evolution of the event horizon, for the model (\ref{modelsingle}) with initial condition 
$\phi_0=16\kappa^{-1}$ and mass $m_{\phi}=12\times10^{-6}\kappa^{-1}$.}
\label{fig22}
\end{center}
\end{figure}
In figure \ref{fig22} we plot the event horizon as a function of the number of e-folds of inflation. While the field $\phi$ rolls down its potential,  $1/H$ remains effectively constant. It only starts to increase when the inflaton approaches its minimum, when inflation ends.

Finally let us comment on the amount of inflation needed to solve the horizon and flatness problem. In order to solve the horizon problem, we need to demand that the observable universe today $1/k_0=1/(a_0H_0)$ was within the horizon at the beginning of inflation $1/k_{in}=1/(a_{in}H_{in})$. Ignoring numerical factors, this means that 
\be
\frac{1}{a_0H_0}\le\frac{1}{a_{in}H_{in}}.  
\ee
Then the minimal amount of inflation can be rewritten by rearranging the above equality
\be 
\frac{a_{fin}}{a_{in}}=\frac{a_{fin}}{a_0}\frac{H_{in}}{H_0}.\label{horizon} 
\ee
The right hand side of this equation can now be evaluated. The value of the Hubble parameter today according to Planck is $H_0=67.3\pm 1.2\ \mathrm{km\ s^{-1}\ Mpc^{-1}}$. As discussed in chapter \ref{Ch1}, the scale factor scales with the temperature of the universe as $a\propto 1/T$. The temperature today is $T_0=2.7255\pm 0.0006\ \mathrm{K}$. As for the value of the Hubble parameter at the beginning of inflation, this can be approximated through the mass scale of inflation $M$ as $H_{in}\sim \kappa M^2$. We will assume the standard value for this mass scale $M\sim10^{16}\ \mathrm{GeV}$. We make the further approximation that all the energy of the inflaton is used for reheating and hence that the temperature at the end of inflation is $T_{fin}\sim M$. At the end of inflation, the energy of the inflaton is converted into radiation due to its oscillations around its minimum and the universe reheats. The efficacy of this process depends on the exact mechanism, however the final temperature must not be as high as the temperature before inflation, otherwise any catastrophic topological defects will be produced again. Hence, the assumption we make here can only give an upper estimate.  Putting everything together we find
\be 
\frac{a_{fin}}{a_{in}}=\frac{T_0}{M}\frac{\kappa M^2}{H_0}=10^{29}\kappa M\sim\mathrm{e}^{60}.
\ee
Hence, the minimum amount of inflation (for our observable universe today to be within the inflationary horizon) is roughly evaluated to $60$ e-folds. This is the number quoted in the literature, although detailed calculations can vary 
between $50-60$ e-folds \cite{Liddle:1993fq}.  

On the other hand, since the curvature density parameter scales like $\Omega_K=K/(a^2H^2)$, demanding that the curvature density parameter in the beginning of inflation was equal to the curvature density parameter today (instead of the many orders of magnitude less predicted in absence of inflation), one finds
\be
\frac{1}{a_0^2H_0^2}=\frac{1}{a_{in}^2H_{in}^2}, 
\ee
that reproduces equation (\ref{horizon}) and hence the $60$ e-folds also solve the flatness problem.

\subsection{The slow-roll approximation}\label{sssr}

In order to study this background analytically and get estimates for its properties in the general case, one needs to employ some kind of assumption to simplify equations (\ref{ee1}) and (\ref{fe1}). Fortunately, these are motivated by the concept of inflation itself. 
As discussed above in order to solve the horizon problem, the universe must expand during inflation by at least the same amount as during all the time after inflation up till now. This points naturally to an exponential type of expansion, which can be achieved by a cosmological constant or in the case of a scalar field dominated universe, by a field that does not roll at all, $\Pi=0$, and hence has reached a minimum of its potential. This vacuum domination is characterised by $\rho=-p=\mathrm{constant}$.

However, if the field does not roll at all inflation would never end. The simplest way to avoid that, is to relax the vacuum dominated universe assumption to a quasi-de Sitter universe $\rho\simeq-p\simeq\mathrm{constant}$. It is exactly this property of a scalar field that makes it the appropriate candidate for an inflaton, i.e.\ the field that drives inflation. The quasi-de Sitter condition is translated to 
\be
%\rho\simeq -p\qquad\mathrm{or}\qquad 
\Pi^2\ll W.\label{cond}
\ee

In that case the field rolls only slowly down a very flat potential. Its kinetic term is not exactly zero, but small, and the potential is quasi-constant, leading to a quasi-exponential expansion. Inflation ends when the field rolls more rapidly while reaching its global minimum, while reheating is achieved by oscillations around this minimum. In the next sections we show that the slow-roll assumption is also in agreement with the almost scale-invariant spectrum of density perturbations and the small bispectrum to power spectrum ratio observed, reinforcing the choice of the slow-roll approximation.

%Assuming that it is the scalar field that dominates the universe and not the curvature, in order to achieve the quasi de-Sitter state the condition is that 
%\be
%\rho\simeq p\qquad\mathrm{or}\qquad \Pi^2\ll|W|.\label{cond}
%\ee
%From the first equation of (\ref{ee1}) one can see that the curvature contribution soon goes to zero, assuming that $W$ remains constant for long enough. The energy density of a scalar field does not scale directly with the scale factor as does the curvature. 

The slow-roll condition (\ref{cond}) can be recast using (\ref{ee1}) into
\be
\e\equiv-\frac{\dot{H}}{NH^2}=\frac{\kappa^2\Pi^2}{2H^2}\ll1.\label{defe}
\ee
This is the first slow-roll assumption. 
To assure that once $\Pi^2$ is much smaller than $W$, it remains so, one should demand that 
\be
\eta\equiv\frac{\dot{\Pi}}{HN\Pi}\ll1,\label{eta} 
\ee 
that is the second slow-roll assumption.
Based on (\ref{eta}) one can construct an infinite hierarchy of slow-roll parameters like \cite{GrootNibbelink:2001qt}
\be
\eta^{(n)}\equiv\frac{1}{H^{n-1}\Pi}\lh\frac{1}{N}\frac{\d}{\d t}\rh^{n-1}\Pi,
\quad\mathrm{with}\quad n>1.\label{defeta}
\ee
For $n=1$ we find the trivial result $\eta^{(1)}=1$, while $\eta\equiv\eta^{(2)}$. 
The second order slow-roll parameter turns out to be
\be
\xi\equiv\eta^{(3)}=\frac{\ddot{\Pi}}{N^2H^2\Pi}-\frac{\dot{N}}{N^2}\eta=-\frac{W_{,\phi,\phi}}{H^2}+3\lh\e -\eta\rh 
\ee
and so on. 

One can view the slow-roll parameters as short-hand notation for the quantities defined above. The Einstein equations and the field equation (\ref{fe1}) can then be rewritten as 
\be
H^2\lh 1-\frac{\e}{3}\rh=\frac{\kappa^2}{3}W,\qquad
\dot{H}=-\e H^2N,\qquad
\eta=-3-\frac{W_{,\phi}}{H\Pi}.\label{fe2} 
\ee
These equations are exact as long as one does not make the slow-roll assumption. When making the slow-roll assumption, one can use expansions in powers of the slow-roll parameters. To zeroth order in this expansion, one recovers the vacuum solution $\rho=-p$. To first order, one can assume the slow-roll parameters as constant. This can be easily shown after calculating the derivatives of the slow-roll parameters. By directly differentiating (\ref{defe}) and using (\ref{eta}) one finds
\be
\dot{\e}=2\e NH(\e+\eta),\label{de} 
\ee
while differentiating (\ref{defeta}) one gets
\be
\dot{\eta}^{(n)}=NH\Big[\lh\e-\eta\rh\eta^{(n)}
+\eta^{(n+1)}\Big].\label{deta}
\ee
The time derivatives of the slow-roll parameters are one order higher than the slow-roll parameters themselves and hence up to first order the slow-roll parameters can be considered as constant, considering of course an appropriate time interval. 

\begin{figure}
\begin{center}
\includegraphics[scale=0.75]{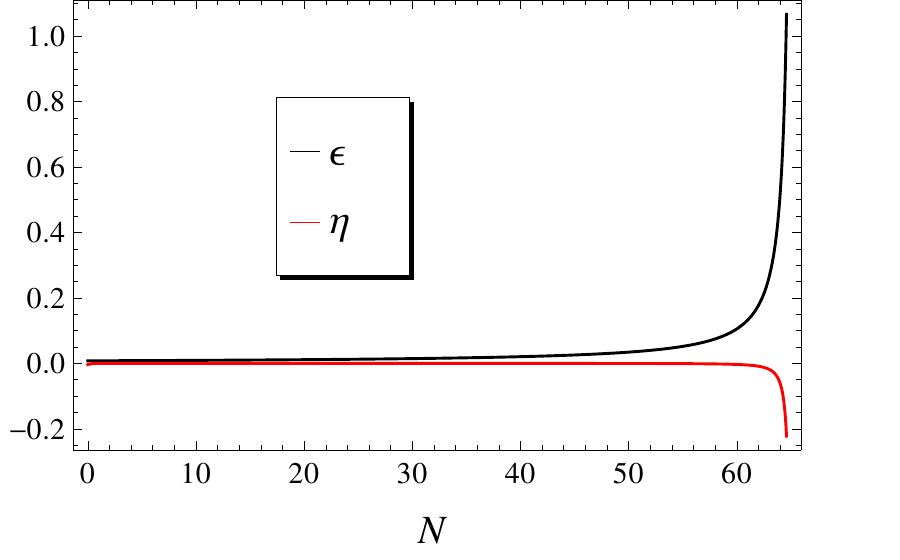}
\caption{The time evolution of the first-order slow-roll parameters, for the model (\ref{modelsingle}) with initial condition 
$\phi_0=16\kappa^{-1}$ and mass $m_{\phi}=12\times10^{-6}\kappa^{-1}$.}
\label{fig23}
\end{center}
\end{figure}

In figure \ref{fig23} we plot the first-order slow-roll parameters for the model (\ref{modelsingle}). Indeed these remain smal during inflation and only increase towards the end of inflation when the fields approaches its minimum. From now on we will define the end of inflation as the time when $\e=1$.

\Section{First-order perturbations}\label{s21}

Apart from solving the horizon and flatness problems, inflation also provides the initial seeds for the anisotropies of the CMB and hence eventually, for the structure in the universe today.  These seeds are related to the quantum perturbations of the inflaton. 
%The quantum perturbations of the scalar field is what makes inflation such a successful scenario. 
Quantum perturbations are created and annihilated in the vacuum. During inflation the superluminal expansion of space-time itself carries the perturbations to scales larger than the horizon before they have time to annihilate and hence they become classical. But in order to understand the details of this process one needs to find the evolution equations of the scalar perturbations. These inevitably produce and in turn are sourced by perturbations of space-time. 

\subsection{The observables}\label{observables}

As will be discussed in subsection \ref{scalar}, slow-roll inflation predicts the creation of almost  Gaussian perturbations frozen once outside the horizon. When these perturbations re-enter inside the horizon they  will give birth 
%to all the structure seen today as well 
to the anisotropies of the CMB, that can be measured to an excellent precision today. Furthermore, the evolution of the CMB anisotropies can be interpolated back in time to be related with the primordial inflationary perturbations. Gaussian statistics predict that the only  quantity that carries all the necessary information about a Gaussian field $f(t,\vc{x})$ is the two-point correlation function. Assuming that such a field can be expanded in Fourier space as
\be
f(t,\vc{x})=\int\frac{\d^3\vc{k}}{(2\pi)^{3/2}}f_{\vc{k}}(t)\mathrm{e}^{\mathrm{i}\vc{k}\cdot\vc{x}},
\ee
the two-point correlation function takes the form \cite{Lyth:1998xn,Bartolo:2004if}
\be
\langle f_{\vc{k}}f^*_{\vc{k'}}\rangle\equiv
\frac{2\pi^2}{k^3}\mathcal{P}_f(k)\delta^{(3)}\lh\vc{k}-
\vc{k'}\rh.\label{powernew} 
\ee
All odd correlation functions are zero, while all even higher order correlation functions are products of the two-point correlation function. The quantity $\mathcal{P}_f$ is called the power-spectrum and it is a measure of the amplitude of the fluctuations at a given scale $k$ (see e.g.\  \cite{Lyth:1998xn,Bartolo:2004if}). Furthermore, if one calculates the mean square value of $f(t,\vc{x})$, one finds 
\be
\langle f^2(t,\vc{x})\rangle=\int\d\mathrm{ln}k\  \mathcal{P}_f(k) 
\ee
and hence the power spectrum is also the contribution to the variance per unit logarithmic interval in $k$, if $\langle f(t,\vc{x})\rangle=0$. 

This implies that one can construct another quantity, apart from the amplitude of the perturbations, that is the slope of the power-spectrum, usually described by a spectral index $n_f(k)$ 
\be
n_f(k)-1\equiv\frac{\d\mathrm{ln}\mathcal{P}_f}{\d\mathrm{ln}k}. 
\ee
In subsection \ref{scalar} we will discuss the form of these observable quantities for perturbations produced during inflation. 

\subsection{The set-up}\label{set}

In the context of perturbation theory around an homogeneous background any quantity $\bar{A}$ that has a non-zero background value, will be decomposed into an homogeneous part $A(t)$ and an infinite series of perturbations as
\be
\bar{A}(t,\vc{x})=A(t)+A_{(1)}(t,\vc{x})+\frac{1}{2}A_{(2)}(t,\vc{x})+\cdots,
\ee 
where the subscripts in the parentheses denote the order of the perturbation. For the scalar field, as was discussed in the previous section, we shall denote the fully non-linear quantity as $\bar{\varphi}(t,\vc{x})$ and its background homogeneous part as $\phi$. 

We will write the fully non-linear metric in the ADM form as \cite{Misner:1974qy}
\be
ds^2=-\bN^2dt^2+\bh_{ij}(dx^i+N^idt)(dx^j+N^jdt),\label{metricexact}.
\ee
The idea of the ADM formalism is to separate time from the rest of the coordinates and split the space-time into a series of spacelike hypersurfaces of constant time. This is done in order to rewrite the action for gravity in a Hamiltonian formulation. The lapse function $N$ is related to the distance between two adjacent hypersurfaces. The spatial metric on each hypersurface  is 
\be
\bh_{ij}=a(t)^2\exp\big[2\alpha\delta_{ij}+2\chi_{ij}
\big], 
\ee
where $\chi_{ij}$ is traceless and $\alpha$ is the scalar curvature perturbation. 
As any tensor, the spatial metric can be split into a scalar, a vector and a tensor part as
\be
\chi_{ij}=D_{ij}F+F_{(i|j)}+\gamma_{ij},
\ee
where $|$ or $\nabla$ is the covariant derivative on the space hypersurface, i.e.\ relative to the  spatial metric $\bh_{ij}$ (being just $\partial_i$ when acting on space scalars) and indices enclosed in parentheses are to be symmetrized. We have also introduced the operator
\be
D_{ij}\equiv\nabla_i\nabla_j-\frac{1}{3}\delta_{ij}\nabla^2. 
\ee 
The vector $F_i$ is divergence-free, while the pure  tensor part $\gamma_{ij}$, that represents gravity waves, is transverse, i.e.\ traceless and divergence-free
%\bea
%\bh_{ij}=a(t)^2\rme^{2\alpha}\rme^{2\gamma_{ij}}\rme^{2E_{,ij}+E_{(i,j)}}=
%a^2\Big[\delta_{ij}+2\alpha_{(1)}\delta_{ij}+2\gamma_{ij(1)}+2E_{,ij}+E_{(i,j)}+\cdots\Big]
%\eea 
\be 
F^i_{|i}=0,\quad 
\gamma^k_{i|k}=0
\quad\mathrm{and}\quad\gamma_i^i=0.
\ee
The shift function $N^i$ is related to the distance between two points on a hypersurface and can be split up again in a scalar part and a divergence-free vector
\be N^i=\partial^i\psi+N^i_{\perp} \quad\mathrm{with}
\quad N^i_{\perp|i}=0.
\ee 
Here we defined $\partial^i\equiv\delta^{ij}\partial_j$. 
Hence we decompose the metric in $4$ scalar contributions $\bar{N},\ \alpha,\ \psi$ and $F$, corresponding to $4$ degrees of freedom, two divergence-free vectors $N_{i\perp}$ and $F_i$, corresponding to another $4$ degrees of freedom and a symmetric, transverse and traceless tensor $\gamma_{ij}$, corresponding to $2$ degrees of freedom. 
\be 
g_{\mu\nu}=
\lh
\begin{array}{cc}
-\bN^2 & \bh_{ij}\partial^j\psi \\
\bh_{ij}\partial^j\psi & a^2\mathrm{exp}\Big[2(\alpha\delta_{ij}+D_{ij}F)\Big]
\end{array}
\rh 
+a^2
\lh
\begin{array}{cc}
0 & N_{i\perp} \\
N_{i\perp} & \mathrm{exp}\lh 2F_{(i|j)}\rh 
\end{array}
\rh 
+a^2
\lh
\begin{array}{cc}
0 & 0 \\
0 & \mathrm{exp}\lh 2\gamma_{ij}\rh 
\end{array}
\rh. 
\ee
Together these account for the $10$ degrees of freedom of gravity in $4$ dimensions. $4$ of these are gauge degrees of freedom related to the choice of coordinates (see next subsection) and can be set explicitly, thus making a gauge choice.

To complete the description of the ADM formalism, we also present here the action for gravity and matter  in this $1+3$ decomposition \cite{Misner:1974qy}:
\be
S=\frac{1}{2}\int\d^4x\sqrt{\bh}\bN\lh \frac{R^{(3)}}{\kappa^{2}}-2\bar{W}-\bh^{ij}\partial_i\varphi
\partial_j\varphi\rh 
+\frac{1}{2}\int\d^4x\frac{\sqrt{\bh}}{\bN}\Big[\frac{1}{\kappa^{2}}\lh
\bE_{ij}\bE^{ij}-\bE^2\rh 
+\lh \dot{\varphi}-N^i\partial_i\varphi\rh^2\Big],  \label{actionexact1}
\ee
where $\bh$ is the determinant of the space metric $\bh_{ij}$, $R^{(3)}$ is the intrinsic 
3-curvature and the tensor $E_{ij}$ (proportional to the extrinsic curvature $\bK_{ij}=-\bN^{-1}\bE_{ij}$) is
\be
\bE_{ij}=\frac{1}{2}\left(\dot{\bh}_{ij}-N_{j|i}
-N_{i|j}\right).\label{excurv}
\ee
The generalised fully non-linear Hubble parameter is defined as the trace of the extrinsic curvature
\be
\bH\equiv\frac{\bE}{3\bN}.\label{genhubble} 
\ee

In the ADM formalism spatial reparametrizations are an explicit symmetry. 
The formalism is constructed so that one can think of $\bh_{ij}$ and $\varphi$ as the dynamical variables and $\bN$ and $N^i$ as Lagrange multipliers giving two constraint equations: the energy and momentum constraints
\bea
&&\kappa^{-2}R^{(3)}-2W-\kappa^{-2}\bN^{-2}(E_{ij}E^{ij}-E^2)-\bar{\Pi}^2-h^{ij}\partial_i\varphi\partial_j\varphi=0,\label{energy}\\
&&\nabla_j\Big[\frac{1}{\bN}(E_i^j-E\delta_i^j)\Big]=\kappa^2\bar{\Pi}\partial_i\varphi,\label{momentum}
\eea
where we have defined the non-linear canonical momentum as
\be
\bar{\Pi}\equiv\bN\frac{\partial\bP_m}{\partial\dvphi}=\frac{\dvphi-N^j\partial_j\varphi}{\bN}.
\ee

%As it will  Hence, it is only natural to 
%make a gauge choice and set $F_i=0$ and $F=0$, namely we will eliminate three degrees of freedom. These are the degrees of freedom related to spatial 
%We choose to set $F=0$. 
%
%Next we can also set to zero the vector modes $N_{i\perp}=0$ and $F_i=0$. Note however, that vector perturbations are anyway irrelevant for inflation governed by a scalar field, since this cannot produce vector perturbations. In addition, any vector perturbations decay because of the expansion of the universe. Thus we are left with 5 degrees of freedom, i.e.\ $\alpha,\ \psi$, $\bN$ and $\gamma_{ij}$. 

%We start by investigating the first-order perturbation predictions. We need to expand the Einstein equations (\ref{stress}) and the field equation (\ref{fe1}) up to first order. It turns out that the remaining scalar degrees of freedom of the metric are the only ones interacting with the scalar field perturbation. The Einstein equations become 

\subsection{Gauge Invariance}\label{ga1}

Since the definition of perturbations depends in general on the gauge choice, a gauge-invariant definition of the 
cosmological perturbations 
is of vital importance to make contact with physical observables, which are obviously gauge-invariant. 
That was investigated in detail in \cite{Bardeen:1980kt} and later in \cite{Mukhanov:1990me}. 
While the unperturbed metric (\ref{frw}) allows for an unambiguous definition of the spatial homogeneous hypersurfaces and a natural time coordinate $t$, this is no longer true in the presence of inhomogeneities. In order to avoid the ambiguity of the choice of coordinates one needs to define gauge-invariant perturbations in any coordinate system by constructing quantities that remain invariant under coordinate transformations. Here we will consider arbitrary coordinate transformations that up to second order take the form  \cite{Bruni:1996im,Malik:2003mv}
\be
\widetilde{x}^{\mu}=x^{\mu}+\beta_{(1)}^{\mu}
+\frac{1}{2}\lh\beta_{(1)|\nu}^{\mu}\beta^{\nu}_{(1)}+\beta_{(2)}^{\mu}\rh\qquad\mathrm{with}\qquad
\beta^0=T
\qquad\mathrm{and}\qquad 
\beta^i=\partial^i\beta+\beta^i_\perp,
\ee
where again we split the space part of the coordinate transformation in a derivative of a scalar and a divergence-free vector. 

Under such coordinate transformations the perturbations of a tensor transform as \cite{Bruni:1996im}
\bea
&&\widetilde{A}_{(1)}=A_{(1)}+L_{\beta_{(1)}}A,\nn\\
&&\widetilde{A}_{(2)}=A_{(2)}+L_{\beta_{(2)}}A+
L_{\beta_{(1)}}^2A+2L_{\beta_{(1)}}A_{(1)},\label{trans}
\eea
where $L_{\beta}$ is the Lie derivative along the vector $\beta$
\be
\lh L_{\beta}A\rh^{\mu_1\mu_2...}_{\nu_1\nu_2...}=\beta^\kappa\partial_\kappa A^{\mu_1\mu_2...}_{\nu_1\nu_2...}
-\partial_\kappa\beta^{\mu_1}
A^{\kappa\mu_2...}_{\nu_1\nu_2...}-\dots +\partial_{\nu_1}\beta^{\kappa}A^{\mu_1\mu_2...}_{\kappa\nu_2...}+\dots .
\ee
Note here that spatial gradients, i.e.\ quantities of the form $\partial_i\bA$, having vanishing background 
values are automatically gauge-invariant at first order, while at second order they transform as
\be
\partial_i\widetilde{A}_{(2)}=\partial_iA_{(2)}
+2L_{\beta_{(1)}}\partial_iA_{(1)}.\label{zerotr}
\ee

From the infinite number of possible gauge-invariant combinations, we choose to work with quantities constructed from the energy density and the logarithm of the spatial curvature $\alpha$. This is the reason why these are referred to as adiabatic cosmological perturbations, since they are related to the perturbation of the total energy density. We will consider a gauge transformation 
%$\beta_{(i)}=(T_{(i)},\vec{q}_{(i)})$ 
between two gauges, 
%from the hatted gauge to some other gauge, 
where for the moment both gauges are taken to be arbitrary. Up to first order we find using (\ref{trans}) that space-time scalar quantities (like the energy density perturbation or the field perturbation) transform as 
\be
\tilde{A}_{(1)}=A_{(1)}+\dot{A}
T_{(1)}
\ee
%\bea
%&&\tilde{\rho}_{(1)}=\hat{\rho}_{(1)}+\dot{\rho}
%T_{(1)},\nn\\
%&&\tilde{\varphi}_{(1)}=\hat{\varphi}_{(1)}+\dphi T_{(1)}.
%\eea
The transformations of the metric functions (\ref{metricexact}) are summarised in table \ref{a1}.  The gradient in the gauge transformation for $\alpha_{(1)}$ is defined as $\partial^2\equiv\partial^i\partial_i$.
%\\[1ex]
\begin{table}[t]
\begin{center}
\begin{tabular}{|l|l|l|}
\hline
Scalar & Vector & Tensor \\[0.5ex]
\hline   & & \\[-0.5ex]
$\tilde{N}_{(1)}=N_{(1)}+\dot{T}_{(1)}N+T_{(1)}\dot{N}$ 
& $\tilde{N}_{i\perp(1)}=N_{i\perp(1)}+\dot{\beta}_{i\perp(1)}$ 
& $\tilde{\gamma}_{ij(1)}=\gamma_{ij(1)}$
\\[0.5ex]
$\tilde{\psi}_{(1)}=\psi_{(1)}+\dot{\beta}_{(1)}-T_{(1)}\frac{N^2}{a^2}$  & $\tilde{F}_{i(1)}=F_{i(1)}+\beta_{i\perp(1)} $ &
\\[0.5ex]
$\tilde{F}_{(1)}=F_{(1)}+\beta_{(1)}$ & &
\\[0.5ex]
$\tilde{\alpha}_{(1)}=\alpha_{(1)}+T_{(1)}HN+\frac{1}{3}\partial^2\beta_{(1)}$ & & \\[0.05ex]
& & \\
\hline
\end{tabular}
%\\[1ex]
\caption{The gauge transformations for the scalar, vector and tensor parts of the metric $g_{\mu\nu}$}.\label{a1}
\end{center}
\end{table}
Out of the $10$ degrees of freedom of the metric, $4$ are gauge degrees of freedom related to the choice of coordinates. 

The tensor $\gamma_{ij}$, that carries two physical degrees of freedom, is gauge invariant.  Actually, its degrees of freedom can be identified with the two polarizations of  gravitational waves directly in the transverse gauge.   
%, i.e.\ with $T,\ \beta$ and $\beta^i_\perp$.
% We choose to set $E=0$ and $E^i=0$. 
We will %make a gauge choice and 
restrict ourselves to gauges with 
%$N_{i\perp}=E_{i\perp}=0$ 
$F_i=0$ and $F=0$ at all orders. 
At first order, this is equivalent to  moving between gauges with $\beta_{(1)}=0$ and $\beta_{i\perp(1)}=0$ as can be checked from table \ref{a1}. The advantage of this  choice, is that the former condition reduces the gauge transformation for $\alpha_{(1)}$ to its form valid on super-horizon scales, where gradients can be ignored. Furthermore, vector perturbations are known to be irrelevant  for scalar field inflation and it turns out that this choice simplifies calculations a lot and implies that $N_{i\perp}=0$ at least up to second order.   
After fixing the spatial reparametrizations (having eliminated $3$ degrees of freedom), the remaining scalar ($3$ metric and $1$ energy density) and vector ($2$ metric) degrees of freedom at first order transform as
\bea
&&\tilde{\rho}_{(1)}=\rho_{(1)}+\dot{\rho}
T_{(1)},\quad
\tilde{\alpha}_{(1)}=\alpha_{(1)}+T_{(1)}HN
,\nn\\
&&\tilde{N}_{i\perp(1)}=N_{i\perp(1)},\quad
\tilde{\psi}_{(1)}=\psi_{(1)}-T_{(1)}\frac{N^2}{a^2},\quad
\tilde{N}_{(1)}=N_{(1)}+\dot{T}_{(1)}N+
T_{(1)}\dot{N}.\label{trans1}
\eea
Since $\bN$ and $N_i$ (accounting for $4$ degrees of freedom) are already known to provide only constraint equations, once $T_{(1)}$ is fixed we will be left with only $1$ scalar dynamical degree of freedom. 
%, meaning practically that we ignore vector modes.
%Indeed vector modes are known to decay during a scalar field dominated era, see section \ref{vec}. Demanding that, 
%for any gauge at first order, is equivalent to choosing $\beta_{i\perp(1)}=0$. 

In order to fix $T_{(1)}$ we will construct a gauge-invariant quantity that reduces to $\alpha_{(1)}$ in the uniform energy-density gauge or equivalently the uniform-field gauge. From now on, we identify the tilded gauge as the uniform-energy gauge, so that $\tilde{\rho}_{(r)}=0$, where the subscript $(r)$ denotes the $r\mathrm{th}$ order in the perturbation expansion. This is equivalent to $\bN\bH=1$ and the time variable coincides with the number of
e-folds $t=\ln{a}+\alpha$. Setting $\tilde{\rho}_{(1)}=0$ in the first equation of (\ref{trans1}) we find 
\be
T_{(1)}
=-\frac{\rho_{(1)}}{\dot{\rho}}
%=-\frac{\hat{\rho}_{(1)}}{\dot{\rho}}
\label{uetr}
\ee
and using the transformation for $\alpha_{(1)}$ in (\ref{trans1}), we obtain
\be
\zeta_{(1)}\equiv\tilde{\alpha}_{(1)}=\alpha_{(1)}
%-\frac{HN}{\dot{\rho}}\hat{\rho}_{(1)}
-\frac{HN}{\dot{\rho}}\rho_{(1)}
%-\frac{H}{\Pi}\varphi_{(1)}
%+\frac{1}{3}\partial^2\beta_{(1)}
.\label{b1}
\ee
% in the super-horizon limit where one can ignore the spatial gradients. We shall set $\beta_{(1)}=0$ in order to avoid this confusion. Note that this way, we also assure our choice $E=0$ at any gauge. 
This is the first-order adiabatic perturbation $\zeta$. It is constructed in such a way, that it remains invariant under first-order gauge transformations. For an arbitrary gauge, the adiabatic perturbation is a combination of the scalar curvature and the energy density perturbation of the scalar field\footnote{Notice that in the general case a term $\partial^2F_{(1)}$ would appear in the definition of $\zeta_{(1)}$,  but since we calculate all our  observables on super-horizon scales, this would not affect our results.}. It is the main quantity for the rest of the calculations in this chapter. Working with a gauge-invariant quantity is the natural choice, since the two scalar degrees of freedom are combined to give the one and only scalar physical degree of freedom in a theory with one scalar field, while at the same time one avoids the confusion of gauge-dependent results. 
Notice that the initial gauge is still arbitrary, but if one was to associate it with the flat gauge 
$\hat{\alpha}_{(r)}=0$ (from now on denoted as hatted), then the time shift would become $T_{(1)}=\zeta_{(1)}/(NH)$.

For a single scalar field it can be shown that the first-order energy density perturbation is proportional to the first-order field perturbation. This is equivalent to showing that the uniform energy density gauge $\rho_{(1)}=0$ coincides with the uniform scalar field gauge $\varphi_{(1)}=0$ at linear order. Indeed using the $0i$ first-order Einstein equation (\ref{aee0i}) derived in appendix \ref{AppA} and perturbing up to first order $NH=\dot{a}/a$, 
\be
N_{(1)}H+H_{(1)}N=\dot{\alpha}_{(1)}, 
\ee
one can show that for an arbitrary gauge 
\be H_{(1)}=-\frac{\kappa^2\Pi}{2}\varphi_{(1)}\ee
and since the energy density perturbation is proportional to the Hubble parameter perturbation (perturbing equation (\ref{ee1})), the above statement is proved, while using (\ref{defe}), the adiabatic perturbation becomes
\be
\zeta_{(1)}=\alpha_{(1)}
%-\frac{HN}{\dot{\rho}}\rho_{(1)}
-\frac{H}{\Pi}\varphi_{(1)}.\label{defz1r} 
\ee 

\subsection{Scalar Perturbations}\label{scalar}

We start by investigating the first-order perturbation predictions for scalars. We need to expand the Einstein equations (\ref{stress}) and the field equation (\ref{fe1}) up to first order. An intermediate step for this calculation where we have kept the field perturbation $\varphi_{(1)}$ (instead of the gauge-invariant perturbation $\zeta_{(1)}$) can be found in appendix \ref{AppA}. It turns out that the remaining scalar degrees of freedom of the metric are the only ones interacting with the scalar field perturbation. The Einstein equations become 
\bea
&&
%\frac{1}{a^2}
00:\quad\partial^2\psi_{(1)}=-\frac{N}{Ha^2}\partial^2\alpha_{(1)}+\e\dot{\zeta}_{(1)}
\equiv-\frac{N}{Ha^2}\partial^2\alpha_{(1)}+\partial^2\lambda%+
%\lh\e-3\rh\lh N_{(1)}-\frac{\dot{\alpha}}{H}+\e N(\zeta-\alpha)\rh H
\label{ee00}\\
&&0i:\quad
N_{(1)}=\frac{\dot{\alpha}_{(1)}}{H}-\e N(\zeta_{(1)}-\alpha_{(1)})\label{ee0i}\\
&&ij:\quad
\ddot{\alpha}_{(1)}+HN\lh3+\e-\frac{\dot{N}}{HN^2}\rh\dot{\alpha}_{(1)}-HN\e\dot{\zeta}_{(1)}-H\dot{N}_{(1)}+H^2N\lh-3+\e+\frac{\dot{N}}{HN^2}\rh N_{(1)}\nn\\
&&
\quad\quad\ -(HN)^2\e\lh3+\e+\eta\rh\lh\zeta_{(1)}-\alpha_{(1)}\rh
=0.\label{eeij}
\eea
Here we have defined for later use 
%the gauge invariant (by construction, since it involves the adiabatic perturbation $\zeta_{(1)}$) quantity 
$\lambda=\e\partial^{-2}\dot{\zeta}_{(1)}$. The gradient will remind us that this quantity vanishes outside the horizon as we will show in subsection \ref{soa}.  Keeping in mind the gauge transformation for $\psi_{(1)}$ (\ref{trans1}) one can perform a consistency check and verify that $\lambda$ is  gauge invariant.
%, for $\beta_{(1)}=0$ or equivalently outside the horizon.  
%After perturbing (\ref{fe}) to first order one finds
The field equation takes the form
\bea
\ddot{\varphi}_{(1)}+\lh3HN-\frac{\dot{N}}{N}\rh\dot{\varphi}_{(1)}-\frac{N^2}{a^2}\partial^2\varphi_{(1)}+N^2W_{,\phi,\phi}\varphi_{(1)}+2NW_{,\phi}N_{(1)}\nn\\
-\dot{\phi}\lh\frac{\dot{N}_{(1)}}{N}-\frac{\dot{N}N_{(1)}}{N^2}-3\dot{\alpha}_{(1)}+
%\frac{1}{a^2}
\partial^2\psi_{(1)}\rh=0.\label{phie}
\eea 
%Equations (\ref{aee00}-\ref{aeeij}) become
%
%where we have take into account that the non-diagonal equations imply that $B_{(1)}=0$. 
Combining equations (\ref{phie}), (\ref{ee00}), (\ref{ee0i}) and (\ref{eeij}) one finds the equation for the first-order gauge-invariant perturbation $\zeta_{(1)}$ to be
\be
\ddot{\zeta}_{(1)}+HN\lh3+2\e+2\eta-\frac{\dot{N}}{HN^2}\rh\dot{\zeta}_{(1)}-\frac{N^2}{a^2}\partial^2\zeta_{(1)}=0\label{zeq}
\ee
or equivalently
\be
\frac{\d}{\d t}\lh\frac{a^3\e}{N}\dot{\zeta}_{(1)}\rh-a\e N\partial^2\zeta_{(1)}=0.
\ee

%%Equation (\ref{zeq}) implies that the adiabatic perturbation behaves like a free massive field and 
%consequently it 
The adiabatic perturbation can be represented in Fourier space by an infinite series of harmonic oscillators $\zeta_\vc{k}$
\be
\zeta(\tau,\vc{x})=\int\frac{\d^3\vc{k}}{(2\pi)^{3/2}}\zeta_\vc{k}(\tau)\rme^{\rmi\vc{k}\cdot\vc{x}}.\label{opz} 
\ee 
For the rest of this subsection, we drop the subscript $(1)$ to lighten the notation and set $N=a$, i.e.\ we choose to work with the conformal time $\tau$. That choice allows the metric to become manifestly conformal to Minkowski space. 
In order to be able to quantize the field, one needs to rewrite equation (\ref{zeq}) in canonical form, namely eliminate the first order derivative.  It is easy to see that this can be achieved by defining a new variable
%\footnote{Notice that for a massless field in de-Sitter space $\e=0$ and $\eta=-3$ and the relevant redefinition is $\zeta=\lh Na^3\rh^{1/2}Z$. For the conformal time choice, $N=a$, $Z_k$ obeys the equation $\ddot{Z}_k+\lh k^2-4\ddot{a}/a\rh=0$, that in the super-horizon regime has the solution $Z_k=c_+(k)a^2+c_-(k)/a^2$. Matching the solution at horizon crossing gives $|Z_k|=$}
\be 
\zeta=-\frac{\kappa}{a\sqrt{2\e}}q.\label{newvar}
\ee
The variable $q$ was first introduced by Sasaki and Mukhanov \cite{Sasaki,Mukhanov}. 
The factor $\kappa/\sqrt{2}$ is needed to assure that the Lagrangian for $q$ has the canonical form (see subsection \ref{soa}). 
After this redefinition, equation (\ref{zeq}) in Fourier space reads 
\be 
%\ddot{Z}_k+\lh\frac{k^2N^2}{a^2}-\sqrt{\frac{N}{a^3}}\frac{\d^2}{\d t^2}\sqrt{\frac{a^3}{N}}+m^2_\zeta\rh Z_k=0.\label{Zeq}
q''_k+\lh k^2-\frac{a''}{a}+m^2_q\rh q_k=0.\label{Zeq}
\ee
with 
\be m^2_q=-
%(aH)^2
\cH^2\lh3\e+3\eta+2\e^2+4\e\eta+\xi\rh.\ee
A prime denotes differentiation with respect to conformal time, while $\cH\equiv a'/a=aH$. Equation (\ref{Zeq}) implies that the redefined adiabatic perturbation behaves like a free massive field. 
Each harmonic oscillator has a time dependent mass $m_q$. 
The Lagrangian associated with this equation of motion is
\be
\mathcal{L}_2=\frac{1}{2}\lh q_\vc{k}'\rh^2-\frac{1}{2}\lh k^2-\frac{a''}{a}+m^2_q\rh q_\vc{k}^2.\label{Lnew}
\ee

Now $q_k$ can be promoted to a quantum operator $q_\vc{k}$, introducing the creation and annihilation operators $a_\vc{k}$ and $a^\dagger_\vc{k}$
\be 
q_\vc{k}(t)=q_k(t)a_\vc{k}+q^*_k(t)a^\dagger_{\mathbf{-}\vc{k}},
\ee
where $q_k$ is a classical solution of (\ref{Zeq}). Note that different choices of of $q_k$ correspond to different choices of vacua for the perturbation. The creation and annihilation operators satisfy the standard commutation relations
\be 
[a_\vc{k},a_\vc{k'}]=0,\qquad
[a^\dagger_\vc{k},a^\dagger_\vc{k'}]=0,\qquad
[a_\vc{k},a^\dagger_\vc{k'}]=\delta^{(3)}(\vc{k}-\vc{k'}),
\ee  
which imply a normalization condition for $q_k$ and its conjugate momentum $q'_k$
\be
q_k^*q_k'-q_kq'^*_k=-\rmi.\label{conj}  
\ee
 
%For the time coordinate choice $N=a$ equation (\ref{Zeq}) becomes
%\be
%\ddot{Z}_\vc{k}+\lh k^2-\frac{\ddot{a}}{a}+m^2_\zeta\rh Z_\vc{k}=0
%\ee
%or equivalently
%\be
%\ddot{Z}_k+\lh k^2+(aH)^2\lh-2+\e+\frac{m^2_\zeta}{(aH)^2}\rh\rh Z_k=0.\label{Zeqa}
%\ee

There exist two regimes where we can study equation (\ref{Zeq}) easily. These regimes can be described in terms of time during inflation or in terms of the scale of the perturbation under study. Let us elaborate on this point since it is crucial for the rest of our discussion. The horizon during inflation $1/H$ is almost constant by construction in the slow-roll approximation. On the other hand perturbations are characterised by their wavelength $a/k$, where $k$ is the comoving scale of the perturbation. Let us consider such a scale $k$. Initially, for early times, the perturbation is well inside the horizon meaning that $k\mg aH$. As time goes by, due to the exponential growth of the scale factor, the wavelength of the perturbation grows and the perturbation crosses the horizon when $k=aH$. At late times, the perturbation is well outside the horizon implying that $k\ll aH$. 

For early times the wavelength of the mode $k$ is much smaller than the Hubble scale $(aH)^{-1}$ and hence it feels it lives in an almost flat space, obeying
\be
q_k''+k^2q_k=0.
\ee 
Hence well within the horizon the modes should approach the plane wave solution of the ordinary flat space-time Minkowski quantum field theory
\be
q_{k}=\frac{1}{\sqrt{2k}}\rme^{-\rmi k \tau}\label{solin}
\ee

At late times, well outside the horizon, we can neglect the $k$ contribution. 
%\be
%\ddot{Z}_k+(aH)^2\lh-2+\e+\frac{m^2_\zeta}{(aH)^2}\rh Z_k=0.\label{Zeqsuper}
%\ee
Assuming also that the slow-roll parameters remain small during inflation, we can find an approximate solution for the perturbations in a de Sitter universe obeying
\be 
q_k''-\frac{a''}{a}q_k=0.
\ee
This is essentially the equation of a massless field in a de Sitter universe. Its solution is 
\be 
q_k=c_+(k)a+\frac{c_-(k)}{a^2}.
\ee
Matching this solution with the sub-horizon solution (\ref{solin}) at horizon crossing $k=aH$ and using (\ref{conj}), we find for the growing mode  
%\be
%|c_{+}(k)|= \frac{1}{a\sqrt{2k}}=\frac{H}{\sqrt{2k^3}}
%\ee
that $|c_+(k)|=H/\sqrt{2k^3}$ and hence outside the horizon the curvature perturbation is frozen 
\be
|\zeta_k|=\frac{\kappa H}{2\sqrt{\e k^3}}. 
\ee
This is an important result, since it implies that the adiabatic perturbation produced during single-field inflation will remain constant as long as it is outside the horizon. 
Actually the operator (\ref{opz}) behaves as
\be
\zeta(\tau,\vc{x})=\int\frac{\d^3\vc{k}}{(2\pi)^{3/2}}\frac{\kappa H}{2\sqrt{\e k^3}}\lh a_\vc{k}+a^\dagger_{\mathbf{-}\vc{k}}\rh\mathrm{e}^{\mathrm{i}\vc{k}\cdot\vc{x}},
\ee
that is, it behaves like a Gaussian variable. That means that outside the horizon quantum operators can be replaced by a stochastic variable with Gaussian statistics:
\bea
&&q_\vc{k}(\tau)
%\rightarrow q_\vc{k}(t)
=q_k(\tau)b(\vc{k}),\nn\\
&&\langle b(\vc{k})\rangle=0,\qquad
\langle b(\vc{k})b^*(\vc{k'})\rangle=\delta^{(3)}(\vc{k}-\vc{k'})
\eea 

The quantity we are interested in is the power-spectrum, discussed in subsection \ref{observables}.  
For our rough calculation for a massless field, the power spectrum is
\be
\mathcal{P}_\zeta=\frac{\kappa^2H^2}{8\pi^2\e}.
\ee
The dependence of $\e$ and $H$ in the above equation on the horizon exit time $\tau_*$ leads to an additional momentum dependence, parametrized by the spectral index 
%\be
%|\zeta_k|\sim k^{-\frac{3}{2}+\frac{n_\zeta-1}{2}}
%\ee
%where
\be
n_\zeta-1=\frac{\d}{\d \mathrm{ln}{k}}\mathrm{ln}\lh\frac{H^2}{\e}\rh =\frac{1}{
(1-\e)aH}\frac{\d}{\d \tau_*}\mathrm{ln}\lh{\frac{H^2}{\e}}\rh  
=-4\e-2\eta.\label{above}
\ee
The second equality in (\ref{above}) comes from the  derivative $(\d\ln{k}/\d\tau_*)^{-1}=(\d\ln(aH)/\d\tau_*)^{-1}$. 
Although this is a rough calculation, one can show that solving equation (\ref{Zeq}) analytically up to first order in slow-roll, one recovers the same result 
%\be
%|\zeta_k|=\frac{H}{\sqrt{2\e k^3}}\lh\frac{k}{aH}\rh^{-2\e-\eta}.
%\ee
%
%The quantity that we can measure is the power spectrum of the perturbations
%\be
%\langle\zeta_\vc{k}\zeta_\vc{k'}\rangle\equiv\frac{2\pi^2}{k^3}\delta^3(\vc{k}-\vc{k'})\mathcal{P}_\zeta 
%\ee
%with 
\be
\mathcal{P}_\zeta=\frac{\kappa^2H_*^2}{8\pi^2\e_*}\lh\frac{k}{k_*}\rh^{-4\e_*-2\eta_*}, 
\ee
where $k_*$ is some pivot scale corresponding to the horizon exit time $\tau_*$. 
The conclusion is that slow-roll inflation predicts an almost scale invariant power-spectrum. Observations are in total agreement with this prediction. Planck has measured the amplitude of the adiabatic inflationary perturbations to be $\mathrm{ln}\lh 10^{10}\mathcal{P}_\zeta(k_*)\rh=3.089^{+0.024}_{-0.027}$, while the spectral index is indeed very close but not equal to $1$,  $n_\zeta(k_*)=0.9603\pm0.0073$, for a pivot scale $k_*=0.05\mathrm{Mpc}^{-1}$ in the middle of the logarithmic scales probed by Planck.

\subsection{Vector Perturbations}\label{vec}

We now turn to vector perturbations. For the class of gauges under study the only vector perturbation at first order is $N_{i\perp(1)}$. 
Vector perturbations are not created during scalar field inflation, since there is no vector component of the stress-energy tensor. However, we present here the first order Einstein equations for a general stress-energy tensor, that simplify to 
\bea
&&\mathrm{0i:}\quad 
%\frac{1}{2a^2N^2}
\frac{1}{2N^2}
\partial^2N_{i\perp(1)}=\kappa^2T^0_{i\perp(1)}\\
&&\mathrm{ij:}\quad 
\frac{\delta^{ik}}
%{a^2N^2}
{N^2}
\Bigg[\lh \frac{\dot{N}}{N}-
3
HN\rh N_{(k\perp(1),j)}-\dot{N}_{(k\perp(1),j)}\Bigg]=\kappa^2
T^i_{j\perp(1)}.
\eea
For scalar field domination the right hand side of these equations is identically zero and hence no vector perturbations are produced, $N_{i\perp(1)}=0$. For the case of a perfect fluid, it can be shown that  $T^i_{j\perp(1)}=0$ and hence the right-hand side of the second equation is still zero. This can be  rewritten as
\be
\frac{\d}{\d t}\lh \frac{a^3}{N}N_{(k\perp(1),j)}\rh=0,
\ee
which implies that even if there were any vector perturbations created at some time, these are diluted away during inflation.

\subsection{Tensor Perturbations}\label{tensor}

The tensor perturbations, although not sourced by the scalar field at first order, evolve on their own and represent the two polarizations of the graviton. The only non-zero Einstein equation is the $ij$, giving
\be
\ddot{\gamma}_{ij}+HN\lh3-\frac{\dot{N}}{HN^2}\rh\dot{\gamma}_{ij}-\frac{N^2}{a^2}\partial^2\gamma_{ij}=0,
\ee
where again we have dropped the $(1)$ subscript. 
We can expand $\gamma_{ij}$ in plane waves
\be
\gamma_{ij}=\int\frac{\d^3k}{(2\pi)^{3/2}}\sum_{s=\pm}\e_{ij}^s(k)
\gamma_\vc{k}^s\rme^{\rmi\vc{k}
\cdot\vc{x}}, 
\ee
where the polarization tensors satisfy $\e_{ii}=k^i\e_{ij}=0$ and $\e^s_{ij}(k)\e^{s'}_{ij}(k)=2\delta_{ss'}$. 
So each polarization mode obeys 
\be
\ddot{\gamma}_k+ HN\lh3-\frac{\dot{N}}{HN^2}\rh\dot{\gamma}_k+k^2\frac{N^2}{a^2}\gamma_k=0. 
\label{tenseq}
\ee
As we did for the adiabatic perturbation, we set the time coordinate to be the conformal time $\tau$ and we perform a redefinition 
\be
\gamma_k=\frac{2\kappa}{a}\Gamma_k 
\ee
to bring this equation to the canonical form
\be
\Gamma_k''+\lh k^2-\frac{a''}{a}\rh\Gamma_k=0. 
\ee
This equation is the same as equation (\ref{Zeq}) in subsection \ref{scalar} for $m_q=0$, so each redefined  polarization mode obeys essentially the equation of a massless scalar field. The procedure remains the same and one can find the same result from  the matching of the growing super-horizon solution at horizon crossing. Hence one finds in the end
\be
|\gamma_k|=\frac{2\kappa H}{\sqrt{2k^3}}
\ee

One can now calculate the power spectrum for the tensor modes
\be 
\langle\gamma_\vc{k}^s\gamma_\vc{k'}^{s'}\rangle=\frac{2\pi^2}{k^3}\delta^3(\vc{k}-\vc{k'})\delta_{ss'}\mathcal{P}_T
\qquad\mathrm{with}\qquad 
\mathcal{P}_T=\frac{2\kappa^2H^2}{\pi^2}.\ee
Comparing to the power spectrum of the adiabatic perturbation we notice that it does not depend on $\e$. Indeed, the amplitude of the tensor perturbations is directly proportional to the energy scale at which inflation occurred. 
The scale dependence of the tensor power spectrum is parametrized by the spectral index $n_T$:
\be
n_T=\frac{\d\mathrm{ln}H^2}{\d \mathrm{ln}k}=-2\e. 
\ee
Therefore, tensor perturbations have an almost scale-invariant power spectrum as does the scalar adiabatic perturbation, but with a different scale dependence.  Furthermore, one can write down a consistency relation between the scalar and the tensor modes by considering the ratio of their power spectra
\be
r=\frac{\mathcal{P}_T}{\mathcal{P}_\zeta}=16\e=-8n_T. 
\ee
The constraint from Planck of the tensor to scalar ratio is $r<0.11$, assuming no running of the spectral indices, i.e.\ a first-order slow-roll  approximation, where the slow-roll parameters are constant and $r<0.26$ allowing for running, both values at $95\%$ CL and a pivot scale $k_*=0.002\ \mathrm{Mpc}^{-1}$.

\subsection{The second-order action}\label{soa}

All of the above results can be reproduced by perturbing the action (\ref{actionexact1}) to second order and solving for the equations of motion for the first order perturbations, instead of perturbing the equations themselves to first order as we have done up to now. To do so we shall work in the ADM formalism described in subsection \ref{set}. 

To that end, we make the same gauge choice made in previous paragraphs, namely set $F_{i\perp}=0$ and $F=0$. In order to find the second order action one first  needs to solve for the constraint equations (\ref{energy}) and (\ref{momentum}) to first order. The resulting equations are just the $00$ and $0i$ Einstein equations (\ref{ee00}) and (\ref{ee0i}) respectively, along with the constraint $N_{i\perp}=0$. It turns out that we do not need to calculate the shift or the lapse function to higher order, since in the action those terms are multiplied by constraint relations and hence vanish.
  
In order to simplify calculations, we will work in the flat gauge $\hat{\alpha}_{(1)}=0$. The only scalar degree of freedom is the field  $\hat{\varphi}_{(1)}=-\Pi/H\zeta_{(1)}$. Perturbing the action (\ref{actionexact1}) to second-order, we find (for details, see appendix \ref{AppB})
\bea
&&\!\!\!\!\hat{S}_2=\frac{1}{2}\int\d^4x\Big\{ a^3\Big[\frac{1}{N}\dot{\hat{\varphi}}_{(1)}^2-NW_{,\phi,\phi}\hat{\varphi}^2_{(1)}
-2\hat{N}_{(1)}(W_{,\phi}\hat{\varphi}_{(1)}+\frac{\dphi
\dot{\hat{\varphi}}_{(1)}}{N^2})+\frac{\hat{N}_{(1)}^2}{N^3}
(-\frac{6}{\kappa^2}(NH)^2+\dphi^2)\Big]\nn\\
&&\qquad\qquad\quad\quad
-aN\lh\partial\hat{\varphi}_{(1)}\rh^2\Big\}.
\eea
Rewriting $\hat{\varphi}_{(1)}$ in terms of $\zeta_{(1)}$ and using the $0i$ constraint (\ref{ee0i}) along with the background equations (\ref{fe2}) and the derivatives (\ref{de}) and (\ref{deta}) to perform integrations by parts, the action can be rewritten in the gauge-invariant form
\be 
S_2=\!\!\int\!\! \d^4x \mathcal{L}_2=\frac{1}{2}
\int\!\! \d^4x\ \Big[2a^3\frac{\e}{\kappa^2N}\dot{\zeta}_{(1)}^2-2a\frac{\e N}{\kappa^2} (\partial\zeta_{(1)})^2
\Big],
\ee
where $\mathcal{L}_2$ is the second-order Lagrangian.  
$\zeta_{(1)}$ would be a pure gauge mode in de-Sitter space and it gets a non-trivial action only to the extent that $\e$ is non-zero.
The equation of motion for $\zeta_{(1)}$, which we will denote by $\delta \mathcal{L}_2/\delta\zeta_{(1)}$, is 
\bea 
&&\!\!\!\!\!\!\!\!\!\frac{\delta \mathcal{L}_2}{\delta\zeta_{(1)}}
\equiv\partial_\mu \frac{\partial\cL_2}{\partial\lh\partial_\mu\zeta_{(1)}\rh }-\frac{\partial\cL_2}{\partial\zeta_{(1)}}
=-2\frac{a^3\e}{N\kappa^2}\Big[\ddot{\zeta}_{(1)}
+NH\lh3+2\e+2\hpa-\frac{\dot{N}}{HN^2}\rh\dot{\zeta}_{(1)}\Big] +2\frac{a\e N}{\kappa^2}\partial^2\zeta_{(1)}\nn\\
&&\qquad\ \!\!\!\!\!\!\!\!=-\frac{\d}{\d t}(2\frac{a^3}{N\kappa^2}\partial^2\lambda)+2\frac{a\e N}{\kappa^2}\partial^2\zeta_{(1)}=0,\label{eq}
\eea
where we remind the reader that $\partial^2\lambda=\e\dot{\zeta}_{(1)}$. 
Thus we have found the evolution equation for the first-order adiabatic perturbation. 
One can show that the first-order energy constraint, which outside the horizon reduces to 
\be
\dot{\zeta}_{(1)}=0,\label{mc}
\ee
is the first integral of the super-horizon part of (\ref{eq}), i.e.\  without the spatial gradient. 

For the conformal time coordinate choice $N=a$, the Lagrangian in terms of the variable $q$, defined in (\ref{newvar}), takes the form (\ref{Lnew}) (after an integration by parts). It now becomes apparent why we included the factor $\kappa/\sqrt{2}$ in (\ref{newvar}): it is necessary for the Lagrangian to attain the canonical form. 

The second-order tensor part of the action takes the form
\be
S_{2\gamma}=\int\d^4x\mathcal{L}_{2\gamma}=\frac{1}{2}\int\d^4x\Big\{\frac{a^3}{4N\kappa^2}(\dot{\gamma}_{ij(1)})^2
-\frac{aN}{4\kappa^2}(\partial_k\gamma_{ij(1)})^2\Big\},
\ee 
where $\mathcal{L}_{2\gamma}$ is the second-order Lagrangian for the tensor modes. To calculate this action there is no need for setting the time reparametrization (relevant to the scalar physical mode), since there is no interaction between tensor and scalar modes. Finally, the equation of motion for the gravitational waves takes the form
\be
\frac{\delta  L_{2\gamma}}{\delta\gamma_{ij(1)}}\equiv-\frac{1}{4}\frac{\d}{\d t}\lh\frac{a^3}{N}\dot{\gamma}_{ij(1)}\rh 
+\frac{aN}{4}\partial^2\gamma_{ij(1)}=0,
\ee
which agrees with equation (\ref{tenseq}).

\Section{Second-order perturbations}\label{S22}

Inflationary first-order perturbation predictions have been verified by observations, making inflation a part of the standard cosmological model despite the theoretical uncertainties related to its initial conditions (see discussion in chapter \ref{Ch1}). 

However, the exact model of inflation remains yet unknown. The observable quantities, i.e.\ the amplitude and the spectral index of the power spectrum, are in agreement with several different models of inflation. As far as single field inflation is concerned, they can be accommodated by different types of potentials and initial conditions for the scalar field. Furthermore, high energy theories of physics predict the existence of various different species of scalar fields during the early universe. This implies that more than one field could be the inflaton(s). In addition, these fields can live in a curved field space or have non-standard kinetic terms, remnants from extra-dimension space-time theories.  

Ideally, one would need to find new observables that could distinguish between these different models. Such a new observable, widely considered lately, is the departure of the inflationary perturbations from the Gaussian distribution. Such a non-Gaussianity would manifest itself through a non-zero odd-point correlation function of the perturbations. The lowest order odd correlation function to study is the three-point correlation function 
\be
\langle\zeta_\mathbf{\vc{k}_1}\zeta_\mathbf{\vc{k}_2}
\zeta_\mathbf{\vc{k}_3}\rangle\equiv (2\pi)^{-3/2}\delta^3\lh\vc{k}_\mathbf{1}+\mathbf{\vc{k}_2}
+\mathbf{\vc{k}_3}\rh B_\zeta(k_1,k_2,k_3).\label{bi}
\ee
The quantity $B_\zeta$ is called the bispectrum. 
The three-point correlation function is computed by definition on a triangle on the sky. The position and orientation of the triangle is irrelevant for the calculation, but its magnitude and shape can parametrize the information we gain. In Fourier space, the three-point correlation function involves three scale vectors $\vc{k_i}$ that are constrained by a $\delta$ function, again forming a triangle. Therefore, the bispectrum depends only on three parameters, that can be chosen among the magnitude of the three scale vectors and the three angles of the triangle.

In order to compute such a correlation function, one needs to move to second-order perturbation theory. In this section, we will present the second-order results for single-field inflation, study gauge invariance and summarize the computation of the bispectrum for scalar perturbations.

\subsection{Gauge invariance}\label{gau12}

It was not until 2003 that Malik and Wands in \cite{Malik:2003mv} defined the gauge-invariant quantity at second order that reduces to the 
curvature perturbation in the uniform energy-density gauge. In \cite{Malik:2005cy} the super-horizon equations of motion of 
these quantities were derived (but see also \cite{Noh:2003yg} for a gauge-ready formulation of the 
perturbations and their equations). 

Here, we start by computing the gauge-invariant curvature  perturbation at second order. This 
corresponds to the second-order curvature perturbation on hypersurfaces orthogonal to comoving wordlines, hence $\tilde{\varphi}_{(r)}=0$, where $(r)$ is the order of the perturbation. This gauge is not necessarily equivalent to the uniform energy density gauge $\tilde{\rho}_{(r)}=0$ beyond first order. Indeed, a space-time scalar transforms to second order as
\be  
%&&
%\tilde{\rho}_{(2)}=\hat{\rho}_{(2)}+\dot{\rho}T_{(2)}+T_{(1)}\lh
%2\dot{\hat{\rho}}_{(1)}+\dot{\rho}\dot{T}_{(1)}
%+\ddot{\rho}T_{(1)}\rh\nn\\&& 
\tilde{A}_{(2)}=A_{(2)}
+\dot{A}T_{(2)}+T_{(1)}\lh
2\dot{A}_{(1)}+\dot{A}\dot{T}_{(1)}
+\ddot{A}T_{(1)}\rh 
\ee 
and using (\ref{uetr}) we find for the second-order time shift assuming $\tilde{\rho}_{(2)}=0$ and $\tilde{\varphi}_{(2)}=0$ respectively
\bea
&&T_{(2)}
%=-\frac{\hat{\rho}_{(2)}}{\dot{\rho}}-T_{(1)}\frac{\dot{\hat{\rho}}_{(1)}}{\dot{\rho}}
=
-\frac{\rho_{(2)}}{\dot{\rho}}+T_{(1)}
\Big[
2\eta\lh\zeta_{(1)}-\alpha_{(1)}\rh
+\frac{1}{NH}\lh\dot{\zeta}_{(1)}-\dot{\alpha}_{(1)}\rh
\Big]\label{t21}\\ 
&&
T_{(2)}
%=-\frac{\varphi_{(2)}}{\dot{\phi}}-T_{(1)}\frac{\dot{\varphi}_{(1)}}{\dot{\phi}}
=
-\frac{\varphi_{(2)}}{\dot{\phi}}+T_{(1)}
\Big[
(\e+\eta)\lh\zeta_{(1)}-\alpha_{(1)}\rh
+\frac{1}{NH}\lh\dot{\zeta}_{(1)}-\dot{\alpha}_{(1)}\rh
\Big]
,\label{t2}
\eea
with $T_{(1)}=(\zeta_{(1)}-\alpha_{(1)})/(HN)$. 
For the above calculation we have taken into account that $\beta_{i(1)}=0$. In order to check the equivalence or not of the two gauge choices we will use the single-field limit of the second-order $0i$ Einstein equation (\ref{r2}) derived in appendix \ref{AppA}
\be 
-\e NH\Big[
-\frac{H}{\Pi}\varphi_{(2)}
+NH\frac{\rho_{(2)}}{\dot{\rho}}
+(\e-\hpa)\lh\zeta_{1(1)}
-\alpha_{(1)}\rh^2\Big]
+\partial^2\cA
=0,
\nn
\ee 
where $\partial^2\cA$ denotes spatial gradients involving $\psi$, vanishing outside the horizon (for the exact expression, see (\ref{r2}) in appendix \ref{AppA}). 
During slow-roll inflation both the first part of this equation, multiplying $\e$, and the gradient $\partial^2\cA$ are small. However the gradient term is exponentially suppressed as $a^{-2}$, while $\e$ remains only constant at a small value. 
%  This equation raises questions about how to use the slow-roll approximation along with the super-horizon approximation. It implies that within the slow-roll approximation, when $\e\ll 1$ the gradient terms and the terms in the brackets can be of the same order.
Indeed, if one was to start from a long-wavelength action the gradient term would not appear at all, as we comment in section \ref{long}.\footnote{Notice that in the flat gauge there is an overall factor $\e$ in $\partial^2\cA$ 
%is no such discrepancy 
since  $\partial_i\psi=\e\dot{\zeta}$ and we do not even need to compare the terms.} 
Hence, the two gauges agree only in the super-horizon limit, where
\be
-\frac{H}{\Pi}\varphi_{(2)}=-NH\frac{\rho_{(2)}}{\dot{\rho}}+(\eta-\e)(\zeta_{(1)}-\alpha_{(1)})^2, 
\ee
where we remind that $\Pi=\dphi/N$. 
Indeed using the above expression in (\ref{t2}), one recovers (\ref{t21}) and the two time shifts become identical. Since we are interested mainly in super-horizon results, we will use the term adiabatic perturbation for $\zeta$ even at second order.

As for the first-order study, we will focus on gauges with $F_{i(2)}\!=\!0\ \mathrm{and}\ F_{(2)}\!=\!0$. Calculating the Lie derivatives and rewriting the first order part of the second order contributions to $\tilde{g}_{ij(2)}$ in terms of the arbitrary gauge, we find

\bea
\tilde{\alpha}_{(2)}\delta_{ij}+\tilde{\gamma}_{ij(2)}
&=&\alpha_{(2)}\delta_{ij}+\gamma_{ij(2)}+NHT_{(2)}
\delta_{ij}
+\partial_{(j}\beta_{i)(2)}
+\partial_{(i}T_{(1)}\partial_{j)}\lh 2\psi_{(1)}-\frac{N^2}{a^2}T_{(1)}\rh\nn\\
&& 
+T_{(1)}\Big[\lh\dot{\alpha}_{(1)}+\dot{\zeta}_{(1)}
\rh\delta_{ij}
  +2\dot{\gamma}_{ij(1)}\Big], 
\eea
where we also used that $N_{i\perp(1)}=0$. The intermediate steps and the result for an arbitrary  $\beta_{i(1)}$ can be found in appendix \ref{AppA}.

%\bea
%&&
%\tilde{\alpha}_{(2)}\delta_{ij}+\tilde{\chi}_{ij(2)}+2\beta_{(i,k)}\beta_{(k,j)}+2\chi_{ik}\beta_{(k,j)} 
%+2\chi_{kj}\beta_{(k,i)}=\alpha_{(2)}\delta_{ij}+\chi_{ij(2)}+NHT_2\delta_{ij}+\beta_{2(i,j)}
%\nn\\
%&&+T_1\Big[\frac{\d}{\d t}\lh T_1NH\rh \Big]\delta_{ij} 
%+NH\beta_k\partial_kT_1\delta_{ij}+T_1\dot{\beta}_{(i,j)}
%+\partial_{(j}T_1\dot{\beta}_{i)}-\frac{N^2}{a^2}\partial_iT_1\partial_jT_1\nn\\
%&&
%+\beta_k\partial_k\partial_{(i}\beta_{j)}
%+\partial_{i}\beta_k\partial_{j}\beta_k
%+\partial_{(i}\beta_k\partial_k\beta_{j)}\nn\\
%&&+2T_1\lh \dot{\alpha}\delta_{ij}+\dot{\chi}_{ij}\rh  
%+2\beta_k\partial_k\lh \alpha\delta_{ij}+\chi_{ij}\rh +2\partial_{(i}T_1N_{j)}+4\partial_{(i}\beta_k\chi_{kj)} 
%\eea

Taking the trace of the above equation we find
%\bea
%\tilde{\alpha}_{(2)}=\alpha_{(2)}+T_2NH+\frac{1}{3}\partial^2\beta_{2}+T_1\lh \dot{\alpha}+\dot{\zeta}
%%+\frac{1}{3}\partial^2\dot{\beta}
%\rh+\beta_k\partial_k\lh \alpha+\zeta\rh 
%+\frac{1}{3}\partial_iT_1\lh -\frac{N^2}{a^2}\partial_iT_1+2N_i+\dot{\beta}_i\rh
%%+\frac{1}{3}\beta_k\partial_k\partial^2\beta
%\eea
\bea
\tilde{\alpha}_{(2)}=\alpha_{(2)}+T_{(2)}NH+\frac{1}{3}\partial^2\beta_{(2)}+T_{(1)}\lh \dot{\alpha}_{(1)}+\dot{\zeta}_{(1)}\rh 
-\frac{N^2}{3a^2}\lh\partial T_{(1)}\rh^2 
+\frac{2}{3}\partial_iT_{(1)}\partial_i\psi_{(1)}
,
\eea
while acting with $D_{ij}$, we find the form of $\partial^2\beta_{(2)}$:
\bea 
\frac{1}{3}\partial^2\beta_{(2)}\!\!\!\!&\equiv&\!\!\!\!
\frac{1}{3}\partial_{i}T_{(1)}\partial_{i}\psi_{(1)}
-\partial^{-2}\partial^i\partial^j\lh \partial_{i}T_{(1)}\partial_{j}\psi_{(1)} \rh 
-\frac{N^2}{6a^2}\lh \partial T_{(1)}\rh^2+\frac{N^2}{2a^2}
\partial^{-2}\partial^i\partial^j\lh \partial_iT_{(1)}\partial_jT_{(1)} \rh\nn\\ 
&&\!\!\!\!-
%\frac{1}{2}
\partial^{-2}\lh\dot{\gamma}_{ij(1)}\partial^i
\partial^j 
T_{(1)}\rh. 
\eea 
Putting everything together we find the second-order gauge transformation for the adiabatic perturbation
%\bea
%\frac{1}{2}\zeta_{(2)}&\equiv& \frac{1}{2}\tilde{\alpha}_{(2)}\nn\\
%&=&\frac{1}{2}\alpha_{2}+
%\frac{1}{2}T_2NH
%+\frac{1}{2}T_1\lh\dot{\alpha}+\dot{\zeta})\rh 
%+\frac{1}{2}\Bigg[\partial_{i}T_1\partial_{i}\psi
%-\partial^{-2}\partial^i\partial^j\lh \partial_{(i}T_1\partial_{j)}\psi \rh \Bigg]
%\nn\\
%&& 
%-\frac{N^2}{4a^2}\Bigg[\lh \partial T_1\rh^2-\partial^{-2}\partial^i\partial^j\lh \partial_iT_1\partial_jT_1 \rh \Bigg]
%-\frac{1}{2}\partial^{-2}\lh\dot{\gamma}_{ij}\partial_i
%\partial_jT_1\rh. 
%\eea

\bea
\frac{1}{2}\zeta_{(2)}\!\!\!\!
&\equiv&\!\!\!\!\frac{1}{2}\tilde{\alpha}_{(2)}\\
&=&\!\!\!\!\frac{1}{2}\alpha_{(2)}
+\frac{1}{2}Q_{(2)}
+\frac{\e+\eta}{2}\lh\zeta_{(1)}-\alpha_{(1)}\rh^2
\!+\!\dot{\zeta}_{(1)}T_{(1)}
-\frac{N^2}{4a^2}\Bigg[\!\lh \partial_i T_{(1)}\rh^2-\partial^{-2}\partial^i\partial^j\lh \partial_iT_{(1)}\partial_jT_{(1)} \rh\! \Bigg]
\nn\\
&&\!\!\!\!
+\frac{1}{2}\Bigg[\partial_{i}T_{(1)}\partial^{i}\psi_{(1)}
-\partial^{-2}\partial^i\partial^j\lh \partial_{i}T_{(1)}\partial_{j}\psi_{(1)} \rh \Bigg]
-\frac{1}{2}\partial^{-2}\lh\dot{\gamma}_{ij(1)}\partial^i
\partial^jT_{(1)}\rh
-\frac{1}{2\e NH}\partial^2\cA\nn
,\label{ga20}
\eea
where the exact expression for $\cA$ can be found in appendix \ref{AppA} and we have introduced the short notation 
\be
Q_{(r)}=-\frac{H}{\Pi}\varphi_{(r)}.
\label{Qs}
\ee
This second-order gauge transformation was computed in our paper \cite{Tzavara:2011hn}. The original computation was done for two fields, but we present here the single-field limit.

If one is to identify the arbitrary gauge with the flat gauge $\hat{\alpha}_{(r)}=0$, the curvature perturbation becomes using (\ref{ee00})
\bea
\frac{1}{2}\zeta_{(2)}\!\!\!\!&\equiv&\!\!\!\!
\frac{1}{2}\hat{Q}_{(2)}
+\frac{\e+\eta}{2}\zeta_{(1)}^2
+\dot{\zeta}_{(1)}\zeta_{(1)}
+\frac{1}{2}\Bigg[\partial_{i}\zeta_{(1)}\partial^{i}\lambda
-\partial^{-2}\partial^i\partial^j\lh \partial_{i}\zeta_{(1)}\partial_{j}\lambda \rh \Bigg]
-\frac{1}{2\e NH}\partial^2\hat{\cA}
\nn\\
&&\!\!\!\!\!\!\!\!
-\frac{N^2}{4a^2}\Bigg[\lh \partial_i \zeta_{(1)}\rh^2
-\partial^{-2}\partial^i\partial^j\lh \partial_i\zeta_{(1)}\partial_j\zeta_{(1)} \rh \Bigg]
-\frac{1}{2}\partial^{-2}\lh\gamma_{ij(1)}
\partial^i\partial^j\zeta_{(1)}\rh,
\label{ga2} 
\eea
where we used (\ref{ee00}) for $\psi_{(1)}$ and the fact that $T_{(1)}=\zeta_{(1)}/(NH)$ in the particular gauge. In the next subsection we will relate this gauge transformation to the action governing the cubic interactions of the adiabatic perturbations.

\subsection{The  action}\label{s21action}

An alternative way to calculate the second-order gauge-invariant perturbation and reconsider its  meaning, is to compute the third-order 
action for the adiabatic perturbation. 
Maldacena \cite{Maldacena:2002vr} was the first to perform that calculation for a single field, in the uniform 
energy-density gauge. 
In this way he managed to find the cubic interaction terms due to non-linearities of the Einstein action as well as 
of the field potential, which among 
other consequences change the ground state of the adiabatic perturbation $\zeta_{1(1)}$. 
This change can be quantified through a redefinition that on super-horizon scales takes the form \cite{Maldacena:2002vr}
\be
\zeta_{1(1)}=\zeta_{1c(1)}+\frac{\e+\hpa}{2}\zeta_{1(1)}^2,
\ee
where $\zeta_{1c}$ is the redefined perturbation. One sees that the correction term of the redefinition coincides with the 
surviving gauge-invariant quadratic term of the transformation (\ref{ga20}), taking into account 
that the super-horizon adiabatic perturbation is constant.
In \cite{Maldacena:2002vr} the curvature perturbation was considered a first-order quantity, 
while the second-order curvature perturbation was not taken into account, since its contribution in the uniform 
energy-density gauge 
is trivial: it introduces a redefinition of the form $\zeta_{1(1)}+\zeta_{1(2)}/2=\zeta_{1c(1)}$ 
(for proof, see the end of this subsection).

In order to compute the third order action and find the cubic interactions that contribute in the bispectrum, we shall work in the flat gauge $\hat{\alpha}_{(r)}=0$. The way to do that is similar to the procedure followed for the second order action. We need to perturb all fields in (\ref{actionexact1}) to second order. It turns out that the overall factor multiplying the second order energy and momentum constraint in the action is zero, so we need not calculate these. The scalar part of the action takes the form (for details on the calculation see appendix \ref{AppB})
\bea
\hat{S}_3\!\!\!\!&=&\!\!\!\!\hat{S}_{3c}+\hat{S}_{3(2)}\nn\\
&&
\!\!\!\!\!\!\!\!\!\!=\int\d^4x\Bigg\{\frac{a^3\e}{\kappa^2} \Bigg[
\frac{\e}{N}\lh1-\frac{\e}{2}\rh\zeta_{(1)}\dot{\zeta}_{(1)}^2
-\frac{2}{N}\dot{\zeta}_{(1)}\partial^i\lambda\partial_i\zeta_{(1)}+
\Big(2\e^2+3\e\eta-\!\eta^2+\gx\Big)H\zeta_{(1)}^2\dot{\zeta}_{(1)}\nn
\\&&\qquad
+\frac{1}{2N}\zeta_{(1)}\partial^i\partial^j\lambda
\partial_i\partial_j
\lambda\Bigg]
%\nn\\
%&&\!\!\!\!\!\!\!\!\!\!
+\frac{a\e^2N}{\kappa^2}\zeta_{(1)}(\partial\zeta_{(1)})^2
\Bigg\}
+\int\d^4x\frac{\delta \mathcal{L}_2}{\delta\zeta_{(1)}}\frac{\hat{Q}_{(2)}}{2}
\label{s31}
\eea
where $\partial^2\lambda=\e\dot{\zeta}_1$ and $\delta \mathcal{L}_2/\delta\zeta$ is the first-order equation of motion. We have split the action in a part $\hat{S}_{3c}$ that comes from the first-order scalar field $\hat{\varphi}_{(1)}$ and is gauge-invariant and a part $\hat{S}_{3(2)}$ coming from the second-order field $\hat{\varphi}_{(2)}$ that turns out to be proportional to the first-order equation of motion. 
One can use $\hat{S}_{3c}$ to easily calculate the non-Gaussianity related to the interaction terms of the action as is explained in detail 
in \cite{Seery:2005wm,Weinberg:2005vy}. This is related to what is known in the literature as $f_{NL}^{(3)}$, the parameter of non-Gaussianity 
related to the three-point correlation function of three first-order perturbations, which is only non-zero in the case of intrinsic 
non-Gaussianity.

The last term in expression (\ref{s31}) comes from the second-order field contribution of  $\hat{\varphi}_{(2)}$ and using (\ref{ga2}) it can be rewritten in terms of gauge invariant quantities
\be
 \frac{\delta \mathcal{L}_2}{\delta\zeta_{(1)}}\frac{\hat{Q}_{(2)}}{2}=
 \frac{\delta \mathcal{L}_2}{\delta\zeta_{(1)}}\lh\frac{\zeta_{(2)}}{2}-f\rh,
\ee
with $f$ corresponding to the quadratic terms of the gauge transformation (\ref{ga2}) 
\bea 
f&=&\frac{\e+\eta}{2}\zeta^2_{(1)}
+\dot{\zeta}_{(1)}\zeta_{(1)}-\frac{1}{4a^2H^2}(\partial\zeta_{(1)})^2
+\frac{1}{4a^2H^2}\partial^{-2}\partial^i\partial^j
(\partial_i\zeta_{(1)}\partial_j\zeta_{(1)})
+\frac{1}{2}\partial^i\zeta_{(1)}\partial_i\lambda\nn\\&&
-\frac{1}{2}\partial^{-2}\partial^i\partial^j
(\partial_{i}\lambda\partial_{j}\zeta_{(1)})
-\frac{1}{2\e NH}\partial^2\cA.
\label{redef1}
\eea
Therefore, $\hat{S}_{3(2)}$ is also gauge invariant. 
A way to think about this term, is that one needs to incorporate  all quadratic first-order terms found by the second-order gauge transformation (\ref{ga2}) directly as a correction to the first-order perturbation
\be
\zeta_{(1)}\rightarrow \zeta_{(1)}+\frac{1}{2}\zeta_{(2)}. 
\ee
The terms proportional to 
$\delta \mathcal{L}_2/\delta\zeta_{(1)}$ can be removed  by a redefinition of $\zeta_{(1)}$
\cite{Maldacena:2002vr} and lead to a change in the ground state of the perturbations. This works as follows.  
The cubic terms of the action (i.e.\  $S_{3c}$) are not affected by the redefinition, because the redefinition always 
involves second-order terms, which would give quartic and not cubic corrections. 
It is only the second-order 
terms (i.e.\  $S_2$) that change. 
Indeed one can show that under a redefinition of the form 
\be 
\zeta_{(1)}=\zeta_{c(1)}+A(\zeta_{c(1)}),\label{red1}
\ee 
the second-order action changes as 
$S_2=S_{2c}+(\delta \mathcal{L}_2/\delta\zeta_{(1)})A$. These new terms cancel out the relevant terms coming from the cubic action and one is left with $S_{3}(\zeta_c)=\hat{S}_{3c}(\zeta_c)$ (remember that the total action up to cubic order is the sum of the second and third-order action). The relevant redefinition in our case is 
\be 
\zeta_{(1)}+\frac{1}{2}\zeta_{(2)}=\zeta_{c(1)}+f(\zeta_{c(1)}),\label{red2}
\ee 
This redefinition also contributes to the non-Gaussianity and it is related to $f_{NL}^{(4)}$, the parameter of non-Gaussianity related to the three-point correlation function of a second-order perturbation (in terms of products of first-order ones) and two first-order perturbations, which reduces to products of two-point functions of the first-order perturbations.  

%Then the last term in the action can be rewritten as
%\be
%\frac{\delta \mathcal{L}_2}{\delta\zeta_{(1)}}\frac{\hat{Q}_{(2)}}{2}=-\frac{\delta \mathcal{L}_2}{\delta\zeta_{(1)}}
%f, 
%\ee 
The cubic action is manifestly gauge invariant. One can check that by computing the action in any other gauge, for example in the uniform-field gauge  $\tilde{\varphi}_{(r)}=0$, to verify that one recovers the same result. In the latter gauge the product of the $f$ term times the first-order field equation occurs already from the first-order field, while the second-order field contribution is trivial:
\be
\tilde{S}_3=\tilde{S}_{3c}+\tilde{S}_{3(2)},\quad\mathrm{with}\quad \tilde{S}_{3c}=\hat{S}_{3c}-\int\d^4x\frac{\delta \mathcal{L}_2}{\delta\zeta_{(1)}}f\quad\mathrm{and}\quad 
\tilde{S}_{3(2)}=\int\d^4x\frac{\delta \mathcal{L}_2}{\delta\zeta_{(1)}}\frac{\zeta_{(2)}}{2}. 
\ee
Therefore, it is of vital importance to incorporate in the calculation the second-order field contributions $S_{3(2)}$, otherwise the action is not gauge invariant as it was shown in our paper \cite{Tzavara:2011hn}.

\subsection{The bispectrum}

The aim of constructing the cubic action (\ref{s31}) for the adiabatic perturbation is to compute the three-point function for $\zeta$. This was done in \cite{Maldacena:2002vr}. Since in this thesis we are mostly interested in the super-horizon non-Gaussianity, we will not elaborate on how to compute the three-point correlation function related to the interaction terms of this action. We are going to summarize the result found in \cite{Maldacena:2002vr} and comment on the part induced by the redefinition of the field $(\ref{red2})$ that is relevant for the super-horizon calculation.    

For a redefined field of the type $\zeta=\zeta_c+A\zeta_c^2$ the three-point correlation function will contain two terms
\be 
\langle\zeta\zeta\zeta\rangle=
\langle\zeta_c\zeta_c\zeta_c\rangle+2A
\lh\langle\zeta\zeta\rangle\langle\zeta\zeta\rangle
+\mathrm{cyclic}\rh. \label{biredef}
\ee
The first term is computed by the interaction terms in the action, while the second one comes from the redefinition. By performing different redefinitions one can reshuffle the contributions between the two terms. In the literature the bispectrum defined in (\ref{bi}) is split in two terms
\be
B_\zeta\equiv B^{(3)}_\zeta +B^{(4)}_\zeta\equiv
-\frac{6}{5}\lh\f^{(3)}+\f^{(4)}\rh 
\lh 2\pi^2\mathcal{P}_{\zeta*}\rh^2\frac{\sum_ik_i^3}
{\prod_i(k_i^3)},\label{fnldef}
\ee
where the preferred redefinition is set by respecting the form of the action (\ref{s31}). The 
first part $B_\zeta^{(3)}$ is related to the interactions in this action and the second part  $B_\zeta^{(4)}$ is related to the redefinition (\ref{red2}). Here we also defined the parameter of non-Gaussianity $\f$ that is roughly the ratio of the three-point correlation function to the product of two two-point correlation functions. Writing this expression we have assumed that the three scales $k_i$ are of the same order and hence the power spectra in the product $\cP_{\zeta*}^2$  are evaluated at the same pivot scale $k_*$ (the general expression for the bispectrum, allowing for arbitrary amplitude of the scales, is given in (\ref{fNL_start})). Since $B_\zeta^{(4)}$ is computed by the redefinitions of the second part of the right-hand side of (\ref{biredef}), its only scale dependence will be in the form of a product of power spectra and hence $\f^{(4)}$ itself should be scale independent.

The bispectrum related to the interaction terms is calculated, within the assumption of slow-roll  at the time when the relevant scales cross the horizon, in \cite{Maldacena:2002vr} and $\f^{(3)}$ is found to be 
%\be
%B_\zeta^{(3)}\equiv-\frac{6}{5}\f^{(3)}
%\lh 2\pi^2\mathcal{P}_{\zeta*}\rh^2\frac{\sum_ik_i^3}
%{\prod_i(k_i^3)}
%%\lh \frac{2\pi^2}{k_1^3}\mathcal{P}_\zeta(k_1)\frac{2\pi^2}{k_2^3}\mathcal{P}_\zeta(k_2)+(k_2\leftrightarrow k_3)+(k_1\leftrightarrow k_3)\rh
%,
%\ee
%with 
\be
-\frac{6}{5}\f^{(3)}=\frac{\e_*}{2}\frac{1}{\sum_ik_i^3}
\Bigg[
\sum_{i\neq j}k_ik_j^2+8\frac{\sum_{i>j}k_i^2k_j^2}{k_1+k_2+k_3}-\sum_ik_i^3
\Bigg]
\ee
where again the three scales $k_1,k_2$ and $k_3$ are assumed to be of the same order, around some pivot scale $k_*$. 

The parameter $\f^{(4)}$ related to the redefinition 
is indeed scale independent and takes the form
\be
%B_\zeta^{(4)}\equiv-\frac{6}{5}\f^{(4)}
%\lh 2\pi^2\mathcal{P}_{\zeta*}\rh^2\frac{\sum_ik_i^3}
%{\prod_i(k_i^3)},\quad\mathrm{with}
%\quad
-\frac{6}{5}\f^{(4)}=\e_*+\eta_*,\label{fnl4s}
\ee
since outside the horizon only the first term in the redefinition (\ref{redef1}) survives. 
Hence, within slow-roll single-field inflation, the non-Gaussianity produced is $\mathcal{O}(\e)$ and thus  negligible; single-field inflation models produce almost Gaussian perturbations. This remains true unless
some non-trivial potential is used \cite{Chen:2006xjb} or higher
derivative contributions are introduced as for the Dirac-Born-Infeld
action
\cite{Alishahiha:2004eh,Silverstein:2003hf,Mizuno:2009cv,Mizuno:2010ag}
or K-inflation \cite{Chen:2006nt,Chen:2009bc}.

The second characteristic of single-field non-Gaussianity is related to the shape of the bispectrum. It turns out that the bispectrum peaks for triangles with $k_1=k_2\equiv k'\mg k_3\equiv k$. Such triangles are called squeezed triangles. In that case one finds
\be
B_\zeta
%=(2\e_{k'}+\eta_{k'})
=-\frac{n_\zeta(k')-1}{2}
\lh \frac{2\pi^2}{k^3}\mathcal{P}_{\zeta}(k)\rh 
\lh \frac{2\pi^2}{k'^3}\mathcal{P}_{\zeta}(k')\rh 
.
\ee 
%where the subscript $k'$ at the slow-roll parameters denotes evaluation at the time $t_{k'}$ when the large scales exit the horizon. \textbf{add maldacena argument} 
In this picture, the scale $k$ crosses the horizon much earlier than the scales $k'$. By the time $t_{k'}$,   the perturbation $\zeta_k$ is already constant and  %one can think of a 
%\be 
%a_n=a(1+\zeta_k)
%\ee
%so that the horizon becomes now 
%\be
%\dot{a}_n=HaN(1+\zeta_k)
%\ee
its only effect on the large scales is that it makes them cross the horizon earlier by a time shift $\delta t_{k'}=-\zeta_k/(NH)$. This time shift changes the amplitude of the perturbations that cross the horizon at a given time, since this depends on the time dependent slow-roll parameters and hence the appearance of the spectral index in the final expression.    
%\be
%\langle\zeta_\mathbf{k_1} \zeta_\mathbf{k_2}
%\zeta_\mathbf{k_3}
%\sim
%-\langle\rangle
%\ee

This type of bispectrum, proportional to a product of two power spectra, is called local and becomes maximum for squeezed triangles. It can be reproduced by a field $\zeta$ departing from Gaussianity as
\be
\zeta=\zeta_g-\frac{3}{5}\f\lh\zeta_g^2-\langle\zeta_g^2\rangle\rh,
\ee    
where $\zeta_g$ is a Gaussian field. 
The factor $-3/5$ is a remnant of the original definition of $\f$ in terms of the gravitational potential  $\Phi$ instead of $\zeta$  \cite{Komatsu:2001rj} 
\be 
\gF = \gF_\mathrm{L} + \f \lh \gF_\mathrm{L}^2 
- \langle \gF_\mathrm{L}^2 \rangle \rh. \label{phiz} 
\ee
During recombination (matter domination) the two are related by $\zeta=-(5/3)\Phi$. Moreover, when computing the ratio of the bispectrum to 
the three permutations of the power spectra squared using expression (\ref{phiz}) for $\gF$, one obtains $2\f$ due to the two ways the 
two $\gF_\mathrm{L}$ inside the second-order solution can be combined with the 
two linear solutions to create the power spectrum. 
Together these two effects
explain the factor $-6/5$ in front of $\f$ in its definition (\ref{fnldef}).

Summarizing, single-field slow-roll inflation models with standard kinetic terms predict local and slow-roll suppressed non-Gaussia\-ni\-ty. In order to discriminate them from other inflation models, one needs to examine the shape dependence and magnitude of $\f$ that these produce. For example, models with more than one inflaton produce a local bispectrum with amplitude that depends on the potential, while models with non-canonical kinetic terms are known to peak for equilateral triangles, with $\f$ that varies depending on the number of fields and their potential. The Planck satellite has constrained the local and equilateral  primordial non-Gaussianity to be $\f^\mathrm{local}=2.7\pm 5.8$ and $\f^\mathrm{equil}=-42\pm 75$ respectively.

%------------------------------ Chapter 3 --------------------------------------

{\Chapter{Two-field Inflation}\label{Ch3}}
%------------------------------------------------------------------------------------------

The simplest extension to the basic single-field model of inflation is to assume the existence of more than one inflaton. Indeed, if one is to associate the process of inflation with high-energy theories, one needs to consider the possible existence of more than one scalar field that can   have potentials flat enough to sustain inflation. This is due to the fact that any high energy extension of the Standard Model of physics, such as GUTs, SUSY and so on, contain several scalar fields that are candidates for the fields driving inflation.   

In this chapter we introduce the tools to study multiple-field inflation. The new feature in these models, is that each field added in the theory introduces an extra physical degree of freedom, which we expect to interact with the adiabatic perturbation studied in the previous chapter. In particular, there exists a combination of the field perturbations that is orthogonal to the energy density perturbation in the flat gauge, usually referred to as the entropic perturbation. This is also known as the isocurvature perturbation, i.e.\ the perturbation corresponding to a compensation of the individual components with each other in order to leave the curvature perturbation constant. The super-horizon interaction of the two perturbations is expected to produce distinctive non-Gaussian features and hence a non-zero bispectrum,  that can in principle distinguish observationally multiple-field models from other models in the literature. In order to understand the physical implications of the isocurvature perturbation, it is enough to introduce one extra field. We shall concentrate on the two-field paradigm, being the easiest to study, but the generalisation to more fields is straightforward although possibly much more cumbersome.  

We will only examine scalar fields with canonical kinetic terms and we will not consider higher-derivative field interactions. Such models motivated by extra-dimensional theories are known to have a completely different phenomenology with regard to inflationary perturbations and are well distinct from the models under study in this thesis (see \cite{Alishahiha:2004eh,Langlois:2008qf,Langlois:2008mn}). We are however dealing with this subject in a paper that is currently in its final stages of preparation, although we choose not to present its results here, since this thesis is dedicated to inflation models with standard kinetic terms which produce non-Gaussianity outside the horizon. Models with non-canonical kinetic terms are known to produce non-Gaussianity before horizon crossing, with different characteristics from those we are studying here.

Furthermore, we will only consider real fields, since complex fields can always be rewritten in terms of two real fields. We will combine the two scalar fields in a vector $\varphi^A$, where the index $A=1,2$ labels the fields. If one is to think of these scalar fields as the coordinates of a real manifold, the concept of a field metric $G_{AB}$ becomes relevant. A generic two-dimensional space is conformally flat, since it is a surface, and its Riemann tensor has only one independent component, the Ricci scalar. In this thesis we will assume that the field space is flat and hence the field metric is trivial (for a study in curved field space see e.g.\ \cite{GrootNibbelink:2001qt}). Although canonical two-field models with a trivial field metric are the simplest models to study, we expect that they still highlight the dominant effects of generic multiple-field inflation.  
%. In our paper  under preparation we  also study models with non-trivial field metrics 
However, we also study such models in our paper  under preparation.
% although we expect that it is the non-canonical kinetic terms that induce the main different features when compared to standard multiple-field models.

In this chapter we will first study the background of two-field models and its characteristics. Next, we will generalise the concept of gauge invariance for the perturbations of the two fields and present the tools for studying these perturbations up to second order (and eventually, in the next chapter, calculate their bispectrum), namely the action that governs the gauge invariant perturbations at all scales and the long-wavelength formalism relevant for their study in the super-horizon regime.

\Section{The two-field background}\label{S31}

The matter properties of a universe dominated by two scalar fields with a potential $W$ are determined from the Lagrangian
\be
\cL_m=\sqrt{-g}\lh-\frac{1}{2}g^{\mu\nu}\partial_\mu\varphi_A\partial_\nu
\varphi^A-W(\varphi^A)\rh,  
\ee
where $\varphi_A$ should be understood as
\be
\varphi_A=G_{AB}\varphi^B,\quad\mathrm{with}\quad
\mathbf{G}=\mathbf{I_2}
\ee
The stress-energy tensor (\ref{stress}) becomes  
\be
T^\mu_\nu
=\partial^\mu\varphi_A\partial_\nu\varphi^A-\delta^\mu_\nu\lh \frac{1}{2}g^{\kappa\lambda}\partial_\kappa\varphi_A\partial_\lambda
\varphi^A+W(\varphi^A)\rh.
\ee  

\begin{figure}[t]
\begin{tabular}{cc}
\includegraphics[scale=0.75]{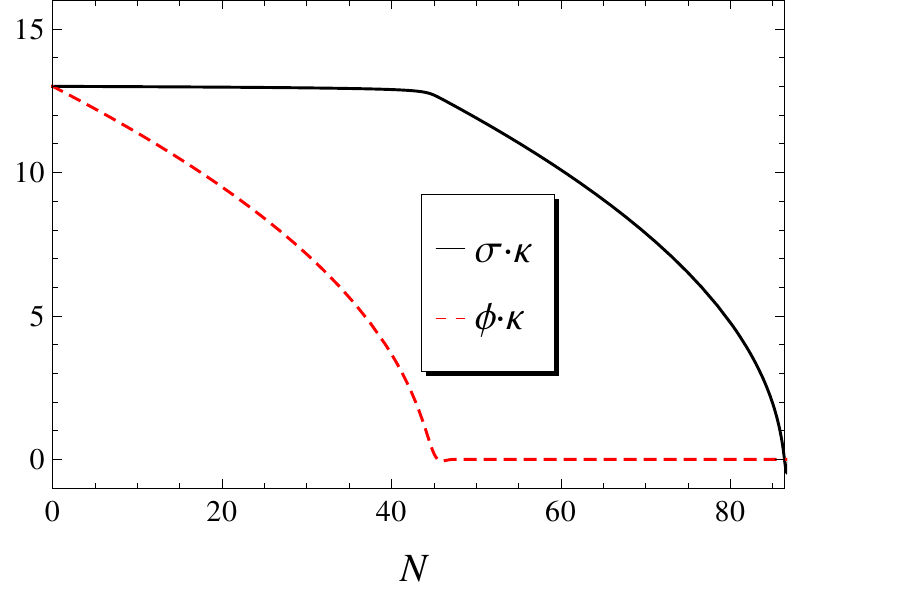}
&\includegraphics[scale=0.75]{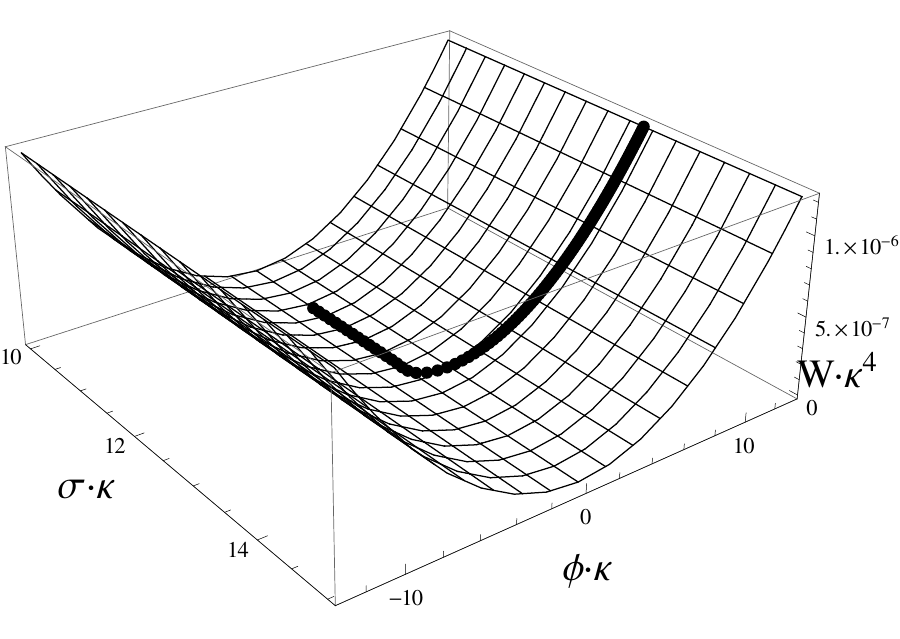}
\end{tabular}
\caption{The time evolution of the fields (left) and the field trajectory (right),
for the model (\ref{model}) with initial conditions 
$\phi_0=\gs_0=13\kappa^{-1}$, $m_\sigma=10^{-5}\kappa^{-1}$ and mass ratio $m_{\phi}/m_{\gs}=12$.}
\label{fig31}
\end{figure}

The background of such a theory is still described by the FLRW metric (\ref{frw}) and the Einstein and field equations take the form
\be
H^2\equiv\lh\frac{\dot{a}}{aN}\rh^2=\frac{\kappa}{3}\lh\frac{\Pi^2}{2}+W\rh,\qquad
\frac{\dot{H}}{N}=-\frac{\kappa^2}{2}\Pi^2,\qquad\dot{\Pi}^A+3NH\Pi^A+NW^{,A}=0,\label{eq2}  
\ee
where we denote $W^{,A_1\cdots A_n}\equiv\partial^n W/(\partial\phi^{A_1}\cdots\partial\phi^{A_n})$. 
The background canonical momentum becomes
\be
\Pi^A=\frac{\dphi^A}{N} 
\ee
while its norm can be found as
\be
\Pi^2\equiv\Pi_A\Pi^A.\label{norm} 
\ee

In order to visualise the inflationary picture described by these equations, we will consider the example of two free massive fields that obey the quadratic potential 
\be 
W_q=\frac{1}{2}m_\phi^2\phi^2+\frac{1}{2}m_\sigma^2\sigma^2. \label{model}
\ee
The only free parameter is the mass ratio of the fields, which we take to be $m_\phi/m_\sigma=12$, hence $\phi$ is a heavy field compared to $\sigma$. For our numerical calculations we use the initial conditions $\phi_0=\sigma_0=13\kappa^{-1}$ at $t=0$. In figure \ref{fig31} we plot the evolution of the fields as a function of the number of e-folds and the field trajectory around the time that the heavy field $\phi$ is approaching zero and oscillates. In the beginning of inflation $\phi$ dominates the expansion while rolling down its potential and about $40$ e-folds after the initial time $t=0$ it starts oscillating around the minimum of its potential. The heavier $\phi$  is, the more persistent are the damped oscillations. During these oscillations the light field $\sigma$ starts driving inflation and rolls down its potential until it also reaches its minimum and starts oscillating at about $85$ e-folds, when inflation ends.  

\begin{figure}[t]
\begin{center}
\includegraphics[scale=0.75]{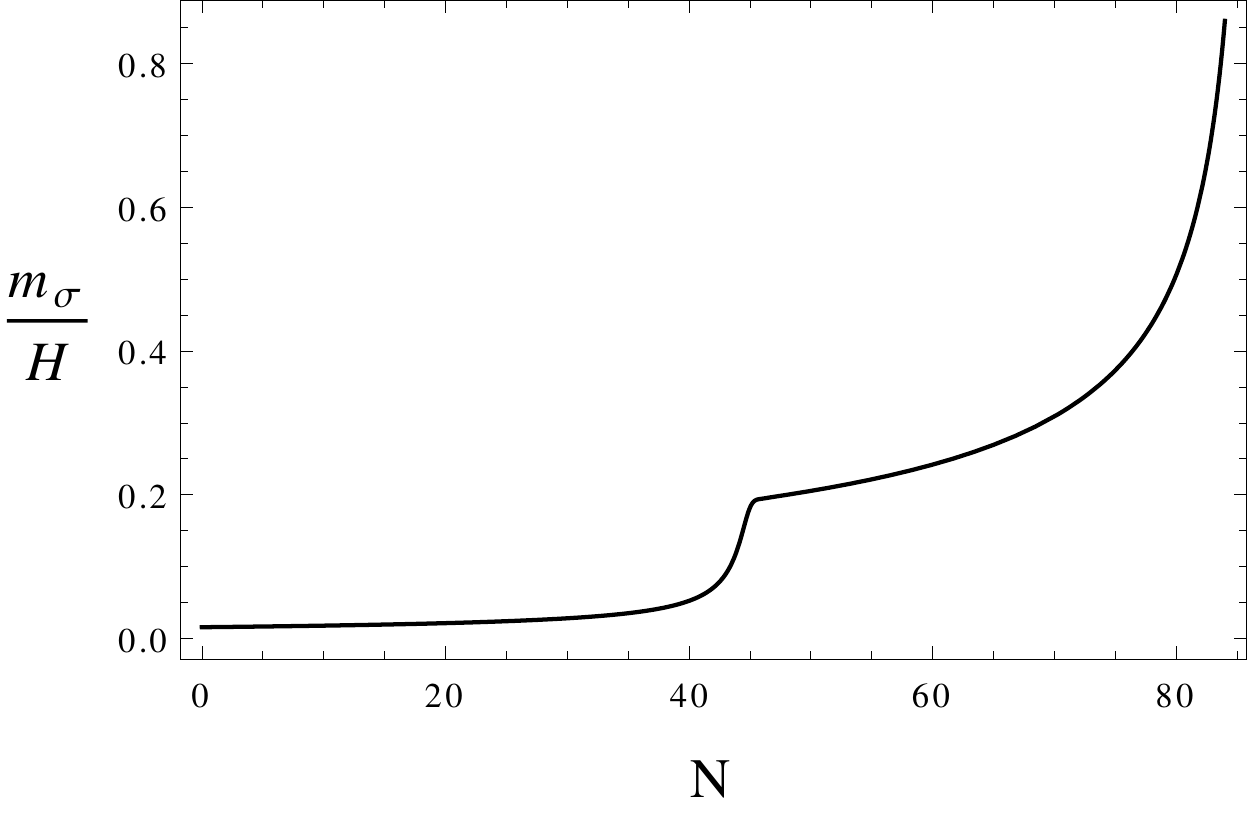}
\caption{The time evolution of the comoving Hubble length scale,
for the model (\ref{model}) with initial conditions 
$\phi_0=\gs_0=13\kappa^{-1}$ and mass ratio $m_{\phi}/m_{\gs}=12$.}
\label{fig32}
\end{center}
\end{figure}

In figure \ref{fig32} we plot the Hubble length scale as a function of the number of e-folds of inflation. There are clearly two different regimes regarding the rate of change of $1/H$. During the domination of the heavy field this remains almost constant, while during the second phase of inflation, dominated by the light field, initially the Hubble length scale increases slowly, while that increase becomes faster when $\sigma$ approaches the minimum of its potential.  

\subsection{The field-space basis}\label{basis}

While working on a two-field space one can introduce an orthonormal basis $\{e_{1A},e_{2A}\}$ that is induced by the dynamics of the system \cite{GrootNibbelink:2001qt,GrootNibbelink:2000vx}. It turns out that such a basis not only simplifies the equations, but also makes it possible to distinguish between effectively single-field effects ($e_{1A}$) and truly multiple-field effects ($e_{2A}$).  

We define the first unit vector $e_{1A}$ as parallel to the canonical momentum of the fields
\be
e_1^A\equiv\frac{\Pi^A}{\Pi}, 
\ee
while the second unit vector is defined as parallel to the component of the time derivative of the momentum orthogonal to the momentum itself
\be
e_2^A\equiv\frac{\dot{\Pi}^A-e_1^Ae_{1B}\dot{\Pi}^B}{|\dot{\Pi}^A-e_1^Ae_{1B}\dot{\Pi}^B|}. 
\ee
The two basis vectors satisfy the orthogonality relation and the completeness relation
\be
e_m^Ae_{nA}=\delta_{mn},\qquad\mathrm{and}\qquad
e_m^Ae_{mB}=\delta^A_B,\label{ortho}
\ee
where low-case indices $m,n=1,2$ denote components in terms of the basis vectors as opposed to capital indices $A,B=1,2$ that label the fields. Moreover, all repeated indices obey the Einstein summation convention. 

Finally, for the two field case under study the components of the unit vector $e_{2A}$ are completely determined in terms of $e_{1A}$ modulo a sign. In order to avoid sudden sign flips of the basis vectors,  and thus make the numerics easier to deal with during the oscillations of the fields, we impose the extra relation \cite{Tzavara:2010ge}
\be
\varepsilon_{A_1A_2}e_1^{A_1}e_2^{A_2}=-1,\label{ant}
\ee
were $\varepsilon_{AB}$ is the fully antisymmetric symbol. 
Indeed, using the orthogonality relation (\ref{ortho}) and (\ref{ant}) one finds
\be 
\mathbf{e_1}=\lh e_1^1,e_1^2\rh,\qquad\mathrm{and}\qquad
\mathbf{e_2}=\lh e_1^2,-e_1^1\rh\label{fbas}
\ee

\begin{figure}
\begin{center}
\includegraphics[scale=0.75]{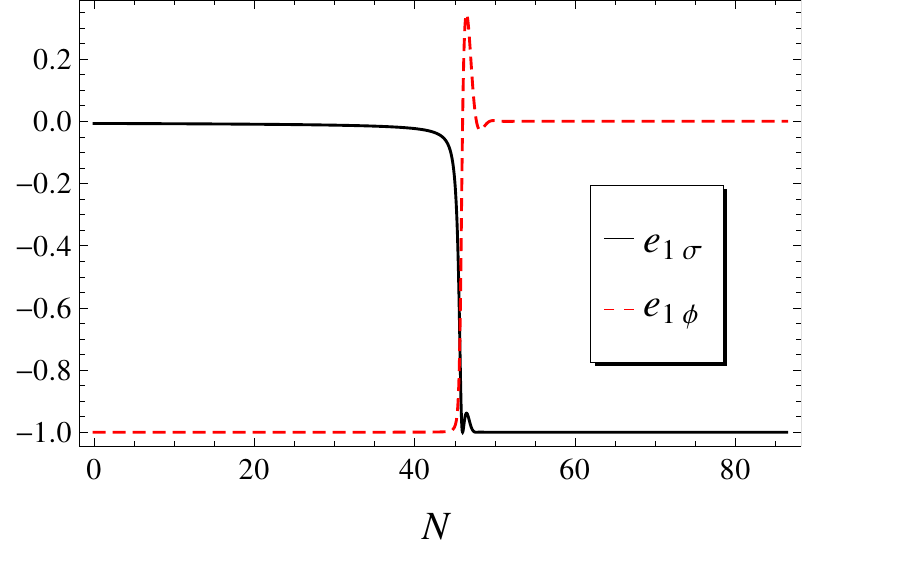}
\caption{The time evolution of the  components of the unit vector $\mathbf{e_1}$,
for the model (\ref{model}) with initial conditions 
$\phi_0=\gs_0=13\kappa^{-1}$ and mass ratio $m_{\phi}/m_{\gs}=12$.}
\label{fig33}
\end{center}
\end{figure}

In figure \ref{fig33} we plot the two components of the unit vector $\mathbf{e_1}$ for the model (\ref{model}). During the domination of the heavy field, inflation is effectively single-field with $e_{1\phi}\simeq1$ and $e_{2\sigma}\simeq0$, while this picture is inverted during the domination of the light field. It is during the turn of the trajectory of the fields, that both fields dominate inflation, and the oscillations of the heavy field are imprinted on the components of the unit vector $e_{1A}$. 

\subsection{The slow-roll parameters}

The concept of slow-roll can be generalised to the case of two-field inflation. The assumption (\ref{cond}) remains the same as for the single field case (\ref{defe})
\be
\e\equiv-\frac{\dot{H}}{NH^2}=\frac{\kappa^2\Pi^2}{2H^2}. 
\ee
The only difference is that now $\Pi$ is the norm of the momentum vector (\ref{norm}). From now on we shall use $\e$ to define the end of inflation, as the time during the domination of the light field that $\e=1$.  

In order to generalise assumption (\ref{eta}), that the time derivative of the momentum is much smaller than the momentum itself, as well as the whole hierarchy of slow-roll parameters defined in (\ref{defeta}), we will make use of the unit vectors defined in the previous subsection. We introduce the vector $\eta^{(n)A}$
\be
\eta^{(n)A}\equiv\frac{1}{H^{n-1}\Pi}\lh\frac{1}{N}\frac{\d}{\d t}\rh^{n-1}\Pi^A.\label{defetanew}
\ee
Note that $\eta^{(1)A}$ is simply $e_1^{A}$. We will use the short-hand notation $\eta^A$ for $\eta^{(2)A}$ and $\xi^A$ for $\eta^{(3)A}$. 
We shall decompose these vectors in components parallel (e.g.\ $\hpa,\xi^\parallel$) and perpendicular (e.g.\ $\hpe,\xi^\perp$) to the momentum as 
\cite{Rigopoulos:2005us} 
(see also \cite{Bernardeau:2002jy}).
\be
\hpa=\eta^Ae_{1A},\quad\xi^\parallel=\xi^Ae_{1A},\cdots,\qquad\hpe=\eta^Ae_{2A},\quad\xi^\perp=\xi^Ae_{2A},\cdots.
\ee
Using the field equation (\ref{fe2}) these can be rewritten as
\be
\hpa=-3-\frac{W_1}{H\Pi},\quad\hpe=-\frac{W_2}{H\Pi},\cdots,\qquad
\xi_m=-\frac{W_{1m}}{H^2}+3(\e-\hpa)\delta_{m1}-3\hpe\delta_{m2},\cdots, 
\label{srpara2}
\ee
where we have defined $W_{mn\cdots}=W^{,AB\cdots}e_{mA}e_{nB}\cdots$.  
In the above relation for $\xi_m$, $m=1$ and $m=2$ should be understood as $\xi^\parallel$ and $\xi^\perp$ respectively. At first order in a  slow-roll expansion, these relations imply that $\hpa=\e-W_{11}/(3H^2)$ and $\hpe=-W_{12}/(3H^2)$.   Notice that in the single field limit all the perpendicular  components are zero by definition (since they are in the $e_{2A}$ direction), while the parallel slow-roll parameters reduce to those defined in (\ref{defeta}).   

Generalising the single-field definition introduced in subsection \ref{sssr}, one can see that the slow-roll condition is related to the parallel components of the vector $\eta^{(n)A}$.  Indeed, the initial idea was that in order to ensure slow-roll one needs to demand $\Pi^2\ll W$ (in other words that the energy density is almost equal to the pressure density of the fields, corresponding to $\e\ll 1$), and in order to preserve this, demand that  $\dot{\Pi}\ll\Pi$. But $\Pi$ is simply $e_{1A}\Pi^A$ and hence the relevant direction for slow-roll is the parallel direction, i.e.\  the component of the momentum parallel to the field trajectory. Strictly speaking, a period during inflation when 
$\hpe$ acquires large values, while $\hpa\ll 1$, is considered  a slow-roll period. We expect that perpendicular components can only become large during the turning of the field trajectory, i.e.\ during the transition time when the fields exchange the domination of the universe. However, in all our calculations hereafter, when we assume slow-roll in order to simplify expressions, we will be taking all slow-roll parameters (parallel and perpendicular) to be small.  Notice though that regarding the calculation of the bispectrum, the slow-roll approximation during inflation will only be used for finding analytical estimates, while for numerical results we will be using the exact, non slow-roll,  expressions. It is only at horizon-crossing that we will be imposing the slow-roll assumption, by demanding that inflation is effectively dominated by one field (as in the example of the quadratic potential) and hence the perpendicular slow-roll parameters will also be small. 

%The system described by (\ref{eq2}) is said to be in the slow-roll regime if
%\be
%\e\ll1\qquad\mathrm{and}\qquad 
%\eta^{(n)A}\ll1,\quad\mathrm{for}\quad n>1.
%\ee

Finally we also define a short-hand notation for the second derivative of the potential projected in the $22$ direction
\be
\chi\equiv\frac{W_{22}}{3H^2}+\e+\hpa. 
\ee
Although we will be referring to it as a slow-roll parameter, $\chi$ can be large during slow-roll inflation (see the discussion below). 

\begin{figure}
\begin{center}
\includegraphics[scale=0.75]{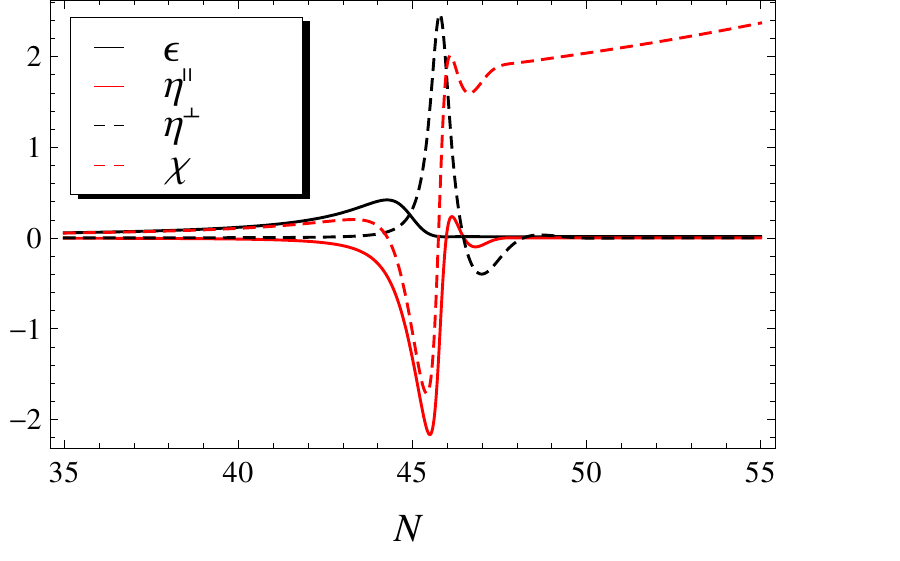}
\caption{The time evolution of the first-order slow-roll parameters and $\chi$ around the turn of the field trajectory, 
for the model (\ref{model}) with initial conditions 
$\phi_0=\gs_0=13\kappa^{-1}$ and mass ratio $m_{\phi}/m_{\gs}=12$.}
\label{fig34}
\end{center}
\end{figure}

In figure \ref{fig34} we plot the first-order slow-roll parameters as a function of the number of e-folds. Notice that solving equations (\ref{eq2}) we have not made any slow-roll assumption and hence the slow-roll parameters should be thought of simply as short-hand notation. The slow-roll parameters remain indeed small during the effectively single-field domination of the heavy or the light field. During the oscillations of the heavy field the slow-roll parameters oscillate too. Although $\e<1$ during the turning of the fields for the specific mass ratio that we use, both $\hpa$ and $\hpe$ vary to values larger than $1$ and hence slow-roll is not a good assumption at this time. 

The behaviour of $\chi$ merits some more discussion. $\chi$ represents effectively the second derivative of the potential in the $22$ direction (see the right-hand side of figure \ref{fig31}). Before the turning of the fields the trajectory goes down the potential in the relatively steep $\phi$ direction, which means that $W_{22}$ then corresponds with the relatively shallow curvature in the direction of the light field $\sigma$ and hence $\chi$ is small. After the turning of the fields the trajectory goes along the bottom of the valley in the $\sigma$ direction and $W_{22}$ corresponds with the large curvature of the potential in the perpendicular direction, leading to large values of $\chi$ (for a model where $\chi$ remains small during all inflation, see for example \cite{Achucarro:2012fd}).

We conclude by deriving a number of results for the derivatives of the slow-roll parameters and the unit vectors that can be computed from their definitions
\bea
&&\dot{\ge} = 2 NH \ge ( \ge + \get^\parallel ),\quad
\dot{\get}^\parallel\!= NH\!\lh\gx^\parallel \!+ (\get^\perp)^2 + (\ge - \get^\parallel) \get^\parallel\rh, \quad
\dot{\get}^\perp\! =NH\!\lh \gx^\perp \!+ (\ge - 2\get^\parallel) \get^\perp\rh,\nn\\
&&\dot{\gc} =NH\lh \ge \get^\parallel + 2 \ge \gc - (\get^\parallel)^2
+ 3 (\get^\perp)^2 + \gx^\parallel + \frac{2}{3} \get^\perp \gx^\perp
+ \frac{\sqrt{2\ge}}{\gk}\frac{W_{221}}{3H^2}\rh,\nn
\\
&&\dot{\xi}^\parallel=NH\lh -\frac{\sqrt{2\ge}}{\gk}\frac{W_{111}}{H^2}+2\hpe\gx^\perp+2\e\gx^\parallel-3\gx^\parallel+9\e\hpa
+3(\hpe)^2+3(\hpa)^2\rh\nn\\
&&\quad\ =NH\lh\eta^{(4)\parallel}+2\e\xi^\parallel
-\hpa\xi^\parallel+\hpe\xi^\perp\rh\nn 
\eea
\bea
&&\!\!\!\!\!\!\!\!\!\!\!\!\!\!\!\!\!\!\!\!\!\!\!\!\!\!\!\!\!\!\!\!\!\!\!\!\!\!\!\!\!\!\!\!\!\!\!\!\!\!\!\!\!\!\!\!\!\!\!\!\!\!\!\dot{\xi}^\perp=NH\lh -\frac{\sqrt{2\ge}}{\gk}\frac{W_{211}}{H^2}-\hpe\gx^\parallel+2\e\gx^\perp-3\gx^\perp
+9\e\hpe+6\hpe\hpa-3\hpe\chi\rh\nn\\
&&\!\!\!\!\!\!\!\!\!\!\!\!\!\!\!\!\!\!\!\!\!\!\!\!\!\!\!\!\!\!\!\!\!\!\!\!\!\!\!\!\!\!\!\!\!\!\!\!\!\!\!\!\!\!\!\!\!\!\!\!\!\!\!\quad\ =NH\lh\eta^{(4)\perp}+2\e\xi^\perp
-\hpa\xi^\perp-\hpe\xi^\parallel\rh
,\nn\\
&&\!\!\!\!\!\!\!\!\!\!\!\!\!\!\!\!\!\!\!\!\!\!\!\!\!\!\!\!\!\!\!\!\!\!\!\!\!\!\!\!\!\!\!\!\!\!\!\!\!\!\!\!\!\!\!\!\!\!\!\!\!\!\!\dot{e}_1^A= NH\hpe e_2^A,\qquad\qquad \dot{e}_2^A=- NH\hpe e_1^A.\label{dere}
\eea
As in the single-field case the derivative of a slow-roll parameter is one order higher in slow-roll. Notice that the derivatives of the parallel slow-roll parameters $\hpa$ and $\xi^\parallel$ involve perpendicular slow-roll parameters. 
% of one order lower. 
This implies that during a slow-roll period with large perpendicular slow-roll parameters, these  derivatives can become large ruining the slow-roll assumption (which as discussed above concerns only the parallel components). However this is only artificial and it is due to our  definition (\ref{defetanew}).  
In \cite{Achucarro:2010da} an alternative  definition was introduced
\be
\eta^{(n)}_m\equiv\frac{1}{H^{n-1}\Pi}\lh\frac{1}{N}\frac{\d}{\d t}\rh^{n-2}\lh e_{mA}\frac{\dot{\Pi}^A}{N}\rh,  
\ee
(only valid for $m=1,2$), which avoids this confusion, in the sense that 
time derivatives of parallel slow-roll parameters no longer contain 
perpendicular components. Although we believe that this new definition 
is an improvement for the physical interpretation of results in the case 
where $\hpa$ remains small and $\eta^\perp$ becomes large, in this thesis 
we keep the definition we used in our original papers, and we will 
always consider the situation where both $\hpa$ and $\eta^\perp$ are either 
small or large at the same time. Note that this definition does not 
change $\hpa$ and $\eta^\perp$, only $\xi$ and higher order slow-roll parameters.

\Section{First-order perturbations}\label{S32}

Given that we now consider a universe filled with two scalar fields, we encounter two types of  perturbations $\varphi_{(r)}^A,\ A=1,2$, one for each field ($(r)$ denotes the order of the perturbation). We expect that a certain combination of the scalar-field perturbations should be equivalent to the adiabatic perturbation introduced for the single-field case. Furthermore, we anticipate the existence of a new type of gauge-invariant perturbation related to the orthogonal combination to the adiabatic one.  It is the interaction of these two scalar modes during the turning of the field trajectory that can give rise to non-linearities, characteristic of multiple-field domination.

In order to study perturbations on the background described in the previous section, one has to extend the set-up described in chapter \ref{Ch2} to include a second scalar field. The gravitational part of the theory remains the same, i.e.\ the gravitational part of the action (\ref{actionexact1}) and hence the left-hand side of the Einstein equations (\ref{stress}). What changes is the field content, so that the action within the ADM formalism that determines this physical system is
\bea
&&S=\frac{1}{2}\int\d^4x\sqrt{\bh}\bN\lh \frac{R^{(3)}}{\kappa^{2}}-2W-\bh^{ij}\partial_i\varphi^A
\partial_j\varphi_A\rh\nn\\
&&\quad 
+\frac{1}{2}\int\d^4x\frac{\sqrt{\bh}}{\bN}\Big[\frac{1}{\kappa^{2}}\lh
\bE_{ij}\bE^{ij}-\bE^2\rh 
+\lh \dot{\varphi}^A-N^i\partial_i\varphi^A\rh
\lh \dot{\varphi}_A-N^i\partial_i\varphi_A\rh\Big]. 
\label{actionexact}
\eea
The fully non-linear canonical momentum becomes
\be 
\bar{\Pi}^A=\frac{\dot{\varphi}^A-N^i\partial_i\varphi^A}{\bN},
\ee
while the energy and momentum constraints take the form
\bea
&&\kappa^{-2}R^{(3)}-2W-\kappa^{-2}\bN^{-2}(E_{ij}E^{ij}-E^2)-\bar{\Pi}^2-h^{ij}\partial_i\varphi_A\partial_j\varphi^A=0,\label{en}\\
&&\nabla_j\Big[\frac{1}{\bN}(E_i^j-E\delta_i^j)\Big]=\kappa^2\bar{\Pi}^A
\partial_i\varphi_A.\label{mo}
\eea
We will not study vector perturbations, since we have shown that vector perturbations decay during inflation. 
From now on we shall focus on scalar perturbations, although in appendix \ref{AppB} we also calculate the tensor part of the perturbed action governing the interactions of the perturbations up to cubic order.   

\subsection{Gauge Invariance}

For this section and the next one, we will work within  gauges with $F=0$ and $F_i=0$, resulting in  $N_{i\perp}=0$ at least up to second order. As explained in subsection \ref{ga1} there exist one scalar, $\alpha$, and two tensor, $\gamma_{ij}$, degrees of freedom for gravity. To these, one has to add the two scalar-field degrees of freedom. After setting the remaining time parametrization, one is left with two scalar degrees of freedom that can be combined to give the adiabatic perturbation and the new gauge-invariant perturbation that we call the isocurvature perturbation. 
%
% However, as in the case of single-field inflation, we are interested in constructing gauge-invariant quantities that permit an unambiguous interpretation of our results. 

\subsubsection{First-order gauge-invariant perturbations}\label{fogip}

As in the case of single-field inflation, we want to find the first-order gauge-invariant quantity that reduces to the curvature perturbation $\tilde{\alpha}_{(1)}$ in the uniform energy density gauge. This was already done in subsection \ref{ga1} for the single-field case and we found the result (\ref{defz1r}) for the linear adiabatic perturbation in terms of the scalar-field perturbation. At the level of the energy density perturbation, the result remains the same. This however is not true for the field perturbation. In order to find the new relation between the energy density perturbation and the field perturbation, we will make use of the first-order momentum constraint (\ref{mo}) or equivalently the $0i$ Einstein equation, which turns out to be
\be
N_{(1)}-\frac{\dot{\alpha}_{(1)}}{H}+\e NQ_{1(1)}=0,
\ee
where generalising the single-field definition (\ref{Qs}) we defined
\be
Q_{m(r)}\equiv
-\frac{H}{\Pi}e_{mA}\varphi^A_{(r)}. 
\ee
Perturbing equations $NH=\dot{a}/a$ and $H^2=(\kappa^2/3)\rho$ to first order we find that the gauge choice $\tilde{\rho}_{(1)}=0$ is equivalent to demanding $e_{1A}\tilde{\varphi}^A_{(1)}=0$ and we find for the adiabatic perturbation (see also \cite{Malik:2003mv,Rigopoulos:2003ak,
Langlois:2005qp})
\be
\zeta_{1(1)}=\alpha_{(1)}-\frac{H}{\Pi}e_{1A}\varphi^A_{(1)}.
\ee
Notice that in order to distinguish the adiabatic perturbation from the second type of perturbation that we are going to define, we will use the subscript $1$,  remnant of the $e_{1A}$ unit vector involved in its definition.

If one was to start from the field transformation
\be
\tilde{\varphi}_{(1)}^A= \varphi_{(1)}^A+\dphi^AT_{(1)},\label{trf1}
\ee
one would recover the above result defining the tilded gauge as $e_{1A}\tilde{\varphi}^A_{(1)}=0$ with $T_{(1)}=Q_{1(1)}/N$. 
Furthermore, multiplying (\ref{trf1}) with $e_{2A}$ the second term in the right-hand side of (\ref{trf1}) vanishes, and one finds a new gauge invariant quantity
\be
\zeta_{2(1)}=-\frac{H}{\Pi}e_{2A}\varphi^A_{(1)}. 
\ee
We identify this quantity as the isocurvature or entropy gauge-invariant perturbation at linear order. It can be defined as the perturbation orthogonal to the adiabatic one, while working in the flat gauge $\hat{\alpha}_{(r)}$, i.e.\ it is independent of the energy density perturbation. 
%It determines the distribution of energy density between the different components of the field 
As promised, the basis we constructed in subsection  \ref{basis} discriminates between effectively single-field effects, here the adiabatic perturbation $\zeta_1$, and purely multiple-field effects as the isocurvature perturbation $\zeta_2$.

Usually the isocurvature perturbation is described in terms of the pressure perturbation of the content of the universe in the uniform-energy density gauge, as for example in \cite{Langlois:2005qp}. Here we choose to characterize it in terms of the fields themselves and the vector $e_{2A}$ (an alternative approach, generalizing for non-canonical scalar fields, was proposed in \cite{Christopherson:2008ry} and in that context a numerical study for the isocurvature perturbation was performed in \cite{Huston:2011fr}).  
The latter indicates we are dealing with a purely multiple field effect and hence it is an appropriate quantity to use 
during the inflationary 
period to describe the non adiabatic perturbations. The pressure for a scalar field is nothing but the matter Lagrangian. It's background value is given in  
(\ref{rp}). 
One can show that the isocurvature perturbation defined in \cite{Langlois:2005qp} is equal to
\be 
\Gamma_{(1)}\equiv p_{(1)}-\frac{\dot{p}}{\dot{\rho}}\rho_{(1)}=
2\Pi^2\lh\eta^{\perp}\zeta_{2(1)}-\frac{1}{NH}\dot{\zeta}_{1(1)}\rh.
\ee 
In the super-horizon regime, where as we will show in (\ref{mch}) there exists a simple relation between the time derivative of the adiabatic perturbation and the isocurvature perturbation,  $\dot{\zeta}_{1(1)}=2NH\hpe\zeta_{2(1)}$, this definition becomes  
\be\Gamma_{(1)}=-2\Pi^2\hpe\zeta_{2(1)},\ee
and $\Gamma_{(1)}$ becomes directly proportional to $\zeta_{2(1)}$. 

\subsubsection{The gradient of the perturbations}\label{grad}

As an alternative to the $\zeta_m$ defined to first order above, one can use the gradient quantity $\zeta_{1i}$ 
along with the isocurvature analogue $\zeta_{2i}$, relevant for working in the super-horizon regime.  
%Both these quantities were defined in \cite{Rigopoulos:2005xx} and later in 
%\cite{Langlois:2005qp} in a covariant way.
%
The use of spatial gradients was first advocated in \cite{Ellis:1989jt} in the
context of the covariant formalism. Later on the authors of
\cite{Rigopoulos:2005xx} constructed invariant quantities under long-wavelength
changes of time-slicing, by considering the
following combination of the spatial gradients of two spacetime scalars $A$ 
and $B$:
\be
\mathcal{C}_i\equiv \partial_i\bA-\frac{\partial_t\bA}{\partial_t\bB}\partial_i\bB,\label{AB}
\ee
%and formed among others the quantity
%\be
%\zeta_i=\partial_i\alpha-\frac{
%NH+\partial_t\alpha}{\partial_t\bar{\rho}}\partial_i\bar{\rho},
%\ee
%which when linearized is just the gradient of the comoving curvature
%perturbation $\zeta$.
%In \cite{Langlois:2005qp} it was argued that this particular
%combination could give rise to a second-order gauge-invariant quantity
%$\gz_{(2)}$ as follows:
%\be
%\partial_i\zeta_{(2)}=\zeta_{i(2)}-\frac{\rho_{(1)}}{\dot{\rho}}
%\dot{\zeta}_{i(1)},
%\label{mw}
%\ee
%where $\zeta_{(2)}$ can be related to yet another gauge-invariant second-order
%quantity defined by Malik and Wands in \cite{Malik:2003mv},
%\be
%\zeta_{(2)}\simeq \zeta_{(2)MW}-\zeta_{(1)MW}^2.
%\ee
%

Hence, in this part we restrict ourselves to super-horizon calculations. 
In the super-horizon regime one can employ the long-wavelength approximation to simplify calculations. 
The latter boils down to ignoring second-order spatial derivatives when compared to time derivatives. 
As a consequence the traceless part of the extrinsic curvature quickly decays and 
can be neglected, see section \ref{long} and   \cite{Salopek:1990jq}. Hence the space part of the metric can be described by 
\be
h_{ij}=a(t)^2e^{2\alpha(t,x)}\chi_{ij}(\vc{x}).
\ee
The field and Einstein equations in that case are identical to (\ref{fe2}), but now the quantities involved are fully non-linear. 
Additionally the momentum constraint (\ref{mo}) can be written as \cite{Rigopoulos:2005xx}
\be
\partial_i\bar{H}=-\frac{\kappa^2}{2}\bar{\Pi}_A\partial_i\varphi^A.\label{fieldeqsu}
\ee 
We will study the long-wavelength approximation  in detail in section \ref{long}.

The gradient quantities that we are introducing here (not gauge-invariant to all orders) are constructed from the curvature perturbation and the field perturbation. Indeed, (\ref{AB}) becomes for that choice
\be
\zeta_{mi}=\delta_{m1}\partial_i\alpha-\frac{\bar{H}}{\bar{\Pi}}\bar{e}_{mA}\partial_i\varphi^A
\label{defgrad},
\ee
where now $\bar{e}_{mA}$ represents the fully non-linear super-horizon version of the orthonormal basis vectors, e.g.\ 
$\bar{e}_{1A}=\bar{\Pi}_A/\bar{\Pi}$, with $\bar{\Pi}_A=\dot{\varphi}_A/\bar{N}$ since we are working in the super-horizon regime. 
Notice that the basis vectors still obey (\ref{ant}) as was shown in 
\cite{Tzavara:2010ge}. 
$\zeta_{mi}$ is by construction gauge-invariant at first order, since it has no background value: it is just the gradient of 
the gauge-invariant $\zeta_{m(1)}$ defined before.

\subsection{The second-order action}\label{saction}

In order to study the evolution of the first-order adiabatic and isocurvature perturbations we will calculate the second-order action governing these perturbations. To that end, we need to perturb the action (\ref{actionexact}) to second order. 
The scalar part of the extrinsic curvature terms appearing in the energy and momentum constraints, as well as in the action itself can be found using the definition (\ref{excurv})
\bea 
&&\bE_{ij}\bE^{ij}-\bE^2=-6\lh NH+\dot{\alpha}\rh^2
+4 \lh NH+\dot{\alpha}\rh\nabla_k\partial^k
\psi+\lh\nabla_i\partial_j\psi \rh^2-\lh\nabla_k\partial^k\psi\rh^2
\nn\\
&&
\bE_i^j-\bE\delta_i^j=
-2\lh NH+\dot{\alpha}\rh\delta_i^j-
\lh 
\nabla_i\partial^j\psi-\nabla_k\partial^k\psi\delta^j_i
\rh .
\eea
In this result we have already set $N_{i\perp}=0$, which we have checked to be true at least up to second order. Assuming that this remains true to all orders, the above expressions are valid for the fully non-linear quantities. The first-order energy (\ref{en}) and momentum constraint (\ref{mo}) equations (or equivalently the first-order $00$ and $0i$ Einstein equations) become
\bea 
&&\partial^2\psi_{(1)}=-\frac{N}{Ha^2}\partial^2\alpha_{(1)}+\e\lh
\dot{\zeta}_{1(1)}-2NH\hpe \zeta_{2(1)}\rh\equiv 
-\frac{N}{Ha^2}\partial^2\alpha_{(1)}+\partial^2\lambda,
\qquad
%\partial_iN_{i\perp(1)}=0,
\nn\\
&&
N_{(1)}=\frac{\dot{\alpha}_{(1)}}{H}
-\e N(\zeta_{1(1)}-\alpha_{(1)}).\label{enmo}
\eea
The term $\partial^2\lambda$, as in the single-field case,  represents a term that vanishes outside the horizon, see (\ref{mch}). This is why we use the gradient in its definition, to remind us that it can be neglected on long wavelengths.

We start by performing our calculation in the gauge $e_{1A}\tilde{\varphi}^A_{(1)}=0$. This constraint is equivalent to working in the uniform energy-density gauge, as was discussed in subsection \ref{fogip}. We give the basic elements of the calculation in appendix \ref{app2}. In the same appendix one can also  find the tensor part of the action. The second-order action takes the form 
\bea
S_2\!\!\!\!&=&\!\!\!\!\!\!\int\!\! \d^4x \cL_2\label{s2}\\
&=&\!\!\!\!\!\!\int\!\! \d^4x\frac{\e a^3}{N\kappa^2} \Bigg\{\!\!
(NH)^2\Big(\!\sqrt{\frac{2\e}{\kappa}}\frac{W_{221}}{3H^2}-2\e^2-(\hpa)^2+3(\hpe)^2+\frac{2}{3}\hpe\gx^\perp
-3\e(\hpa-\chi)+2\hpa\chi\Big) \zeta_{2(1)}^2
 \nn\\
&&\qquad\quad+
\dot{\zeta}_{1(1)}^2+\dot{\zeta}_{2(1)}^2
 -4NH \hpe \dot{\zeta}_{1(1)}\zeta_{2(1)}
+2NH\chi \dot{\zeta}_{2(1)}\zeta_{2(1)}
-\frac{N^2}{a^2}\Big( (\partial\zeta_{1(1)})^2+(\partial\zeta_{2(1)})^2\Big)
\Bigg\}\nn,
\label{l2f}
\eea
where $\cL_2$ is the second-order Lagrangian. 
While we have started from an action describing the evolution of the fields $\tilde{\alpha}_{(1)}$ and $e_{2A}\tvarphi^A_{(1)}$ we have now 
constructed an action 
in terms of the adiabatic and isocurvature perturbations $\zeta_{1(1)}$ and $\zeta_{2(1)}$.
The equations of motion that $\zeta_{1(1)}$ and $\zeta_{2(1)}$ obey are ($\delta \cL_2/\delta\zeta_m$ being a short-hand notation for the 
relevant variations of the Lagrangian)
\bea
&&\!\!\!\!\!\!\!\!\!\!\!\!\!\frac{\delta \cL_2}{\delta\zeta_1}=-2\frac{a^3\e}{N\kappa^2}\Bigg[\ddot{\zeta}_{1(1)}+
NH\lh3+2\e+2\hpa-
\frac{\dot{N}}{HN^2}
\rh \dot{\zeta}_{1(1)}
-2(NH)^2(3\hpe+2\e\hpe+\gx^\perp)\zeta_{2(1)}
\nn\\
&&\qquad\qquad\qquad\quad\!\!\!\!\!\!\!\!\!\!\!\!\!
-2 NH\hpe \dot{\zeta}_{2(1)}
-\frac{N^2}{a^2}\partial^2\zeta_{1(1)}\Bigg]
=-\frac{\d}{\d t}(2\frac{a^3}{N\kappa^2}\partial^2\lambda)+2\frac{a\e N}{\kappa^2}\partial^2\zeta_{1(1)}=0,
\nn
\eea
\bea
&&\!\!\!\!\!\!\!\!\!\!\!\!\!\frac{\delta \cL_2}{\delta\zeta_2}=-2\frac{a^3\e }{N\kappa^2} \Bigg[\ddot{\zeta}_{2(1)}+\!NH\lh
3+2\e+2\hpa-\frac{\dot{N}}{HN^2}\rh \dot{\zeta}_{2(1)}+(NH)^2
(3\chi+2\e^2+4\e\hpa+\gx^\parallel)
\zeta_{2(1)}
\nn\\
&&\!\!\!\!\!\!\!\!\!\!\!\!\!\qquad\qquad\qquad\quad\!+2NH\hpe  \dot{\zeta}_{1(1)}-\frac{N^2}{a^2}
\partial^2\zeta_{2(1)}\Bigg]=0.\label{eql2}
\eea
Thus we have found the evolution equations for the first-order adiabatic and isocurvature perturbations. In the next two subsections we are going to study these equations, considering their slow-roll limit, as well as their behaviour deep inside or far outside the horizon.
%where $\partial^2\lambda=\e\dot{\zeta}_1-2\e\hpe NH\zeta_2$ (for the reason of introducing $\lambda$ see appendix \ref{app2}).

While working in the flat gauge we find the same action (\ref{s2}) (see appendix \ref{app2}). So the curvature perturbations 
$\zeta_{m(1)}$ 
satisfy to first order the same equations in both gauges as expected, due to the gauge invariance of $\zeta_{m(1)}$ 
(or equivalently the gauge invariance of the action).

\subsection{The equations of motion}\label{tem}

In order to study the general properties of the equations of motion (\ref{eql2}), let us move to Fourier space, where
\be
\zeta_m(t,\vc{x})=\int\frac{\d^3\vc{k}}{(2\pi)^{3/2}}\zeta_{m\vc{k}}(t)\rme^{\rmi\vc{k}\cdot\vc{x}}.\label{fourierm}
\ee 
We start by defining the time derivative of the perturbations $\theta_m\equiv\dot{\zeta}_m$, in order to split equations (\ref{eql2}) in a set of first-order differential equations.  To simplify the form of the equations we also set the time variable. In order to make connection with the long-wavelength formalism, introduced in section \ref{long}, we choose to work with a time coordinate that coincides with the number of e-folds, i.e.\ $N=1/H$. Then, equations (\ref{eql2}) for the Fourier components of the perturbations simplify to
\bea 
&&\dot{\theta}_1+\lh3+\e+2\hpa\rh\theta_1-2\hpe\theta_2
-2\lh 3\hpe+2\e\hpe+\gx^\perp\rh\zeta_2+\frac{k^2}{a^2H^2}\zeta_1=0,\nn\\
&&\dot{\zeta}_1=\theta_1,\nn\\
&&\dot{\theta}_2+\lh3+\e+2\hpa\rh\theta_2+2\hpe\theta_1
+\lh 3\chi+2\e^2+4\e\hpa+\gx^\parallel\rh\zeta_2+\frac{k^2}{a^2H^2}\zeta_2=0,\nn\\
&&\dot{\zeta}_2=\theta_2,\label{eql3}
\eea
where we dropped the $\vc{k}$ index denoting the Fourier component and the $(1)$ subscript denoting the order of the perturbations, to lighten the notation. These equations for the first-order adiabatic and isocurvature perturbations can also  be derived by perturbing the background field and Einstein equations, as was done in \cite{Gordon:2000hv}. Notice that in the general case, each type of time derivative of the perturbations $\theta_m$ is sourced by both $\theta_1$ and $\theta_2$ and the isocurvature perturbation $\zeta_2$. 

In the super-horizon limit, one can ignore the contribution of the $k^2$ term when working up to zeroth order in a gradient expansion  and the set of equations  
coincides with that derived in \cite{Rigopoulos:2005xx} for the gradient of the perturbations (see section \ref{long} for details), since up to first 
order $\zeta_{mi(1)}=\partial_i\zeta_{m(1)}$.  Then one can show (using the time derivative of $\hpe$ derived in (\ref{dere})) that the constraint equation 
\be
\theta_1=2\hpe\zeta_2 \label{newcon}
\ee
is a solution of the super-horizon part of the first equation of (\ref{eql3}). This constraint equation is just the super-horizon limit of the first-order energy constraint (\ref{enmo}). Indeed outside the horizon gradients can be ignored and hence we can set $\partial^2\psi_{(1)}=0$. Since $\partial^2\lambda$ is a gauge invariant quantity, it has to be zero independently of $\partial^2\alpha_{(1)}$ and we find    
\be
\partial^2\lambda=\e\lh\dot{\zeta}_{1(1)}-2NH\hpe
\zeta_{2(1)}\rh=0.\label{mch}
\ee
In fact it was shown in \cite{Rigopoulos:2005us} that this is the case at all orders for the gradient of the adiabatic perturbation, where now $\hpe,\ N$ and $H$  generalise to the fully non-linear quantities. Rewriting the equations of motion for the time derivatives of the perturbations in the super-horizon limit, using (\ref{newcon}), we find
\bea
&&\theta_1=2\hpe\zeta_2,\quad\dot{\zeta}_1=\theta_1\nn\\
&&\dot{\theta}_2+\lh3+\e+2\hpa\rh\theta_2
+\lh 3\chi+2\e^2+4\e\hpa+4(\hpe)^2+\gx^\parallel\rh\zeta_2
=0,\quad\dot{\zeta}_2=\theta_2.\label{eq4}
\eea
One can see that outside the horizon the evolution of the isocurvature perturbation is decoupled from the adiabatic one and hence the adiabatic perturbation does not source the isocurvature perturbation on these scales.  

Working up to first order in a slow-roll expansion, we can further simplify the equation for the isocurvature perturbation
\be 
\dot{\theta}_2+3\theta_2
+3\chi\zeta_2=0.
\ee
One can check that this equation has the simple solution
\be
\theta_2=-\chi\zeta_2,  
\ee
where we neglected the $\dot{\theta}_2$ contribution since it is second order in slow-roll.  
In paper \cite{Rigopoulos:2005us} it was found that for the fully non-linear gradient of the perturbation  $\zeta_{2i}$ this remains true outside the horizon at all orders   assuming slow-roll. In the next chapter, we will use the super-horizon slow-roll expressions for $\theta_m$, in order to calculate the non-Gaussianity immediately after horizon crossing (where we will assume that slow-roll holds), as well as to study the evolution of non-Gaussianity during slow-roll inflation.
%\be 
%\dot{\zeta}_{2(1)}=-NH\chi\zeta_{2(1)}\label{xsr}
%\ee 
%is the super-horizon first integral of 
%the equation for $\zeta_{2(1)}$, which can be easily verified by ignoring the second-order time derivative and gradient, along with the next to leading order slow-roll parameters in (\ref{eql2}). 

\subsection{The scalar perturbations at horizon-crossing}\label{sphc}

In order to solve for the perturbations and promote them to quantum operators as we did for the single-field case in subsection \ref{scalar},  
we shall work with the conformal time coordinate $N=a$ and we will define the new variables
\be
\zeta_{m}=-\frac{\kappa}{a\sqrt{2\e}}q_m.\label{redefzeta}
\ee
For the rest of this subsection we drop the subscript $(1)$ to lighten the notation. 
Now the Lagrangian (\ref{l2f}) for the $k$ mode vector $\mathbf{q}_{k}^T=(q_{1k},q_{2k})$ becomes
%\be
%\cL_2=\frac{1}{2}\lh\dot{Z}_{mk}-\varepsilon_{mn}aH\hpe Z_{nk}\rh^2
%-\frac{1}{2}\lh k^2-\frac{\ddot{a}}{a}+M_n^2\rh Z_{nk}^2 
%-V_{12}Z_{1k}Z_{2k},
%\ee
\be
\cL_2=\frac{1}{2}\lh\mathbf{q}_k'+\mathbf{S}\mathbf{q}_k\rh^T
\lh\mathbf{q}_k'+\mathbf{S}\mathbf{q}_k\rh 
-\frac{1}{2}\lh k^2-\frac{a''}{a}\rh 
\mathbf{q}_k^T\mathbf{q}_k   
-\frac{1}{2}\mathbf{q}_k^T\mathbf{\Omega}\mathbf{q}_k,
\ee
where the matrices $\mathbf{S}$ and $\mathbf{\Omega}$ take the form
\be
%S_{mn}=-aH\hpe\varepsilon_{mn}
\mathbf{S}=-\cH\hpe
\lh\begin{array}{cc}
0 &1\\
-1 & 0
\end{array}\rh
\quad\mathrm{and}\quad
\mathbf{\Omega}=-\cH^2
\lh\begin{array}{cc}
3\e +3\hpa+2\e^2+4\e\hpa+\xi^\parallel  & \lh3+2\e\rh\hpe+\xi^\perp\\
\lh3+2\e\rh\hpe+\xi^\perp &
-3\lh\e+\hpa-\chi\rh 
\end{array}\rh  
\ee
%where the interaction parameters $M_n^2$ and $V_{12}$ take the values
%\bea
%&&M_1^2=-(aH)^2\lh 3\e +3\hpa+2\e^2+4\e\hpa+\xi^\parallel\rh
%,\qquad 
%M_2^2=-3(aH)^2\lh\e+\hpa-\chi\rh ,\nn\\
%&&V_{12}=-(aH)^2\lh\lh3+2\e\rh\hpe+\xi^\perp\rh. 
%\eea
and we remind that $\cH=aH$. 
This Lagrangian has the canonical form, while the canonical momentum of the perturbation $q_m$ is
\be
\mathbf{\Pi_{k}}=\mathbf{q_k}'
-\cH\hpe \boldsymbol{\varepsilon}\mathbf{q_k}, 
\ee 
where $\boldsymbol{\varepsilon}$ is the fully antisymmetric Levi-Civita matrix.
The equations of motion turn out to be
\be
\mathbf{q}_k''+2\mathbf{S}\mathbf{q}_k' 
+\lh \lh k^2-\frac{a''}{a}\rh \mathbf{I_2}+\dot{\mathbf{S}}+
\mathbf{S}^2+\mathbf{\Omega}\rh\mathbf{q}_k=0.
\ee
%\bea
%&&\ddot{Z}_{1k}-2aH\hpe \dot{Z}_{2k}+2\lh(aH)^2\lh 1+\e+\hpa\rh\hpe+V_{12}\rh Z_{2k}+\lh k^2-\frac{\ddot{a}}{a}+m_1^2\rh  Z_{1k}=0\nn\\
%&&\ddot{Z}_{2k}+2aH\hpe \dot{Z}_{1k}-2(aH)^2\lh 1+\e+\hpa\rh\hpe  Z_{1k}+\lh k^2-\frac{\ddot{a}}{a}+m_2^2\rh  Z_{2k}=0,
%\eea
%where the mass terms take the values
%\be
%m_1^2=M_1^2-(aH)^2\hpe,\qquad\mathrm{and}\qquad
%m_2^2=M_2^2-(aH)^2\hpe
%\ee
In order to remove the first order derivatives and study analytically the system of equations, we perform a rotation of the perturbations $q_m=R_{mn}\widetilde{q}_n$, where $\mathbf{R}$ is a rotation matrix, hence real and with unit determinant \cite{GrootNibbelink:2001qt}. The equations are then rewritten as 
\be 
\widetilde{\mathbf{q}}_{k}''
+\lh \lh k^2-\frac{a''}{a}\rh \mathbf{I_2}
+
\mathbf{\Omega}\rh\widetilde{\mathbf{q}}_k=0
,\qquad
\mathbf{R}'+\mathbf{S}\mathbf{R}=0,\qquad
\mathbf{\tilde{\Omega}}\equiv \mathbf{R^{-1}\Omega R}.\label{maeq}
\ee 
%with
%\be 
%\mathbf{\Omega}=
%\lh 
%\begin{array}{cc}
%k^2-\frac{\ddot{a}}{a}+M_1^2 & V_{12}\\
%V_{12} & k^2-\frac{\ddot{a}}{a}+M_2^2
%\end{array}
%\rh.  
%\ee 
The canonical momentum in terms of the rotation matrix becomes  $\Pi_m=R_{mn}\tilde{q}_n$. 
One can study the matrix equation (\ref{maeq}) the same way we did for the single-field case. 

We can now promote $q_m$ 
and $\Pi_m$ to quantum operators
\be
\mathbf{q}_\vc{k}=\mathbf{q}_k
\mathbf{a}_\vc{k}
+\mathbf{q}^*_k\mathbf{a}^\dagger_\vc{k}=
\mathbf{R\tilde{q}}_k\mathbf{a}_\vc{k}
+\mathbf{R\tilde{q}}^*_k\mathbf{a}^\dagger_\vc{k}
\qquad\mathrm{and}\qquad
 \mathbf{\Pi}_\vc{k}=\mathbf{\Pi}_k
 \mathbf{a}_\vc{k}^\dagger
+\mathbf{\Pi}^*_k\mathbf{a}_\vc{k}.
\ee
$\mathbf{q}_k$ is now a $2\times 2$ matrix coupling to the creation and annihilation operators which are vectors with components in the $e_{m}$ basis. They satisfy
\be
[\mathbf{a}_\vc{k},\mathbf{a}_\vc{k'}]=
[\mathbf{a}^\dagger_\vc{k},\mathbf{a}^\dagger_\vc{k'}]=\mathbf{0},\qquad[\mathbf{a}_\vc{k},\mathbf{a}^\dagger_\vc{k'}]=
\delta^{(3)}(\vc{k}-\vc{k'})\mathbf{I_2}. 
\ee
As a result the perturbations and their canonical momentum satisfy the equation 
\be
\mathbf{q}^*_k\mathbf{\Pi}_k^T-\mathbf{q}_k
\mathbf{\Pi}^{*T}_k=-\mathrm{i}\mathbf{I_2} 
\ee

Following the study of the single-field case we shall solve this equation in the two limits of the sub-horizon $k\mg aH$ and super-horizon $k\ll aH$ regimes. In the sub-horizon limit equation (\ref{maeq}) becomes
\be
\widetilde{\mathbf{q}}_{k}''
+k^2\widetilde{\mathbf{q}}_{k}=0,\quad\mathrm{
with\ solution}\quad 
\mathbf{\tilde{q}}_k=\frac{1}{\sqrt{2k}}\mathrm{e}^{-\mathrm{i}k\tau}\mathbf{I_2},\qquad \mathbf{\Pi}_k=\frac{\mathrm{i}\sqrt{k}}{\sqrt{2}}\mathrm{e}^{-\mathrm{i}k\tau}\mathbf{R}.
\ee
In the super-horizon regime and assuming that the interaction terms (of order at least $\mathcal{O}(\e)$) remain small during inflation, (\ref{maeq}) reduces to  
\be 
\widetilde{\mathbf{q}}_{k}
-\frac{a''}{a}\widetilde{\mathbf{q}}_{k}=0,\quad\mathrm{with\ solution}\quad 
\widetilde{\mathbf{q}}_k=\mathbf{C_+}a+\frac{1}{a^2}\mathbf{C_-}
,\qquad
\mathbf{\Pi}_k=aH\mathbf{R}
\lh\mathbf{C_+}a-\frac{2}{a^2}\mathbf{C_-}
\rh
\ee
Matching the two solutions at horizon crossing we find $
\mathbf{C_+}=H/\sqrt{2k^3}\mathbf{I_2}$ 
and hence using the redefinition (\ref{redefzeta}) we find for the first-order perturbation at horizon crossing
\be
\boldsymbol{\zeta}_k=-\frac{\kappa H}{2\sqrt{\e k^3}}\mathbf{R}.\label{Xam} 
\ee
The operator $\boldsymbol{\zeta}_{\vc{k}}$ behaves like a classical random variable on super-Hubble scales, i.e.\ 
\be
\boldsymbol{\zeta}_{\vc{k}}=\int\frac{\d^3
\vc{k}}{(2\pi)^{3/2}} \frac{\kappa H}{2\sqrt{\e k^3}}\mathbf{R}\mathbf{b}_\vc{k}\mathrm{e}^{\mathrm{i}
\vc{k}\cdot\vc{x}},
\quad\mathrm{with}\quad
\langle\mathbf{b_\vc{k}}
\mathbf{b}_\vc{k'}^{*T}\rangle=
\delta^{(3)}(\vc{k}-\vc{k'})\mathbf{I_2}.
\ee
The $\mathbf{R}$ matrix is evaluated at horizon crossing. When computing observable quantities, for example the power spectrum, the $\mathbf{R}$ matrix cancels out with its inverse
\be
\langle\boldsymbol{\zeta}_\vc{k}
\boldsymbol{\zeta}^{*T}_\vc{k'}\rangle=
\frac{\kappa^2H^2}{4\e k^3}
\langle \mathbf{R}
\mathbf{b_\vc{k}}\mathbf{b}_\mathbf{\vc{k}'}^{*T}
\mathbf{R}^T
\rangle
=
\frac{\kappa^2H^2}{4\e k^3}
\mathbf{R}\langle 
\mathbf{b_\vc{k}}\mathbf{b}_\mathbf{\vc{k}'}^{*T}
\rangle\mathbf{R}^T=\frac{\kappa^2H^2}{4\e k^3}
\delta^{(3)}(\vc{k}-\vc{k'})\mathbf{I_2}.
\label{2cor}
\ee 
Hence, we can set the value of $\mathbf{R}=\mathbf{I_2}$ at horizon crossing, since it does not carry any relevant information for measurable quantities.

Although we computed this solution for a massless field around horizon-crossing, it turns out to be in agreement with the detailed calculation done in \cite{GrootNibbelink:2001qt}. Given this one can compute the power spectrum for the adiabatic perturbations, the isocurvature perturbations or the mixed power spectrum at horizon crossing from equation (\ref{2cor}) as
\be
\mathcal{P}_{\zeta_1}=\frac{\kappa^2H_*^2}{8\pi^2\e_*},\qquad
\mathcal{P}_{\zeta_2}=\frac{\kappa^2H_*^2}{8\pi^2\e_*},\qquad 
\mathcal{P}_{\zeta_1\zeta_2}=0.
\ee 
However, unlike the single-field case where the adiabatic perturbation once outside the horizon remains constant, both the adiabatic and the isocurvature perturbations evolve in the super-horizon regime and interact with each other. Hence, these values of the power-spectra are only the initial values around horizon-crossing. In order to find the value of the power spectra at the end of inflation, one needs to solve the evolution equations for $\zeta_1$ and $\zeta_2$ (\ref{eql2}) outside the horizon. We are going to do that in section \ref{long} within the long-wavelength formalism.

\Section{Second-order perturbations}\label{S33}

We move next to studying second-order perturbations, since in the end of the day we want to study the non-linearities produced during two-field inflation and calculate the bispectrum of the adiabatic perturbations. 
Again, we shall not consider vector perturbations.   
Nevertheless, it is known that  at second order,  isocurvature perturbations source vorticity and hence vector
perturbations are generated. However, this is likely to be a subdominant contribution and hence we shall assume that it does not affect the observables we calculate in this thesis  (for details see  \cite{Christopherson:2009bt}). 

In this section we first define the second-order gauge invariant quantities. Next we compute the cubic action, since we already know from the single-field case that the redefinitions of the fields occurring in this action can be compared to the second-order gauge-invariant definitions and contribute to  the super-horizon $\f^{(4)}$. 
%Finally we will describe the basic elements of the long-wavelength formalism that we will use in the next chapters to calculate the super-horizon evolution of the perturbations. 

\subsection{Gauge invariance}

As for the first-order gauge invariant quantities, the second-order gauge-invariant quantities can be calculated fairly easily given that we have already done so for the single-field case. The gravitational part of the calculation remains the same. What changes is the matter content which can be easily generalised to two fields. This subsection is based on our paper \cite{Tzavara:2011hn}.

\subsubsection{Second-order gauge invariant perturbations}\label{gic}

Already in chapter \ref{Ch2} we found the second order gauge transformation of the adiabatic perturbation at second order
\bea
\frac{1}{2}\zeta_{(2)}\!\!\!\!&\equiv&\!\!\!\!\frac{1}{2}\tilde{\alpha}_{(2)}\nn\\
&=&\!\!\!\!\frac{1}{2}\alpha_{(2)}
+\frac{1}{2}NHT_{(2)}
+\frac{1}{2}\lh\dot{\alpha}_{(1)}+\dot{\zeta}_{1(1)}\rh T_{(1)}
-\frac{N^2}{4a^2}\Bigg[\lh \partial T_{(1)}\rh^2-\partial^{-2}\partial_i\partial_j\lh \partial_iT_{(1)}\partial_jT_{(1)} \rh \Bigg]
\nn\\
&&\!\!\!\!
+\frac{1}{2}\Bigg[\partial_{i}T_{(1)}\partial_{i}\psi_{(1)}
-\partial^{-2}\partial_i\partial_j\lh \partial_{(i}T_{(1)}\partial_{j)}\psi_{(1)} \rh \Bigg]
-\frac{1}{2}\partial^{-2}\lh\dot{\gamma}_{ij(1)}\partial_i
\partial_jT_{(1)}\rh,
\eea
where the tilded gauge is the uniform energy density gauge. The second order time shift in terms of the energy density perturbation was calculated in (\ref{t2}) for the single-field case and for two fields it generalises to 
\be 
T_{(2)}=
-\frac{\rho_{(2)}}{\dot{\rho}}+T_{(1)}
\Big[
2\hpa\lh\zeta_{1(1)}-\alpha_{(1)}\rh
+\frac{1}{NH}\lh\dot{\zeta}_{1(1)}-\dot{\alpha}_{(1)}\rh
\Big],\label{t2r}\ee
where the first-order time shift is  $T_{(1)}=(\zeta_{1(1)}-\alpha_{1(1)})/NH$.

As for the single-field case, the uniform energy density gauge is not equivalent to the uniform field gauge at second order and as a consequence the gauge invariant quantities defined in each gauge have different expressions. 
Indeed at second order the fields transform as
\be
\tilde{\varphi}^A_{(2)}=
\varphi_{(2)}^A+T_{(2)}\dphi^A+T_{(1)}\lh 
\dot{T}_{(1)}\dphi^A+T_{(1)}\ddot{\phi}^A
+2\dot{\varphi}^A_{(1)}
\rh \label{phitr} 
\ee
and multiplying with $e_{1A}$ and using expressions (\ref{difz}) one can find the  second-order time shift for the uniform-field gauge $e_{1A}\tilde{\varphi}^A_{1(2)}=0$ in terms of the first-order gauge-invariant perturbations:
\be
T_{(2)}
=
%-\frac{1}{NH}\tilde{Q}_{1(2)}
\frac{1}{NH}Q_{1(2)}
+T_{(1)}
\Big[
(\e+\hpa)\lh\zeta_{(1)}-\alpha_{(1)}\rh
+\frac{1}{NH}\lh\dot{\zeta}_{(1)}-\dot{\alpha}_{(1)}\rh
-2\hpe\zeta_{2(1)}\Big].\label{tt2}
\ee

In the single-field case the two time shifts were found to be equal outside the horizon  and hence in this regime one can identify the uniform-field gauge with the uniform energy density gauge. This also remains true for the multiple-field case. The relation between $\rho_{(2)}$ and $Q_{1(2)}$ is calculated in  (\ref{r2}) for two fields
\bea
&&\!\!\!\!\!\!\!\!\!\!\!\!\!\!\!\!\!\!\!\!\!\!\!\!
-\e NH\Big[
Q_{1(2)}+NH\frac{\rho_{(2)}}{\dot{\rho}}
+(\e-\hpa)\lh\zeta_{1(1)}
-\alpha_{(1)}\rh^2
-(\e+\hpa)\zeta_{2(1)}^2
-2\hpe\zeta_{2(1)}\lh\zeta_{1(1)}
-\alpha_{(1)}\rh 
\nn\\&&\!\!\!\!\!\!
-\frac{2}{NH}\partial^{-2}\partial_i\dot{\zeta}_{2(1)}
\partial_i\zeta_{2(1)}\Big]
+\partial^2\cA
=0,
\label{r2n}
\eea
where $\partial^2\cA$ are terms involving gradients  $\partial^2\psi$ and hence they can be ignored on super-horizon scales. 
%Therefore,  there is no discrepancy in ignoring them during slow-roll inflation, when $\e\ll1$. 
Once the super-horizon terms are neglected, substitution of (\ref{r2n}) in (\ref{t2r}) shows that the two time shifts become the same also in the multiple field case and therefore the uniform energy density gauge coincides with the uniform-field gauge outside the horizon.
%in the uniform energy density gauge where $\zeta_{(1)}=\tilde{\alpha}_{(1)}$
%\be
%\tilde{Q}_{1(2)}=
%(\e+\hpa)\zeta^2_{2(1)}
%-\frac{2}{NH}\partial^{-2}\partial_i\lh\dot{\zeta}_{2(1)}
%\partial_i\zeta_{2(1)}\rh
%\ee

%To summarize, from now the tilded gauge will be the gauge 
%\be
%e_{1A}\tilde{\varphi}^A_{(1)}=0,\qquad
%\zeta_{(1)}=\alpha_{(1)}
%\qquad
%\mathrm{and}\qquad
%\tilde{Q}_{1(2)}=(\e+\hpa)\zeta^2_{2(1)}-\frac{2}{NH}\partial^{-2}\partial_i\lh\dot{\zeta}_{2(1)}
%\partial_i\zeta_{2(1)}\rh, 
%\ee
%the latter found from the super-horizon limit of (\ref{r2}) in the uniform energy density gauge. 
%This gauge at first order is simply the uniform  energy density gauge. At second order it coincides with the uniform energy density gauge only in the super-horizon regime, while in general it is a gauge where
%\be
%-\e N^2H^2\frac{\tilde{\rho}_{(2)}}{\dot{\rho}}
%-\partial^{-2}\partial_i\Bigg[
%\frac{1}{NH}\partial_k\dot{\zeta}_{(1)}\partial_k
%\partial_i\psi_{(1)}
%+2\frac{a^2H}{N}\partial_k
%\psi_{(1)}\partial_i\partial_k\psi_{(1)}
%+\partial^2\zeta_{(1)}\partial_i\psi_{(1)}
%+\partial_i\partial_k\zeta_{(1)}\partial_k\psi
%\Bigg]
%=0,
%\ee
%
%
%\be 
%\partial^2\psi_{(1)}=
%-\frac{N}{Ha^2}\partial^2\zeta_{(1)}+\partial^2\lambda
%\ee 

Now the second-order gauge invariant perturbation becomes
\bea
\frac{1}{2}\zeta_{1(2)}\!\!\!\!&=&\!\!\!\!\frac{1}{2}\alpha_{(2)}
+\frac{1}{2}Q_{1(2)}
+\frac{\e+\hpa}{2}\Big[\lh\zeta_{1(1)}-\alpha_{(1)}\rh^2
-\zeta_{2(1)}^2
\Big]
+\lh\frac{\dot{\zeta}_{1(1)}}{NH}-\hpe \zeta_{2(1)}\rh \lh\zeta_{1(1)}-\alpha_{(1)}\rh
\nn\\&&\!\!\!\!\!
-\frac{1}{NH}\partial^{-2}\partial_i\lh\dot{\zeta}_{2(1)}
\partial_i\zeta_{2(1)}\rh 
-\frac{N^2}{4a^2}\Bigg[\lh \partial T_{(1)}\rh^2-\partial^{-2}\partial_i\partial_j\lh \partial_iT_{(1)}\partial_jT_{(1)} \rh \Bigg]
-\frac{1}{2\e NH}\partial^2\cA
\nn\\
&&\!\!\!\!
+\frac{1}{2}\Bigg[\partial_{i}T_{(1)}\partial_{i}\psi_{(1)}
-\partial^{-2}\partial_i\partial_j\lh \partial_{(i}T_{(1)}\partial_{j)}\psi_{(1)} \rh \Bigg]
-\frac{1}{2}\partial^{-2}\lh\dot{\gamma}_{ij(1)}\partial_i
\partial_jT_{(1)}\rh
,\label{zgen}
\eea
%
%In the tilded gauge this becomes
%\bea
%\frac{1}{2}\zeta_{(2)}\!\!\!\!&=&\!\!\!\!\frac{1}{2}\tilde{\alpha}_{(2)}
%-\frac{1}{2}NH\frac{\tilde{\rho}_{(2)}}{\dot{\rho}}
%\eea
In the flat gauge $\hat{\alpha}_{(2)}=0$ this takes the form
\bea
\frac{1}{2}\zeta_{1(2)}\!\!\!\!&=&\!\!\!\!
\frac{1}{2}\hat{Q}_{1(2)}
+\frac{\e+\hpa}{2}\lh\zeta_{1(1)}^2
-\zeta_{2(1)}^2\rh 
+\lh\frac{\dot{\zeta}_{1(1)}}{NH}-\hpe \zeta_{2(1)}\rh\zeta_{1(1)}
-\frac{1}{NH}\partial^{-2}\partial_i\lh
\dot{\zeta}_{2(1)}
\partial_i\zeta_{2(1)}\rh 
\nn\\&&\!\!\!\!\!
-\frac{N^2}{4a^2}\Bigg[\lh \partial \zeta_{1(1)}\rh^2-\partial^{-2}\partial_i\partial_j
\lh \partial_i\zeta_{1(1)}\partial_j\zeta_{1(1)} \rh \Bigg]
-\frac{1}{2\e NH}\partial^2\hat{\cA}
-\frac{1}{2}\partial^{-2}\lh\dot{\gamma}_{ij(1)}\partial_i
\partial_j\zeta_{1(1)}\rh 
\nn\\
&&\!\!\!\!
+\frac{1}{2}\Bigg[\partial_{i}\zeta_{1(1)}\partial_{i}
\lambda
-\partial^{-2}\partial_i\partial_j\lh \partial_{(i}\zeta_{1(1)}\partial_{j)}\lambda\rh \Bigg].
\label{flat2} 
\eea
The super-horizon limit of this is just the first line of the expression. 
Notice that unlike in the original definition of $\zeta_{1(2)}$ in terms of $\rho_{(2)}$, a non-local term appears in (\ref{flat2}) in the super-horizon limit  
%when one uses the fields 
%instead of the energy density 
(because of (\ref{r2n})). The time derivatives of the first-order perturbations for super-horizon scales were found in subsection \ref{saction} to be 
\be
\dot{\zeta}_{1(1)}=2NH\hpe \zeta_{2(1)}
\ee   
for the adiabatic perturbation and
\be 
\dot{\zeta}_{2(1)}=-NH\chi \zeta_{2(1)}\label{xsr}
\ee    
for the isocurvature perturbation, the latter valid only in the slow-roll regime. Using these expressions  we find that just after horizon-crossing, when we assume that slow-roll holds
\be 
\frac{1}{2}\zeta_{1(2)}=\frac{1}{2}\hat{Q}_{1(2)}
+\frac{\e+\hpa}{2}\zeta_{1(1)}^2-\frac{\e+\hpa-\chi}{2}\zeta_{2(1)}^2+\hpe\zeta_{1(1)}\zeta_{2(1)},
\ee
i.e.\  without a non-local term.

Multiplying (\ref{phitr}) with $e_{2A}$ we find 
the second-order transformation for the field perturbation in the $\mathbf{e}_2$ direction 
\be 
\frac{1}{2}\tilde{Q}_{2(2)}=\frac{1}{2}Q_{2(2)}+
\lh\zeta_{1(1)}-\alpha_{(1)}\rh\Big[
(\e+\hpa)\zeta_{2(1)}+\frac{1}{HN}\dot{\zeta}_{2(1)}+\frac{\hpe}{2}\lh\zeta_{1(1)}-\alpha_{(1)}\rh 
\Big].\label{Q2tr}
\ee
Notice how the $Q_{2(2)}$ gauge transformation does not involve any gradient terms and hence remains the same for all scales. Although $Q_{2(2)}$ is gauge invariant we cannot identify it with the isocurvature perturbation, since  it is not orthogonal to the second-order adiabatic perturbation in the flat gauge (that is the original definition of the isocurvature perturbation). We will find the second-order isocurvature perturbation in the next paragraph. If we identify the arbitrary gauge with the flat gauge, the transformation takes the form 
\be 
\frac{1}{2}\tilde{Q}_{2(2)}=\frac{1}{2}\hat{Q}_{2(2)}+
\zeta_{1(1)}\Big[
(\e+\hpa)\zeta_{2(1)}+\frac{1}{HN}\dot{\zeta}_{2(1)}+\frac{\hpe}{2}\zeta_{1(1)} 
\Big].
\ee
%Let us discuss this result. Using the basis vectors $e_{1A}$ and $e_{2A}$ to decompose the fields is equivalent to renaming the two scalar field degrees of freedom from $\varphi^A\ (A=1,2)$, to $e_{1A}\varphi^A$ and $e_{2A}\varphi^A$. These degrees of freedom have to be supplemented with the scalar degree of freedom of gravity. From these three, only two are independent as was discussed in \ref{ga1}. Our starting point was to make the natural choice to to define the adiabatic perturbation, i.e.\ one of the two dynamical degrees of freedom, from the combination of the $e_{1A}$ scalar-field degree of freedom (corresponding to the perturbation of the energy density) and the curvature one, since it is the energy density that sources the curvature of space-time. Hence, the $e_{2A}$ perturbation of the scalar field is the second degree of freedom. 
%Indeed, the transformation of $Q_{2(r)}$ does not involve the perturbations $\alpha_{(r)}$ or $Q_{1(r)}$  of the same order. The fact that $Q_{2(2)}$ does not couple directly to gravity (except from the first-order scalar gauge-invariant perturbation $\zeta_{1(1)}$) is translated to the absence of any other metric elements, e.g.\  $\psi$ in its transformation. 
%There are also no gradients in its transformation and hence this is exact at any scale, another proof that it does not interact directly with gravity. 

%However, based on this discussion we do not expect that $\zeta_{2(1)}$ will involve terms  other that $\zeta_{1(1)},\ \zeta_{2(1)}$ and $Q_{2(2)}$. 

\subsubsection{The gradient of the perturbations}
\label{grad2}

Expanding to second order the gradient of the perturbations (\ref{defgrad}) we find for the adiabatic perturbation
\bea
&&\frac{1}{2}\zeta_{1i(2)}=\frac{1}{2}\partial_i\lh\alpha_{(2)}+Q_{1(2)}\rh
-\frac{\dot{\alpha}_{(1)}}{N\Pi}e_{1A}\partial_i\varphi^A_{(1)}
-\frac{H}{N\Pi^2}\dvphi^A_{(1)}\partial_i\varphi_{A(1)}
-2\frac{1}{N\Pi}e_{1A}\dvphi^A_{(1)} \partial_iQ_{1(1)}\nn\\&&\qquad\quad\ 
=\frac{1}{2}\partial_i\zeta_{1(2)}
-\frac{1}{HN}\lh\zeta_{1(1)}-\alpha_{(1)}\rh\partial_i\dot{\zeta}_{1(1)}
,\label{akoma}
\eea
where we used the basis completeness relation and (\ref{difz}) to rewrite the terms. The second order adiabatic perturbation appearing in the second line is the super-horizon limit of (\ref{zgen}).
In the uniform energy-density gauge this gives
\be
\frac{1}{2}\tilde{\zeta}_{1i(2)}=\frac{1}{2}\partial_i\tilde{\alpha}_{(2)},\label{zai}
\ee
while in the flat gauge where $\partial_i\ha=0$ and $\zeta_{1(1)}=\hat{Q}_{1(1)}$, we find 
\bea 
&&\!\!\!\!\!\frac{1}{2}\hat{\zeta}_{1i(2)}
=\partial_i\Bigg[\frac{1}{2}\hat{Q}_{1(2)}+\frac{\e+\hpa}{2}\lh\zeta_{1(1)}^2-\zeta_{2(1)}^2\rh-\hpe\zeta_{1(1)}\zeta_{2(1)}
+\frac{1}{HN}\dot{\zeta}_{1(1)}\zeta_{1(1)}\Bigg]-
\frac{1}{HN}\dot{\zeta}_{2(1)}\partial_i\zeta_{2(1)}
\nn\\&&\qquad\qquad\!\!\!\!\!
-\frac{1}{HN}\zeta_{1(1)}\partial_i\dot{\zeta}_{1(1)},
\label{z1fg}
\eea
$\tilde{\zeta}_{1i(2)}$ in the uniform energy-density gauge (\ref{zai}) coincides with the gradient of the 
gauge-invariant second-order adiabatic perturbation. 
However, by comparing (\ref{flat2}) to (\ref{z1fg}) we see 
that in the flat gauge $\hat{\zeta}_{1i(2)}$ is the gradient 
of the gauge-invariant curvature perturbation $\zeta_{1(2)}$ expressed in the flat gauge plus a new non-local term. 
This is in agreement 
with the findings in \cite{Langlois:2005qp}. This new term is nothing else but the
gauge transformation of $\zeta_{1i(2)}$. A quantity with zero background value as $\zeta_i$ is, transforms as (\ref{zerotr}). 
One can check, using the gauge transformations (\ref{trans}) for $\rho$ and for $\alpha$ 
and requiring that $\tilde{\rho}_{(1)}=0$, 
that 
\be
\frac{1}{2}\tilde{\zeta}_{mi(2)}=\frac{1}{2}\zeta_{mi(2)}
+\frac{1}{NH}\lh\zeta_{1(1)}-\alpha_{(1)}\rh \dot{\zeta}_{mi(1)},
\ee
where we replaced $T_{(1)}=\lh\zeta_{1(1)}-\alpha_{(1)}\rh/(NH)$. If we identify the second gauge with the flat gauge we find
\be
\frac{1}{2}\tilde{\zeta}_{mi(2)}=\frac{1}{2}\hat{\zeta}_{mi(2)}
+\frac{1}{NH}\zeta_{1(1)} \dot{\zeta}_{mi(1)}.\label{gat}
\ee

Next we try to find the second-order gauge-invariant part of the isocurvature perturbation $\zeta_{2i}$ by expanding (\ref{defgrad}) and using  (\ref{ant}) to express $\vc{e}_{2}$ in terms of $\vc{e}_1$,
\be
\frac{1}{2}\zeta_{2i(2)}=\frac{1}{2}\partial_iQ_{2(2)}
-\frac{\dot{\alpha}_{(1)}}{N\Pi}e_{2A}\partial_i\varphi^A_{(1)}
+\frac{H}{N\Pi^2}\varepsilon^{BA}\dvphi_{B(1)}
\partial_i\varphi_{A(1)}-2\frac{1}{N\Pi}e_{1A}\dvphi^A_{(1)} \partial_i\zeta_{2(1)},
\ee
or equivalently using (\ref{difz}) to rewrite the two last terms 
\bea
&&\!\!\!\!\!\!\!\!\!\!\!\!\!\!
\frac{1}{2}\zeta_{2i(2)}
%=\frac{1}{2}\partial_iQ_{2(2)}
%-\frac{\dot{\alpha}_{(1)}}{N\Pi}e_{2A}\partial_i\varphi^A_{(1)}
%-\frac{1}{N\Pi}e_{2A}\dvphi^A_{(1)} \partial_iQ_{1(1)}-\frac{1}{N\Pi}e_{1A}\dvphi^A_{(1)} \partial_i\zeta_{2(1)}
%\nn\\&&
%\qquad\qquad\!\!\!\!
=\partial_i
\Bigg[\frac{1}{2}Q_{2(2)}
+\frac{\hpe}{2}\Big[\lh\zeta_{1(1)}-\alpha_{(1)}\rh^2
-\zeta_{2(1)}^2\Big]
+(\e+\hpa)\lh\zeta_{1(1)}-\alpha_{(1)}\rh \zeta_{2(1)}
\label{another}
\\&&\!\!\!\!\!\!\!\!\!\!\!\!\!\!
\qquad\qquad\!\!
+\frac{1}{NH}\dot{\zeta}_{2(1)}\lh
\zeta_{1(1)}-\alpha_{(1)}\rh 
+\frac{1}{NH}\partial^{-2}\partial_j\lh\dot{\zeta}_{1(1)}
\partial_j\zeta_{2(1)}\rh \Bigg]
-\frac{1}{NH}\lh\zeta_{1(1)}-\alpha_{(1)}\rh\partial_i
\dot{\zeta}_{2(1)}
.\nn
\eea
In the uniform energy density gauge where $\tilde{\alpha}_{(2)}=\zeta_{1(2)}$ we find 
\be 
\frac{1}{2}\tilde{\zeta}_{2i(2)}=\partial_i\Bigg[\frac{1}{2}\tilde{Q}_{2(2)}-\frac{\hpe}{2}\zeta_{2(1)}^2
+\frac{1}{NH}\partial^{-2}\partial_j\lh\dot{\zeta}_{1(1)}
\partial_j\zeta_{2(1)}\rh \Bigg]
%+\dot{\zeta}_{1(1)}\partial_i\zeta_{2(1)}
\label{z2u},
\ee
For the flat gauge $\ha_{(r)}=0$ we find
% using (\ref{difzflat})
\bea 
&&\!\!\!\!\!\!\!\!\frac{1}{2}\hat{\zeta}_{2i(2)}=\partial_i\Bigg[\frac{1}{2}\hat{Q}_{2(2)}
+\frac{\hpe}{2}\lh\zeta_{1(1)}^2-\zeta_{2(1)}^2\rh
+(\e+\hpa)\zeta_{1(1)}\zeta_{2(1)}
+\frac{1}{NH}\dot{\zeta}_{2(1)}\zeta_{1(1)}
\nn\\
&&\qquad\qquad\!\!\
+\frac{1}{NH}\partial^{-2}\partial_j\lh\dot{\zeta}_{1(1)}
\partial_j\zeta_{2(1)}\rh 
\Bigg]
%+\dot{\zeta}_{1(1)}\partial_i\zeta_{2(1)}
-\frac{1}{NH}\zeta_{1(1)}\partial_i\dot{\zeta}_{2(1)}\label{z2f}.
\eea
We notice that the last term in the second line corresponds again to a gauge transformation familiar from the curvature perturbation case 
studied earlier (\ref{gat}). Using the gauge transformation (\ref{Q2tr}) we verify that the rest of the expression is a gauge-invariant quantity 
corresponding to the one in (\ref{z2u}). Indeed this expression is gauge-invariant beyond the long-wavelength approximation 
\bea
&&\!\!\!\!\!\!\!\!\!\!\!\frac{1}{2}\zeta_{2(2)}
=
\frac{1}{2}Q_{2(2)}
+\frac{\hpe}{2}\Big[\lh\zeta_{1(1)}-\alpha_{(1)}\rh^2
-\zeta_{2(1)}^2\Big]
+(\e+\hpa)\lh\zeta_{1(1)}-\alpha_{(1)}\rh \zeta_{2(1)}
 \nn\\&&
\qquad\qquad\!\!\!\!\!\!\!\!\!\!\!\!
+\frac{1}{NH}\dot{\zeta}_{2(1)}\lh
\zeta_{1(1)}-\alpha_{(1)}\rh 
+\frac{1}{NH}\partial^{-2}\partial_j\lh\dot{\zeta}_{1(1)}
\partial_j\zeta_{2(1)}\rh 
,
\eea

We conclude that the gradients of the perturbations allow not only to define the adiabatic gauge-invariant perturbation, but also the isocurvature one, which otherwise could not be defined using the gauge-transformation procedure followed in the previous subsection. 
%are in some sense equivalent to the perturbations themselves
%, since both 
%allow for the definition  of gauge-invariant second-order adiabatic and isocurvature quantities. 
In addition, as will be shown in the next section (and originally shown in \cite{Rigopoulos:2005xx}), 
%Furthermore, 
since the 
gradients are defined using fully non-linear quantities, they allow for an easy treatment of the second-order perturbation system. 
%, as was shown in \cite{Rigopoulos:2005xx}.

\subsection{The cubic action}\label{cub}

%\subsubsection{The third-order action}

As we did for the single-field case, we want to compute the cubic action of the perturbations in order to find the interactions that contribute to the intrinsic non-Gaussianity, related to $\f^{(3)}$,  and find the redefinitions related to the second-order gauge-invariant quantities and contributing to $\f^{(4)}$. 
Seery and Lidsey \cite{Seery:2005gb} performed this calculation for the multiple-field case in the flat gauge 
in terms of the scalar fields $\varphi^A$ and not of the adiabatic and isocurvature perturbations $\zeta_m$. They found 
no redefinitions, but their results would have to be supplemented 
by the $\delta N$ formalism (with its associated slow-roll approximation at horizon-crossing) to say anything about the non-Gaussianity 
of the gauge-invariant perturbations $\zeta_m$.

In this section we present our findings in \cite{Tzavara:2011hn} where we generalized this calculation  to second order in the expansion of the curvature perturbation in both the uniform 
energy-density gauge and the flat gauge. Doing so we computed the full form of the third-order action. The latter not only 
consists of the cubic interactions of the first-order curvature perturbations, but also of lower order interaction terms of the second 
order quantities. Here, we first perform the calculation relevant to the first-order quantities and then add the second-order 
effects. In this subsection we only present the scalar part of the action, but in appendix \ref{app3} the 
tensor part can be found as well. We emphasize that in this subsection we no longer make the long-wavelength approximation, so that 
the results are valid at any scale. 
We present here only the final results, while in appendix \ref{app3} we give the intermediate steps of the 
calculation.  
The scalar cubic action in the uniform energy-density gauge due to the first-order perturbations $\zeta_m$ takes the form
\bea
\tilde{S}_{3(1)}\!&=&S_{3(1)}-\int\d^4x\frac{\delta L_2}{\delta\zeta_m}f_m
\eea
with 
\bea
&&f_1=\frac{\e+\hpa}{2}\zeta_1^2-\hpe\zeta_1\zeta_2
+\frac{1}{NH}\dot{\zeta}_1\zeta_1-\frac{N^2}{4a^2}\Big((\partial\zeta_1)^2
-\partial^{-2}\partial^i\partial^j(\partial_i\zeta_1\partial_j\zeta_1)\Big)
\nn\\
&&\qquad
+\frac{1}{2}\Big(\partial^i\lambda\partial_i\zeta_1-\partial^{-2}
\partial^i\partial^j(\partial_i\lambda\partial_j\zeta_1)\Big),\nn\\
&&f_2=(\e+\hpa)\zeta_1\zeta_2
+\frac{1}{NH}\dot{\zeta}_2\zeta_1+\frac{\hpe}{2}\zeta_1^2,\label{rede1}
\eea
where we remind that $\partial^2\lambda=\e\lh \dot{\zeta}_1-2NH\hpe\zeta_2\rh$ and we have dropped the $(1)$ subscript of the first-order perturbations. 
The exact form of $S_{3(1)}$ can be found in appendix \ref{app3} or equivalently it is the cubic part of (\ref{fin}). 
The reason for introducing $S_{3(1)}$ without the tilde will become clear below. 

The terms proportional to 
$\delta L_2/\delta\zeta_m$, i.e.\  the first-order equations of motion, 
can be removed  by a redefinition of $\zeta_m$
\cite{Maldacena:2002vr} and lead to a change in the ground state of the perturbations. 
%This works as follows.  
As explained already in subsection \ref{s21action}, the cubic terms of the action (i.e.\  $\tilde{S}_{3(1)}$) are not affected by the redefinition, because the redefinition always 
involves terms proportional to $\zeta_m^2$, which would give quartic and not cubic corrections. 
It is only the second-order 
terms (i.e.\  $\tilde{S}_2$) that change. 
Indeed one can show that under a redefinition of the form $\zeta_m=\zeta_{mc}+f_m$, the second-order action changes as 
$S_2=S_{2c}+(\delta L_2/\delta\zeta_m)f_m$. These new terms cancel out the relevant terms coming from the cubic action 
(remember that the total action up to cubic order is the sum of the second and third-order action) and we are left 
with
\be
\tilde{S}_{3(1)}=S_{3(1)}(\zeta_{mc}).\label{ss3}
\ee

If we repeat the same calculations for the flat gauge (see appendix \ref{app3}), 
performing several integrations by part, we find that 
\be
\hat{S}_{3(1)}=S_{3(1)}(\zeta_{m}).
\ee 
This is a consequence of the action staying invariant under a gauge transformation. 
Nevertheless if one associates the redefinition appearing in the uniform energy-density gauge to a change in the ground state 
of $\zeta_m$, it would mean that directly after horizon crossing, when super-horizon effects have not yet 
been switched on, the second-order contribution to $\zeta_1$ would be zero for the flat gauge and non-zero for the uniform energy-density 
gauge. In terms of non-Gaussianity, this can be restated as: the non-Gaussianity present after horizon-crossing is different for the 
two gauges. 
Indeed if one was to calculate the three-point functions for the above action, one would need to perform 
two steps. First, change to the interaction picture, where it can be proved that the interaction Hamiltonian up to and 
including cubic order is just $H_{int}=-L_{int}$, where $L_{int}$ are the cubic terms of the Lagrangian, and compute the 
expectation value $\langle\zeta_c\zeta_c\zeta_c\rangle$ as in \cite{Maldacena:2002vr}. Second, take into account that the 
fields have been redefined as $\zeta=\zeta_c+A\zeta_c^2$. Then the three-point correlation function can be 
written as
\be
\langle\zeta\zeta\zeta\rangle=\langle\zeta_c\zeta_c\zeta_c\rangle
+2A[\langle\zeta_c\zeta_c\rangle
\langle\zeta_c\zeta_c\rangle+\mathrm{cyclic}].\label{cyclic}
\ee
These new terms, products of the second-order correlation functions, are only present in the uniform energy-density gauge 
if we restrict ourselves to $S_{3(1)}$. 

In order to cure this bad behavior we need to add to the above results the effect of the second-order fields. 
We find (see appendix \ref{app3})
\bea
\tilde{S}_{3(2)}=\int\d^4x\Bigg\{\frac{\delta L_2}{\delta\zeta_1}\lh\frac{\tilde{Q}_{1(2)}}{2}+\frac{\zeta_{1(2)}}{2}\rh+
\frac{\delta L_2}{\delta\zeta_2}\frac{\tilde{Q}_{2(2)}}{2}\Bigg\}.
\eea
Since all terms in $\tilde{S}_{3(2)}$ are proportional to $\delta L_2/\delta\zeta_m$, $\tilde{S}_{3(2)}$ only contains 
redefinitions of $\zeta_m$. 
Notice that the second-order lapse and shift functions do not appear in the final action, since these two 
are multiplied by a factor equal to the energy and momentum constraint equations (\ref{n1ni}). 
On the other hand, the second-order field perturbations are dynamical variables that obey second-order equations of motion that cannot 
be set to zero in the action. The single-field limit of this 
action is just the term proportional to $\zeta_{1(2)}$, since $\tilde{Q}_{m(i)}=0$ identically in that case for 
the uniform energy-density gauge. 
The term proportional to $\zeta_{1(2)}$ in $\tilde{S}_{3(2)}$, along with the terms proportional to $\lambda$ in $\tilde{S}_{3(1)}$, 
originate from the contribution of $N^i$ in the action. The latter vanish outside the horizon since then $\partial^2\lambda$ 
coincides with the super-horizon energy constraint and hence is identically zero. 
So if we were to study only the quadratic contributions of the first-order perturbations outside the horizon, 
we would be allowed not only to ignore the tensor parts of the metric \cite{Salopek:1990jq}, 
but also work in the time-orthogonal gauge $N^i=0$.
 
Coming back to the redefinition, its final form, including the tensor parts (see appendix \ref{app3}), is
\bea
&&\!\!\!\!\!\!\!\!\zeta_1=\zeta_{1c}-\frac{\zeta_{1(2)}}{2}-\frac{\tilde{Q}_{1(2)}}{2}+\frac{1}{NH}\dot{\zeta_1}\zeta_1+\frac{\e+\hpa}{2}\zeta_1^2
-\hpe\zeta_1\zeta_2
-\!\frac{N^2}{4a^2}\Big((\partial\zeta_1)^2
-\partial^{-2}\partial^i\partial^j(\partial_i\zeta_1\partial_j\zeta_1)\Big)\nn\\
&&\qquad\!\!\!\!\!\!\!\!
+\frac{1}{2}\Big(\partial^i\lambda\partial_i\zeta_1-\partial^{-2}\partial^i\partial^j(\partial_i\lambda\partial_j\zeta_1)\Big)
-\frac{1}{4}\partial^{-2}(\dot{\gamma}_{ij}\partial^i\partial^j\zeta_1),\nn\\
&&\zeta_2=\zeta_{2c}-\frac{\tilde{Q}_{2(2)}}{2}+\frac{1}{NH}\dot{\zeta_2}\zeta_1+\frac{\hpe}{2}\zeta_1^2+(\e+\hpa)\zeta_1\zeta_2.\label{rede2}
\eea

Finally we perform the above calculations for the flat gauge and find the action
\bea
\hat{S}_{3(2)}=\int\d^4x\Bigg\{\frac{\delta L_2}{\delta\zeta_1}\frac{\hat{Q}_{1(2)}}{2}
+\frac{\delta L_2}{\delta\zeta_2}\frac{\hat{Q}_{2(2)}}{2}\Bigg\}.
\eea
The redefinitions in the flat gauge take the simple form
\bea
&&\zeta_1=\zeta_{1c}-\frac{\hat{Q}_{1(2)}}{2}\nn\\
&&\zeta_2=\zeta_{2c}-\frac{\hat{Q}_{2(2)}}{2}.\label{red}
\eea

We want to write these redefinitions as well as the action itself in terms of gauge-invariant quantities and compare them. We would also like 
to compare with the definitions of the second-order gauge-invariant perturbations found in subsection \ref{gic}. 
After using the second-order constraint (\ref{r2}) and the uniform energy gauge definition of $\zeta_{2(2)}$ 
(\ref{z2u}) we can rewrite (\ref{rede2}) as
\bea
&&\!\!\!\!\!\!\!\!\!\!\!
\zeta_1+\frac{\zeta_{1(2)}}{2}=\zeta_{1c}+\frac{1}{NH}\dot{\zeta_1}\zeta_1+\frac{\e+\hpa}{2}\lh\zeta_1^2-\zeta_2^2\rh-\hpe\zeta_1\zeta_2
-\frac{1}{NH}\partial^{-2}\partial^i\lh\!\dot{\zeta}_2\partial_i\zeta_2\!\rh
-\frac{1}{4}\partial^{-2}(\dot{\gamma}_{ij}\partial^i\partial^j\zeta_1)
\nn\\
&&\qquad\qquad\quad\!\!\!\!\!\!\!\!\!\!\!
-\frac{N^2}{4a^2}\Big((\partial\zeta_1)^2
-\partial^{-2}\partial^i\partial^j(\partial_i\zeta_1\partial_j\zeta_1)\Big)
+\frac{1}{2}\Big(\partial^i\lambda\partial_i\zeta_1
-\partial^{-2}\partial^i\partial^j
(\partial_i\lambda\partial_j\zeta_1)\Big)
-\frac{1}{2\e NH}\partial^2\hat{\cA}\nn\\
&&\!\!\!\!\!\!\!\!\!\!\!
\zeta_2+\frac{\zeta_{2(2)}}{2}=\zeta_{2c}
+\frac{1}{NH}\dot{\zeta_2}\zeta_1+\frac{1}{NH}\zeta_2\dot{\zeta}_1+\frac{\hpe}{2}\lh\zeta_1^2-\zeta_2^2\rh+(\e+\hpa)\zeta_1\zeta_2
-\frac{1}{NH}\partial^{-2}\partial^i\lh\zeta_2\partial_i
\dot{\zeta}_1\rh.\nn\\ \label{finu}
\eea
When comparing the first equation of (\ref{finu}) with (\ref{flat2}), we see that we recover (\ref{red}). The same is true for the isocurvature 
part of the redefinition: comparing the second equation of (\ref{finu}) with (\ref{z2f}), we recover the redefinition 
for $\zeta_2$ (\ref{red}). 
Hence the two redefinitions are the same, as is necessary for the action to be gauge-invariant. 
Notice that the single-field limit of (\ref{finu}) is 
$\zeta_1+\zeta_{1(2)}/2=\zeta_{1c}+(\e+\hpa)\zeta_1^2/2$ 
in agreement with the total redefinition found in the uniform energy-density gauge. 

Equation (\ref{finu}) is the implicit definition of the redefined, gauge-invariant $\zeta_{mc}$. One can see that up to 
and including second order, it is a function of only the combination $\zeta_{m(1)} + \zeta_{m(2)}/2$. One can also notice 
that the purely second-order perturbation $\zeta_{m(2)}$ does not occur explicitly in the cubic action (see e.g.\  (\ref{fin}) below). 
Hence once could in principle
consider the quantities $\zeta_{m(2)}$ (and similarly $Q_{m(2)}$) as auxiliary quantities and try to avoid introducing them in the 
first place, but consider the quadratic first-order terms directly as a correction to the first-order perturbations, as is done for 
the single-field case in \cite{Rigopoulos:2011eq}. While the calculations would be roughly equivalent, we have chosen not to follow this route 
for two reasons. In the first place it seems conceptually simpler to us to expand the perturbations and the action consistently up to 
the required order, and more logical to view quadratic first-order terms as a correction to a second-order quantity than to a first-order 
one. Secondly, in the multiple-field case (as opposed to the single-field case), one would have to introduce the second-order quantities at 
some intermediate steps anyway in order to find the correct non-linear relation between the $Q_m$ and $\zeta_m$ (which is derived from the 
second-order gauge transformation performed in subsection \ref{gic}). 

So in the end we have managed to find the source of the non-Gaussianity present at horizon crossing due to first-order perturbations 
and identify it with the quadratic 
terms of (\ref{finu}). With source here we mean the second-order perturbation that, when contracted with two first-order 
perturbations, gives the bispectrum. 
Equation (\ref{finu}) is gauge-invariant, as it should be. 
Additionally, the redefinition of the perturbations  
that we perform is essential not only to simplify calculations but also to find the gauge-invariant form of the 
action itself. 
We clearly see that the quadratic corrections in the flat gauge seem to be zero if one takes into account only the first-order fields. 
In that gauge all of the second-order contributions are hidden in the second-order fields as opposed to the uniform energy-density 
gauge where part of the quadratic contributions is attributed to the redefinition of the first-order $\zeta_m$ and the rest of them lie 
in the second-order field. 

\subsubsection{Summary}\label{summary}

Let us summarize our results for the cubic action. Cosmological gauge-invariant perturbations 
should obey a gauge-invariant action. Using first-order perturbations the action up to third order is the same in the uniform 
energy-density gauge and the flat gauge only after a redefinition of $\zeta_m$ in the uniform energy-density gauge 
$\zeta_m=\zeta_{mc_1}+f_{m1}$ (\ref{rede1}) (the subscript $1$ indicating the use  of only first-order perturbations)  and takes the form 
\bea
S&=&\hat{S}(\zeta_{m})=\tilde{S}(\zeta_{mc_1})=\!\!\int\!\!\d^4x \frac{a\e N}{\kappa^2}
(\e\zeta_1-1)\Big((\partial\zeta_1)^2+(\partial\zeta_2)^2\Big)
\nn\\
&+&\!\!\int\!\!\d^4x\Bigg\{\!\frac{a^3\e}{N\kappa^2}\Bigg[\lh1+\e\zeta_1\rh\lh\dot{\zeta}_1^2+\dot{\zeta}_2^2\rh
-2\partial^i\lambda\lh\dot{\zeta}_2\partial_i\zeta_2+\dot{\zeta}_1\partial_i\zeta_1\rh\nn\\
&&\qquad\qquad\quad
-2(\e+\hpa)\zeta_2\partial^i\lambda\partial_i\zeta_2+4\hpe\zeta_2\partial^i\lambda\partial_i\zeta_1
+\frac{1}{2}\zeta_1\lh\partial^i\partial^j\lambda\partial_i\partial_j\lambda-\lh\partial^2\lambda\rh^2\rh\nn\\
&&\qquad\qquad\quad
+2NH\dot{\zeta}_2\lh\chi\zeta_2+\e\zeta_1\lh(\e+\hpa)
\zeta_2+\hpe\zeta_1\rh\rh
+NH\dot{\zeta}_1\lh-4\hpe\zeta_2+\e\zeta_1^2(3\hpa+2\e)
\rh\nn\\
&&\qquad\qquad\quad
+(NH)^2\zeta_1^2\lh \lh\sqrt{\frac{2\e}{\kappa}}\frac{W_{211}}{H^2}-2\e\lh\e\hpe+\hpa\hpe+\gx^\perp+3\hpe\rh\rh\zeta_2\right.\nn\\
&&\qquad\ \left.\qquad\qquad\quad\quad\quad\  
+\lh\sqrt{\frac{2\e}{\kappa}}\frac{W_{111}}{3H^2}-\e\lh\gx^\parallel+3\hpa-(\hpe)^2-(\hpa)^2\rh\rh\zeta_1
\rh\nn\\
&&\qquad\qquad\quad+
(NH)^2\zeta_2^2\lh
\sqrt{\frac{2\e}{\kappa}}\frac{W_{221}}{3H^2}-2\e^2-(\hpa)^2+3(\hpe)^2+\frac{2}{3}\hpe\gx^\perp
-3\e(\hpa-\chi)+2\hpa\chi\right.\nn\\
&&\qquad\qquad\ \ \qquad\quad\quad\quad\  
+
\Bigg(\sqrt{\frac{2\e}{\kappa}}\frac{W_{221}}{H^2}+\e\lh-3(\hpe)^2+(\e+\hpa)^2\rh+3\e(\chi-\e-\hpa)\Bigg)\zeta_1\nn\\
&&\left.\qquad\qquad\ \ \qquad\quad\quad\quad\  
+\sqrt{\frac{2\e}{\kappa}}\frac{W_{222}}{3H^2}\zeta_2\rh\Bigg]\Bigg\}
\label{fin}
\eea
where we have kept the notation $\partial^2\lambda=\e\lh\dot{\zeta}_1-2NH\hpe \zeta_2\rh$ in order to mark clearly the terms 
that vanish outside the horizon, namely the terms proportional to $\lambda$ along with the terms involving second-order space derivatives. 
This is one of our main results. 
We managed to compute the cubic action for adiabatic and isocurvature perturbations in the exact theory, beyond any super-horizon or 
slow-roll approximation. 
Its single-field limit coincides with the action computed by Maldacena in \cite{Maldacena:2002vr} or by Rigopoulos in 
\cite{Rigopoulos:2011eq}, namely expression (\ref{s31}). 
Let us examine the implications of this action. 
Forgetting about the redefinition of the perturbations in the uniform energy-density gauge, the form of the action is gauge-invariant. 
One can use it to easily calculate the non-Gaussianity related to the interaction terms as is explained in detail 
in \cite{Seery:2005wm,Weinberg:2005vy}. This is known in the literature as $f_{NL}^{(3)}$, the parameter of non-Gaussianity 
related to the three-point correlation function of three first-order perturbations, which is only non-zero in the case of intrinsic 
non-Gaussianity.

However, taking into account the need for a redefinition of the perturbations in the uniform energy-density gauge, one might worry that 
the action is not actually gauge-invariant. 
The action in the uniform energy-density gauge before the redefinition  
has extra terms that are proportional to the second-order equations that the perturbations obey. 
This means that when calculating the non-Gaussianity in the uniform energy-density gauge, one not only has contributions due to the 
interaction terms in the cubic action, but also ones due to the redefinition of $\zeta_m$, which contribute as explained in (\ref{cyclic}). 
They are part of what is known in the literature as $f_{NL}^{(4)}$, the parameter of non-Gaussianity related to the 
three-point correlation function of a second-order perturbation (in terms of products of first-order ones) and two first-order perturbations, 
which reduces to products of two-point functions of the first-order perturbations. 

This would mean that the non-Gaussianity calculated in the two gauges would not be the same due to the lack of any redefinition in the flat 
gauge. 
However, if one takes into account only the corrections coming from first-order perturbations, the redefinition associated to 
the second-order 
perturbation is not complete as one can check by comparing the super-horizon version of the adiabatic part of (\ref{rede1}) with (\ref{flat2}).   
As we showed, the solution of this issue is to include second-order fields since they also contribute to the cubic 
action. 
As one would expect these do not change the action itself, so that 
(\ref{fin}) still holds. 
The effect of the new terms is to redefine the perturbations in both gauges. 
It should be noted that, if one had incorporated all quadratic first-order terms (found by a second-order gauge transformation 
as in subsection \ref{gic}) directly as a correction to the first-order perturbations, one would have found the two contributions 
$S_{3(1)}$ and $S_{3(2)}$ together and hence there would have been no initial discrepancy between the two gauges. 
However, we explained before our reasons for proceeding in this way. 
So in any case we finally obtain
\be
S=\hat{S}(\zeta_{mc})=\tilde{S}(\zeta_{mc}),\label{finale}
\ee
where $\zeta_{mc}$ is given in (\ref{finu}).  
Now the two redefinitions as well as the action itself are the same for the two gauges, 
hence the action is truly gauge-invariant and the $f_{NL}^{(4)}$, related to the products of first-order $\zeta_m$ in the redefinitions, 
is the same in the two gauges. In the next chapter we will use these redefinitions to calculate the related super-horizon $\f^{(4)}$ and incorporate them as the horizon-crossing source of non-Gaussianity in the long-wavelength formalism. 

This exact action allows one to compute $f_{NL}^{(4)}$ without the need for 
the slow-roll approximation at horizon crossing that is essential for both the long-wavelength formalism and the $\delta N$ formalism: 
the long-wavelength 
formalism needs slow-roll at horizon crossing in order to allow for the decaying mode to vanish rapidly, while the $\delta N$ 
formalism requires it in order to ignore the derivatives with respect to the canonical momentum. 
Additionally, up to now only 
the slow-roll \textit{field} action \cite{Seery:2005gb}  (and not the action of the $\zeta_m$ themselves) was known, 
so in order to compute the non-Gaussianity at horizon crossing one had to use the long-wavelength 
or $\delta N$ formalism to transform to $\zeta_m$ 
and hence one was in any case required to  make the assumption of slow-roll, 
even if the exact action for the fields would have been known. It will be interesting to investigate models that do not satisfy 
the conditions for the long-wavelength or $\delta N$ formalism using the action (\ref{fin}).

%In order to connect the redefinitions to some previously derived results in the literature we assume the super-horizon and slow-roll approximations. The 
%super-horizon approximation is already assumed in (\ref{finu}) and it can be supplemented by the condition $\dot{\zeta}_1=2\hpe\zeta_2$. 
%The slow-roll assumption translates into $\dot{\zeta}_2=-\chi\zeta_2$. Then 
%the quadratic part of the redefinitions, relevant to $f_{NL}^{(4)}$, takes the form 
%\bea
%&&\zeta_1=\zeta_{1c}+\frac{\e+\hpa}{2}\zeta_1^2+\hpe\zeta_2\zeta_1-\frac{\e+\hpa-\chi}{2}\zeta_2^2\nn\\
%&&\zeta_2=\zeta_{2c}+\frac{\hpe}{2}\zeta_1^2+(\e+\hpa-\chi)\zeta_1\zeta_2+\frac{\hpe}{2}\zeta_2^2.
%\eea
%The redefinitions in this form were used in \cite{Tzavara:2010ge} to find the second-order source term of the evolution equations for the 
%super-horizon perturbations. 
It is worth noting that the super-horizon limit of the tensor part of the cubic action, found in appendix \ref{AppB}, contains interactions only of the tensor and the adiabatic perturbations. The isocurvature perturbations do not interact with the gravity waves outside the horizon. 
%This is in agreement with the discussion in subsection \ref{gic}, where we concluded that the isocurvature perturbation does not interact directly with gravity, but only through its interaction with the adiabatic perturbation. 
Although interesting, we will not further elaborate on this subject, since the study of tensor perturbations is beyond the scope of this thesis.  

\Section{The long-wavelength formalism}\label{long}

Up until now, we distinguished our study of perturbations in first and second order. Indeed, we found the action and gauge-invariant expressions for both cases and managed to solve the equations of motion for the first-order perturbations around horizon crossing (notice that the study of the super-horizon evolution of the first-order perturbations was done in \cite{GrootNibbelink:2000vx,GrootNibbelink:2001qt}). However, studying the evolution of the  second-order system turns out to be non-trivial and one needs to simplify the calculation by using assumptions. Since we are mainly interested in the super-horizon evolution of the perturbations, this is the first assumption to make. Furthermore, it turns out that it is easier to first find the super-horizon equations of motion for the fully non-linear quantities and then perturb up to second order.    

The above procedure is called the long-wavelength formalism and was originally developed by Rigopoulos, 
Shellard and
Van Tent \cite{Rigopoulos:2005xx, Rigopoulos:2005ae, Rigopoulos:2005us}, 
hereafter refered to as RSvT, based on the pioneer work of \cite{Salopek:1990jq}. 
The long-wavelength formalism corresponds to the leading-order 
approximation of the spatial gradient expansion 
(see \cite{Salopek:1990jq,Tanaka:2006zp} and references therein). 
In this expansion all quantities are expanded in terms of a small parameter 
$1/(HL)$, where $L$ is the characteristic physical length scale of the 
perturbations (i.e.\  proportional to $a$). The leading-order approximation 
of the spatial gradient expansion is equivalent to neglecting the $k^2$ term 
(which comes from the second-order spatial gradient and is of order 
$\cO(1/(HL)^2)$) with respect to the $\cO(1)$ terms in the 
equation for the perturbation modes. Because of the very rapid growth of $a$ 
during inflation, this is in principle a well-justified approximation from 
just a few e-folds after horizon crossing of the perturbation mode under 
consideration, when the decaying mode will have disappeared. 
However, as pointed out in \cite{Leach:2001zf,Takamizu:2010xy},
if slow roll is broken at horizon crossing and for some e-folds afterwards,
a cancellation of the $\cO(1)$ terms can cause the decaying mode to remain 
important during this period. This is easily seen even in the single-field case 
\be 
\ddot{\zeta}_1+NH\lh3+2\e+2\hpa-\frac{\dot{N}}{HN^2}\rh\dot{\zeta}_1 -\frac{N^2}{\alpha^2}\partial^2\zeta_1=0.\nn
\ee 
When the slow-roll parameters become of order $\mathcal{O}(1)$  the factor of the first derivative of the perturbation can be of the same order as the gradient term. 
 In those papers it was shown that 
for single-field inflation there may be an enhancement of
the curvature perturbation both at first and second order due to the effect 
of the $k^2$ term even on super-horizon scales, if the decaying mode has 
not yet vanished. Hence, in order for the long-wavelength 
approximation to be valid on super-horizon scales (our first assumption), we also need to assume slow roll to hold around horizon crossing (our second assumption),
for the decaying mode to quickly disappear.  
%Hence, the two assumptions needed to study the super-horizon evolution of the perturbations within the long-wavelength formalism  neglecting the $k^2$ terms is our first assumption.  

\subsection{The set-up}

We start by considering the form of the metric (\ref{metricexact}) relevant for super-horizon calculations and how can this be simplified. We shall choose a gauge where $N_i=0$, hence we set $3$ degrees of freedom,  $N_{i\perp}=0$ and $\partial_i\psi=0$. The trace $\bE$ and traceless part 
$\widehat{E}$ of the extrinsic curvature become with this gauge choice
\be
\bE=3\lh NH+\dot{\alpha}\rh=\partial_t\ln\sqrt{\bh},\qquad
\widehat{E}_{ij}\equiv \bE_{ij}-\frac{1}{3}\bE\delta_{ij}=\frac{1}{2}a^2\mathrm{e}^{2\alpha\delta_{ij}}\partial_t
\mathrm{e}^{\chi_{ij}}.\label{long1}   
\ee
Within the long-wavelength approximation, where  we can ignore spatial gradients, the action (\ref{actionexact1}) becomes 
\be
S=\frac{1}{2}\int\d^4x\sqrt{\bh}\Bigg\{
%\bN\Big[\frac{R^{(3)}}{\kappa^2}-2W\Big] 
-2\bN W
+\frac{1}{\bN}\Big[
\frac{1}{\kappa^2}\lh \bE_{ij}\bE^{ij}-\bE^2\rh
+\dot{\varphi}^2\Big]\Bigg\},\label{actionlong} 
\ee
where we ignored all gradient terms, i.e.\ the intrinsic curvature and gradients of the fields. We can find the evolution of the traceless part of the extrinsic curvature varying this action with respect to the spatial metric \cite{Salopek:1990jq}
\be
\frac{\partial}{\partial t}\ln\lh\frac{\widehat{E}^i_j}{N}\rh=-\bE,\quad
\mathrm{with\ solution}\quad
\widehat{E}^i_j=C^i_j(\vc{x})\frac{1}{a^3}.
\ee
Hence, $\widehat{E}^i_j$ decays exponentially fast during inflation. From now on we will set it to zero. From equation (\ref{long1}) we see that this is equivalent to demanding that $\chi_{ij}$ depends on $\vc{x}$ only. In other words, $\chi_{ij}$ does not participate in the long-wavelength dynamics. We have already checked at first order in subsection \ref{tensor} and appendix \ref{AppB} that gravity waves freeze on such scales.
Given the above, the metric (\ref{metricexact}) reduces to 
\be 
\d s^2=-\bN^2\d t^2+a^2\mathrm{e}^{2\alpha}\chi_{ij}(\vc{x})\d x^i\d x^j.
\ee
To make the connection with the previous paragraphs, the above arguments are in agreement with our findings so far assuming $N^i=0$. 
%Hence the confusion due to equation (\ref{r2}) is avoided and the gradient terms in this equation are by definition zero within the long-wavelength approximation. 

The generalised fully non-linear Hubble parameter (\ref{genhubble}) takes the form 
\be
\bH =\frac{1}{\bN}\lh\dot{\alpha}+HN\rh . 
\ee
In order to simplify super-horizon calculations we choose to work in a flat gauge 
\be\bN\bH=1,\label{flatga}\ee
that is we choose time slices in which the expansion of the 
universe is homogeneous and the time variable coincides with the number of
e-folds $t=\alpha+\ln{a}$. From now on we will drop the hat for quantities in the flat gauge to lighten the notation. 

 The energy and momentum constraints become for the fully non-linear fields
\be 
\bH^2=\frac{\kappa^2}{3}\lh\frac{1}{2}\bar{\Pi}^2+\bW\rh ,\qquad
\partial_i\bH=-\frac{\kappa^2}{2}\bar{\Pi}_A\partial_i\varphi^A.\label{eqlw1}
\ee
Variation of the action (\ref{actionlong}) with respect to $\bh_{ij}$ gives the dynamical equation for the evolution of the Hubble parameter
\be 
\dot{\bH}=-\frac{\kappa^2}{2\bH}\bar{\Pi}^2.
\ee
Finally the field equation becomes
\be
\cD_t\bar{\Pi}^A=-3
%\bN\bH
\bar{\Pi}^A-\frac{1}{
\bH}\bar{W}^{,A},
\ee
where a superscript $,A$ denotes a derivative $W^{,A}\equiv\partial W/\partial\varphi_A$. In the context of the separate universe picture, these fully non-linear equations represent the FLRW separate evolution of each point of the long-wavelength universe, related with each other by the second equation in (\ref{eqlw1}).

Combining these equations one can construct the long-wavelength evolution equation for the gradient of the perturbations $\zeta_{im}$
\bea
&&\dot{ \theta}_{i1}+\lh3+
%2
\bar{\e}+
2\bar{\eta}^\parallel
%-\frac{\dot{\bN}}{\bN^2\bH}
\rh 
%\bN\bH
\theta_{i1}-2\lh3\bar{\eta}^\perp
+2\bar{\e}\bar{\eta}^\parallel+\bar{\xi}^\perp\rh
%(\bN\bH)^2
\zeta_{i2}-2\bar{\eta}^\perp 
%\bN\bH
\theta_{i2}=0,\nn\\
&&
\dot{\theta}_{i2}+
2\bar{\eta}^\perp 
%\bN\bH
\theta_{1i}
+\lh 3\bar{\chi}+2\bar{\e}^2+4\bar{\e}
\bar{\eta}^\parallel+\ 
\bar{\xi}^\parallel
 \rh
% (\bN\bH)^2
\zeta_{i2}
+\lh3+
%2
\bar{\e}+
2\bar{\eta}^\parallel
%-\frac{\dot{\bN}}{\bN^2\bH}
\rh 
%\bN\bH
\theta_{i2}=0,\label{eqlw}
%\cD_t^2{\zeta}_i^A+NH\lh3+2\bar{\e}+
%2\bar{\eta}^\parallel-
%\frac{\dot{N}}{N^2H}\rh\cD_t{\zeta}_i^A
%+(\bN\bH)^2\bar{\Xi}^A_B\zeta_i^B=0.
\eea
%The matrix $\Xi^A_B$ takes the form
%\be
%\Xi^A_B=\frac{\bar{W}^A_B}{\bH^2}+ 
%\lh 3\bar{\e}+3\bar{\eta}^\parallel+2\bar{\e}^2+4\bar{\e}
%\bar{\eta}^\parallel+(\bar{\eta}^\perp)^2+\bar{\xi}
%^\parallel\rh
%\delta^A_B-
%2\bar{\e}\lh(3+\bar{\e})\bar{e}_1^A\bar{e}_{1B}
%+\bar{e}_1^A\bar{\eta}_B+\bar{\eta}^A
%\bar{e}_{1B}\rh 
%\ee
where we defined the time derivative of the perturbations $\theta_{im}\equiv\dot{\zeta}_{im}$.  No slow-roll assumption has been made, the slow-roll parameters can be thought of as short-hand notation.  
Notice that the slow-roll parameters and the unit vectors appearing in this equations are the fully non-linear quantities
\bea
\dot{\bar{H}}\equiv-\bar{\e}\bN\bH^2,\quad
\bar{\eta}^{(n)A}\equiv\frac{1}{\bH^{n-1}\bar{\Pi}}\lh
\frac{1}{\bN}\frac{\d}{\d t}
\rh^{n-1}\bar{\Pi},\quad
\bar{e}_{1A}\equiv\frac{\bar{\Pi}^A}{\bar{\Pi}}.
\eea 
Their time derivatives retain the form (\ref{dere}) where now all quantities should be substituted by the fully non-linear ones. Eventually, we are interested in expanding the equation of motion (\ref{eqlw}) for the adiabatic perturbation up to second order (in order to calculate the bispectrum), so one also needs expressions for the gradients of the slow-roll parameters and the unit vectors. The gradients involved in this calculation can be found using the four equations (\ref{eqlw1}-\ref{eqlw}) 
\bea
&&
\partial_i\bar{\e}=
-2\bar{\e}\lh \bar{\e}+\bar{\eta}^\parallel\rh\zeta_{i1}
-2\bar{\eta}^\perp\zeta_{i2},
\nn\\
&&
\partial_i\bar{e}_{1A}=-\bar{e}_{2A}\lh\bar{\eta}^\perp\zeta_{i1}
-\lh\bar{\e}+\bar{\eta}^\parallel\rh\zeta_{i2}
-\theta_{i2}\rh,\quad
\partial_i\bar{e}_{2A}=\bar{e}_{1A}\lh\bar{\eta}^\perp\zeta_{i1}
+\lh\bar{\e}+\bar{\eta}^\parallel\rh\zeta_{i2}
+\theta_{i2}\rh ,
\nn\\
&&\partial_i\bar{\eta}^\parallel=\lh-\bar{\e}\bar{\eta}^\parallel
+(\bar{\eta}^\parallel)^2
-(\bar{\eta}^\perp)^2-\bar{\xi}^\parallel\rh
\zeta_{i1} 
-\lh\bar{\e}\bar{\eta}^\perp+\bar{\xi}^\perp
\rh\zeta_{i2}-\bar{\eta}^\perp\theta_{i2},\nn\\
&&
 \partial_i\bar{\eta}^\perp=
\lh-\bar{\e}\bar{\eta}^\perp+2\bar{\eta}^\parallel
\bar{\eta}^\perp-\bar{\xi}^\perp
\rh\zeta_{i1}
+\lh
 \bar{\e}\bar{\eta}^\parallel
+(\bar{\eta}^\parallel)^2
+(\bar{\eta}^\perp)^2+3\bar{\chi}\rh
\zeta_{i2} 
+\lh3+\bar{\eta}^\parallel\rh \theta_{i2},
\nn\\
&&
\partial_i\bar{\chi}=-\lh
W_{221}-(\bar{\eta}^\parallel)^2
+3(\bar{\eta}^\perp)^2+\bar{\xi}^\parallel
+\frac{2}{3}\bar{\eta}^\perp\bar{\xi}^\perp
+\bar{\e}\lh\bar{\eta}^\parallel+2\bar{\chi} \rh
\rh\zeta_{i1}\nn\\
&&\qquad\quad\!
-\lh W_{222}+3\bar{\e}\bar{\eta}^\perp+\bar{\xi}^\perp
+\frac{2}{3}\lh\bar{\e}+\bar{\eta}^\parallel\rh\lh 3\bar{\eta}^\perp+\bar{\xi}^\perp\rh \rh\zeta_{i2}
-\lh 3\bar{\eta}^\perp+
\frac{2}{3}\bar{\xi}^\perp\rh\theta_{i2}
\nn\\
&&
\partial_i\bar{\xi}^\parallel=
3\lh
W_{111}-3\bar{\e}\bar{\eta}^\parallel
-(\bar{\eta}^\parallel)^2-(\bar{\eta}^\perp)^2
+\bar{\xi}^\parallel-\frac{2}{3}\lh\bar{\e}\bar{\xi}^\parallel+\bar{\eta}^\perp
\bar{\xi}^\perp\rh 
\rh\zeta_{i1}\nn\\
&&\qquad\quad\!
+3\lh
W_{211}-3\bar{\e}\bar{\eta}^\perp
-2\bar{\eta}^\parallel\bar{\eta}^\perp
+\bar{\xi}^\perp-\frac{2}{3}\lh\bar{\e}+\bar{\eta}^\parallel\rh
\bar{\xi}^\perp
\rh\zeta_{i2}
-\lh 3\bar{\eta}^\perp+2\bar{\xi}^\perp\rh\theta_{i2}
%\nn\\
%&&
%\partial_i\bar{\xi}^
\label{partialder}
\eea

Two comments are in order regarding the first-order limit of the equations (\ref{eqlw}). First, comparing to (\ref{eql3}) one can check that in the long-wavelength limit these agree with the first order version of (\ref{eqlw}). Indeed at linear order, the slow-roll parameters reduce to their background values and $\zeta_{im(1)}=\partial_i\zeta_{m(1)}$, so that the gradient can be removed. 
Second, it was shown in (\ref{mch}) that at first order in the perturbation expansion  $\dot{\zeta}_{1(1)}=2NH\hpe \zeta_{2(1)}$. One can check by direct substitution in equations (\ref{eqlw}) that this relation is generalised for the gradient of the perturbations to the fully non-linear case \cite{Rigopoulos:2005us}
\be
\theta_{i1}=2\bN\bH\bar{\eta}^\perp \zeta_{i2},
\ee
so that we do not need to consider $\theta_{i1}$ as an independent variable. 

Hence, using the above constraint equation, we can rewrite equations (\ref{eqlw}) as a matrix equation for the vector $v_{ia}\equiv\lh\zeta_{i1},\zeta_{i2},\theta_{i2}\rh^T$
\be
\dot{v}_{ia}(t,\vc{x})+A_{ab}(t,\vc{x})v_{ib}(t,\vc{x})=0, \label{mas} 
\ee
where the matrix $A_{ab}$ has the form
\be 
\mathbf{A}=
\lh 
\begin{array}{ccc}
0 & -2\bar{\eta}^\perp
%(\bN\bH)^2 
& 0\\
0 & 0 & -1
%\bN\bH
\\
0 & \lh3\bar{\chi}+2\bar{\e}^2
+4\bar{\e}\bar{\eta}^\parallel
+4(\bar{\eta}^\perp)^2
+\bar{\xi}^\parallel\rh
%(\bN\bH)^2 
& 
\lh3+
%2
\bar{\e}+
2\bar{\eta}^\parallel
%-\frac{\dot{N}}{N^2H}
\rh 
%\bN\bH
\end{array}
\rh.  \label{Amat}
\ee

\subsection{Solving the equations}

%The third assumption of the long-wavelength formalism is related to solving 
Equation (\ref{mas}) is a long-wavelength equation. To first order it is the super-horizon limit of equations (\ref{eql3}), i.e.\ ignoring the gradient of the perturbations. Hence, it needs to be supplemented with initial conditions at horizon crossing, that will act as source terms in the right-hand side of the equation. Expanding to second order, we find
\bea
&&\dot{v}_{ia}^{(1)}+A_{ab}^{(0)}(t)v_{ib}^{(1)}=
b_{ia}^{(1)}(t,\vc{x}),\label{eqlw11}\\ 
&&\dot{v}_{ia}^{(2)}+A_{ab}^{(0)}(t)v_{ib}^{(2)}=
-A_{ab}^{(1)}(t,\vc{x})v_{ib}^{(1)}
+b_{ia}^{(2)}(t,\vc{x}).\label{eqlw2}
\eea 
In order to understand how the source term  $b_{ia}^{(1)}$ arises, let us consider the full equations of motion at linear order (\ref{eql3}). We then define smoothed long-wavelength variables, corresponding to the perturbations we are solving for within the long-wavelength formalism, using a window function $\cW$ with smoothing length $R\equiv c/(aH)$ and $c$ a constant of the order of a few. In Fourier space this means $\zeta_m(k)=\cW(kR)\zeta_{m,\mathrm{lin}}(k)$, where $\zeta_{m,\mathrm{lin}}(k)$ are the solutions of the exact equations (\ref{eql3})  , and an identical expression for $\theta_m$. Rewriting equations (\ref{eql3}) in terms of the smoothed variables, one is left with terms that depend on the exact variable $\zeta_{m}$. These terms form the source $b^{(1)}_{ia}$.

The window function is chosen so that it ensures that short wavelengths are cut out, making our long-wavelength approximation applicable, $\cW(kR)\rightarrow 0$ for $k\mg 1/R$  and ensuring that at sufficiently late times the solution of the linear-order equation (\ref{eqlw11}) does not depend on the exact shape of $\cW$, $\cW(kR)\rightarrow 1$ for $k\ll 1/R$. These conditions imply that $\dot{\cW}(kR)$ peaks around a time just after horizon crossing and hence we only need to know $\zeta_{m,\mathrm{lin}}$ around this time. 
The source $b_{ia}^{(1)}$ can then be expressed in terms of the horizon-crossing solutions $X_{am}$ as
\be
b_{ia}^{(1)}=\int\frac{\d^3\vc{k}}{(2\pi)^{3/2}}\dot{\mathcal{W}}(k)X_{am}^{
(1)}(k)
a_{m\vc{k}}^{\dagger}\rmi k_i \rme^{\rmi\vc{k}\cdot\vc{x}}+\mathrm{c.c.}.
\ee
The quantum creation ($a^{\dagger}_{m\vc{k}}$) and conjugate annihilation 
($a_{m\vc{k}}$) operators satisfy the usual commutation relations. The linear
horizon-crossing solutions can be determined exactly numerically, or analytically within
the slow-roll approximation (which, as observations indicate, seems to be a
very good approximation at horizon crossing). Within the long-wavelength formalism we have to assume slow-roll during horizon crossing, as was explained in the beginning of this section, and hence we will use the analytical expression for the linear perturbation at horizon crossing derived in subsection \ref{sphc} (see (\ref{x01})).

The matrix $A_{ab}^{(1)}$ is found by perturbing the exact $\mx{A}$ matrix, using (\ref{partialder}), giving
$A_{ab}^{(1)}(t,\vc{x})=\bar{A}^{(0)}_{abc}(t)v_c^{(1)}(t,\vc{x})$. 
The explicit form of $\mx{\bar{A}}$ is given in (\ref{Abar}), where we
have dropped the superscript $(0)$ for notational convenience.
We have also defined
$v_c^{(1)}\equiv \partial^{-2}\partial^iv_{ic}^{(1)}$.

The effect of the source term $b_{ia}^{(2)}$ on $\f^{(4)}$ is expected to be small, since it is a second-order horizon-crossing contribution. Nevertheless, we compute it 
explicitly in order to allow for an exact comparison with known results in the
literature. Identically to the first-order case, $b_{ia}^{(2)}$ is the second-order exact perturbation  smoothed by the window function $\cW$. As was  explained above, because of the form of the window function we only need the second-order exact solution around horizon crossing. 
These contributions can be found from the cubic action, found in section \ref{cub} and they are just the quadratic terms in the super-horizon redefinitions of the  perturbations (\ref{finu}),  or equivalently the quadratic terms of the second-order gauge transformation that survive outside the horizon and therefore can be computed analytically within the slow-roll assumption. Therefore, the second-order source term is a convolution integral
\bea\label{secondsource}
 b_{ia}^{(2)}= && \int\frac{\d^3\vc{k}}{(2\pi)^{3/2}}\int\frac{\d^3\vc{k'}}{(2\pi)^{3/2}}
\dot{\mathcal{W}}(\mathrm{max}(k',k))\nn\\
&&\times \Bigg\{L_{abc}(t)X_{bm}^{(1)}(k',t)X_{cn}^{(1)}(k,t)
a_{m\vc{k'}}^{\dagger}
a_{n\vc{k}}^{\dagger}\rmi (k'_i+k_i)
\rme^{\rmi(\vc{k'}+\vc{k})\cdot\vc{x}}\nn\\
&&\ \ \ \ \  +N_{abc}(t)X_{bm}^{(1)}(k',t)X_{cn}^{(1)}(k,t)
a_{m\vc{k'}}^{\dagger}
a_{n\vc{k}}^{\dagger}\rmi k_i\rme^{\rmi(\vc{k'}+\vc{k})\cdot\vc{x}}
+\mathrm{c.c.}\Bigg\},
\eea
where the derivative of the window function peaks at the scale that exits 
the horizon last. 
We have split $b_{ia}^{(2)}$ into a local part proportional to $L_{abc}$ and 
a non-local part proportional 
to $N_{abc}$. The explicit components of $L_{abc}$ and $N_{abc}$ are given in (\ref{nl}) and appendix \ref{appSecondSource}. 
%The explicit expressions for the components of $L_{abc}$ and $N_{abc}$ are given in section~\ref{secGreen}.

We are ready now to solve (\ref{eqlw11}) and  (\ref{eqlw2}) in order to compute the evolution of the super-horizon observable quantities. These are the power spectrum and bispectrum of the adiabatic primordial perturbations, along with their scale dependence encoded through their spectral indices.

%------------------------------ Chapter 4 --------------------------------------

{\Chapter{The super-horizon evolution of the perturbations}\label{Ch4}}
%------------------------------------------------------------------------------------------

Here we find the expression for the super-horizon bispectrum within the long-wavelength formalism. We first derive an exact result, assuming slow-roll only at horizon crossing. We compute this for an equilateral triangle, effectively assuming that the three scales of the triangle are of the same order and hence $\f$ can be computed at a single pivot scale $k_*$. We also provide the formula for a squeezed triangle. Next we apply the slow-roll assumption during all the inflationary period in order to simplify the expressions and eventually, in the next chapter, deduce qualitative conclusions about the behaviour of $\f$. This chapter is based on our findings in \cite{Tzavara:2010ge}. 

\Section{The set-up}\label{S41}

In this section we discuss the tools needed to solve equations (\ref{eqlw11}) and (\ref{eqlw2}) and make contact between the solution and the bispectrum of the adiabatic perturbations. We also discuss the properties of the  $\delta$N formalism, another formalism to compute the super-horizon non-Gaussianity. 

\subsection{Green's functions}
\label{secGreen}

Equations (\ref{eqlw11}) and (\ref{eqlw2}), together with the initial condition
$v_{ia}\left(t\rightarrow-\infty\right)=0$, can be solved using a simple Green's
function $G_{ab}(t,t')$. In matrix notation it satisfies
\cite{Rigopoulos:2005us,Rigopoulos:2005ae}
	\be \label{Greeneqmot}
	\frac{\d}{\d t} \mx{G}(t,t') + \mx{A}(t) \mx{G}(t,t') = \mx{0},
	\qquad\qquad
	\mx{G}(t,t) = \mx{1}.
	\ee 
Starting from this equation, to lighten the notation, when we write $\mx{A}$ 
we actually mean $\mx{A}^{(0)}$, i.e.\ the matrix in (\ref{Amat}) with all 
local slow-roll parameters replaced by their background version that depends 
on time only.
Looking at this equation of motion and its initial condition, we see that the
solution can be written as
	\be \label{GisFFinv}
	\mx{G}(t,t') = \mx{F}(t) \mx{F}\inv(t'),
	\ee 
where $\mx{F}(t)$ satisfies the same equation of motion (\ref{Greeneqmot}) as
$\mx{G}(t,t')$ with an arbitrary initial condition. From this we immediately
derive that
	\be 
	\frac{\d}{\d t'} \mx{G}(t,t') - \mx{G}(t,t') \mx{A}(t') = \mx{0}.
\label{grstar}
	\ee 

The solution of (\ref{eqlw11}) and (\ref{eqlw2}) can now be written as the 
time integral of $G_{ab}$ contracted with the terms on
the right-hand side of these equations:
\be\label{via1sol}
v_{ia}^{(1)}(t,\vc{x})=\int\frac{\d^3\vc{k}}{(2\pi)^{3/2}}v_{am}(k,
t)\hat { a }
_m^{\dagger}(\vc{k})\rmi k_i \rme^{\rmi\vc{k}\cdot\vc{x}}+\mathrm{c.c.},
\ee
with
\be
v_{am}(k,t)=\int_{-\infty}^t\d
t'G_{ab}(t,t')\dot{\mathcal{W}}(k,t')X_{bm}^{(1)}(k,t')
\ee
and
\be
\label{u2hor}
 v_{ia}^{(2)}(t,\vc{x})=-\int_{-\infty}^t\!\!\!\d
t'G_{ab}(t,t')\bar{A}_{bcd}(t')v_{ic}^{(1)}(t',\vc{x})v_{d}^{(1)}(t',\vc{x})+
\int_{-\infty}^t\!\!\!\d t'G_{ab}(t,t')b_{ib}^{(2)}(t',\vc{x}).
\ee
As before, $v_d^{(1)}\equiv \partial^{-2}\partial^iv_{id}^{(1)}$.

Written in components (\ref{Greeneqmot}) gives the following equations
for the Green's functions:
	\bea\label{Greeneqmot2f}
	\frac{\d}{\d t} G_{1x}(t,t') & = & 2 \get^\perp(t) G_{2x}(t,t'),
	\non\\
	\frac{\d}{\d t} G_{2x}(t,t') & = & G_{3x}(t,t'),
	\\ 
	\frac{\d}{\d t} G_{3x}(t,t') & = & - A_{32}(t) G_{2x}(t,t')
	- A_{33}(t) G_{3x}(t,t'),
	\non\\
	G_{ab}(t,t) & = & \gd_{ab}.
	\non
	\eea
We can also rewrite this as a second-order differential equation for $G_{2x}$:
	\be \label{G22eq}
	\frac{\d^2}{\d t^2} G_{2x}(t,t') + A_{33}(t) \frac{\d}{\d t}
	G_{2x}(t,t') + A_{32}(t) G_{2x}(t,t') = 0.
	\ee 
For the derivatives with respect to $t'$ we find:
	\bea\label{Greeneqmot2ftp}
	\frac{\d}{\d t'} G_{x2}(t,t') & = & -2 \get^\perp(t') \gd_{x1}
	+ A_{32}(t') G_{x3}(t,t'),
	\non\\ 
	\frac{\d}{\d t'} G_{x3}(t,t') & = & - G_{x2}(t,t') 
	+ A_{33}(t') G_{x3}(t,t').
	\eea

The solutions for the $x=1$ components of (\ref{Greeneqmot2f}) are simple: 
$G_{11} = 1$, $G_{21} = G_{31} = 0$. 
To find the solutions for the $x=2,3$ components 
we assume that we have found a solution $g(t)$ that satisfies (\ref{G22eq}). 
Then a second, independent, solution is given by
	\be \label{deffY}
	f(t) = g(t) \int^t \d\bt \: Y(\bt),
\qquad\qquad 
	Y(t) \equiv \frac{1}{g^2(t)}e^{- \int^t \d\bt A_{33}(\bt)}
	= \frac{1}{g^2(t)} \frac{\mathrm{e}^{-3t}}{H(t) \ge(t)}.
	\ee 
Hence
	\bea
	 &&G_{23}(t,t') =\frac{1}{g(t') Y(t')} f(t)
	- \frac{f(t')}{g^2(t') Y(t')}g(t),
	\\
	 &&G_{22}(t,t')= \lh \frac{\dot{g}(t') f(t')}
	{g^3(t') Y(t')} + \frac{1}{g(t')} \rh g(t)- \frac{\dot{g}(t')}{g^2(t') Y(t')}f(t)
=\frac{g(t)}{g(t')}-\frac{\dot{g}(t')}{g(t')}G_{23}(t,t'),\qquad
	\eea
and
	\be
	 G_{33}(t,t')= \frac{\dot{g}(t)}{g(t)} \, G_{23}(t,t')
	+ \frac{g(t) Y(t)}{g(t') Y(t')},\quad
	G_{32}(t,t') = \frac{\dot{g}(t)}{g(t)} \, G_{22}(t,t')
	- \frac{\dot{g}(t')}{g(t')} \frac{g(t) Y(t)}{g(t') Y(t')}.
	\ee	
Of course $G_{13}(t,t') = 2 \int_{t'}^t \d\bt \, \get^\perp(\bt)
G_{23}(\bt,t')$ and $G_{12}(t,t') = 2 \int_{t'}^t \d\bt \, \get^\perp(\bt)
G_{22}(\bt,t')$.
For exact calculations the Green's functions will be determined numerically,
but in an approximate slow-roll treatment we can sometimes find analytic
solutions, see section~\ref{secSlowRoll}.

For the linear mode solutions at horizon crossing, $X_{am}^{(1)}$, we will
assume the analytic slow-roll solutions. Observations of the spectral index indicate
that slow roll is a good approximation {\em at} horizon crossing. Note 
however that, with the exception of section~\ref{secSlowRoll}, we do not 
assume slow roll to hold {\em after} horizon crossing. Moreover, the assumption
of slow roll at horizon crossing is not a requirement to compute these
linear solutions, we could just as well numerically compute the linear mode 
solutions exactly.
For the window function used in the
calculation of the linear solution we take a step function, see
\cite{Rigopoulos:2005xx,Rigopoulos:2005us}, so that its time
derivative is a delta function: $\dot{\cW} = \gd(k c/(a H \sqrt{2}) - 1)$, where
$c$ is a constant of the order of a few, e.g.\ $c=3$. Then
\be 
v_{am}(t) = G_{ab}(t,t_*) X_{bm}^{(1)}(t_*) \gTh(t-t_*),
\ee 
where the step function $\gTh(x)$ equals 1 for $x\geq 0$ and 0 for $x<0$. 
The time $t_*$ is defined by $aH=kc/\sqrt{2}$, i.e.\ a time slightly after
horizon crossing when we have entered the long-wavelength regime. 
While results right at $t_*$ of course depend on the details of the window
function, a few e-folds later any dependence on $\cW$ has disappeared.
Moreover, under the assumption of slow roll at horizon crossing, all
quantities change very little between horizon crossing and $t_*$, so that
final results do not depend on the choice of $c$ and $t_*$ can be taken
equal to the horizon-crossing time determined from $k=aH$ in the final
expressions. For a detailed study of the impact of the first few e-folds after horizon crossing on the evaluation of observable quantities, see \cite{Nalson:2011gc}.  
Defining $\gamma$ as $\gamma\equiv -
\gk H/(2 k^{3/2} \sqrt{\ge})$ the matrix $\mx{X}^{(1)}(t_*)$ is given by
\be
\mathbf{X}^{(1)}(t_*)=\gamma_*
\lh 
\begin{array}{cc}
1 & 0\\
0 & 1\\
0 & -\chi_*
\end{array}
\rh,\label{x01}
\ee
where the subscript $*$ means evaluation 
at $t=t_*$. 
The $2\times 2$ upper part of this matrix is just the result we found in (\ref{Xam}), corresponding to the values of $\zeta_{1(1)k}$ and $\zeta_{2(1)k}$, coupling initially only to the quantum operators $a_{1\mathbf{k}},a_{1\mathbf{k}}^\dagger$ and $a_{2\mathbf{k}},a_{2\mathbf{k}}^\dagger$ respectively. The third line, corresponding to the time derivative of the isocurvature perturbation $\theta_{2(1)}$, can be found by the slow-roll version of (\ref{eql2}) immediately after horizon crossing (hence ignoring the space gradient), equation (\ref{xsr}).
%\be 
%3\dot{\zeta}_{2(1)}+3\chi\zeta_{2(1)}=0.\label{xsr}
%\ee
%
%
% 
%$X_{11}^{(1)}(t_*) = X_{22}^{(1)}(t_*) = \gamma_*$,
%$X_{32}^{(1)}(t_*) = -\gc_* \gamma_*$, the other components being zero
%\cite{Rigopoulos:2005us}. 
Hence we have
\bea
&v_{11} =  \gamma_* \gTh(t-t_*), &	v_{12}(t) =  \gamma_* \lh G_{12}(t,t_*) 
- \gc_* G_{13}(t,t_*) \rh \gTh(t-t_*),\nn\\
	&v_{21} =  0, &  v_{22}(t) =  \gamma_* \lh G_{22}(t,t_*) 
- \gc_* G_{23}(t,t_*) \rh \gTh(t-t_*),\nn\\
	&v_{31} =  0, & 	v_{32}(t) =  \gamma_* \lh G_{32}(t,t_*) 
- \gc_* G_{33}(t,t_*) \rh 	\gTh(t-t_*).\label{vamGreenrel}
\eea
We also define the short-hand notation $\bv_{am}$ by 
$v_{am}(t) = \gamma_* \gTh(t-t_*)\bv_{am}(t)$.

For the second-order horizon-crossing solutions we find from (\ref{z1fg}) and
(\ref{z2f}) (see appendix \ref{appSecondSource}) that the slow-roll matrices $L_{abc}$ 
and $N_{abc}$ have elements satisfying
\bea
&&L_{111*} =\e_*+\hpa_*,\qquad\qquad\qquad\qquad L_{122*}=-\lh\e_*+\hpa_*-\chi_*\rh,\nn\\
&&L_{211*} =\hpe_*,\qquad\qquad\qquad\qquad\ \ \ \ \ \ L_{222*}=\hpe_*,\nn\\
&&L_{112*}+L_{121*}=2\hpe_*,
\ \ \ \ \qquad\qquad\ \  N_{112*}+N_{121*}=-2\hpe_*,\nn\\
&&L_{212*}+L_{221*}=2\lh\e_*+\hpa_*-\chi_*\rh,
\  N_{212*}+N_{221*}=\chi_*,\label{nl}
\eea
with the other elements of $N_{abc}$ being zero. As explained in the appendix,
the slow-roll approximation which expresses $\theta_{2(1)}$ in terms of $\zeta_{2(1)}$ (\ref{xsr}) has 
been used. This means in particular that the subscripts $a,b,c$ only take the 
values 1 and 2, but not 3. However, for consistency in the notation, we
will define here all entries of $L_{abc}$ and $N_{abc}$ to be equal to zero 
if one or more of the indices are equal to 3.

\subsection{Two and three point statistics}
\label{secCorrelators}

%. On such slices 
%the adiabatic perturbation variable has the simple form
%\be
%\tilde{\zeta}_{1i}=\partial_i\tilde{\alpha}.
%\label{redefinedzetagauge}
%\ee
%At first and second order the relation between the adiabatic component
%of $\gz_i$ in the two gauges is (see \ref{grad} and \ref{grad2})
%\bea
%\label{gaugetrans}
%\tilde{\zeta}_{1i(1)}&=&\zeta_{1i(1)},\\
%\tilde{\zeta}_{1i(2)}&=&\zeta_{1i(2)}
%+\zeta_{1(1)}\dot{\zeta}_{1i(1)}=\zeta_{1i(2)}
%+2\hpe\zeta_{1(1)}\zeta_{2i(1)},
%\eea
%where again $\zeta_{1(1)}\equiv \partial^{-2}\partial^i\zeta_{1i(1)}$ and the second relation for $\tilde{\zeta}_{1i(2)}$ is valid on super-horizon scales. 
%Indeed one can show that not only do we end up with a total gradient through 
%this gauge transformation, we also 

%:
%in the flat gauge and for superhorizon scales one can show that \cite{Tzavara:2011hn} 
%(see also \cite{Langlois:2005qp} for 
%the energy density definition of $\zeta_i$)
%\be
%\zeta_{1i(2)}=\partial_i\zeta_{1(2)}
%-\zeta_{1(1)}\dot{\zeta}_{1i(1)}.
%\ee 
%The second term on the right-hand side cancels exactly the 
%gauge transformation term and we are left with the space gradient of the 
%gauge-invariant quantity.

Having now expressed the first and second order  perturbations outside the horizon in terms of their value at horizon crossing and the Green's functions, we move to computing the quantities that can be measured observationally, namely the power spectrum and the bispectrum of the perturbations. 

We start by computing the power spectrum for the adiabatic perturbation introduced in (\ref{powernew}). 
From equation (\ref{via1sol}) we see that the Fourier coefficients of 
$\gz_{1(1)}(\vc{x},t)$ 
are given by $\gz^{(1)1}_{\vc{k}}(t) = v_{1m}(k,t)\lh a_{m\vc{k}}^{\dagger}
+a_{m-\vc{k}}\rh $. Hence the power spectrum for the adiabatic perturbation becomes 
%,  that is the two-point
%correlator of the Fourier coefficients, is
%\be
%\langle\zeta_{1(1)\vc{k_1}} \zeta_{1(1)\vc{k_2}}\rangle
% = \delta^3(\vc{k_1}+\vc{k_2}) v_{1m}(k_1, t)v_{1m}(k_1, t).
%\ee
%The power spectrum of the perturbation $\gz$ (see \ref{observables} and \cite{Lyth:1998xn,Bartolo:2004if}), defined to remove the overall delta function and 
%the factor $1/k^3$ coming from the $v_{1m}$ (see (\ref{vamGreenrel})), is:
\be 
\mathcal{P}_{\zeta}(k,t) \equiv \frac{k^{3}}{2\pi^2} \, 
v_{1m}(k, t) v_{1m}(k, t).
\label{power}
\ee 
The scalar spectral index is then defined as
\be
n_{\zeta} - 1 \equiv\frac{\d\ln{\mathcal{P}_{\zeta}}}{\d\ln{k}}
= \frac{\d\ln{\mathcal{P}_{\zeta}}}{\d t_*}\frac{\d t_*}{\d\ln{k}}
=\frac{\d\ln{\mathcal{P}_{\zeta}}}{\d t_*}\frac{1}{1-\e_*},
\label{spectralind}
\ee
where we used that $k=aH\sqrt{2}/c$ and $\dot{H}=-\e H$ for the time coordinate choice $NH=1$.

For our calculations performed so far within the long-wavelength formalism we have used time slices on which the expansion of the universe is
homogeneous, $\der_i\alpha = 0$, since it simplifies super-horizon 
calculations (see (\ref{flatga})). However, to make contact with the proper gauge-invariant
expression for the second-order $\gz_1$ it turns out to be 
necessary to change to uniform energy density time slices, $\partial_i\rho=0$
in order 
to obtain the gauge-invariant quantity 
corresponding to the curvature perturbation $\zeta_1$. This is done by the gauge transformation (\ref{gat}). 

By combining the different permutations of
$\langle\zeta_{1(2)\vc{k_1}}
\zeta_{1(1)\vc{k_2}}\zeta_{1(1)\vc{k_3}}\rangle$ of the Fourier components of the 
linear and second-order 
adiabatic solutions (first subtracting the average of $\tgz_{1(2)}(\vc{x},t)$ 
to get rid of the divergent part), we find the 
bispectrum\footnote{In the literature (e.g.\  \cite{Maldacena:2002vr}) one 
often sees a factor $(2\pi)^3$ in front of the bispectrum (as well as in 
front of the power spectrum). This is due to a different definition of the 
Fourier transform. We use the convention where both the Fourier transform
and its inverse have a factor $(2\pi)^{-3/2}$.}
\cite{Rigopoulos:2005us}
\bea
\langle \zeta_{1\vc{k_1}}\zeta_{1\vc{k_2}}\zeta_{1\vc{k_3}}\rangle^{(2)} & = & 
(2\pi)^{-3/2}\delta^3(\sum_s\vc{k_s})\left[f(k_1,k_2)+f(k_1,k_3)+f(k_2,k_3)
\right]\non\\
& \equiv & (2\pi)^{-3/2}\delta^3 (\sum_s\vc{k_s})B_{\zeta}(k_1,k_2,k_3),
\eea
where
\bea
 f(k,k') \equiv v_{1m}(k) v_{1n}(k') \Bigg(&&
 \!\!\!\!\!\!\!\!\!\!\!\! \!\get^{\perp} 
v_{2m}(k) v_{1n}(k')+
\frac{1}{2}G_{1a}(t,t_{k'})\lh L_{abc}+N_{abc}\rh(t_{k'})X_{bm}(k,t_{k'})
X_{cn}(k',t_{k'})\non\\
&& \!\!\!\!\!\!\!\!\!\!\!\!- \frac{1}{2}\int_{-\infty}^t\d t'\,G_{1a}(t,t')\bA_{abc}v_{bm}(k)v_{cn}(k')\Bigg)
+ {k} \leftrightarrow {k}',
\label{fkk}
\eea
where $k'$ refers to the scale that exits the horizon last. The first term in the parenthesis comes from the gauge transformation (\ref{gat}). 
Finally we introduce the parameter $\f$, basically defined as the bispectrum
divided by the power spectrum squared, which gives a relative measure of the
importance of non-Gaussianities of the bispectral type (see 
e.g.\ \cite{Maldacena:2002vr, Vernizzi:2006ve}):
\bea\label{fNL_start}
-\frac{6}{5}\f & \equiv & \frac{B_{\zeta
} (k_1,k_2,k_3)} {\frac{2\pi^2}{k_1^3}\mathcal{P}_{\zeta}(k_1)\frac{2\pi^2}{k_2^3}\mathcal{P}_{\zeta}(k_2)
+(k_2\leftrightarrow k_3) + (k_1\leftrightarrow k_3)} \non\\
& = & \frac{f({k}_1,{k}_2) + f({k}_1,{k}_3)
+ f({k}_2,{k}_3)}
{v_{1m}(k_1) v_{1m}(k_1) v_{1n}(k_2) v_{1n}(k_2)+\mathrm{2\ perms.}}.\label{bisp}
\eea
This is essentially $\f^{(4)}$, but we suppress from now on the superscript $(4)$. Notice that unlike definition (\ref{fnldef}), here we assume no pivot scale for our computation and $\f$ depends on all three relevant scales.

%The quotient is called $-\frac{6}{5}\f$ and not simply $\f$ because it was
%originally defined in terms of the gravitational potential $\gF$ and not $\gz$
%as $\gF = \gF_\mathrm{L} + \f \lh \gF_\mathrm{L}^2 
%- \langle \gF_\mathrm{L}^2 \rangle \rh$ \cite{Komatsu:2001rj}. 
%During recombination (matter domination) the two are 
%related by $\gz = - \frac{5}{3} \gF$. Moreover, when computing the bispectrum
%divided by the three permutations of the power spectrum squared using this
%expression of $\gF$ one obtains $2\f$ due to the two ways the 
%two $\gF_\mathrm{L}$ inside the second-order solution can be combined with the 
%two linear solutions to create the power spectrum. Together these two effects
%explain the factor $-6/5$.\footnote{In \cite{Rigopoulos:2005us} and
%earlier papers we used a slightly different definition of $\f$ which was
%larger by a factor of -18/5. Here we conform to the definition that is now 
%generally accepted in the literature.}

\subsection{\texorpdfstring{ $\delta N$}{DN}-formalism}
\label{dnform}

An alternative formalism to compute $\f$ is the so-called $\delta N$-formalism
\cite{Starobinsky:1986fxa,Sasaki:1995aw,Sasaki:1998ug,Lyth:2004gb,Lyth:2005fi}.
In order to compare our results of the next sections to those obtained
using the $\gd N$ formalism, for those cases where the latter are available,
we give here a brief overview.

The $\gd N$ formalism uses the fact that the adiabatic perturbation $\zeta_1$ 
on large scales is equal to the perturbation of the number of e-folds
$\delta N(t,t_*)$ between an initial flat hypersurface at $t=t_*$, which
is usually taken to be the horizon crossing time, and a final uniform
density hypersurface at $t$. One can then expand the number of e-folds
in terms of the perturbations of the fields and their momenta on the
initial flat hypersurface 
\be \delta N(t,t_*)=\frac{\partial
  N}{\partial\phi^A_*}\delta\phi^A_*+\frac{\partial
  N}{\partial\Pi^A_*}\delta\Pi^A_*
+\frac{1}{2}\frac{\partial^2N}{\partial\phi^A_*\partial\phi^B_*}
\delta\phi_*^A\delta\phi_*^B+\ldots .
\ee 
So instead of integrating the evolution of $\zeta_1$ through
equations (\ref{eqlw11}) and (\ref{eqlw2}) one can evaluate the
derivatives of the number of e-folds at horizon crossing and thus
calculate $\zeta_1$.

Because of the computational difficulty associated with the derivatives
with respect to $\Pi^A$, slow roll is assumed at horizon exit so that
the terms involving the momentum of the fields can be ignored. 
This is a crucial assumption for the $\gd N$ formalism.
The final formula then reads 
\be 
\delta N(t,t_*)=\frac{\partial N}{\partial\phi^A_*}\delta\phi^A_*
+\frac{1}{2}\frac{\partial^2N}{\partial\phi^A_*\partial\phi^B_*}
\delta\phi_*^A\delta\phi_*^B,
\ee 
up to second order. From it one finds the following expression for the 
bispectrum:
\be 
\langle  \zeta_{1\vc{k_1}}
\zeta_{1\vc{k_2}}\zeta_{1\vc{k_3}}\rangle^{(2)}
=
\frac{1}{2}N_{,A}N_{,B}N_{,CD}\langle \delta\phi^A_{k_1}
\delta\phi^B_{k_2}(\delta\phi^C\star\delta\phi^D)_{k_3}\rangle+\mathrm{perms.},
\ee
where $\star$ denotes a convolution and the average of 
$(\delta\phi \star \delta\phi)$ has been subtracted to avoid divergences. 
$N_{,A}$ denotes the derivative of $N$ with respect to the field $\phi^A_*$. 
Using Wick's theorem this can be rewritten as products 
of two-point correlation functions to yield finally
\be
-\frac{6}{5}f_\mathrm{NL,\delta N} = \frac{N^{,A}N^{,B}N_{,AB}}
{\left(N_{,C}N^{,C}\right)^2}.
\label{dnresult}
\ee 
Notice that this result is momentum independent and local in real space, 
although attempts to generalize to a scale-dependent situation have 
recently been made in \cite{Byrnes:2009pe,Byrnes:2010ft}.
This formula can be used numerically or analytically to calculate $\f$. 
However, for any analytical results and insight one must assume
the slow-roll approximation to hold at all times after horizon exit (see
for example \cite{Vernizzi:2006ve,Choi:2007su,Battefeld:2007en}),
except for the special case of a separable Hubble parameter
\cite{Byrnes:2009qy,Battefeld:2009ym}.

\Section{General analytic expression for \texorpdfstring{$\f$}{fNL} 
for two fields}\label{general}

In this section we will further work out the exact long-wavelength expression 
for $\f$ given in (\ref{fNL_start}). 
No slow-roll approximation is used on super-horizon scales in this section. 
In particular this means the formalism can deal with sharp turns in the
field trajectory after horizon crossing during which slow roll temporarily
breaks down. In the first subsection
we restrict ourselves to the case where $k_1=k_2=k_3$ to lighten the 
notation. In the second subsection we show how the result for $\f$ changes 
in the case of arbitrary momenta.

\subsection{Equal momenta}

In the case of equal momenta, equation (\ref{fNL_start})
reduces to
\bea\label{fNL}
-\frac{6}{5}\f =  \frac{-v_{1m}(t) v_{1n}(t)}{\lh v_{1m}(t) v_{1m}(t) \rh^2}
\Bigg\{&&
\!\!\!\!\!\!\!\!\!
\int_{-\infty}^t\!\!\!\!\!\d t' G_{1a}(t,t') 
\bA_{abc}(t') v_{bm}(t') v_{cn}(t')
- 2 \get^\perp(t) v_{2m}(t) v_{1n}(t)\nn\\
&&\!\!\!\!\!\!\!\!
-G_{1a}(t,t_*)M_{abc*} v_{bm}(t_*)v_{cn}(t_*)\Bigg\}.
\eea
We remind the reader that indices $l, m, n$ take the values 1 and 2 (components
in the two-field basis), while indices $a,b,c,\ldots$ take the values 1, 2, 
and 3 (labeling the $\gz_1$, $\gz_2$, and $\gth_2$ components).
To make the expressions a bit shorter, we will drop the
time arguments inside the integrals, but remember that for the Green's
functions the integration variable is the second argument. Using the result
(\ref{timederA}) proved in \ref{appTimeder} we can write $\bA_{ab1}$ as a 
time derivative and do an integration by parts, with the result
	\be\label{fNLint}
	\int_{-\infty}^t\!\!\!\!\!\d t' G_{1a} \bA_{abc}
	v_{bm} v_{cn}
	= 2 \get^\perp v_{2m} v_{1n}
	+ \int_{-\infty}^t\!\!\!\!\! \d t' A_{ab} \frac{\d}{\d t'} 
	[ G_{1a}v_{bm} v_{1n} ]+ 
	\int_{-\infty}^t \!\!\!\!\!\d t' G_{1a} \bA_{ab\bc}
	v_{bm} v_{\bc n},
	\ee
where the index $\bc$ does not take the value 1.
Here we used that the linear solutions $v_{am}$ are zero at $t=-\infty$ 
(by definition), that the Green's function $G_{1a}(t,t) = \gd_{1a}$, and that 
$A_{1b} = -2\get^\perp \gd_{b2}$ (exact).
We see that the first term on the right-hand side exactly cancels 
with the gauge correction (the second term in (\ref{fNL})) that is necessary to 
create a properly gauge-invariant second-order result. 

We start by working out the second term on the right-hand side of
(\ref{fNLint}). We find
\bea
 I && \equiv \int_{-\infty}^t\d t' A_{ab} \frac{\d}{\d t'} 
	\left [ G_{1a} v_{bm} v_{1n} \right ]
         =\gamma_*^2 \int_{-\infty}^t \d t' A_{ab} \frac{\d}{\d t'} 
	\left [ G_{1a} \bv_{bm} \bv_{1n} \gTh(t'-t_*) \right ]\nn\\
	&& = A_{ab*} G_{1a}(t,t_*) v_{bm*} v_{1n*}
	+ \gamma_*^2 \gTh(t-t_*) \int_{t_*}^t \d t' A_{ab} \left [ G_{1d} 
	A_{da} \bv_{bm} \bv_{1n}- G_{1a} A_{bd} \bv_{dm} \bv_{1n}\right.\nn\\ 
	&&\qquad\qquad\qquad\qquad\qquad\qquad\qquad\qquad\qquad\qquad\qquad\left.- G_{1a} \bv_{bm} A_{1d} \bv_{dn} \right ]\nn\\
	&&=A_{ab*} G_{1a}(t,t_*) v_{bm*} v_{1n*}
	- \gamma_*^2 \gTh(t-t_*) \int_{t_*}^t \d t' A_{ab} A_{1d} G_{1a} 
	\bv_{bm} \bv_{dn},
\label{Ainteg}	
\eea
where, as before, a subscript $*$ means that a quantity is evaluated at $t_*$.
Using the explicit form of the matrix $\mx{A}$ (\ref{Amat}) and the solutions 
$v_{am}$ (\ref{vamGreenrel}) this becomes
	\bea
	I & = & \gamma_*^2 \gTh(t-t_*) \gd_{m2} \gd_{n1} 
	\lh -2 \get^\perp_* + \gc_* G_{12}(t,t_*) + A_{32*} G_{13}(t,t_*)
- \gc_* A_{33*} G_{13}(t,t_*) \rh\\
	&& + \gamma_*^2 \gTh(t-t_*) \gd_{m2} \gd_{n2} 
	\int_{t_*}^t \d t' 2\get^\perp \bv_{22} 
	\left [ -2\get^\perp \bv_{22}-G_{12} \bv_{32} 
	+ A_{32} G_{13} \bv_{22} + A_{33} G_{13} \bv_{32} \right ].\non
	\eea
From now on we will drop the overall step function, which just encodes the
obvious condition that $t\geq t_*$.
Realizing that $A_{32}\bv_{22} + A_{33}\bv_{32} = - \frac{\d}{\d t'} \bv_{32}$ 
we can do an integration by parts:
	\bea\label{Ires}
	I = & \gamma_*^2 \gd_{m2} & \Bigg[ \gd_{n1} 
	\lh -2 \get^\perp_* + \gc_* G_{12}(t,t_*) + A_{32*}G_{13}(t,t_*)
	- \gc_* A_{33*} G_{13}(t,t_*) \rh \non\\ 
 && -\delta_{n2}2 \get^\perp_* \gc_*G_{13}(t,t_*)\Bigg]\\
	+\gamma_*^2 & \gd_{m2} \gd_{n2} & \int_{t_*}^t \d t' 2\get^\perp 
\left [ -2\get^\perp (\bv_{22})^2+ 
	\lh - 2 G_{12} + A_{33} G_{13} 
	+ \frac{\dot{\get}^\perp}{\get^\perp} G_{13} \rh \bv_{22} \bv_{32} 
	+ G_{13} (\bv_{32})^2
	\right ].\non
	\eea
To this result we have to add the final term on the right-hand side of 
(\ref{fNLint}). Using the explicit expression for the matrix $\mx{\bA}$ 
and doing
some more integrations by parts this can be worked out further, as can be found
in appendix \ref{derivation}. The final result for $\f$ in the
equal momenta limit is (including also the final term of (\ref{fNL}))
\be\label{fNLresult}
-\frac{6}{5}\f=\frac{-2\bv_{12}^2}{[1+(\bv_{12})^2]^2}
\Bigg( g_{iso}+g_{sr}+g_{int}\Bigg), 
\ee
where
\bea
 g_{iso}&=&(\ge+\get^\parallel) (\bv_{22})^2 + \bv_{22} \bv_{32},
\qquad
g_{sr}=-\frac{\es+\hpas}{2\bv_{12}^2}+\frac{\get^\perp_* \bv_{12}}{2}
-\frac{3}{2}\lh\ge_*+\get^\parallel_*-\gc_*+\frac{\hpes}{\bv_{12}}\rh,\nn\\
 g_{int}&=&- \int_{t_*}^t \d t' \Biggl[ 2 (\get^\perp)^2 (\bv_{22})^2 
	+ (\ge+\get^\parallel) \bv_{22} \bv_{32} + (\bv_{32})^2
- G_{13} \bv_{22} \lh C \bv_{22} + 9 \get^\perp \bv_{32} \rh\Biggr].
\label{gisosrint}
\eea
Here we have defined
\be 
C \equiv 12 \get^\perp \gc - 6 \get^\parallel \get^\perp
	+ 6 (\get^\parallel)^2 \get^\perp + 6 (\get^\perp)^3
- 2 \get^\perp \gx^\parallel- 2 \get^\parallel \gx^\perp
	- \frac{3}{2}(\tW_{211} + \tW_{222}),
\label{Ci}
\ee 
where $\tW_{lmn} \equiv (\sqrt{2\ge}/\gk)W_{lmn}/(3H^2)$. 
We should add that although no slow-roll approximation has been used on
super-horizon scales, we did assume slow roll to hold at horizon crossing,
in order to use the analytic linear short-wavelength 
solutions~(\ref{vamGreenrel}) and to remove any dependence on the window
function $\cW$. Observations of the scalar spectral
index seem to indicate that slow roll is a good approximation at horizon 
crossing. In a numerical treatment
we could use the exact numerical solutions instead.

Looking at (\ref{fNLresult}), which is one of our main results, 
we can draw a number of important conclusions.
In the first place there is a part of $\f$, namely the first term in $g_{sr}$,
that survives in the single-field limit. It corresponds to the single-field
non-Gaussianity produced at horizon crossing and comes from the $b_{ia}^{(2)}$
source term. It agrees with the single-field result of Maldacena 
\cite{Maldacena:2002vr} for $\f^{(4)}$ (\ref{fnl4s}).
The rest of the result is proportional to $\bv_{12}$, 
which describes the contribution of the isocurvature mode to the adiabatic
mode. In the single-field case it is identically zero, so that there is no
super-horizon contribution to $\f$ in that case. Moreover, since 
$\gth_1 = 2 \get^\perp \gz_2$, such a contribution only builds up when
$\get^\perp$ is non-zero, i.e.\ when the field trajectory makes a turn in 
field space. We also see that there are three different sorts of terms in
the expression for $\f$. 
The $g_{sr}$ terms are proportional to a slow-roll parameter evaluated at $t_*$ 
and thus are always small because we assume slow roll to hold at 
horizon-crossing. 
Although the terms proportional to $\bv_{12}$ and $1/\bv_{12}$ in $g_{sr}$ are 
time varying, one can easily show that neither $\bv_{12}/(1+\bv_{12}^2)^2$ nor 
$\bv_{12}^3/(1+\bv_{12}^2)^2$ are ever bigger than $0.33$. 
The $g_{iso}$ terms are proportional to
$\bv_{22}$, the pure isocurvature mode. These terms can be big, in particular
during a turn in field space, but in the models that we consider, where the
isocurvature mode has disappeared by the end of inflation, they become
zero again and cannot lead to observable non-Gaussianities. The reason that
we do not consider models with surviving isocurvature modes is that in that
case the evolution after inflation is not clear. In the presence of 
isocurvature modes the adiabatic mode is not necessarily constant (indeed,
that is the source of the non-Gaussianities we are considering here), which
means that the final results at recombination might depend on the details 
of the transition at the end of inflation and of (p)reheating. Hence we will
make sure that in all models we consider the isocurvature modes have
disappeared by the end of inflation, which means in particular that the turn
of the trajectory in field space has to occur a sufficient number of e-folds
before the end of inflation. Note, however, that this is a constraint we
impose voluntarily to simplify the evolution after inflation, it is in no
way a necessary condition for the validity of our formalism during inflation.
Finally, the third type of term in 
(\ref{fNLresult}) is the integral in $g_{int}$. It is
from this integrated effect that any large, persistent non-Gaussianity 
originates.

For completeness we also calculate the power spectrum, which according to 
equation~(\ref{power}) takes the simple form
\be
\mathcal{P}_{\zeta}=\frac{\kappa^2 H_*^2}{8\pi^2\e_*}(1+\bv^2_{12}),
\label{powerspectrum}
\ee
and the spectral index, calculated analytically using 
equations~(\ref{spectralind}), (\ref{grstar}) and (\ref{dere}), 
\bea
\label{spectralindex}
n_{\zeta} - 1 = &&\!\!\!\!\!\!\!\!
 \frac{1}{1-\es} \Bigg[-4\e_*-2\hpa_*
+2\frac{\bv_{12}}{1+\bv^2_{12}}\Big(-2\hpe_*+\chi_*\bv_{12}\\
 && +G_{13}(t,t_*)\lh -\tilde{W}_{221*}+2\e^2_* +\eta^{\parallel 2}_*
+\eta^{\perp 2}_*  +3\e_*(\hpa_*-\chi_*)-2\hpa_*\chi_*
+\chi_*^2 \rh \Big)\Bigg].\nn
\eea

\subsection{General momenta}
\label{Secgenmom}

We turn now to the more general case where each scale exits the
horizon at a different time $t_{k_i}$, defined by $aH=k_ic/\sqrt{2}$,
where $c \approx 3$ is a constant allowing for some time to pass after horizon
exit so that the long-wavelength approximation is valid (see the discussion
in section~\ref{secGreen}). It is
important to realize that it is not the momentum dependence of the
bispectrum that we are discussing here, but of $\f$.  The momentum
dependence of the local bispectrum is dominated by the momentum
dependence of the power spectrum squared, leading to the well-known
result (see e.g.\  \cite{Babich:2004gb}) that it peaks on squeezed
triangles where one of the momenta is much smaller than the other
two. Here we are discussing the momentum dependence of $\f$, so one
has divided by the power spectrum squared. This $\f$, that is really $f_\mathrm{NL}^{(4)}$ in the literature, is usually assumed to be
momentum-independent. However, as we will show this is not true and
its momentum dependence can lead to relative effects of order $10\%$ even 
within the range of momenta that are observable by Planck.

Assuming $k_1 \geq k_2$, i.e.\  $t_{k_1} \geq t_{k_2}$, we find that 
(\ref{fkk}) reduces to
\bea
 f(k_1,k_2) &=& -\frac{\gamma^2_{k_1}\gamma^2_{k_2}}{2}
\bv_{1mk_2}(t)\bv_{1nk_1}(t)
\Bigl[\int_{t_{k_1}}^t \d t' G_{1a}(t,t') \left[ \bA_{a\bb\bh}
+ \bA_{a\bh\bb}\right]\bv_{\bb mk_2}(t') \bv_{\bh n k_1}(t')   \non\\
 && 
    -\int_{t_{k_1}}^t \d t' G_{1a}(t,t') \left[ A_{ab}A_{1e}
+ A_{ae}A_{1b}\right]\bv_{bmk_2}(t') \bv_{enk_1}(t')\nn\\
 &&
+ A_{ab}(t_{k_1}) G_{1a}(t,t_{k_1}) \left[ \bv_{bmk_2}(t_{k_1}) 
\delta_{1n}
+ \bv_{1mk_2}(t_{k_1})\delta_{bn} \right]\nn\\
 &&-G_{1a}(t,t_{k1})M_{abc}(t_{k_1})
\left[\bv_{cmk_2}(t_{k_1})\delta_{bn}+\bv_{bmk_2}(t_{k_1})\delta_{cn}\right]
\Bigr],
\eea
where again we have used the result (\ref{timederA}) and have done an
integration by parts that cancels the gauge correction term as in
(\ref{fNLint}). The indices $\bb$ and $\bh$ do not take the value
1. We have introduced the notation $\bv_{i1k_l}\equiv\delta_{i1}$ and
$\bv_{i2k_l}\equiv G_{i2}(t,t_{k_l})-\chi_{k_l}G_{i3}(t,t_{k_l})$,
where $\chi_{k_l}$ is evaluated at $t_{k_l}$.  We notice that due to
the step functions the integral's lower limit corresponds to the time
when both scales have entered the long-wavelength regime, i.e.\  the
time when the larger $k_1$ (smaller wavelength) exits the horizon. The 
expression has become more complicated as compared to (\ref{fNLint}) and
(\ref{Ainteg}) since the $\bv_{bmk_i}$ refer to a different initial
value depending on the horizon crossing time of each scale $k_i$.
Following the same procedure as in the previous section we find that
\bea\label{fNLgen}
-\frac{6}{5}\f(k_1,k_2,k_3) = \frac{f({k}_1,{k}_2)+f({k}_2,{k}_3)
+f({k}_1,{k}_3)}
{\gamma^2_{k_1}\gamma^2_{k_2}[1+(\bv_{12k_1})^2][1+(\bv_{12k_2})^2]
+\mathrm{2\ perms.}},
\eea
where
\bea
	&&f(k_1,k_2)=-2\gamma^2_{k_1}\gamma^2_{k_2}(\widetilde{v}_{12})^2
\Bigg(g_{iso}({k}_1,{k}_2)+g_{sr}({k}_1,{k}_2)+g_{int}({k}_1,{k}_2)
%+g_{k}({k}_1,{k}_2)
\Bigg),
\label{fNLresultgenmom}
\eea
with
\bea
 g_{iso}({k}_1,{k}_2)\!&=&\!(\e+\hpa)(\widetilde{v}_{22})^2
+\widetilde{v}_{22}\widetilde{v}_{32},\nn\\	
g_{sr}({k}_1,{k}_2)&=&\hpe_{k_1}\Bigg(\frac{G_{22k_1k_2}\bv_{12k_1}}{2}   
-\frac{1}{\bv_{12k_2}}-\frac{G_{22k_1k_2}}{2\bv_{12k_1}} 
\Bigg)+\frac{3\chi_{k_2}}{4}G_{33k_1k_2}
-\frac{3}{2}(\e_{k_1}+\hpa_{k_1})G_{22k_1k_2}\nn\\
&&
+\frac{\chi_{k_1}}{4}\Bigg(2\frac{\bv_{12k_1}}{\bv_{12k_2}}+
G_{22k_1k_2}\Bigg)-\frac{\e_{k_1}+\hpa_{k_1}}{2(\widetilde{v}_{12})^2}
\nn\\
&&
+\frac{G_{13}(t,t_{k_1})}{2}\Bigg[\frac{3\lh\chi_{k_1}G_{22k_1k_2}
-\chi_{k_2}G_{33k_1k_2}\rh}{2\bv_{12k_1}}
+G_{32k_1k_2}\lh\frac{3+\e_{k_1}+2\hpa_{k_1}}{2\bv_{12k_1}}
+\hpe_{k_1}\rh\Bigg]\nn\\
&&
-\frac{3}{4}G_{32k_1k_2}
-\frac{1}{2}G_{12k_1k_2}\lh
\e_{k_1}+\hpa_{k_1}+2\hpe_{k_1}-\frac{\chi_{k_1}}{2}\lh1+\bv_{12k_1}\rh
+\frac{\e_{k_1}+\hpa_{k_1}}{\bv_{12k_1}}\rh,
\nn\\
g_{int}({k}_1,{k}_2)&=&\!-\!\int_{t_{k_1}}^t\!\!\!\!\d t'\Big[2(\hpe)^2
(\widetilde{v}_{22})^2
\!+\!(\e+\hpa)\widetilde{v}_{22}\widetilde{v}_{32}\!+\!(\widetilde{v}_{32})^2
\!-\!G_{13}\widetilde{v}_{22}(C\widetilde{v}_{22}+9\hpe\widetilde{v}_{32})
\Big],\label{comp}
\eea
for $k_1\geq k_2\geq k_3$ and $C$ was defined in (\ref{Ci}). We introduced 
the notation
\begin{displaymath}
\begin{array}{l}
(\widetilde{v}_{12})^2\equiv\bv_{12k_1}\bv_{12k_2},\qquad 
(\widetilde{v}_{22})^2\equiv\bv_{22k_1}\bv_{22k_2},\qquad
(\widetilde{v}_{32})^2\equiv\bv_{32k_1}\bv_{32k_2},\\  
\widetilde{v}_{22}\widetilde{v}_{32}\equiv\frac{1}{2}(\bv_{22k_1}\bv_{32k_2}
+\bv_{22k_2}\bv_{32k_1}),
\end{array}
\end{displaymath}
and also $G_{ijk_1k_2}\equiv G_{ij}(t_{k_1},t_{k_2})$, while the
subscript on the slow-roll parameters denotes evaluation at the
relevant time that the scale exits the horizon. The Green's functions
that appear without arguments denote $G(t,t_{k_1})$ outside or
$G(t,t')$ inside the integral.

Although this expression is quite a bit longer than
(\ref{fNLresult}), there are many similarities between the two
results. The whole expression is again proportional to $\bv_{12k_i}$,
except for the single-field horizon-crossing result, so that there is
no super-horizon contribution to $\f$ for the single-field case. 
In the $g_{iso}$ term and the first two lines of  $g_{sr}$  we
recognize the familiar terms of the equal-momenta case, i.e.\  the
isocurvature contributions proportional to $\bv_{22k_i}$ as well as
the horizon crossing terms now evaluated at $t_{k_1}$ and $t_{k_2}$
(note that for $k_1=k_2$, $G_{iik_1k_2}=1$ identically and we regain
the expressions of (\ref{gisosrint})). The integral has also retained
its form. The rest of the terms in $g_{sr}$, namely those in the third and fourth line, are terms
arising due to the different horizon-crossing times of the scales and
are identically zero for the equal-momenta case $k_1=k_2$ where
$G_{ijk_1k_2}=\delta_{ij}$. All these terms  are proportional
to a slow-roll parameter evaluated at horizon crossing (using the fact
that $G_{13}=G_{12}/3$ up to slow-roll corrections, see (\ref{GsolSR})),
except for the first term on the fourth line. 
However, $G_{32k_1k_2}$ is expected to be
quite small: for $k_1=k_2$ it is zero, and for $k_1 \mg k_2$ it becomes 
the linear solution for the isocurvature velocity $\gth_2$ 
(see (\ref{vamGreenrel})). 
Hence we do not expect these terms to give a large 
contribution, which is confirmed numerically. We will further study the scale dependence of $\f$ and the contribution of each term in chapter \ref{Ch6}.

\Section{Slow-roll approximation}
\label{secSlowRoll}

While the exact result for $\f$, equation (\ref{fNLresult}) or 
(\ref{fNLgen}), is an extremely
useful starting point for an exact numerical treatment, the integral cannot
be done analytically. In order to find explicit analytic results that will be 
very useful to gain insight and draw generic conclusions, we need to simplify
the problem by making the slow-roll approximation. In 
subsection~\ref{secSRgenexpr} we further work out (\ref{fNLresult}) under this 
approximation. Even then the integral can only be done analytically for 
certain specific classes of inflationary potentials, which are treated in the next chapter.

\subsection{General expressions}
\label{secSRgenexpr}

Considering the slow-roll version of equation (\ref{G22eq}) we find
that $g(t)$ (as defined above equation (\ref{deffY})) 
satisfies
	\be \label{gSReq}
	\dot{g} + \gc \, g = 0.
	\ee 
We see that $Y(t) \propto \exp(-3t)$ so that $f(t)$ is a rapidly 
decaying solution that can be neglected (see (\ref{deffY}) for definitions). 
After the decaying mode has vanished the solutions for the Green's 
functions simplify to
	\bea\label{GsolSR}
	&&G_{22}(t,t') = \frac{g(t)}{g(t')},\qquad\qquad 
	G_{12}(t,t') = \frac{2}{g(t')} \int_{t'}^t\d\bt\ \hpe(\bt)g(\bt),\nn\\
	&&G_{32}(t,t') = - \gc(t) G_{22}(t,t'),\qquad\qquad 
	G_{x3}(t,t') = \frac{1}{3}G_{x2}(t,t').
	\eea

\subsubsection{Equal momenta}

Using the last two relations in (\ref{GsolSR}) and dropping higher-order terms 
in slow roll, (\ref{gisosrint}) reduces to
\bea
\label{srtot}
 g_{iso}&=&(\e +\hpa-\chi)(\vb_{22})^2,\qquad
g_{sr}=-\frac{\es+\hpas}{2\bv_{12}^2}+\frac{\get^\perp_* \bv_{12}}{2}
-\frac{3}{2}\lh\ge_*+\get^\parallel_*-\gc_*+\frac{\hpes}{\bv_{12}}\rh,\\
 g_{int}&=&\!\!\int_{t_*}^t\!\!\!\!\d t' (\bv_{22})^2 \Bigg[ 2 \get^\perp
\!\lh\! -\! \get^\perp\! +\! \frac{(\ge\!+\!\get^\parallel\!-\!\gc)\gc}
{2\get^\perp}\rh
\!+G_{12} \lh \get^\perp \gc\! -\! 2 \get^\parallel 
\get^\perp\!-\!\frac{1}{2} (\tW_{211} \!+\! \tW_{222}) \rh \Bigg].\non
\eea
Inserting these terms into (\ref{fNLresult}) we find an expression that can 
be considered the final expression for $\f$ in the slow-roll 
approximation, and is the one that will be used in section~\ref{intsection}. 
It also proves useful, however, to rewrite it in a different way using 
integration by parts.

We use the slow-roll version of relation (\ref{Greeneqmot2ftp}),
$2\get^\perp = - \frac{\d}{\d t'} G_{12}(t,t') + \gc G_{12}(t,t')$, to do an
integration by parts, leading to
	\bea\label{Keq}
	g_{int} & = & \bv_{12} \lh -\get^\perp_* + 
	\frac{(\ge_* + \get^\parallel_* - \gc_*)\gc_*}{2 \get^\perp_*} \rh
+ \int_{t_*}^t \d t' G_{12} (\bv_{22})^2 
        \Biggl[ 2 \get^\perp \gc
	- \frac{(\ge+\get^\parallel-\gc)\gc^2}{2\get^\perp} \non\\
	&&- 2 \get^\parallel \get^\perp
	-\frac{1}{2} (\tW_{211} + \tW_{222}) 
	+ \frac{\d}{\d t'} \lh - \get^\perp 
	+ \frac{(\ge+\get^\parallel-\gc)\gc}{2\get^\perp} \rh
	\Biggr].
	\eea
Using the slow-roll version of the relations (\ref{W111W211rel}),
	\be \label{xiparxiperpSR}
	\gx^\parallel = 3 \ge \get^\parallel + (\get^\parallel)^2
	+ (\get^\perp)^2 - \tW_{111}
	\ \ \ \ \ \ \ \mbox{and}\ \ \ \ \ \ \  
	\gx^\perp = 3 \ge \get^\perp + 2 \get^\parallel \get^\perp
	- \get^\perp \gc - \tW_{211},
	\ee 
as well as the time derivatives of the slow-roll parameters in (\ref{dere}), 
we can derive that
	\bea
	\frac{\d}{\d t}\lh-\get^\perp +\frac{(\ge+\get^\parallel-\gc)\gc}
{2\get^\perp} \rh
	&=& \frac{1}{2\get^\perp} \Bigg[ -\gc^3 + (\ge+\get^\parallel)\gc^2
- 4 \lh \ge\get^\parallel+(\get^\perp)^2 \rh \gc\\
	&&\!\!\!\!+ 4 \lh \ge^2 \get^\parallel + \ge(\get^\parallel)^2
	- \ge(\get^\perp)^2 + \get^\parallel(\get^\perp)^2 \rh \!-\!(\ge+
\get^\parallel-\gc)\tW_{111}\nn\\
	&&\!\!\!\!+ \lh 2\get^\perp + \frac{(\ge+\get^\parallel-\gc)\gc}
{\get^\perp}\rh \tW_{211} 
	+ (\ge+\get^\parallel-2\gc)\tW_{221}\Bigg].\non
\eea
Inserting this into expression (\ref{Keq}) for $g_{int}$ and including the 
remaining terms in the expression for $\f$ we finally obtain
\bea\label{fNLresultSR}
-\frac{6}{5}\f(t) &=& \frac{-2(\bv_{12})^2}{[1+(\bv_{12})^2]^2}
\Biggl\{ (\ge+\get^\parallel-\gc) (\bv_{22})^2
-\frac{\es+\hpas}{2\bv_{12}^2}-\frac{\get^\perp_* \bv_{12}}{2}+\! \frac{(\ge_*\! +\! \get^\parallel_* \!-\! \gc_*)
\gc_*}{2 \get^\perp_*}\, \bv_{12} \non\\ 
&& \qquad\qquad\qquad
-\frac{3}{2}\lh\ge_*+\get^\parallel_*-\gc_*+\frac{\hpes}{\bv_{12}}\rh 
\nn\\
&&+ \int_{t_*}^t \d t' G_{12} (\bv_{22})^2 
\Bigg[2\frac{\ge\get^\parallel}{\get^\perp} 
	\lh\! -\gc\! +\! \ge\! +\! \get^\parallel\! -\!
	\frac{(\get^\perp)^2}{\get^\parallel} \rh\!+\!\frac{1}{2}(\tW_{211}\! 
-\! \tW_{222}\! -\! \frac{\gc}{\get^\perp} \tW_{221})
\nn\\
&& \qquad\qquad\qquad\qquad
- \frac{\ge+\get^\parallel-\gc}{2\get^\perp}
\lh \tW_{111} \!-\! \tW_{221}\! -\! \frac{\gc}{\get^\perp} \tW_{211} \rh
\Bigg] \Biggr\}.
\eea
This is the alternative final result for $\f$ in the slow-roll approximation.

Equation (\ref{fNLresultSR}), as well as (\ref{srtot}), is characterized by the
same features as the result of the exact formalism. We can easily
distinguish the pure isocurvature $\bv_{22}$ term, which we
assume to vanish before the end of inflation in order for the
adiabatic mode to be constant after inflation, as well as the terms evaluated
at the time of horizon crossing, which are expected to be small. Any
remaining non-Gaussianity at recombination has to originate from the integral.
In subsections \ref{secEqPow} and
\ref{intsection} we will further work out the expressions of this section
for the case of certain classes of potentials to gain insight into their 
non-Gaussian properties. But first we look at
the momentum dependence of $\f$.

\subsubsection{Squeezed limit}
\label{slowsq}

In this section we will calculate the slow-roll expression for $\f$ in the
case where $k \equiv k_3 \ll k_1 = k_2 \equiv k'$, what is usually refered to as 
the squeezed limit. Note that we assume $k_1 = k_2$
for simplicity, to keep the expressions manageable, it is not a necessary
condition.
We start with equation (\ref{fNLgen}) and
follow the procedure of the previous subsection, that is we use the slow-roll
approximations (\ref{GsolSR}) for the Green's functions and drop
higher-order terms in slow roll. Since there are only two relevant
scales the expression simplifies to give
\bea
	-\frac{6}{5}\f&=&\frac{-2\bv_{12k'}/[1+(\bv_{12k'})^2]}
{\gamma^2[1+(\bv_{12k'})^2]+2[1+(\bv_{12k})^2]}
	\Bigg[\gamma^2\bv_{12k'}\!\Bigg(\!g_{iso}(k',k')\!
	+\!g_{sr}(k',k')\!+\!g_{int}(k',k')\!\Bigg) \non\\
	&&+2\bv_{12k}\Bigg(g_{iso}(k',k)\!+\!g_{sr}(k',k)
\!+\!g_{int}(k',k)\Bigg)\Bigg],
\label{fNLgeni}
\eea
where $\gamma\equiv\gamma_{k'}/\gamma_{k}$ and
\bea
 g_{iso}(k',k)&=&(\e+\hpa-\chi)\bv_{22k}\bv_{22k'},\\
 g_{sr}(k',k)\!&=&\!\hpe_{k'}\Bigg(\frac{G_{22k'k}\bv_{12k'}}{2}   
-\frac{1}{\bv_{12k}}-\frac{G_{22k'k}}{2\bv_{12k'}}) 
\Bigg)\!+\!\frac{3\chi_{k}}{4}G_{33k'k}
-\frac{3}{2}(\e_{k'}+\hpa_{k'})G_{22k'k}
\nn\\
&&\!\!\!
+\frac{\chi_{k'}}{4}\Bigg(2\frac{\bv_{12k'}}{\bv_{12k}}+G_{22k'k}\Bigg)-\frac{\e_{k'}+\hpa_{k'}}{2\bv_{12k}\bv_{12k'}},\nn\\
&&\!\!\!+\frac{1}{4}(\chi_{k'}G_{22k'k}\!-\!\chi_{k}G_{33k'k})
+\frac{1}{12}G_{32k'k}(-6+\e_{k'}\!+\!2\hpa_{k'}+2\hpe_{k'}\bv_{12k'})\nn\\
&&\!\!\!-\frac{1}{2}G_{12k'k}\lh 
\e_{k'}+\hpa_{k'}+ 2\hpe_{k'}-\frac{\chi_{k'}}{2}(1+\bv_{12k'})+\frac{\e_{k'}+\hpa_{k'}}{\bv_{12k'}}\rh,
\nn\\
 g_{int}(k',k)\!&=&\!\!\!\int_{t_{k'}}^t\!\!\!\!\d
t'\bv_{22k}\bv_{22k'}\!\Bigg[ 2 \get^\perp\!\!
	\lh\!\!-\get^\perp\!\!
+\!\frac{(\ge\!+\!\get^\parallel\!-\!\gc)\gc}{2\get^\perp}\!\rh
\!+\!G_{12}\!\lh\! \get^\perp\! \gc\! -\! 2 \get^\parallel
\get^\perp\!\!-\! \frac{1}{2} (\tW_{211}\!
	+\! \tW_{222})\! \rh\!\! \Bigg].\nn
\eea
The first line of (\ref{fNLgeni}), proportional to $\gamma^2$, comes
from the $f(k',k')$ term and it is identical to expression
(\ref{srtot}). The difference is that now it occurs with a weight
$\gamma^2$ compared to the terms originating from $f(k',k)$ that come
with a weight 2. Obviously, in the case of equal momenta where
$G_{ijk_1k_2}=\delta_{ij}$, the expression reduces to equation (\ref{srtot}). 
The $\gamma$ terms can be safely neglected in the squeezed limit because
$\gamma^2$ scales as $e^{-3\Delta t}$, where $\gD t$ is the number of e-folds
between horizon exit of the two scales. If for example the two scales
exit the horizon with a delay $\Delta t\sim 7$, which corresponds to
$k'\sim 1000k$, approximately the resolution of the Planck satellite, we
find that $\gamma^2\sim10^{-9}$. We will extensively  investigate the scale dependence of $\f$ in chapter \ref{Ch6}.

%The functions $\bv_{12}$ and $\bv_{22}$ increase and decrease
%respectively (from their initial values 0 and 1) only a little until
%the turning of the fields. The later the relevant scale exits the
%horizon, the less time there is available for $\bv_{i2}$ to evolve, so
%the smaller is the value of $\bv_{12}$ (and the larger for $\bv_{22}$)
%during this period.  During the turning of the fields isocurvature
%effects turn on. Both $\bv_{12}$ and $\bv_{22}$ vary wildly during
%this period. $\bv_{12}$ grows and reaches a constant value afterwards,
%while $\bv_{22}$ varies and reaches zero when isocurvature effects
%cease.  In the models we studied we found that while during this
%period $\bv_{22}$ continues to behave in the same way, i.e. being
%larger for the scale that exits last, $\bv_{12}$ changes behaviour and
%also becomes larger for the scale that exits last. In the end we
%observe that $\f$ in the squeezed limit is smaller than in the
%equal-momenta case that was treated in the previous subsection. The
%effect is particularly pronounced during the turn of the field
%trajectory, mainly due to $g_{iso}$.  As we will show in
%section~\ref{secNumQuadrPot}, these effects can reduce the value of
%$\f$ during the turn of the field trajectory by $10\%$ on scales that
%are within the resolution of Planck. 

%------------------------------ Chapter 5 --------------------------------------

{\Chapter{Concrete examples}\label{Ch5}}
%------------------------------------------------------------------------------------------

In this chapter we will study several classes of field potentials that allow for the integral in (\ref{srtot}) or in (\ref{fNLresultSR}) to be solved analytically, presenting the results of our paper \cite{Tzavara:2010ge}. Although our exact result (\ref{fNLresult}) can treat numerically any type of potential, we are still interested in obtaining analytical slow-roll estimates in order to derive qualitative conclusions about the behaviour of $\f$. The types of potentials we are going to study do not exhaust the large variety of models in the literature but are still representative of the more common models. 
Next, we present numerical results in order to compare with the analytical slow-roll estimates, and study a model found in \cite{Tzavara:2010ge} able to produce $\f$ of order $\cO(1)$.  

Various multiple field models have been studied in the literature. Non-standard kinetic term models producing non-Gaussianity at horizon-crossing were studied for example in
\cite{Langlois:2008qf,Arroja:2008yy,Mizuno:2009cv,
Cai:2009hw,Senatore:2010wk}. 
Large non-Gaussianity can also be produced at the end of inflation
\cite{Lyth:2005qk,Bernardeau:2004zz,Barnaby:2006km,Enqvist:2004ey,
Enqvist:2005qu,Jokinen:2005by}
or after inflation, in models with varying inflaton decay rate
\cite{Zaldarriaga:2003my} and in curvaton models
\cite{Bartolo:2003jx,Enqvist:2005pg,Ichikawa:2008iq,Malik:2006pm,Sasaki:2006kq,
Huang:2008zj}. Within the $\delta N$
formalism several authors have investigated the bispectra of specific
multiple field inflation models
\cite{Seery:2005gb,Kim:2006te,Battefeld:2006sz,Battefeld:2007en,
Langlois:2008vk,Cogollo:2008bi}. Two-field models, being easier to deal with, 
have gained popularity
though. Vernizzi and Wands studied the double field sum potential
\cite{Vernizzi:2006ve}, while the double product potential was studied
in \cite{Choi:2007su}. Conditions for large non-Gaussianity were found in
\cite{Byrnes:2008wi}.

\Section{Potentials with equal powers}
\label{secEqPow}

We start with studying sum potentials of equal powers. The lowest order such potential is of course the quadratic potential (\ref{model}), which is already known to produce no substantial non-Gaussianity. We will relate this characteristic of the quadratic potential with the the integral (\ref{fNLresultSR}). Next we will derive a general conclusion for potentials of any power.

\subsection{Quadratic potential}

The quadratic potential has been widely examined in the past and it is
known that it cannot produce large non-Gaussianity (see for example
\cite{Vernizzi:2006ve}).  Here we use our results to analytically
explain why. While the quadratic potential is a special case of the
more general sum potential treated later on, it is still
interesting to discuss it separately in a different way. We start by
deriving the result that for a quadratic two-field potential within
slow roll,
	\be \label{chirel}
	\gc = \frac{\d}{\d t} \ln \frac{\ge \get^\perp}{\get^\parallel}.
	\ee 
Working out the right-hand side, using (\ref{dere}), we find
	\be 
	\gc
	= 2\ge + \get^\parallel - \frac{(\get^\perp)^2}{\get^\parallel}
	- \frac{\gx^\parallel}{\get^\parallel} 
	+ \frac{\gx^\perp}{\get^\perp}.
	\ee 
Inserting the relations (\ref{xiparxiperpSR}) (with the third derivatives of the
potential equal to zero, since we have a quadratic potential) this reduces to
	\be 
	\gc = \ge + \get^\parallel - \frac{(\get^\perp)^2}{\get^\parallel}.
	\ee 
It can be checked that this result does indeed satisfy the general equation 
for the time derivative of $\gc$ (\ref{dere}) within the approximations made,
and the remaining integration constant is fixed by realizing that this result
has the proper limit in the single-field case. This concludes the proof of
(\ref{chirel}).

Since the third-order potential derivatives as well as the first term of the
integral in (\ref{fNLresultSR}) are identically zero, we find that for
a quadratic potential the integral completely vanishes in the slow-roll
approximation and no persistent large non-Gaussianity is produced. 
Numerically we find that even for large mass ratios, when during the turn
of the field trajectory slow roll is broken, the integral is still
approximately zero, see subsection~\ref{secNumQuadrPot}.

Using this result (\ref{chirel}) for $\gc$ we can also solve (\ref{gSReq}):
	\be 
	g(t) = \frac{\get^\parallel}{\ge \get^\perp},
	\ee 
and hence find that
	\be\label{GsolSRquadr}
	G_{22}(t,t')= \frac{\ge(t') \get^\perp(t')}{\get^\parallel(t')}
	\frac{\get^\parallel(t)}{\ge(t) \get^\perp(t)},\ \ \ \ \ \ \ 
	G_{12}(t,t') = - \frac{\ge(t') \get^\perp(t')}{\get^\parallel(t')}
	\lh\! \frac{1}{\ge(t)} \!+\! 2 t\! -\! \frac{1}{\ge(t')} \!-\! 2t'\!\rh.
	\ee
Note that even though $g(t)$ is a large quantity, of order inverse slow roll, it
is still slowly varying, as we have shown, with its time derivative an order of
slow roll smaller.

\subsection{Potentials of the form \texorpdfstring{ $W=\alpha \phi^p+\beta \gs^q$}{}}\label{pots}

For a potential of the form
\be
W(\phi,\gs)=\alpha \phi^p+\beta \gs^q
\label{polynomtype}
\ee
we can work out explicitly the form of the integrand in equation
(\ref{fNLresultSR}). We have to use the slow-roll version of equations
(\ref{eq2}) and (\ref{srpara2}) to easily find after substitution
that
\be
 g_{int}\!=\!-\!\!\int_{t_*}^t\!\!\frac{\alpha\beta p^4 
(y\!-\!1)\phi^{p-3}\gs^{py-3}\left(y(p y\!-\!1)\phi^2
+(p\!-\!1)\gs^2\right)
\left(\alpha^2\phi^{2 p}\gs^2\!+\!\beta^2 y^2\phi^2\gs^{2 p y}\right)^2\!\!}
{2\kappa^4\left(\alpha\phi^p+\beta\gs^{p y}\right)^4
\left(\alpha(p-1)\phi^p\gs^2-\beta y (py-1)\phi^2\gs^{p y}\right)^2}
\d t',
\ee
where $y\equiv q/p$. 

From this expression we can derive an important
result: for $y=1$, i.e.\ $p=q$, we immediately
see that the integral is zero. This means that no persistent
non-Gaussianity can be produced after horizon exit for potentials
of the form $W(\phi,\gs)=\alpha \phi^p+\beta\gs^p$, at least within
the slow-roll approximation. This generalizes the result for the 
two-field quadratic potential of the previous subsection to any potential 
with two equal powers. We will come back to the case $p\neq q$ in subsection \ref{uvn} where we will study the predictions for $\f$ produced by general sum potentials.

\Section{Other integrable forms of potentials}\label{intsection} 

In general, the first step of finding an analytical expression for
the integral $g_{int}$ is to solve the differential equation
(\ref{gSReq}) for $g$ in order to determine the Green's functions.
To do that, one tries to express $\chi$ as a time derivative
of some other quantity. In the slow-roll limit 
\be
\label{etaslow}
\tW_{11} = \ge - \hpa, \qquad\qquad
\tW_{21} = - \hpe,
\ee
so that $\chi$ can be written as
\be
\label{chislow}
\chi=2\e +\tilde{W}_{22}-\tilde{W}_{11}.
\ee
which, as can be checked, cannot be expressed as a derivative of a known 
quantity for a general potential. 
Here we introduced the notation $\tilde{W}_{mn}=W_{mn}/(3H^2)$. 
Thus we are forced to examine special 
classes of potentials.

\subsection{Product potentials}

First we consider potentials of the form
\be
W(\phi,\gs)=U(\phi)V(\gs),
\ee
inspired by the analytical study done in 
\cite{Choi:2007su,Byrnes:2008wi}. 
From our point of view, the advantage of these potentials 
is that their mixed second derivative $\tilde{W}_{\phi\gs}$ can be
expressed in terms of the first derivatives to finally give for
the second-order derivatives of the potential in the adiabatic
and isocurvature directions:
\bea
\label{mixed}
 \tilde{W}_{11}&=&\tilde{W}_{\phi\phi}\ef^2+\tilde{W}_{\gs\gs}\ex^2
+4 \epsilon\ef^2 \ex^2,\qquad\qquad 
\tilde{W}_{22}=\tilde{W}_{\phi\phi}\ex^2+\tilde{W}_{\gs\gs}\ef^2
-4\e\ef^2\ex^2,\nn\\
 \tilde{W}_{21}&=&\left(\tilde{W}_{\phi\phi}-\tilde{W}_{\gs\gs}
+2\e(\ex^2-\ef^2)\right)\ef\ex,
\eea
where we used (\ref{fbas}) to eliminate the unit vector $\vc{e}_{2}$
in terms of $\vc{e}_1$. It is straightforward to show that the second-order
derivatives in the directions of the basis vectors are
related:
\be
\label{w21slow}
\frac{2\e+\tilde{W}_{22}-\tilde{W}_{11}}{\tilde{W}_{21}}=\frac{\ex}{\ef}
-\frac{\ef }{\ex},
\ee
so that only two of them are independent. 
Now we can use (\ref{srpara2}) and the above results to write $\chi$ as
\be
\chi=\tilde{W}_{21}\left(\frac{\ex}{\ef}-\frac{\ef}{\ex}\right)
=-\frac{\mathrm{d}}{\mathrm{d}t}\ln\left(\ef\ex\right),
\ee
where the derivatives of the unit vectors are given in (\ref{dere}). 
Hence looking at equation (\ref{gSReq}) we can identify the 
Green's function $g$ to be
\be
g(t)=\ef(t)\ex(t).
\ee
After a few more manipulations the integrand of $G_{12}(t,t')$ 
in (\ref{GsolSR}) takes the form
\be
\hpe(t)g(t)=\frac{1}{4}\frac{\mathrm{d}S}{\mathrm{d}t},
\ee
where $S\equiv\ef^2-\ex^2$, so that the analytical form of the two 
independent linear perturbation solutions in the slow-roll approximation is (the same results were obtained 
in \cite{Peterson:2010mv} for the transfer functions $T_\mathcal{{RS}}$ and 
$T_\mathcal{{SS}}$ of product and sum potentials, 
which turn out to coincide with $\bv_{12}$ and $\bv_{22}$).
\be
\bv_{12}=\frac{S-S_*}{2 e_{1\phi *}e_{1\gs*}},\qquad\qquad 
\bv_{22}=\frac{e_{1\phi}e_{1\gs}}{e_{1\phi *}e_{1\gs *}}.
\ee

The final step is to write the integrand of $g_{int}$ in (\ref{srtot}) 
in terms of the potential's derivatives and rearrange terms to form
time derivatives. One can prove that then the integrand can be rewritten as
\be
 g_{int}=\frac{1}{1-S_*^2}\int_{t_*}^t \frac{\d}{\d t'}
\Bigg[\Big(S(t)-S(t')\Big)
\Big(\tilde{W}_{\sigma\sigma}(t')\ef^2(t')-\tilde{W}_{\phi\phi}(t')
e^2_{1\sigma}(t')\Big)\Bigg]\d t'.
\ee
After performing the integration and adding the rest
of the terms we find that
\be
-\frac{6}{5}\f=\frac{2(S-S_*)^2(S_*^2-1)}{(1+S^2-2SS_*)^2}
\left(g_{iso}+g_{sr}+g_{int}\right),
\label{fpro}
\ee
where now
\bea
 g_{iso}&=&
\frac{S^2-1}{S_*^2-1}\left(\e+\hpa-\chi\right),
\nn\\ 
 g_{sr}&=&
-\frac{1}{2(S-S_*)}\Bigg[\lh\es+\hpas\rh\frac{1+3S(S-2S_*)+2S_*^2}
{S-S_*}-\chi_*\frac{-3+S^2+4SS_*-2S_*^2}{2S_*}\Bigg],\nn\\
 g_{int}&=&
-\frac{S_*(S-S_*)}{S_*^2-1}
\left(\es+\hpas-\chi_*\frac{S_*^2+1}{2S_*^2}\right).
\eea
Comparing
to the results of \cite{Choi:2007su,Byrnes:2008wi} we find complete 
agreement. 

Looking at the result for $\f$ for the product potential we can draw a number
of conclusions. The only time-dependent
slow-roll parameters appear in $g_{iso}$. These terms and consequently 
$\f$ can vary significantly during a turn of the field trajectory but, as 
explained before, in the models we consider isocurvature modes have
disappeared by the end of inflation so that the adiabatic mode will be 
constant after inflation, which means $g_{iso}$ will disappear again and
cannot give any persistent non-Gaussianity.
The rest of the terms involve slow-roll parameters evaluated at horizon
crossing, which are small.
Hence we conclude that any large non-Gaussianity will have to come from
the denominator becoming very small (since $|S| \leq 1$ the numerator cannot
become large) to compensate for the small slow-roll parameters at horizon 
crossing. We see that this can only happen when $S,S_* \rightarrow \pm 1$.
In the remainder of this section we will study the two different cases
that satisfy this condition: a $90^\circ$ turn in the field trajectory
($S=-S_*$), or the same field dominating both at the beginning and at the end
($S=S_*$).

First we study the case where the field trajectory makes a $90^\circ$ turn.
The field $\phi$ is dominant right after horizon crossing, which means 
$|\exs|\ll 1$, $|\efs| \approx 1$ and hence $S_*\rightarrow 1$. 
Later on a turn in the field trajectory occurs  and afterwards $\gs$ dominates 
inflation, so that $|\ef|\ll 1$, $|\ex|\approx 1$ and $S\rightarrow-1$.  
Then we find that both $g_{sr}$ and $g_{iso}$ go to zero, which means 
in particular that we satisfy the condition on the disappearance of the 
isocurvature mode that allows us to directly extrapolate the results at 
the end of inflation to the time of recombination. 
The non-zero term comes as expected from $g_{int}$
and it is given by:
\be
-\frac{6}{5}\f = \ge_* + \hpa_* - \gc_* = - \tW_{\gs\gs*}, 
\ee
since 
$\tW_{\gs\gs*}=\tW_{22*}=\gc_*-\ge_*-\hpa_*$.
Hence we see that for any product potential where the field trajectory
makes a $90^\circ$ turn no significant non-Gaussianity will be produced,
at least within the slow-roll assumptions used to derive this analytic
result.

Next we look at the opposite limit, where one of the fields, $\phi$,
is dominant both at horizon crossing and at the end of inflation.
This means $|\exs|\ll 1$, $|\efs| \approx 1$ and $|\ex|\ll 1$, $|\ef|
\approx 1$, so that $S_*\rightarrow 1$ and $S\rightarrow 1$. This
includes the case where we have a perfectly straight field trajectory,
i.e.\  an effectively single-field situation, where obviously no
super-horizon non-Gaussianity is produced. However, we find that even
more generally in this limit the contributions from $g_{iso}$ and
$g_{int}$ go to zero and we are left with only the single-field result
from $g_{sr}$:
\be
-\frac{6}{5}\f =\ge_* + \hpa_*.
\ee
Hence no significant non-Gaussianity is produced in this limit.

We conclude that if we impose the condition of the disappearance of the
isocurvature mode by the end of inflation, to simplify the evolution
afterwards, the product potential can never give large non-Gaussianity, 
at least within the slow-roll approximation.

\subsection{Potentials of the form \texorpdfstring{ $W(\phi,\gs)=(U(\phi)+V(\gs))^{\nu}$}{(U+V)n}}\label{uvn}

Next we consider potentials of the form
\be
W(\phi,\gs)=(U(\phi)+V(\gs))^{\nu},
\ee
first studied for general 
$\gn$ in \cite{Tzavara:2010ge} and \cite{Meyers:2010rg}.  
While of course not the most general two-field potential, 
it can accommodate potentials with coupling terms of the form 
$\alpha^2\phi^2+\beta^2\gs^2+2\alpha\beta\phi\gs$ or higher-order 
combinations.
Note that in the case of $\gn=1$ the potential becomes the simple sum potential,
which has been studied before \cite{Vernizzi:2006ve,Byrnes:2008wi}.

Just as for the product potential, we find that mixed second derivatives 
of the potential can be expressed in terms of the other derivatives:
\bea
&&\tilde{W}_{11}\!=\!\tilde{W}_{\phi\phi}\ef^2+\tilde{W}_{\gs\gs}\ex^2
+\frac{4\e(\nu -1)\ef^2\ex^2}{\nu },\ \ 
\tilde{W}_{22}\!=\!\tilde{W}_{\gs\gs}\ef^2+\tilde{W}_{\phi\phi}\ex^2
-\frac{4\e(\nu -1)\ef^2\ex^2}{\nu },\nn\\
&&\tilde{W}_{21}\!=\!\left(\tilde{W}_{\phi\phi}-\tilde{W}_{\gs\gs}
\right)\ef\ex+\frac{2\e(\nu -1)\ef\ex(\ex^2-\ef^2)}{\nu }.
\eea
Again there are only two independent second derivatives of the potential
in our basis:
\be
\label{w21sum}
\tilde{W}_{21}=\left(\tilde{W}_{22}-\tilde{W}_{11}+\frac{2\e(\nu-1)}{\nu}\right)\frac
{\ex\ef} {\ex^2-\ef^2}.
\ee
Following the procedure of the previous section we rewrite $\chi$ as
\be
\chi=\frac{2\e}{\nu} +\tilde{W}_{21}\left(\frac{\ex}{\ef}-\frac{\ef}{\ex}\right)
=-\frac{\mathrm{d}}{\mathrm{d}t} \ln\left(H^{2/\nu}\ef\ex\right),
\ee
and then find an analytical expression for $g$,
\be
g(t)=H^{2/\nu}(t)\ef(t)\ex(t).
\ee
The integrand of $G_{12}(t,t')$ is now written as
\be
\hpe(t)g(t)=\frac{1}{2}\left(\frac{\kappa^2}{3}\right)^{1/\nu}\frac
{\mathrm{d}
Z}{\mathrm{d}t},
\ee
where $Z\equiv V\ef^2-U\ex^2$. Finally we find that
\be
\bv_{12}=\frac{Z-Z_*}{W_*^{1/\nu} e_{1\phi *}e_{1\gs*}},\qquad\qquad
\bv_{22}=\frac{W^{1/\nu} e_{1\phi}e_{1\gs}}{W_*^{1/\nu}
e_{1\phi *}e_{1\gs *}}
\label{G12}.
\ee

Rewriting the integrand of $g_{int}$ in terms of the potential's derivatives yields after a few manipulations
\bea
 g_{int}=\frac{W_*^{-2/\nu}}{\efs^2\exs^2}&&
\!\!\!\!\!\!\!\!\!
 \int_{t_*}^t\!\!\!\d t'
\Bigg\{\frac{\d}{\d t'}\Big[\frac{W^{2/\nu}(t')\e(t')\ef^2(t')
e_{1\sigma}^2(t')}{\nu}\Big]\nn\\ 
 &&
\!\!\! 
 +\frac{\d}{\d t'}\Big[W^{1/\nu}(t')\frac{Z(t)-Z(t')}{2}
\Big(\tilde{W}_{\sigma\sigma}(t')\ef^2(t')-\tilde{W}_{\phi\phi}(t')
e^2_{1\sigma}(t')\Big)\Big]\Bigg\}
\eea
and adding the rest of the terms results in
\be
-\frac{6}{5}\f\!=\!-\frac{2W_*^{2/\nu}(Z-Z_*)^2\efs^2\exs^2}
{\left(\exs^2(Z+U_*)^2\!+\!\efs^2(Z-V_*)^2\right)^2}
\Bigg(g_{iso}+g_{sr}+g_{int}\Bigg),
\label{nuv}
\ee
where
\bea
 g_{iso}&=&
\left(\frac{W^{1/\nu}\ef\ex}{W_*^{1/\nu}\efs\exs}\right)^2
\Bigg(\e+\hpa-\chi\Big),\nn\\
 g_{sr}&=&
-\frac{3}{2}(\es+\hpas-\chi_*)+\frac{Z-Z_*}{W_*^{1/\nu}(\efs^2-\exs^2)}
(-\frac{\es}{\nu}+\frac{\chi_*}{2})
\Bigg[1-\frac{3W_*^{2/\nu}\efs^2\exs^2}{(Z-Z_*)^2}\Bigg]\nn\\
 &&\!\!\!\!-(\es+\hpas)\frac{W_*^{2/\nu}\efs^2\exs^2}{2(Z-Z_*)^2}\nn\\
 g_{int}&=&
\frac{Z-Z_*}{2W_*^{1/\nu}}\left(\frac{1}{\exs^2}-\frac{1}{\efs^2}\right)
\Bigg(\es+\hpas-\frac{\chi_*}{2}\Big(1+\frac{1}{(\efs^2-\exs^2)^2}\Big)
\Bigg)\nn\\
&&\!\!\!\!+\frac{\e}{\nu}\left(\frac{W^{1/\nu}\ef\ex}{W_*^{1/\nu}\efs\exs}
\right)^2-\frac{\es}{\nu}
\Big(1-\frac{2(Z-Z_*)}{W_*^{1/\nu}(\efs^2-\exs^2)}\Big).
\label{nuvt}
\eea
Note that the first term on the second line of $g_{int}$ is also
related to the pure isocurvature mode (the term in the parentheses is just $\bv_{22}$ (\ref{G12})), but we have not incorporated it
in $g_{iso}$ in order to remind the reader that it originates from the
integral. 

As in the case of the product potential we will study two limiting cases,
to get some insight into the behaviour of $\f$. First is the limit where
the field trajectory makes a $90^\circ$ turn. We assume that $\phi$ 
dominates inflation at horizon exit, that is $|\exs| \ll 1$,
$|\efs| \approx 1$ and $Z_* \rightarrow V_*$. At late times, after the 
turn of the field trajectory, the second field $\gs$ is dominant and the 
contribution of $\phi$ is negligible, so that $|\ef|\ll 1$,
$|\ex| \approx 1$ and $Z \rightarrow -U$. 
Then we find that $g_{sr}$ and $g_{iso}$ go to zero, while the remaining
contribution to $\f$ comes from $g_{int}$, as expected,
\be
-\frac{6}{5}\f=-\frac{U_*+V_*}{U+V_*}\tW_{\gs\gs *}
= \frac{U_*+V_*}{U+V_*} \lh \ge_* + \hpa_* - \gc_* \rh. 
\label{lim}
\ee
So we see that we need a significant decrease in $U$ between horizon crossing
and the end of inflation, as well as a relatively small value of $V_*$, 
to get a large $\f$. Of course we cannot increase
$U_*/U$ too much without breaking slow roll.
In section~\ref{secNewPot} we investigate numerically the properties of a 
model with a sum potential and confirm the validity of the above limit.

Let us consider the consequences of this limit for a sum of monomial potentials like the models studied in subsection \ref{pots}. Because of the form of its potential, the final value of the initially dominant field $\phi$ will be zero and hence $U=0$. Assuming slow-roll at horizon crossing we approximate $3H^2_*\approx \kappa^2\lh U_*+V_*\rh $ (we remind that $\tilde{W}_{\sigma\sigma*}=W_{\sigma\sigma*}/(3H_*^2)$ ) and for $V=\beta\sigma^q$, (\ref{lim}) becomes
\be
-\frac{6}{5}\f=-\frac{V_{\gs\gs *}}{\kappa^2 V_*}=-\frac{q(q-1)}{\kappa^2 \sigma_*^2}.
\ee
Let us study this simple result. $\f$ can be large either if $\sigma_*$ is small or if $q$ is a big integer. However, if $\sigma_*$ is small, the $\sigma$ field will rapidly reach its minimum  before the heavy field $\phi$ has come to rest, and hence at the end of inflation both fields will perform oscillations and our assumptions $|\ef|\ll 1$, $|\ex| \approx 1$ will not hold any more. 
On the other hand, if $q\mg1$, $\sigma$ will take over the domination of the universe before $\phi$ has reached its minimum, resulting again in both fields oscillating at the end of inflation and hence non-vanishing isocurvature modes  at the time. We conclude that within our assumptions a sum potential of the form (\ref{polynomtype}) cannot produce large non-Gaussianity. 

In the opposite limit $\gf$ dominates both at horizon crossing
and at the end of inflation, i.e.\  $|\exs|\ll 1$, $|\efs| \approx 1$ and 
$|\ex|\ll 1$, $|\ef| \approx 1$ so that $Z_* \rightarrow V_*$ and
$Z \rightarrow V$. Then the expression reduces to
\be
-\frac{6}{5}\f = -\frac{U_*+V_*}{V-V_*} \lh \ge_* + \hpa_* - \gc_* \rh,
\ee
which comes from $g_{int}$. Note that we have assumed here that $V \neq V_*$. 
In the (effectively)
single-field case this is not valid; in that case we find that $g_{int}$
and $g_{iso}$ are zero and $g_{sr}$ goes to the single-field result,
$\ge_* + \hpa_*$. 
We remark that in this limit $g_{iso}$ is zero, so that the
adiabatic mode is conserved after inflation.
In order to make $\f$ large, one might be tempted to take $V$ close to
$V_*$. However, that means $\sigma$ does not evolve and we are in an
effectively single-field situation, where the above limit is not
valid. Instead the situation is somewhat similar to the previous
limit: we need a large value of $U_*$ and relatively small values of
$V_*$ and $V$ to overcome the small values of the slow-roll parameters
at horizon crossing. It might not be simple to satisfy these
conditions together with the requirements of this limiting case that
$\phi$ dominates both at horizon crossing and at the end of inflation,
with a period of $\sigma$ domination in between; we did not further
study those types of models.

As a final remark we point out that the power $\nu$ of the potential
does not appear explicitly in the limits for $\f$. Of course its value
will play a role in determining the field trajectory and the values
of the slow-roll parameters, but that is only a relatively small effect.
We have verified this result numerically for several values of the power
$\nu$ of sum potentials of the form (\ref{polynomtype}).

\Section{Numerical results}
\label{numerical}

The formalism we have developed so far provides a tool to calculate
the exact amount of non-Gaussianity produced during inflation driven
by a general two-field potential, beyond the slow-roll
approximation. While we assumed slow roll in the previous sections,
in order to derive analytical results, we return here to the
exact formalism for a numerical treatment.
In the following subsections we investigate the
properties of the quadratic potential as well as a potential of the
sum type that can produce an $\f$ of the order of a few, and compare our
results to those of the $\delta N$-formalism.

\subsection{Comparison with \texorpdfstring{ $\delta N$}{DN} for the quadratic 
potential}
\label{secNumQuadrPot}

We investigate the quadratic potential (\ref{model})
\be
W_q=\frac{1}{2}m_{\phi}^2\phi^2+\frac{1}{2}m_{\gs}^2\gs^2
\ee
choosing our parameters as follows: $m_{\phi}/m_\gs=20$, 
$m_{\gs}=10^{-5}\kappa^{-1}$ and the initial conditions 
$\phi_0=\gs_0=13\kappa^{-1}$ at $t=0$ for a total of about 85
e-folds of inflation. From now on we will denote the heavy field as $\phi$. 
We choose to present this particular mass ratio because 
%the fields oscillate 
%wildly during the turn and 
slow roll is badly broken during the turn and hence it provides a 
serious check both of our formalism and the $\gd N$ one. Of course we have
also run tests with smaller mass ratios when slow roll is unbroken and verified
our analytical slow-roll results.

\begin{figure}[h]
\begin{tabular}{cc}
\includegraphics[scale=0.7]{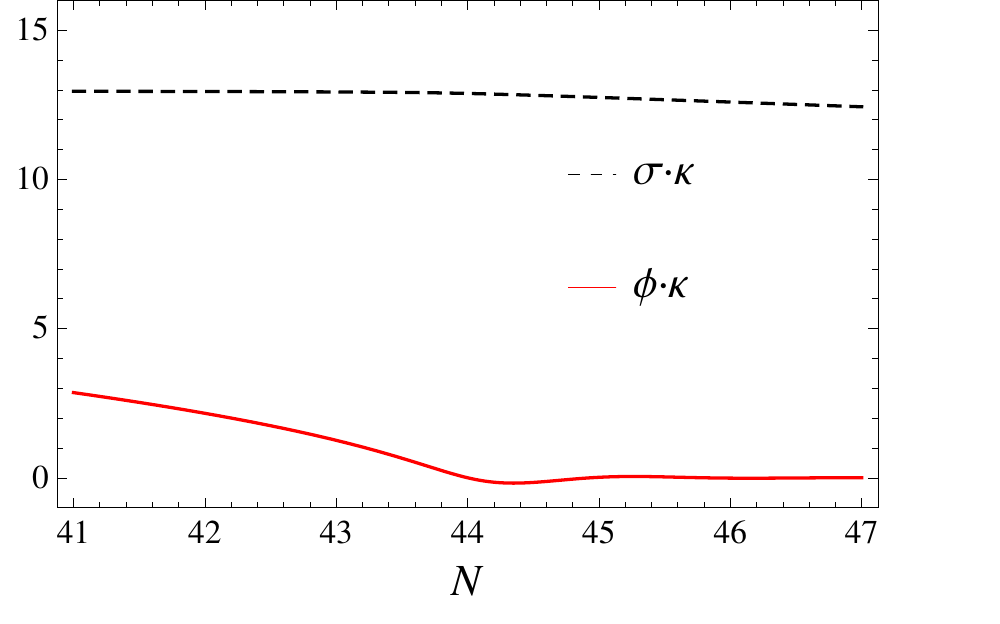}
&\includegraphics[scale=0.65]{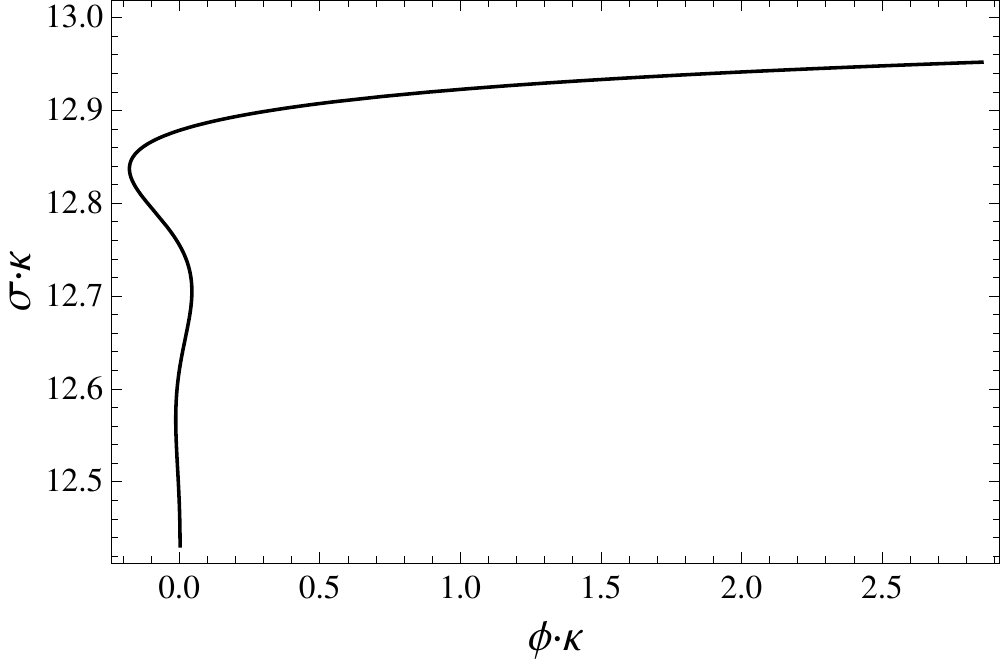}
\end{tabular}
\caption{The time evolution of the fields (left) and the field trajectory
(right) during the period of the turn of the field trajectory,
for the model (\ref{model}) with initial conditions 
$\phi_0=\gs_0=13\kappa^{-1}$ and mass ratio $m_{\phi}/m_{\gs}=20$.}
\label{fig51}
\end{figure}

\begin{figure}

\begin{tabular}{cc}
\includegraphics[scale=0.65]{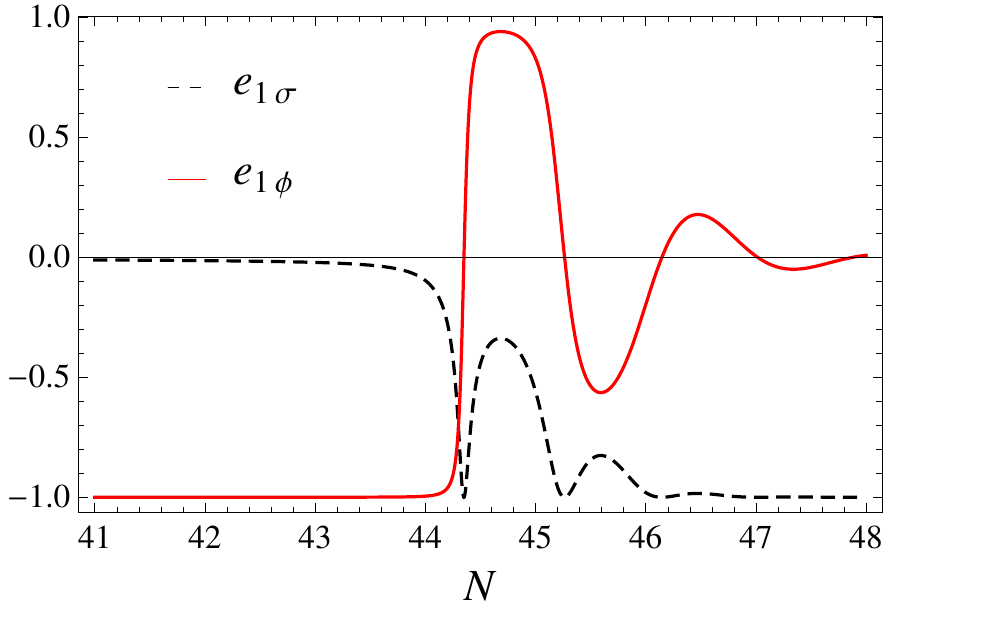}
&\includegraphics[scale=0.65]{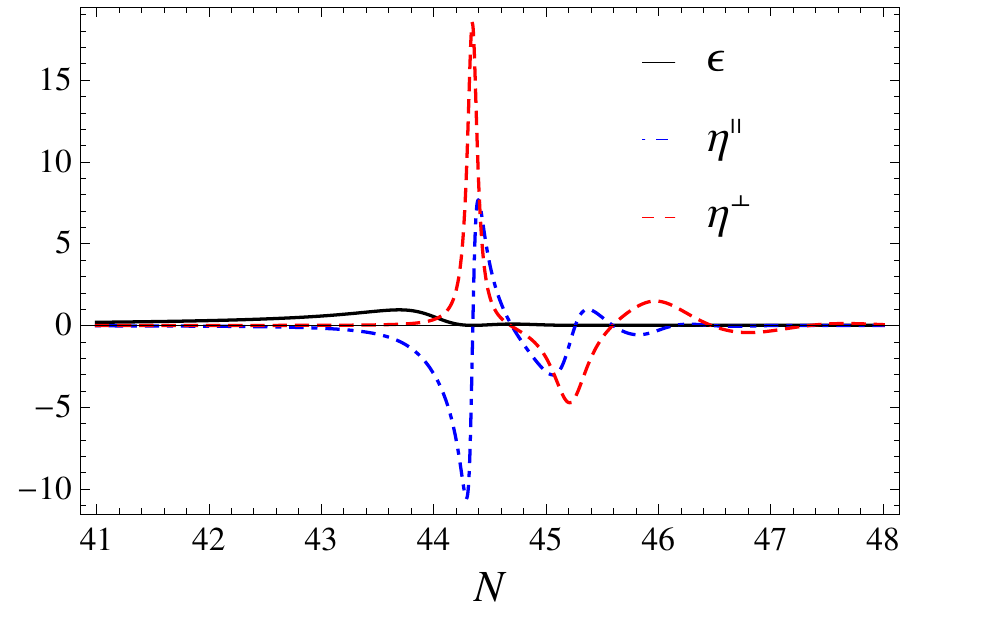}
\end{tabular}
\caption{The unit vectors (left) and the slow-roll parameters 
$\e,\hpa$ and $\hpe$ (right) as a function of time during the turn of the
field trajectory, for the same model as in figure~\ref{fig51}.}
\label{fig52}
\end{figure}

We solve the field equations (\ref{eq2}) numerically and in
figures~\ref{fig51} and \ref{fig52} we plot the values of the fields,
the unit vectors, and the slow-roll parameters as a function of time
during the range of e-folds where the heavy field $\phi$ is approaching zero
and starts oscillating. In
the beginning of inflation $\phi$ dominates the expansion while
rolling down its potential and about 40 e-folds after the initial
time $t=0$ it starts oscillating around the minimum of its
potential. The heavier $\phi$ is, the more persistent are the damped
oscillations. During the period of oscillations the unit vectors, as
well as the slow-roll parameters $\e$, $\hpa$, and $\hpe$, oscillate
too. For $m_{\phi}/m_\gs=20$ the maxima of
the slow-roll parameters are much larger than unity and slow roll is
temporarily broken. 
During these oscillations the light field $\gs$ starts driving 
inflation and rolls down its potential until it also reaches its minimum and
starts oscillating. We take the end of inflation when $\e=1$ during this
second period of oscillations.  The situation is
similar to the limiting case we studied in subsection \ref{uvn} with
$|\exs|\ll1$ and $|\ef|\ll1$.

\begin{figure}[h]
\begin{center}
\begin{tabular}{ll}
\includegraphics[scale=0.8]{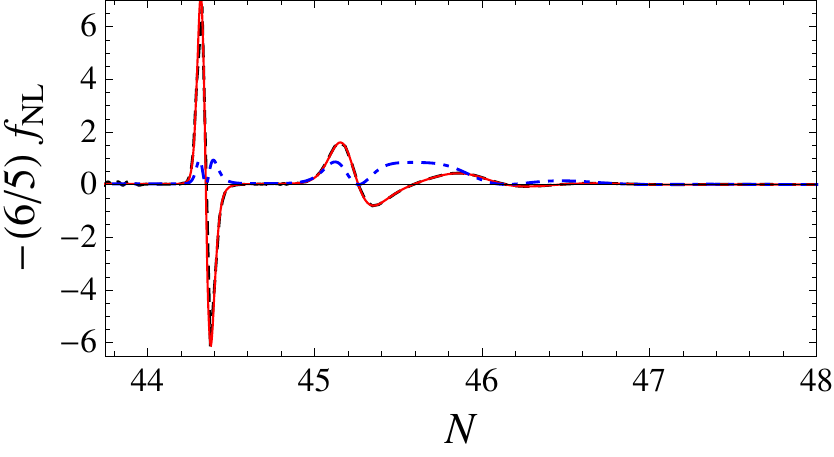}
&\includegraphics[scale=0.8]{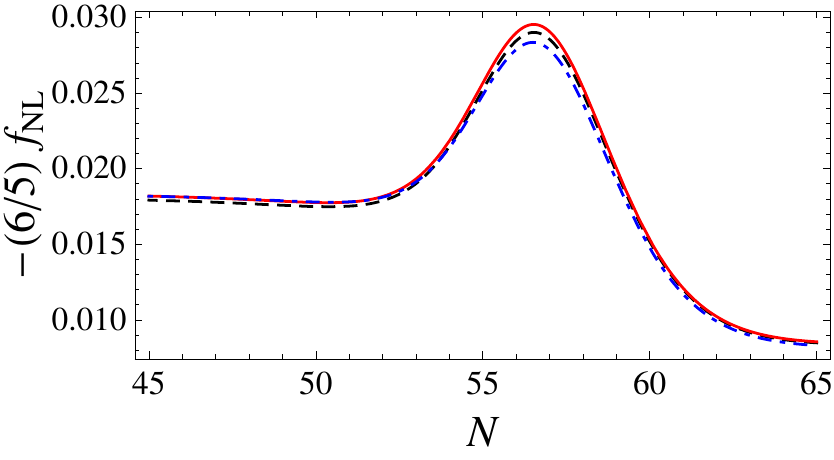}
\end{tabular}
\caption{We plot the $\f$ parameter for the model (\ref{model}) with initial 
conditions $\phi_0=\gs_0=13\kappa^{-1}$ and mass ratio $m_{\phi}/m_{\gs}=20$ 
(left) and $m_{\phi}/m_{\gs}=4$ (right).
The red line is the exact numerical result, while the blue dot-dashed line 
shows 
the slow-roll analytical approximation (but using the exact background).
We also show the numerical $\delta N$ result as the black dashed line, which
lies practically on top of our red result.}
\label{fig53}
\end{center}
\end{figure}

In figure \ref{fig53} we plot the $\f$ parameter as calculated in
our formalism, both the numerical exact version (\ref{fNLresult}) 
and the slow-roll 
analytical approximation (\ref{fNLresultSR}) (but using the 
exact background), as well as the result computed numerically
in the context of the $\delta N$-formalism.  The horizon-crossing
time is defined as 60 e-folds before the end of inflation. We do not
expect any large non-Gaussianity to be produced in this model,
since we have shown that the integral of (\ref{fNLresultSR}) is equal
to zero in the slow-roll approximation. The final
value of $\f$ calculated in all three cases is $\mathcal{O}(10^{-2})$. 
Our results coincide with those of
the $\delta N$-formalism, thus reinforcing the validity of both
formalisms. Any deviation between the results of the two formalisms is attributed to the different accuracy of our numerics.  
We also show $\f$ for a much smaller mass ratio, $m_\gf/m_\gs = 4$,
where slow roll remains valid throughout the turn of the field trajectory,
verifying our analytical slow-roll result.

The peak of the $\f$ parameter during the turning of the fields is due
to the isocurvature terms $g_{iso}$ in the slow-roll analytical formula.
As expected this effect is transient and disappears
when the isocurvature mode $\bv_{22}$ has been fully converted to the 
adiabatic one. There is no surviving isocurvature mode in this model.
The higher is the mass ratio, the larger is the magnitude of
the peak as a consequence of the more violent oscillations.

\begin{figure}[h]
\begin{center}
\begin{tabular}{cc}
\includegraphics[scale=0.8]{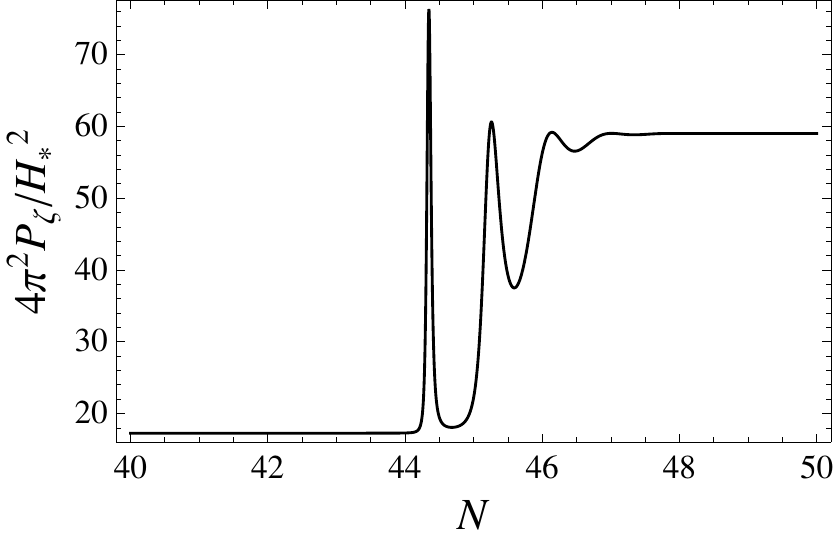}
&\includegraphics[scale=0.8]{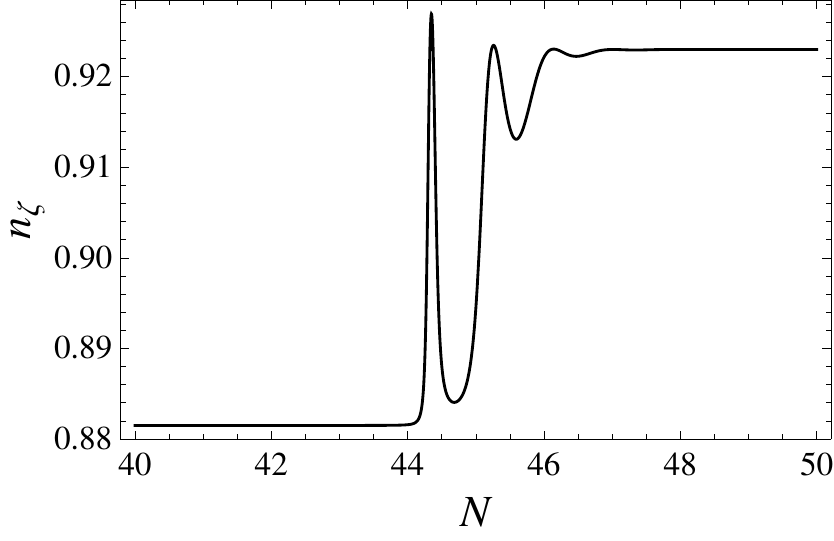}
\end{tabular}
\caption{The exact numerical power spectrum (left) and the spectral 
index (right) for the same model as in figure~\ref{fig51}.}
\label{fig54}
\end{center}
\end{figure}

For completeness, we plot in figure~\ref{fig54} the power spectrum 
(\ref{powerspectrum}) and the spectral index (\ref{spectralindex}) 
of this model. We see there is a jump in both of them
during the oscillatory period of the heavy field, but afterwards they
become constant again.

%Finally in figure~\ref{fig55} we plot the exact numerical $\f$ in
%the squeezed limit and in the equal momenta limit. As mentioned in 
%section~\ref{slowsq}, 
%we see that the $\f$ parameter in the squeezed limit is smaller than 
%in the equal-momenta one. 
%From figure \ref{fig55} we can see that for $k'=1000k$
%(roughly corresponding to the Planck resolution) the peak value of
%$\f$ is more than $10\%$ smaller than for $k'=k$, for this particular model.

%\begin{figure}
%\begin{center}
%\begin{tabular}{cc}
%\includegraphics[scale=0.8]{figure59}
%&\includegraphics[scale=0.8]{figure510}
%\end{tabular}
%\caption{In the first plot we depict the $\f$ parameter for the equal 
%momenta limit in red and the squeezed limit 
%result for $k'/k=1000$ in dashed black. In the second plot we show the 
%dependence of the discrepancy of the first peak value of $\f$ in the two 
%limits on the ratio $k'/k$. We used the same model as in figure~\ref{fig51}
%and $t_{k'}=25$ (60 e-folds before the end of inflation).}
%\label{fig55}
%\end{center}
%\end{figure}

\subsection{A simple model producing large non-Gaussianity}
\label{secNewPot}

In this section we introduce a model that produces an $\f$ of the order of 
a few, which is two orders of magnitude larger than the single-field 
slow-roll result. So in that sense we can call it large. From the point of 
view of observations with the Planck satellite it is probably still a little 
bit too small, but we have taken this particular model to be able to make
the connection with our analytical results.

\begin{figure}[h]
\begin{center}
\begin{tabular}{cc}
\includegraphics[scale=0.65]{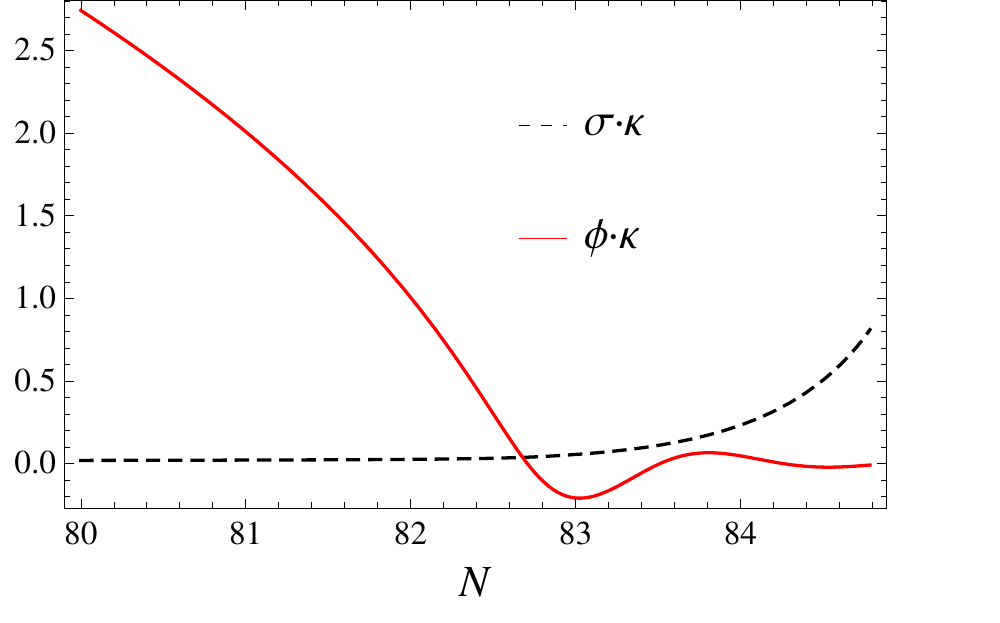}
&\includegraphics[scale=0.75]{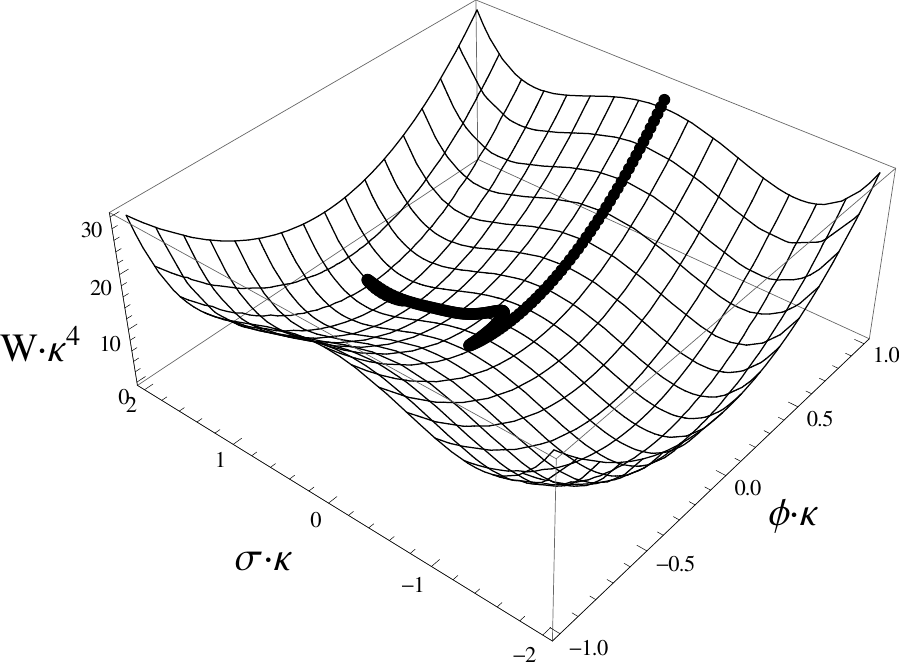}
\end{tabular}
\caption{The time evolution of the fields (left) and the field trajectory
(right), for the model (\ref{newpot}) with initial conditions 
$\phi_0=18\kappa^{-1},\gs_0=0.01\kappa^{-1}$ and parameters 
$a_2=20\kappa^{-2},b_2=7\kappa^{-2}$ and $b_4=2$. Only the time interval
during the turn of the field trajectory is shown.}
\label{fig56}
\end{center}
\end{figure}

The $\f$ limit (\ref{lim}) that we calculated in subsection \ref{uvn} can
be simplified for the sum potential ($\nu=1$) to give
\be
-\frac{6}{5}\f=-\frac{V_{\gs\gs *}}{\kappa^2(U+V_*)},
\ee
where we used the definition of $\tilde{W}_{mn}$ and the slow-roll
version of (\ref{eq2}) for $H$.  We can easily infer that in order
to obtain a large value for $\f$, the heavy field $\phi$ should end up
with a small value at the end of inflation, while $\gs$ should obey a
potential characterized by a large second derivative and a small value
at horizon crossing. Such properties can be accommodated by
a potential of the form
\bea
\label{newpot}
U(\phi)&=&a_2\phi^2,\nn\\
V(\gs)&=&b_0-b_2\gs^2+b_4\gs^4,
\eea
with $b_0=b_2^2/(4b_4)$ so that the minimum of the potential has $W=U+V=0$.

%\begin{figure}
%\begin{center}
%\begin{tabular}{cc}
%\includegraphics[scale=0.7]{figure510a}&
%\includegraphics[scale=0.7]{figure510a}
%\end{tabular}
%\caption{The potential (\ref{newpot}) and the the field trajectory
%(right), with initial conditions 
%$\phi_0=18\kappa^{-1},\gs_0=0.01\kappa^{-1}$ and parameters 
%$a_2=20\kappa^{-2},b_2=7\kappa^{-2}$ and $b_4=2$. Only the time interval
%during the turn of the field trajectory is shown.}
%\label{fig56a}
%\end{center}
%\end{figure}

\begin{figure}
\begin{center}
\begin{tabular}{cc}
\includegraphics[scale=0.65]{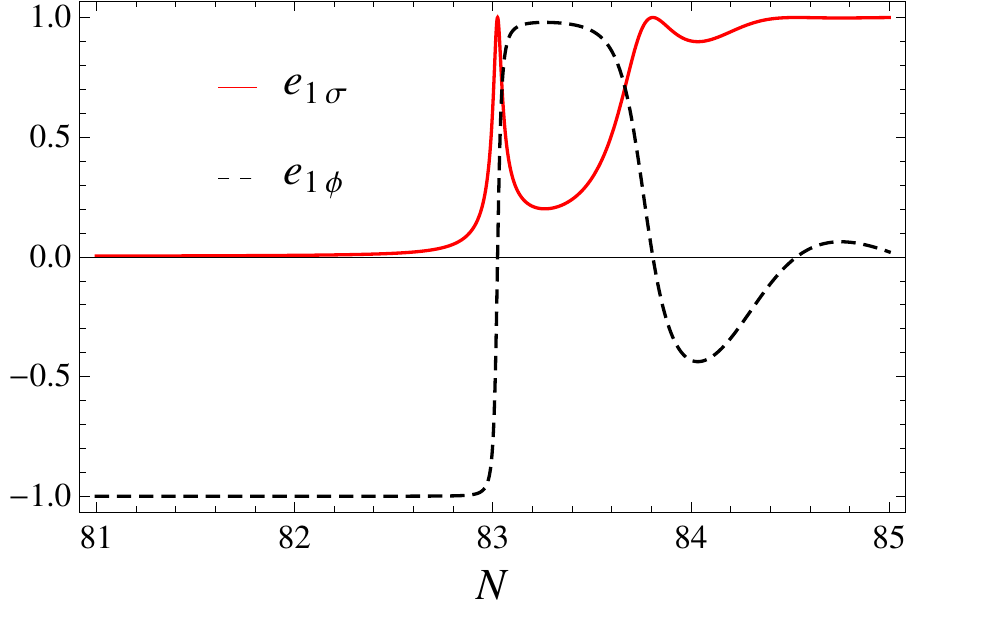}
&\includegraphics[scale=0.7]{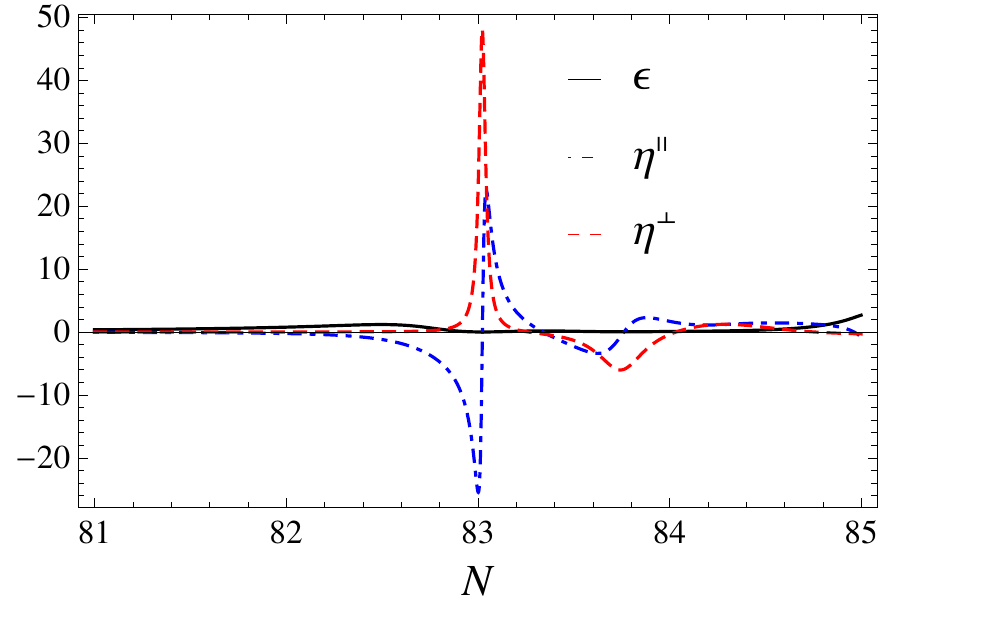}
\end{tabular}
\caption{The unit vectors (left) and the slow-roll parameters 
$\e,\hpa$ and $\hpe$ (right) as a function of time during the turn of the
field trajectory, for the same model as in figure~\ref{fig56}.}
\label{fig57}
\end{center}
\end{figure}

To illustrate the above we investigate a model with
$a_2=20\kappa^{-2},b_2=7\kappa^{-2}$, $b_4=2$, and initial
conditions $\phi_0=18\kappa^{-1}$ and $\gs_0=0.01\kappa^{-1}$, so
that the light field is standing on the local maximum of its
potential, for a total amount of 85 e-folds of inflation. 
This type of effective potential might be realized in the
early universe during second-order phase transitions.  We solve
the field equations (\ref{eq2}) numerically and in figures
\ref{fig56} and \ref{fig57} we plot the evolution of the fields and the 
unit vectors, as well as the slow-roll parameters. The situation is
qualitatively the same as in the case of the quadratic potential: in
the beginning $\phi$ dominates inflation while rolling down its
potential, then there is a period of violent oscillations around $\gf=0$, 
and $\gs$ takes over and starts rolling down towards the minimum 
of its potential.

The behaviour of the unit vectors is that of the limiting case we
studied in subsection \ref{uvn}, that is $|\exs|\ll1$ and $|\ef|\ll1$. We will next obtain an analytical estimate for the magnitude of $\f$. The final value of 
$\f$ is reached when the fields have rolled down to
their minima, that is at $\phi=0$ and $\gs=\sqrt{b_2/(2b_4)}$ for a positive
initial condition for $\gs$. Then $\f$ becomes
\be
-\frac{6}{5}\f=
8\frac{b(1-6b\gs_*^2)}{\kappa^2(1-2b\gs_*^2)^2},
\ee
where $b\equiv b_4/b_2$. Note that within our approximation $\f$
depends only on the value of $\sigma_*$ at horizon crossing once we
have fixed the ratio $b$.  Since the turning of the fields occurs
only a few e-folds before the end of inflation, we will explicitly
assume $W\simeq U$ is a good approximation for nearly all the period
of inflation.  Then we can solve the field equations in the slow-roll
approximation to find
\be
 \phi(t)=\phi_0\sqrt{1-\frac{4t}{\kappa^2\phi_0^2}},\qquad\qquad
\gs(t)=\Bigg[2b-\left(2b-\frac{1}{\gs_0^2}\right)
\left(1-\frac{4t}{\kappa^2\phi_0^2}\right)^{r}\Bigg]^{-1/2},
\ee
where $r\equiv b_2/a_2$.

The time of horizon crossing $t_*=t_{fin}-60$ can be approximately
found from the final time $t_{fin}\simeq\kappa^2\phi_0^2/4$ and thus
we calculate the values of the fields at horizon exit as functions of
the initial conditions $\phi_0$ and $\gs_0$.  Using these results in
$\f$ we find 
\be
-\frac{6}{5}\f=-\frac{8b}{\kappa^2\left(1-2b\gs_0^2\right)^2}
\Bigg[2\left(-1+\tilde{\phi}_0^{2r}\right)b\gs_0^2+1\Bigg]
\Bigg[2\left(1+2\tilde{\phi}_0^{2r}\right)b\gs_0^2-1\Bigg], 
\ee
where $\tilde{\phi}_0=\phi_0/(2\sqrt{60}/\kappa)=\phi_0/\phi_*$.

We now check the dependence of the above expression on the initial 
condition $\gs_0$. Since we assumed that $|\exs|\ll1$ 
and $W\simeq U$ we examine the case $\gs_0\ll 1$ where 
\be
-\frac{6}{5}\f=\frac{8b}{\kappa^2}(1-2b\gs_0^2\tilde{\phi}_0^{2r})
\label{an}
\ee 
up to second order with respect to $\gs_0$. The parameter $\f$ becomes 
maximal if $b=\tilde{\phi}_0^{-2r}/(4\gs_0^2)$ and its 
value is then
\be
-\frac{6}{5}\f=\frac{\tilde{\phi}_0^{-2r}}{\kappa^2\gs_0^2}.
\ee
Since $\tilde{\phi}_0>1$, the smaller the ratio $r$ and the smaller the 
initial value of the field $\gs$, the higher is the value of $\f$.

Nevertheless one has to assure that the turn of the field trajectory
does not occur too late (too close to the end of inflation), so that the
isocurvature mode will have had the time to disappear before the end of
inflation (so that we can directly extrapolate the results at the end of
inflation to the time of recombination and do not have to take further
evolutionary effects into account) and the oscillations of the heavy field 
do not coincide with those of the light field. 
The higher is the ratio $b$, the larger is $\f$, but
then the minimum of the potential approaches $\gs_0$ and consequently
there is less time available for $\bv_{12}$ and thus for the
adiabatic perturbation to become constant. This turns out to be a
non-trivial requirement: although we do not claim to have scanned the
whole parameter space of the model, we could not find parameter
values that passed the above test and at the same time yielded a very
large $\f$. The values we have chosen to work with respect the above condition and using expression (\ref{an}) we expect
to find $-(6/5)\f\sim 2$. Note that unlike the monomial potentials studied in subsection \ref{pots} and further discussed in subsection \ref{uvn}, it is the tuning of the extra parameter $b$ that allows to produce significant non-Gaussianity by the end of inflation without both fields oscillating at the same time.  

If one were to take $b_4=5$ instead of $2$, one would find $-(6/5)\f\sim 4$, 
but in that case the turn of the fields occurs too near the end of inflation so
that the isocurvature mode will not have disappeared
completely by the end of inflation. Looking at the contributions of $g_{iso}$ 
and $g_{int}$ separately, we see that even in that case $g_{int}$ has already 
gone to a constant while $g_{iso}$ is still decreasing towards zero, 
so that we feel reasonably confident that the estimate is good even
for that model, but we cannot be absolutely certain without a better
treatment of the end of inflation, which is beyond the scope of this thesis. 

\begin{figure}
\begin{center}
\begin{tabular}{cc}
\includegraphics[scale=0.85]{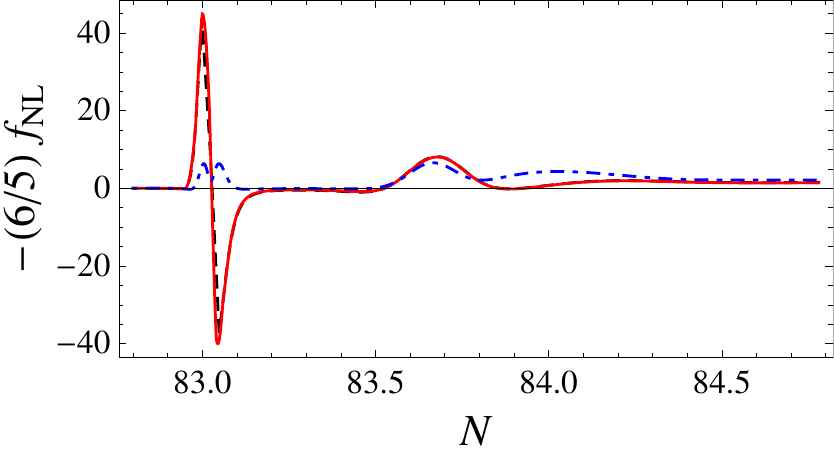}
&\includegraphics[scale=0.8]{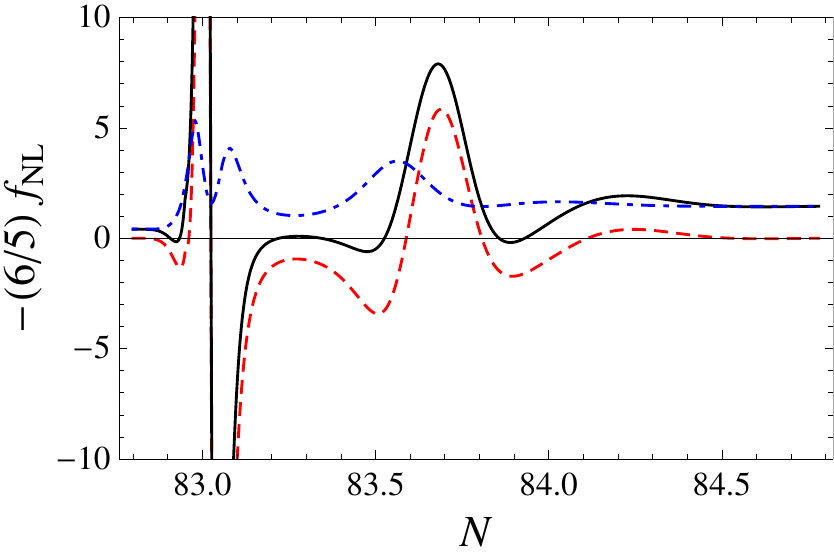}
\end{tabular}
\caption{In the first plot we show the non-Gaussianity parameter $\f$
  as calculated in our formalism exactly (red line) and within the
  analytical slow-roll approximation (blue dot-dashed line) as well as
  numerically in the context of the $\delta N$ formalism (black dashed
  line) for the model (\ref{newpot}). In the second plot we show again
  the total $\f$, now as the black solid line, and split it up into
  the isocurvature contribution proportional to $g_{iso}$ (red dashed
  line) and the integral contribution proportional to $g_{int}$ (blue
  dot-dashed line).  We use the same model as in figure~\ref{fig56}.}
\label{fig58}
\end{center}
\end{figure}

In figure~\ref{fig58} we plot $\f$ for the model (\ref{newpot}) with
the parameter values described above.  Again a notable feature comes
up during the turn of the fields. It comes from the isocurvature term
of (\ref{nuv}) that gets very big during the turn of the field
trajectory, but as soon as the fields relax it vanishes again. We do
not plot $g_{sr}$ separately since it turns out to be negligible. Note
how the final value of $\f$ depends only on the integrated effect, as
the isocurvature contribution has vanished. The final slow-roll
analytical value is calculated to be $-(6/5)f_{\mathrm{NL},sr}=2.15$
while the values obtained numerically by our formalism and the $\delta
N$ formalism are $-(6/5)f_{\mathrm{NL}}=1.43$ and
$-(6/5)f_{\mathrm{NL},\delta N}=1.48$, respectively. the difference of the two values is due to the different numerical methods used.  
We see excellent
agreement between the exact numerical result of our formalism and the
$\gd N$ one, within the numerical accuracy. 
The slow-roll analytical result does very badly during
the turn of the field trajectory, when slow roll is badly broken, but
gives a reasonable estimate (within $50\%$) of the final value.

\begin{figure}
\begin{center}
\includegraphics[scale=1]{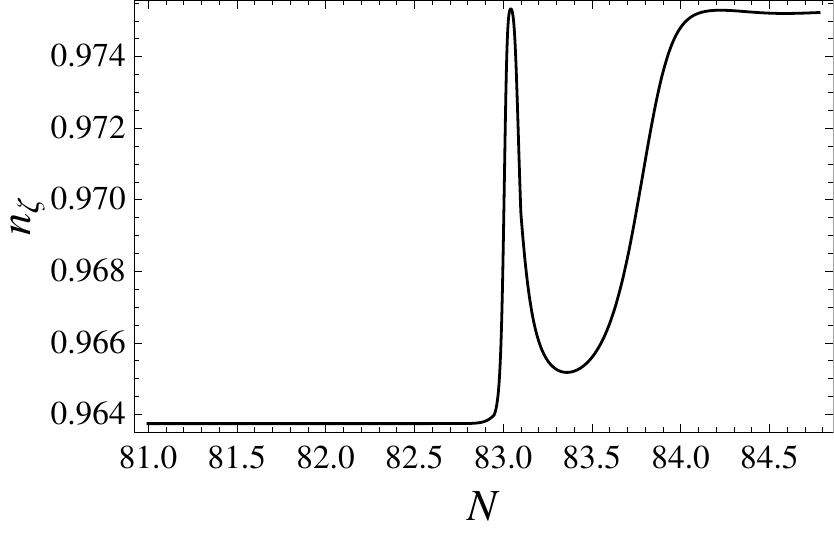}
\caption{The exact numerical spectral index for the same model as in 
figure~\ref{fig56}.}
\label{fig510}
\end{center}
\end{figure}

Finally in figure~\ref{fig510} we plot the spectral index for this
model. Its value is in the range of the Planck observations $n_\zeta=0.9603\pm0.0073$ \cite{Ade:2013uln}, but lies near
the upper limit.

%------------------------------ Chapter 6 --------------------------------------

{\Chapter{Scale dependence}\label{Ch6}}
%------------------------------------------------------------------------------------------

In this chapter we will study and investigate the sources  the scale dependence of the bi\-spectrum (\ref{bisp}), presenting our results found in \cite{Tzavara:2012qq}. Non-Gaussianity produced at horizon-crossing is known to be momentum-dependent. 
The scale dependence of the equilateral $\f$ produced for example from DBI inflation \cite{Langlois:2008qf,Arroja:2008yy,Mizuno:2009cv,Cai:2009hw,Senatore:2010wk}, has been examined both theoretically 
\cite{Chen:2005fe,Khoury:2008wj,Byrnes:2009qy,Leblond:2008gg} and in terms of observational forecasts \cite{LoVerde:2007ri,Sefusatti:2009xu}. 
In this chapter we are going to study the scale dependence of local-type models that has not been studied as much. 
Squeezed-type non-Gaussianity, produced outside the horizon, is usually associated with a parameter of non-Gaussianity 
$\f^\mathrm{local}$ that is local in real space, and therefore free of any explicit momentum dependence, defined through 
$\zeta(x)=\zeta_{L}(x)+(3/5)\f^\mathrm{local}(\zeta_{L}(x)^2-\langle\zeta_L(x)\rangle^2)$, where $\zeta_{L}$ is the linear Gaussian part. 
Nevertheless, calculations of $\f$ for several types of multiple-field models (see e.g.\  \cite{Vernizzi:2006ve,Byrnes:2008wi,
Tzavara:2010ge}) show that there is always a momentum dependence inherited from the horizon-crossing era, which can in principle result in 
a tilt of $\f$.  
When a physical quantity exhibits such a tilt one usually introduces a spectral index, as for example in the case of the power spectrum.  
The observational prospects of the detection of this type of scale dependence of local $\f$ were studied in \cite{Sefusatti:2009xu}.  
Only recently spectral indices for $\f$ were defined in \cite{Byrnes:2010ft,Byrnes:2009pe,Byrnes:2012sc}, keeping constant the shape of the 
triangle or two of its sides, within the 
$\delta N$ formalism.  
Note, however, that most theoretical predictions have considered equilateral triangles for simplicity, even though the local-type 
configuration is maximal on squeezed triangles. 
If one were to calculate a really squeezed triangle, then $\f^\mathrm{local}$ acquires some intrinsic momentum dependence 
due to the different relevant scales, as was shown in subsection \ref{Secgenmom} and \cite{Tzavara:2010ge}.

\begin{figure}
\begin{center}
\includegraphics[scale=0.2]{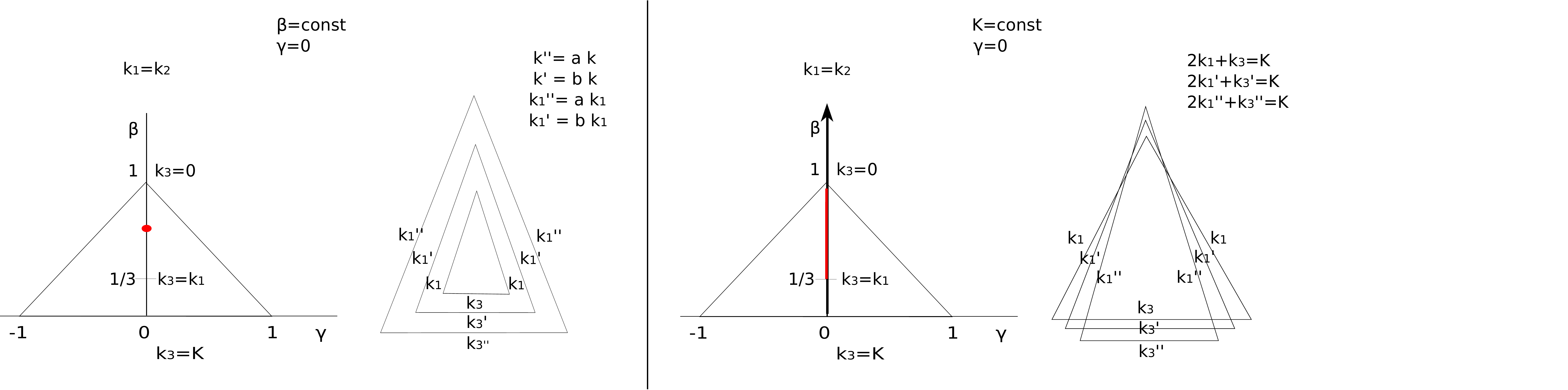}
\end{center}
\caption{The types of deformations of the momentum triangle we are 
considering. Left: Conformal transformation of the 
triangle. In the $\gamma,\beta$ plane this corresponds to a 
constant (red) point. Right: Keeping the perimeter of the triangle 
$K$ constant we change the shape of an isosceles 
$\gamma=0$ triangle, moving along the bold red line for 
$k_1=k_2\geq k_3$.}
\label{fig01}
\end{figure}

\Section{Sources of scale dependence}\label{sour}

In this section, we start by reminding the reader the basic elements of the long-wavelength calculation for $\f$ and also introduce the new scale variables that we are going to use in order to study the dependence of $\f$ on the momentum triangle. Next we discuss the main sources of the scale dependence of $\f$ and define spectral indices in order to parametrize this.
 
\subsection{The key variables}
%
%In order to study the scale dependence of the non-Gaussianity produced in two-field models of inflation We parametrize the bispectrum as 
%\be
%B_{\zeta}(k_1,k_2,k_3)=-\frac{6}{5}\f\lh \frac{2\pi^2}{k_1^3}
%\mathcal{P}_{\zeta}(k_1)\frac{2\pi^2}{k_2^3}\mathcal{P}_{\zeta}(k_2)
%+\lh k_2\leftrightarrow k_3\rh
%+\lh k_1\leftrightarrow k_3\rh\rh 
%\ee

%\subsection{Long-wavelength results}

%In this thesis we use the long-wavelength formalism to study the 
%parameter of 
%non-Gaussianity $\f$ and its scale dependence. 
%The long-wavelength formalism consists of doing a perturbative expansion
%of the exact differential equations for the non-linear cosmological 
%perturbations on super-horizon scales, which were obtained by neglecting
%second-order spatial gradient terms
%(see \cite{Rigopoulos:2005xx,Rigopoulos:2005us,Tzavara:2010ge}).
%Slow-roll solutions for the first-order perturbations at horizon-crossing
%are used as initial conditions.\footnote{The non-Gaussianity produced 
%at horizon-crossing is not included in the long-wavelength formalism, but can 
%be computed in another way, see \cite{Tzavara:2011hn}, and 
%added by hand. It is negligibly small in models with standard kinetic terms.}
%The super-horizon 
%assumption is equivalent to the leading order of the spatial gradient expansion and 
%requires the slow-roll assumption to be satisfied at horizon-crossing (but not 
%afterwards, see \cite{Leach:2001zf,
%Takamizu:2010xy,Tzavara:2010ge}). 
The non-Gaussianity parameter for an isosceles triangle of 
the form $k_1=k_2\equiv k'\geq k_3\equiv k$ was found in (\ref{fNLgeni}) (see also our paper  \cite{Tzavara:2010ge}) to be
\bea
	-\frac{6}{5}\f=\frac{-2\bv_{12k'}/[1+(\bv_{12k'})^2]}
{1+(\bv_{12k'})^2+2\frac{\gamma_k^2}{\gamma_k'^2}[1+(\bv_{12k})^2]}
	\!&\!\!\Bigg[&\!\!\bv_{12k'}\!\Bigg
(\!g_{sr}
	(k',k')\!
	+\!g_{iso}(k',k')\!+\!g_{int}(k',k')\!\Bigg) \non\\	
&&\hspace{-1.5cm}+2\frac{\gamma_k^2}{\gamma_k'^2}\bv_{12k}\Bigg(g_{sr}(k',k)\!+\!g_{iso}(k',k)
\!+\!g_{int}(k',k)\!\Bigg)\Bigg],
\eea
where $\f=\f(t;t_{k'},t_k)$ depends on $t_{k'}$ and $t_k$, denoting 
the horizon-crossing times of the two scales $k'$ and $k$ of the triangle,
respectively. We remind that $\gamma_k\equiv -
\gk H/(2 k^{3/2} \sqrt{\ge})$. 
This result is exact and valid beyond the slow-roll approximation after horizon-crossing. 
%All quantities appearing in this formula will be explained below.

The quantity $\bv_{12}$ is a key quantity for our study in this chapter. It is essentially a  
transfer function showing how the isocurvature mode (denoted by the subscript $2$) sources  
the adiabatic component $\zeta_1$. 
Two more transfer functions appear in $g_{sr},\ g_{iso}$ and $g_{int}$, namely $\bv_{22}$ and $\bv_{32}$, showing how the isocurvature mode sources the isocurvature 
component $\zeta_2$ and the velocity of the isocurvature component 
$\theta_2$, respectively. $\bv_{a2}$ is a function of the 
horizon-exit time $t_k$ of the relevant perturbation of scale $k$ and it 
also evolves with time $t$, at least during inflation. In (\ref{fNLgeni}) as 
well as in the formulas that follow, $\bv_{a2k}\equiv\bv_{a2}(t,t_k)$.  
The indices $a,b$ take the values $1,2,3$, indicating respectively the 
adiabatic perturbation $\zeta_1$, the isocurvature perturbation $\zeta_2$,
and the isocurvature velocity $\theta_2$.\footnote{Due to the exact relation
$\theta_1 = 2 \eta^\perp \zeta_2$, there is no need to consider the velocity
of the adiabatic perturbation $\theta_1$ as an additional variable
\cite{Rigopoulos:2005us}.} We remind the reader that 
$\bv_{a2}$ comes from the 
combination of the Green's functions $G_{a2}$ and $G_{a3}$ of the system of equations for the 
super-horizon perturbations (\ref{vamGreenrel}), satisfying 
\be
G_{ab}(t,t)=\delta_{ab}.
\label{Green_norm}
\ee

The contributions $g_{iso}$, $g_{sr}$ and $g_{int}$ are given in (\ref{comp}).
%  is a term that survives as long as the isocurvature modes are alive,
%\be
%g_{iso}({k}_1,{k}_2)=(\e+\hpa)(\widetilde{v}_{22})^2
%+\widetilde{v}_{22}\widetilde{v}_{32}.\label{giso}
%\ee
%If at the end of inflation these are non-zero, $\f$ can still evolve 
%afterwards and we cannot be sure that its value survives until today. Finally, 
%$g_{int}$ is given by
%\be
%g_{int}({k}_1,{k}_2)=\!-\!\int_{t_{k_1}}^t\!\!\!\!\d t'\Big[2(\hpe)^2
%(\widetilde{v}_{22})^2
%\!+\!(\e+\hpa)\widetilde{v}_{22}\widetilde{v}_{32}\!+\!(\widetilde{v}_{32})^2
%\!-\!G_{13}(t,t')\widetilde{v}_{22}(C\widetilde{v}_{22}+9\hpe\widetilde{v}_{32})
%\Big]\label{gint}
%\ee
%with   
%\be 
%C \equiv 12 \get^\perp \gc - 6 \get^\parallel \get^\perp
%	+ 6 (\get^\parallel)^2 \get^\perp + 6 (\get^\perp)^3
%- 2 \get^\perp \gx^\parallel- 2 \get^\parallel \gx^\perp
%	- \sqrt{\frac{\e}{2}}\frac{1}{\kappa H^2}(W_{211} + W_{222}),
%\ee 
%where $W_{mnl}=W^{,ABC}e_{mA}e_{nB}e_{lC}$. 
%It is from this integrated effect that any large, persistent non-Gaussianity 
%originates, if we consider only models where the isocurvature modes have
%vanished by the end of inflation.
For the analytical approximations that we will provide (in addition to the
exact numerical results), it is useful to note that within the slow-roll
approximation $g_{int}$ can be rewritten as 
\be
g_{int}(k_1,k_2) = \bv_{12k_1} G_{22 k_1 k_2} \left( -\eta^\perp_{k_1} 
+ \frac{(\epsilon_{k_1}
+\eta^\parallel_{k_1} -\chi_{k_1})\chi_{k_1}}{2\eta^\perp_{k_1}} \right) 
+ \tg_{int}(k_1,k_2),
\label{gint}
\ee
where $\tg_{int}$ is the integral in (\ref{fNLresultSR}) that is identically zero for the two-field
quadratic model, or even more generally for any two-field equal-power sum model (see also section \ref{secEqPow}).

In order to study the dependence of the non-Gaussianity on the 
shape of the triangle, instead 
of using $k_1,k_2$, and $k_3$ we will use the variables introduced in 
\cite{Rigopoulos:2004ba,Fergusson:2008ra}, 
\bea
K=\frac{k_1+k_2+k_3}{2},\qquad\gamma=\frac{k_1-k_2}{K},
\qquad\beta=-\frac{k_3-k_1-k_2}{2K},\label{newvar1}
\eea
which correspond to the perimeter of the triangle and two scale ratios 
describing effectively the angles of the triangle. They have the following 
domains: $0\leq K\leq \infty,\ 0\leq\beta\leq1$ and 
$-(1-\beta)\leq\gamma\leq1-\beta$, see figure~\ref{fig01}. 
As one can check from 
the above equations,  
the local bispectrum becomes maximal for $\beta=1$ and $\gamma=0$,   
or $\beta=0$ and $\gamma=\pm 1$, i.e.\  for a squeezed triangle. 
In this chapter we always assume 
$k_1=k_2$, dealing only with equilateral or isosceles triangles
(note that the relation $k_1=k_2$ is satisfied by definition for both 
equilateral and squeezed triangles).  
The two scales of the triangle $k_3 \equiv k \le k' \equiv k_1 = k_2$ can be expressed 
in terms of the new parameters $\beta$ and $K$ as 
\be
k=(1-\beta)K
\qquad\mathrm{and}
\qquad k'=\frac{1+\beta}{2}K,\label{par}
\ee
while $\gamma=0$. The condition $k\le k'$ means that we only have to study acute isosceles triangles $1/3\leq\beta\leq1$.

\subsection{Discussion}

\begin{figure}
\begin{tabular}{cc}
\includegraphics[scale=0.75]{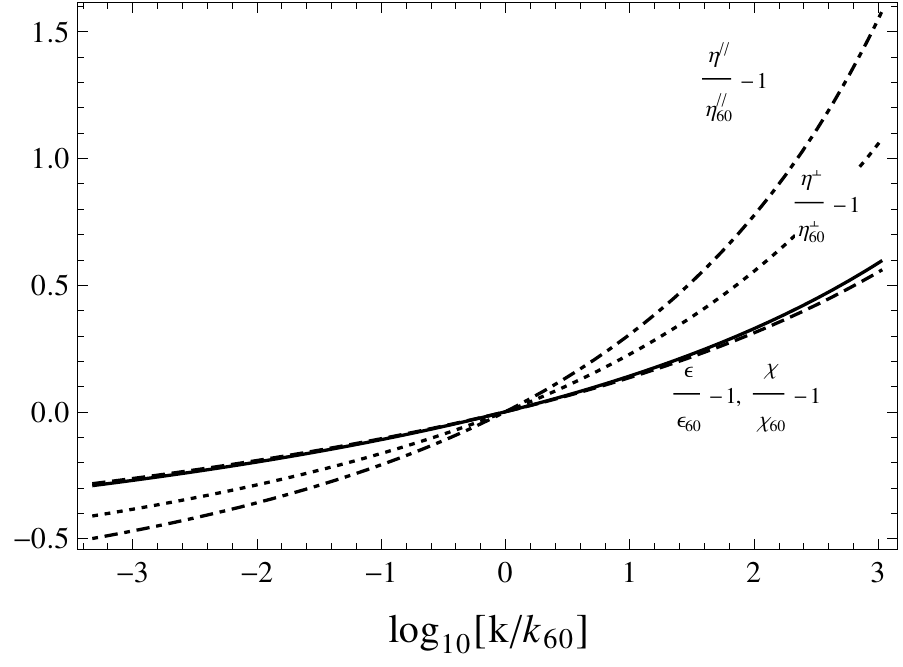}
&\includegraphics[scale=0.75]{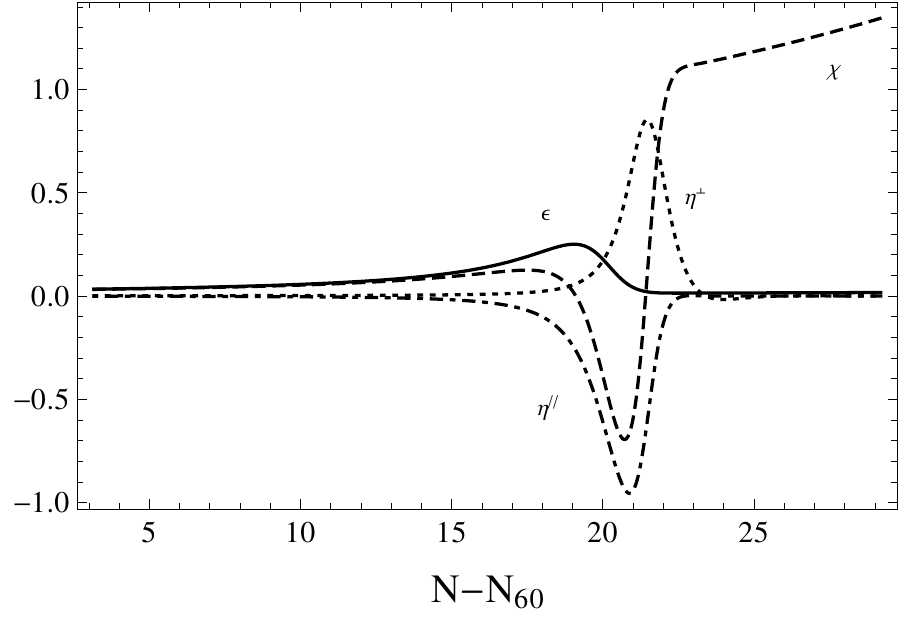}
\end{tabular}
\caption{Left: The relative change of the horizon-crossing first-order slow-roll 
parameters $\e$ (solid curve), $\hpa$ (dot-dashed curve), $\hpe$ (dotted curve) and $\chi$ (dashed curve) 
at $t_k$ as a function of the 
ratio $k/k_{60}$ of the horizon-exit scale to the scale that left the horizon $60$ e-folds 
before the end of inflation, 
for the model (\ref{qua}) with mass ratio $m_{\phi}/m_{\gs}=9$. 
Right: The evolution of the first-order slow-roll parameters $\e$ (solid curve), $\hpa$ (dot-dashed curve), $\hpe$ (dotted curve) and $\chi$ (dashed curve) as a function of the number of e-foldings $N-N_{60}$ for the time interval around the turning of the fields, for the same model.
}
\label{fig2}
\end{figure}

Inspecting (\ref{fNLgeni}) one sees that there are two sources of momentum 
dependence for $\f$: the slow-roll parameters 
at horizon-crossing and the Green's functions $G_{ab}$ or their combinations 
$\bv_{a2}$. In order to study their impact 
we shall use the quadratic model 
\be
W_q=\frac{1}{2}m_{\phi}^2\phi^2+\frac{1}{2}m_{\gs}^2\gs^2,\label{qua}
\ee
with $m_{\phi}/m_{\gs}=9$. The procedure to follow is to solve for the 
background quantities (\ref{eq2}) and then for the Green's functions (see subsection \ref{secGreen}) 
in order to apply the formalism. The quadratic model's Green's functions can be found 
numerically, or even analytically within the slow-roll approximation, which is valid for a 
small mass ratio like the one we chose here. However, all our calculations in this chapter 
are numerical and exact, without assuming the slow-roll approximation after horizon 
crossing. We only use the slow-roll approximation after horizon crossing for the
analytical approximations that we provide (e.g.\ eq.~(\ref{fnlfin})) and 
sometimes to clarify the physical interpretation of results (e.g.\ the use of 
(\ref{difu}) below to explain the behaviour of $\bv_{12}$).
Inflation ends at $t_f$ defined as the time when $\e_f=1$. 
From now on a subscript $f$ 
will denote quantities evaluated at the end of inflation. 
We also define the scale that exited the 
horizon $60$ e-folds before the end of inflation as $k_{60}$ and use it as a reference 
scale, around which we perform our computations ($k_{60}$ being the scale that 
corresponds to the text books' minimal necessary amount of inflation). 

In figure \ref{fig2} we plot the first-order slow-roll parameters for a 
range of horizon-crossing times around $k_{60}$. While the heavy field rolls 
down its potential the slow-roll parameters increase, reflecting the 
evolution of the background. This implies that 
$\f$, which is in general proportional to the 
slow-roll parameters evaluated at $t_k$ and $t_{k'}$, 
should increase as a function of $k$ and $k'$. 
This can easily be verified for the initial value of $f_{\mathrm{NL},in}$ at 
$t=t_{k'}$, which according to (\ref{fNLgeni}) with $\bv_{12k'}=0$ takes 
the value 
\be
-\frac{6}{5}f_{\mathrm{NL},in}=\e_{k'}+\hpa_{k'}+\frac{2\frac{\gamma_k^2}{\gamma^2_{k'}}G_{12k'k}}{1+2\frac{\gamma_k^2}{\gamma^2_{k'}}
[1+(G_{12k'k})^2]}\hpe_{k'}G_{22k'k}.
\label{fnlin}
\ee

\begin{figure}
\begin{tabular}{cc}
\includegraphics[scale=0.75]{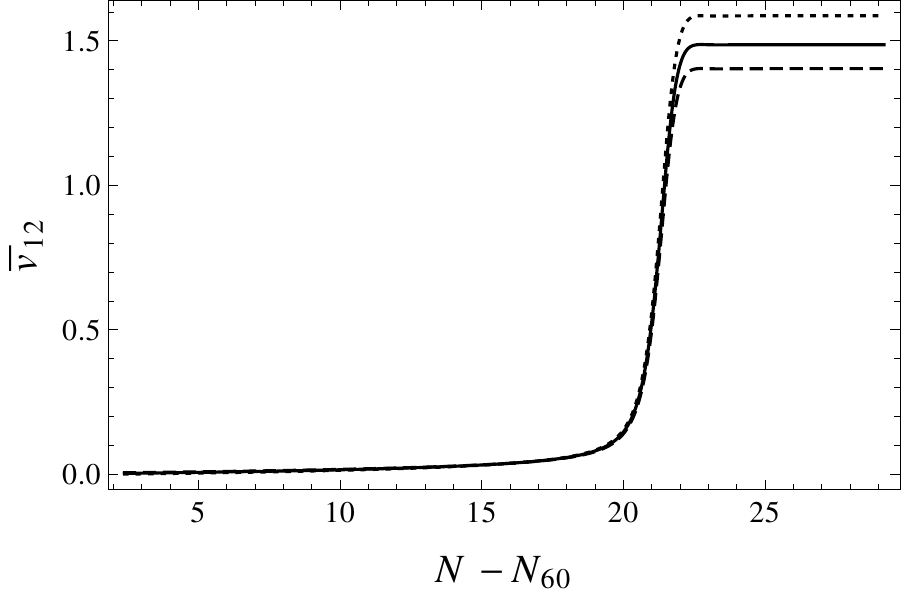}
&\includegraphics[scale=0.75]{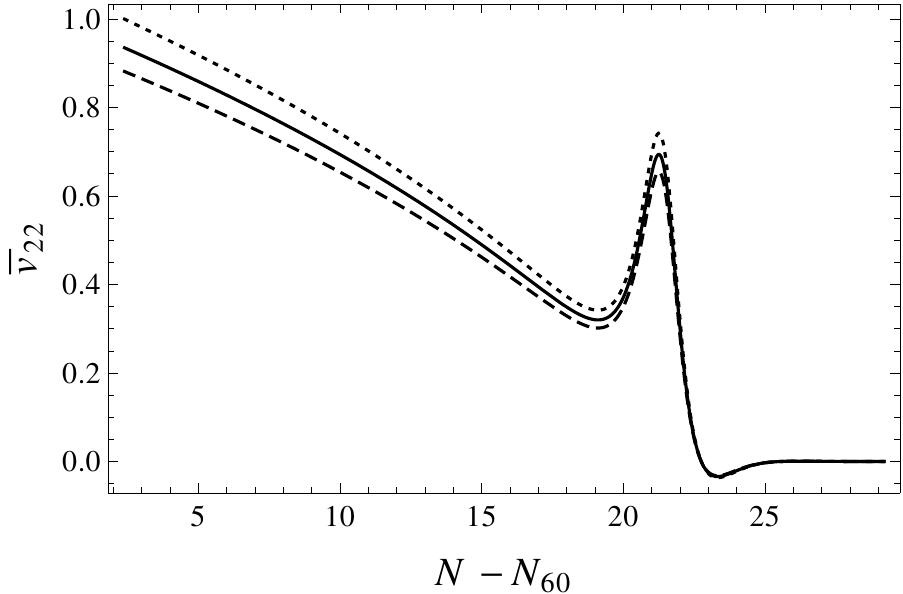}
\end{tabular}
\caption{The evolution of the transfer functions $\bv_{12}$ (left) and 
$\bv_{22}$ (right) as a function of the number 
of e-foldings $N-N_{60}$ for the time interval around the turning of the fields 
and for different horizon exit scales, varying 
from top to bottom as $k_{60}\times10$ (dotted curve), $k_{60}$ (solid curve) and $k_{60}/10$ (dashed curve), 
for the model (\ref{qua}) with mass ratio $m_{\phi}/m_{\gs}=9$.}
\label{fig1}
\end{figure}

Apart from the slow-roll parameters the other source of momentum dependence for $\f$ lies in the Green's functions and 
particularly how their time evolution depends on the relevant horizon-crossing scale. The two main quantities that we 
need to study in order to understand their impact on $\f$ are the transfer functions $\bv_{12}$ and $\bv_{22}$. This is due to the fact 
that $\bv_{32}$ is slow-roll suppressed and the rest of the Green's functions appearing in (\ref{fNLgeni}) can 
be rewritten in terms of $\bv_{12}$ and $\bv_{22}$ within the slow-roll approximation (for details, see section  \ref{secSlowRoll}).  
In particular  $G_{a3}=G_{a2}/3$, $G_{32}(t,t_k)=-\chi(t) G_{22}(t,t_k)$ and hence $G_{a2} \approx \bv_{a2}$. Note that except for the era of 
the turning of the fields, the slow-roll assumption is a good approximation during inflation in this particular model.   
The slow-roll evolution equations for $\bv_{12k}$ and $\bv_{22k}$ are
\be
\frac{\d}{\d t}\bv_{12k}=2\hpe\bv_{22k}\qquad\mathrm{and}\qquad\frac{\d}{\d t}\bv_{22k}=-\chi\bv_{22k}.\label{difu}
\ee

As was discussed above, $\bv_{12}$ describes how the isocurvature mode sources the 
adiabatic one, while $\bv_{22}$ describes how the 
isocurvature mode sources itself.  
By definition $\bv_{12}(t_k,t_k)=0$ and $\bv_{22}(t_k,t_k)=1$ at horizon crossing, 
since no interaction of the different modes has yet occurred (see also 
(\ref{vamGreenrel}) and (\ref{Green_norm})). 
For the transfer functions of the adiabatic mode one finds that 
\be 
\bv_{11}=1\quad\mathrm{and}\quad\bv_{21}=0,\label{lastid}
\ee
since the curvature perturbation is conserved for purely 
adiabatic perturbations and adiabatic perturbations cannot source entropy 
perturbations on super-horizon scales. 
In order to better understand the 
role of the transfer functions, we can use the Fourier transformation of the 
perturbations \cite{Tzavara:2010ge} along with (\ref{lastid}), to find
\bea
&&\zeta_{1}(t)=\int\frac{\d^3\vc{k}}{(2\pi)^{3/2}}\gamma_k\bv_{1m}\hat{a}^{\dagger}_m(\vc{k})
 e^{\rmi \vc{k}\cdot \vc{x}}=\zeta_{1}(t_k)+\bv_{12}(t,t_k)\zeta_{2}(t_k),\nn\\
&&\zeta_{2}(t)=\int\frac{\d^3\vc{k}}{(2\pi)^{3/2}}\gamma_k\bv_{2m}\hat{a}^{\dagger}_m(\vc{k})
 e^{\rmi \vc{k}\cdot \vc{x}}=\bv_{22}(t,t_k)\zeta_{2}(t_k),\label{ze}
\eea
where $\zeta_{m}$ with $m=1,2$ are the first-order adiabatic and isocurvature 
perturbation. 
 
Let us start by discussing the time evolution of $\bv_{12}$. Each one of the 
curves on the left-hand side of figure \ref{fig1} corresponds to the 
time evolution of $\bv_{12}$ for a different horizon-exit scale. 
At $t=t_k$, i.e.\  when the relevant mode $k$ exits the horizon, $\bv_{12k}=0$   
since the isocurvature mode has not had time to affect the adiabatic one. Outside 
the horizon and well in the slow-roll regime of the sole dominance of the heavy field, the
isocurvature mode sources the adiabatic one and the latter slowly increases. 
As time goes by, the heavy field rolls down its potential 
and the light field becomes more important. 
During this turning of the field trajectory, the slow-roll parameters suddenly
change rapidly, with important consequences for the evolution of the adiabatic
and isocurvature mode. The transfer function
$\bv_{12k}$ grows substantially during that era because of the increasing 
values of $\hpe$ in (\ref{difu}) as well as the growing contribution of $\bv_{22k}$, 
to become constant afterwards  
when the light field becomes dominant in an effectively single-field universe. 

Note that the earlier the mode exits the horizon, the smaller is the final $\bv_{12k}$. This is opposite to the behaviour of the initial 
value, just after horizon-crossing, 
when the earlier the scale exits the horizon the more has its adiabatic mode been sourced by the isocurvature one at a given time $t$, and hence the 
larger is its $\bv_{12k}$. 
This can be understood by the evolution equations of $\bv_{12k}$ and $\bv_{22k}$ 
in (\ref{difu}), 
showing that $\bv_{12k}$ is sourced by $\bv_{22k}$, which itself is a decreasing function 
of time, 
at least during eras when the universe is dominated by a single field (see 
the right-hand side of figure \ref{fig1}). If the equation (\ref{difu}) for $\bv_{12k}$ did not depend on $\hpe$, the $\bv_{12k}$ curves 
would never cross each other since they would be similar and only boosted by their horizon-crossing time shift. It is the increasing 
value of $\hpe$ 
%or in other words the fact that the background is two-field dominated (and hence induces isocurvature effects) 
that results in the larger values of $\bv_{12k}$ for larger $k$.

\begin{figure}
\begin{tabular}{cc}
\includegraphics[scale=0.75]{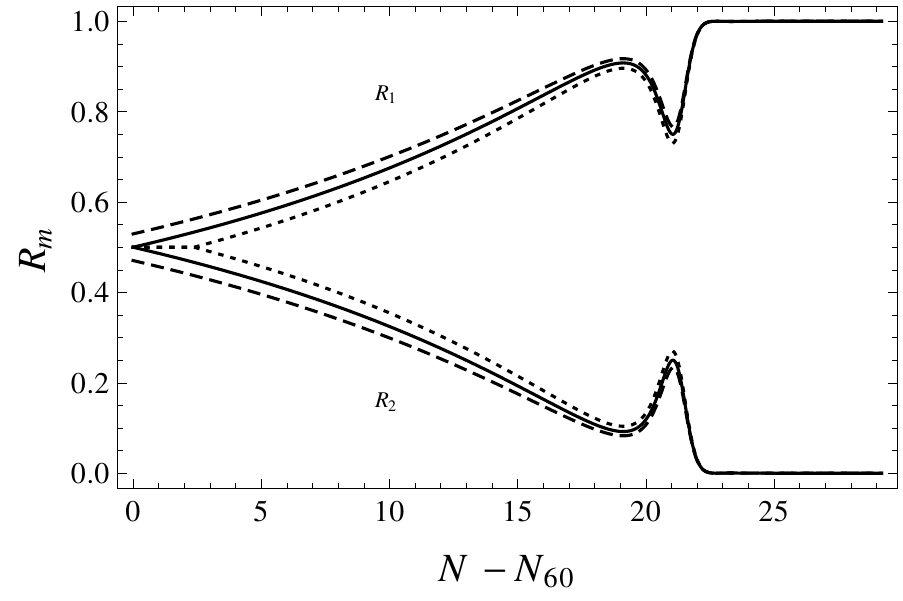}
&\includegraphics[scale=0.75]{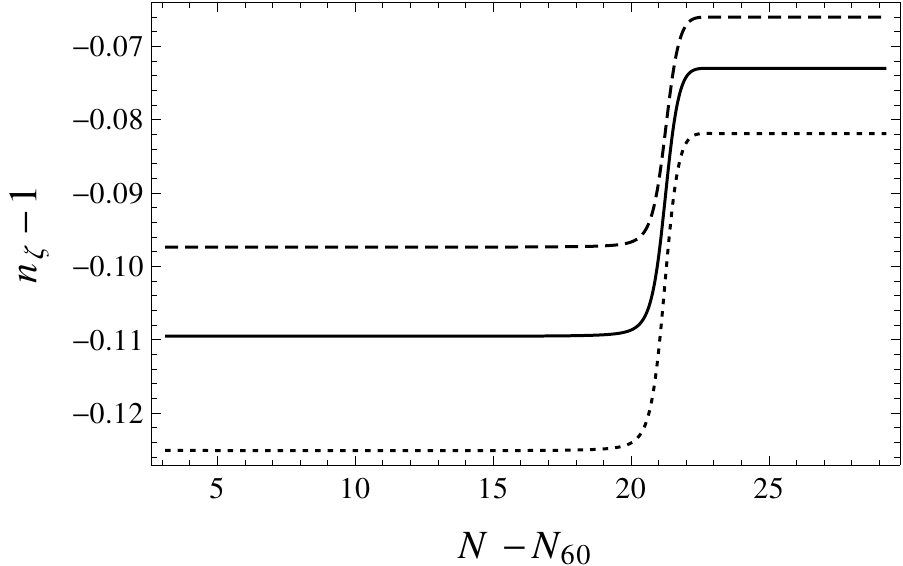}
\end{tabular}
\caption{Left: The time evolution of the adiabatic $R_1$ and the isocurvature 
$R_2$ ratios of power spectra as a function of the number of e-foldings $N-N_{60}$ 
for scales $k_{60}/10$ (dashed line), $k_{60}$ (solid line) and 
$k_{60}\times10$ (dotted line). Right: The time evolution of the spectral index (\ref{spectralind}) 
as a function of the number of e-foldings $N-N_{60}$ 
for scales $k_{60}/10$ (dashed line), $k_{60}$ (solid line) and 
$k_{60}\times10$ (dotted line). Both plots are made 
for the model (\ref{qua}) with mass ratio $m_{\phi}/m_{\gs}=9$.}
\label{fig11}
\end{figure}

On the right-hand side of figure \ref{fig1} we show the evolution of $\bv_{22}$. 
According to (\ref{difu}) $\bv_{22}$ and hence the isocurvature mode evolves independently from the adiabatic mode. 
At horizon-crossing $t=t_k$, the transfer function $\bv_{22k}=1$.
%, i.e.\  the isocurvature mode has not yet interacted with the adiabatic mode. 
Once outside the horizon, the isocurvature mode decays due to the small but positive values of $\chi$.
%Once outside the horizon and due to the dominance of the universe by a single field (but nonetheless an $\eta^\perp$ that is not quite zero), isocurvature modes are diluted while sourcing the adiabatic mode. 
During the turning of the fields the slow-roll parameters evolve rapidly, thus leading to first an enhancement of $\bv_{22k}$ and then a diminution due to the varying value of $\chi$ in (\ref{difu}). 
As can be seen from the right-hand side plot in figure \ref{fig2}, during the turning $\chi$ first becomes negative and then positive. 
After the turning of the fields, the remnant isocurvature modes again decay and (for this model) at the end of inflation none are left.
%After the turning of the fields, the remnant isocurvature modes are again diluted to adiabatic modes and (for this model) at the end of inflation none are left.   
The parameter $\chi$ plays a crucial role in the evolution of the isocurvature mode. 
It represents effectively the second derivative of the potential in the $22$ direction. 
Before the turning of the fields the trajectory goes down the potential in the 
relatively steep $\phi$ direction, which means that $W_{22}$ then corresponds
with the relatively shallow curvature in the direction of the light field 
$\sigma$ and hence $\chi$ is small. After the turning the trajectory goes along
the bottom of the valley in the $\sigma$ direction and $W_{22}$ corresponds with
the large curvature of the potential in the perpendicular direction, leading 
to large values of $\chi$. The negative values of $\chi$ during the turn come
from the contribution of $\hpa$.

Instead of looking at the tranfer functions, using (\ref{ze}) one can also 
construct more physical quantities from the operators $\zeta_{m}$ 
and hence from the $\bv_{m2}$, namely the ratios 
of the adiabatic and isocurvature power spectrum to the total power spectrum:
\bea
&&R_1\equiv\frac{\langle\zeta_{1}\zeta_{1}\rangle}{\langle\zeta_{1}\zeta_{1}\rangle+
\langle\zeta_{2}\zeta_{2}\rangle}=\frac{1+\lh\bv_{12}\rh^2}{1+\lh\bv_{12}\rh^2+\lh\bv_{22}\rh^2},\nn\\
&&R_2\equiv\frac{\langle\zeta_{2}\zeta_{2}\rangle}{\langle\zeta_{1}\zeta_{1}\rangle+
\langle\zeta_{2}\zeta_{2}\rangle}=\frac{\lh\bv_{22}\rh^2}{1+\lh\bv_{12}\rh^2+\lh\bv_{22}\rh^2}.
\eea
These are plotted on the left-hand side of figure \ref{fig11} as a function 
of the number of e-foldings for different scales. 
One can clearly see that both 
ratios start as equal to $1/2$ when the scale exits the horizon, while 
afterwards the adiabatic ratio $R_1$ increases to reach $1$ at the end 
of inflation and the isocurvature $R_2$ decreases to reach $0$, for this 
particular model. During the turning of the fields we see that the temporary
increase in the isocurvature mode due to the negative value of $\chi$ is 
reflected in $R_2$, while the adiabatic $R_1$ necessarily has the opposite 
behaviour. 

On the right-hand side of figure \ref{fig11} we plot the time evolution of the spectral index (\ref{spectralind}) of the power spectrum. 
The spectral index measures by construction the tilt of the power spectrum for different horizon-crossing scales and hence it 
depends on the horizon-crossing slow-roll parameters. For multiple-field models the power spectrum evolves during inflation even after horizon-crossing, 
and so does the spectral index.   
During the turning of the fields the spectral index increases, to remain constant afterwards. 
The earlier 
a scale exits the horizon the less negative is its spectral index $n_{\zeta}-1$. This implies that the power spectrum itself decreases faster 
for larger horizon-crossing scales. This is due to the fact that except for the factor $1+(\bv_{12k})^2$ in the expression for the power 
spectrum there is also an inverse power of $\e_k$ (see (\ref{spectralind})).

\subsection{Spectral indices}

Finally let us discuss the scale dependence of the local $\f$ in terms of the relevant spectral indices. 
Equation (\ref{fNLgeni}) for an isosceles triangle implies that 
\bea
-\frac{6}{5}\f&=&\frac{1}{(2\pi^2)^2}\frac{f(k',k')+2(\frac{k'}{k})^3f(k',k)}
{\mathcal{P}_\zeta(k')^2+2(\frac{k'}{k})^3\mathcal{P}_\zeta(k')\mathcal{P}_\zeta(k)},
\eea
where 
\be
f(k',k) = -2 (k'k)^3 \gamma_{k'}^2 \gamma_k^2 \, \bv_{12k'} \bv_{12k}
\left( g_{sr}(k',k) + g_{iso}(k',k) + g_{int}(k',k) \right).
\ee 
For an arbitrary triangle configuration this is generalized as
\bea
-\frac{6}{5}\f&=&\frac{1}{(2\pi^2)^2}\frac{k_3^3f(k_1,k_2)+\mathrm{perms.}}
{k_3^3\mathcal{P}_\zeta(k_1)\mathcal{P}_\zeta(k_2)+\mathrm{perms.}}.
\eea
The local $\f$ depends on a two-variable function $f(k_1,k_2)$, with $k_1\geq k_2$. This is due to its super-horizon origin, which 
yields classical non-Gaussianity proportional to products of two power spectra. 
Hence one expects that the scale dependence of 
$\f$ can be expressed in terms of only two spectral indices, characterizing the function $f$. 
Notice that this is particular to the local case. In general the bispectrum 
cannot be split as a sum of two-variable functions and one anticipates that three spectral indices would be needed. 

The next issue to be resolved is which are the relevant spectral indices for $f$. The naive guess would be 
$f(k_1,k_2)=f(k_{1,0},k_{2,0})(k_1/k_{1,0})^{\tn_{k_1}}(k_2/k_{2,0})^{\tn_{k_2}}$. 
We tested this parametrization and we did not find good agreement with the exact value of $f$. 
Instead, we found that $f$ is best approximated by keeping either the shape or the magnitude of the triangle constant. 
This statement can be expressed as 
\be
f(k_1,k_2)=f_0\lh\frac{K}{K_0}\rh^{\tn_{K}}\lh\frac{\omega}{\omega_0}\rh^{\tn_{\omega}},
\label{ind}
\ee
where
\be
\tn_{K}\equiv\frac{\d\ln f}{\d\ln K}\qquad\mathrm{and}\qquad \tn_{\omega}\equiv\frac{\d\ln f}{\d\ln\omega}\label{inddef}
\ee
and
\be
\omega\equiv\frac{k_1}{k_2}=\frac{1+\beta}{2(1-\beta)}. 
\ee
The last equality is valid only for the isosceles case $\gamma=0$ (see
(\ref{newvar1})). We dropped 
the $-1$ of the power spectrum spectral index definition to follow the 
definitions in \cite{Byrnes:2009pe,Byrnes:2010ft}. 
We added a tilde to indicate that these spectral indices are defined for the
function $f$, not yet for the full $\f$.
In the next two sections we are going to examine the scale-dependence of $\f$, 
changing the magnitude and the shape of the triangle separately, 
and verify assumption (\ref{ind}).

\Section{Changing the magnitude of the triangle}\label{conf}

\begin{figure}
\begin{center}
\includegraphics[scale=0.8]{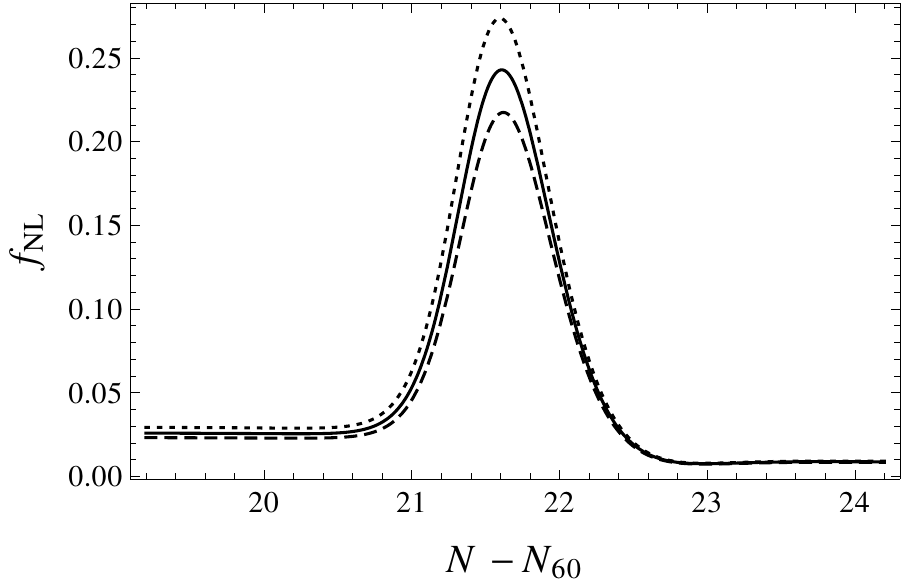}
\end{center}
\caption{The time evolution of $\f$ in terms of the number of e-foldings 
$N-N_{60}$ around the time of the turning of the fields,
for equilateral ($\omega=1$) triangles with $K=(3/2)k_{60}$ (solid curve), 
$K=(3/2)k_{60}/10$ (dashed curve) and $K=(3/2)k_{60}\times10$ 
(dotted curve),
for the model (\ref{qua}) with mass ratio $m_{\phi}/m_{\gs}=9$.}
\label{fig66}
\end{figure}

\begin{figure}
\begin{tabular}{cc}
\includegraphics[scale=0.75]{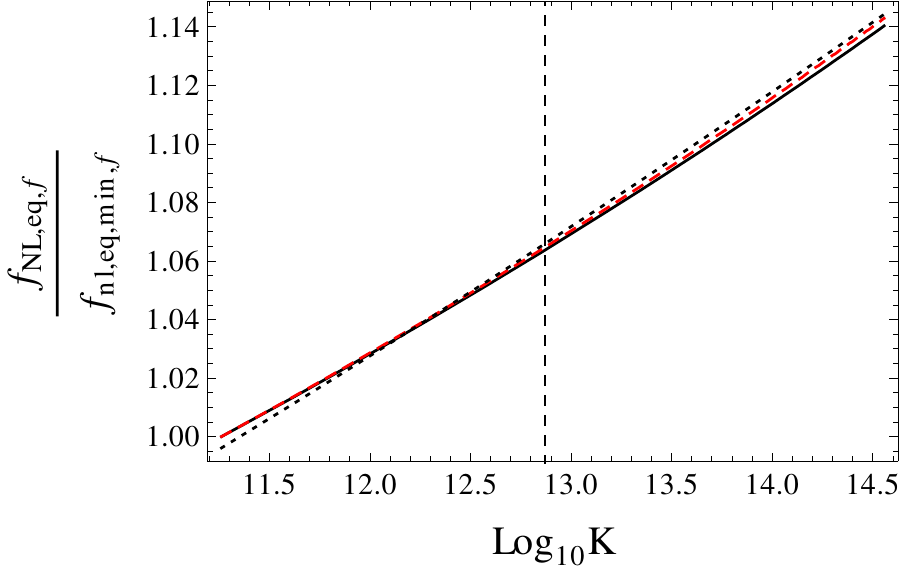}
&\includegraphics[scale=0.75]{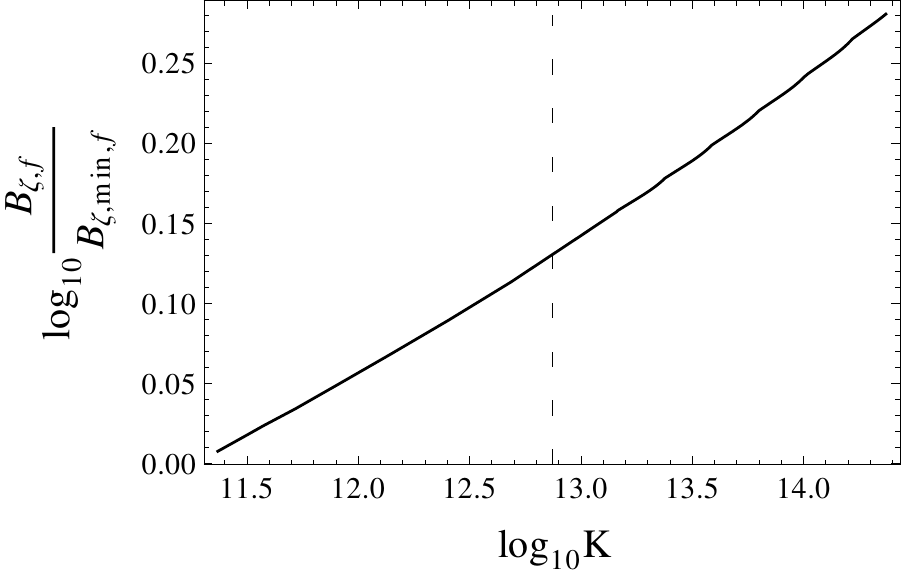}
\end{tabular}
\caption{Left: The relative change of the final value of $\f$ arbitrarily normalized to one at the smallest value of $K$ on the figure, 
as a function of $K$ for equilateral triangles ($\omega=1$), calculated exactly (solid curve), using the analytical approximation (\ref{fnlfin}) (dashed red curve) 
and using the shape index (\ref{nkeq}) (dotted black curve).  
Right: The logarithm of the final value of the exact bispectrum, similarly normalized,  
as a function of 
$K$ for equilateral triangles ($\omega=1$).  
Both figures are for the quadratic model (\ref{qua}) with mass ratio 
$m_{\phi}/m_{\gs}=9$. The vertical dashed line corresponds to $K=(3/2)k_{60}$.}
\label{fig631}
\end{figure}

In this section we shall study the behaviour of $\f$ for triangles 
of the same shape but different size, see the left-hand side of 
figure~\ref{fig01}. In figure 
\ref{fig66} we plot the time evolution of $\f$ for equilateral 
triangles (the result would remain qualitatively the same for any isosceles 
triangle) of perimeter $K=(3/2)k_{60}\times10$ (top curve), 
$K=(3/2)k_{60}$ (middle curve) and $K=(3/2)k_{60}/10$ (bottom curve). The later 
the relevant scale exits the horizon the larger is its initial $\f$ as 
explained in the previous section. $\f$ grows during the turning of the 
fields due to isocurvature effects as described by (\ref{comp}), but by the end of inflation, when isocurvature modes vanish, it 
relaxes to a small, slow-roll suppressed value (see e.g.\  \cite{Vernizzi:2006ve,Tzavara:2010ge}). 
In figure \ref{fig631} we plot the final value of $\f$ (left) and the final 
value of the bispectrum (right) for equilateral triangles, varying $K$ for 
values around $K=(3/2)k_{60}$, within the Planck satellite's resolution 
($k'/k\sim1000$). The later the scale exits the horizon, i.e.\  the larger
$K$, the larger is the final value of $\f$ and of the bispectrum. 

The final value of $\f$ can be found analytically for the quadratic model 
within the slow-roll approximation. 
By the end of inflation $\bv_{22,f}=0$ so that $g_{iso}$
vanishes, while (\ref{gint}) can be further 
simplified to give some extra horizon-crossing terms and a new integral 
$\tg_{int}$ that is identically zero for the quadratic potential 
(see section \ref{secEqPow}) and \cite{Tzavara:2010ge}). 
For simplicity we give here the final value of 
$\f$ for equilateral triangles, 
\be
f_{\mathrm{NL},eq,f}(k)=\frac{
3\lh\bv_{12k}\rh^2\lh\e_k+\hpa_k-\chi_k+\frac{\hpe_k}{\bv_{12k}}\rh
+\lh\bv_{12k}\rh^3\lh\hpe_k-
\frac{\lh\e_k+\hpa_k-\chi_k\rh\chi_k}{\hpe_k}\rh+\e_k+\hpa_k}
{\lh1+\lh\bv_{12k}\rh^2\rh^2}.\label{fnlfin}
\ee
This formula is actually valid for any two-field model for which isocurvature modes
vanish at the end of inflation and for which $\tg_{int}=0$, 
like for example equal-power sum models. 
Inspecting the various terms it turns out that although $\bv_{12k}$ tends to decrease 
the value of $f_{\mathrm{NL},f}$ as a function of $k$, it is the contribution of the horizon-crossing  
slow-roll parameters that wins and leads to an increase of the parameter 
of non-Gaussianity for larger horizon-crossing scales.
Note that for equilateral triangles $K$ is simply $3k/2$.

\begin{figure}
\begin{tabular}{ll}
\includegraphics[scale=0.75]{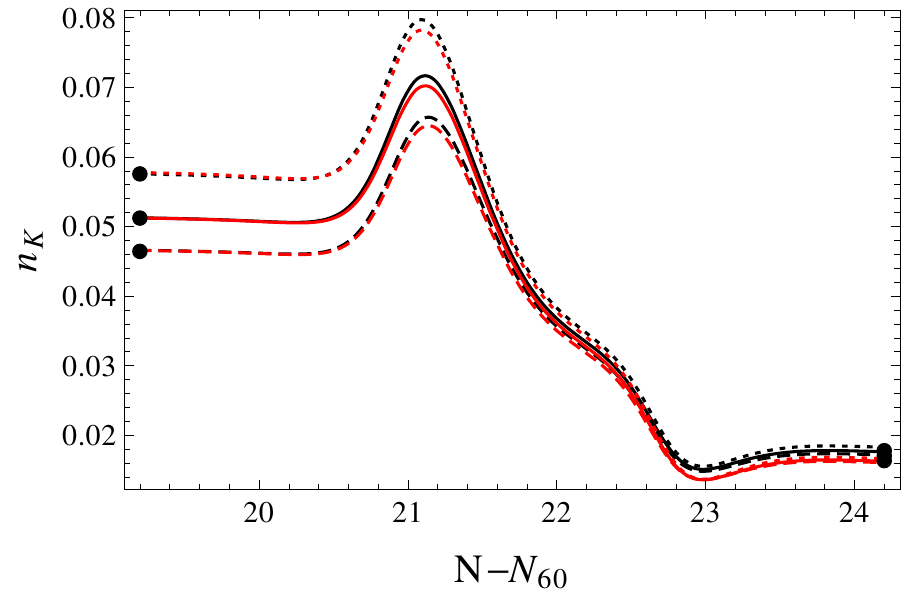}
&\includegraphics[scale=0.75]{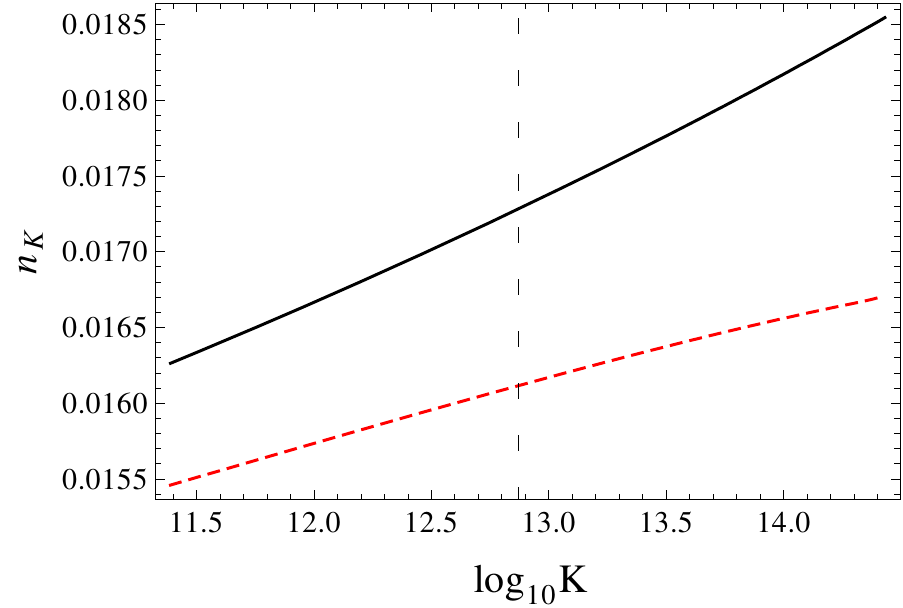}
\end{tabular}
\caption{Left: The time evolution of the conformal index 
$n_K$ (\ref{con}) around the time of the turning of the fields,
for triangles with $\omega=1$ (black curves) and 
$\omega=5/2$ (red curves, below the $\omega=1$ curves), 
with perimeter $K=(3/2)k_{60}$ (solid curve), 
$K=(3/2)k_{60}/10$ (dashed curve) and $K=(3/2)k_{60}\times10$ (dotted curve). 
The three points on the left and on the right correspond to the analytical values 
of the index as calculated from (\ref{ind1}) and (\ref{A1}) respectively. 
Right: The final value of the conformal index $n_K$ 
(\ref{con}) for triangles with $\omega=1$ (black curve) and $\omega=5/2$ (red dashed curve)  
as a function of $K$. 
Both figures are for the model (\ref{qua}) with mass 
ratio $m_{\phi}/m_{\gs}=9$.}\label{fig651}
\end{figure}

We turn now to the spectral index $n_{K}$. Using (\ref{inddef}) with (\ref{par}) and assuming 
that $\gamma=0$ and $\beta=\mathrm{const.}$, we can express $\tn_{K}$ in terms of the horizon-crossing time derivatives as 
\be 
\tn_{K}(t;t_{k_1},t_{k_2})=\frac{\partial\ln{f}}{\partial t_{k_1}}\frac{1}{1-\e_{k_1}}
+\frac{\partial\ln{f}}{\partial t_{k_2}}\frac{1}{1-\e_{k_2}}.
\label{con}
\ee 
Then $\f$ takes the form
\be
-\frac{6}{5}\f=\frac{1}{(2\pi^2)^2}\frac{f(k'_0,k'_0)\lh\frac{K}{K_0}\rh^{\tn_{K}(t_{k'_0},t_{k'_0})}+
2\omega^3f(k'_0,k_0)\lh\frac{K}{K_0}\rh^{\tn_{K}(t_{k'_0},t_{k_0})}}
{\mathcal{P}_\zeta(k'_0)^2\lh\frac{k'}{k'_0}\rh^{2(n_\zeta(t_{k'_0})-1)}
+2\omega^3\mathcal{P}_\zeta(k'_0)\mathcal{P}_\zeta(k_0)\lh\frac{k'}{k'_0}\rh^{n_\zeta(t_{k'_0})-1}\lh\frac{k}{k_0}\rh^{n_\zeta(t_{k_0})-1}}.
\label{nkf}
\ee
Note that the ratios $k'/k'_0=k/k_0=K/K_0$, since $\beta=\mathrm{const}$. 

The above formula can be simplified in the limit of squeezed-triangle configurations, as well as in the equilateral limit. 
When one takes the squeezed limit $\omega^3\mg1$ (note that this would be true for $\beta\gtrsim2/3$), one finds:
\bea
-\frac{6}{5}\f&=&\frac{1}{(2\pi^2)^2}\frac{f(k'_0,k_0)}{\mathcal{P}_\zeta(k'_0)\mathcal{P}_\zeta(k_0)}
\lh\frac{K}{K_0}\rh^{\tn_{K}(t_{k'_0},t_{k_0})-n_\zeta(t_{k'_0})-n_\zeta(t_{k_0})+2}\nn\\
&\equiv&-\frac{6}{5}f_{\mathrm{NL},0}
\lh\frac{K}{K_0}\rh^{n_{K}(t_{k'_0},t_{k_0})},
\eea
where
\be
n_K(t;t_{k'},t_k)\equiv\frac{\d \ln\f}{\d\ln K}=\frac{\partial\ln{\f}}{\partial t_{k'}}\frac{1}{1-\e_{k'}}
+\frac{\partial\ln{\f}}{\partial t_{k}}\frac{1}{1-\e_{k}}.\label{nkt}
\ee
For the equilateral case $\omega=1$, (\ref{nkf}) becomes
\bea
-\frac{6}{5}\f&=&\frac{1}{(2\pi^2)^2}\frac{f(k'_0,k'_0)}{\mathcal{P}_\zeta(k'_0)^2}
\lh\frac{K}{K_0}\rh^{\tn_{K}(t_{k'_0},t_{k'_0})-2n_\zeta(t_{k'_0})+2}\nn\\
&\equiv&-\frac{6}{5}f_{\mathrm{NL},0}
\lh\frac{K}{K_0}\rh^{n_{K}(t_{k'_0},t_{k'_0})}.\label{nkeq}
\eea

The conformal spectral index $n_K$ measures the change of $\f$ due to the 
overall size of the triangle, namely due to a conformal transformation of the triangle. 
For an isosceles triangle this is conceptually sketched on the left-hand side of 
figure \ref{fig01}, but it can be generalized for any shape. $n_K$ coincides with the 
$n_{\f}$ of \cite{Byrnes:2009pe,Byrnes:2010ft} and grossly speaking it describes the 
tilt of $\f$ due to the pure evolution of the inflationary background (note that for 
an equilateral triangle this statement would be exact, since all three scales are equal and there is no relative scale-dependent evolution outside the horizon).

On the left-hand side of figure \ref{fig651} we plot the time evolution of the conformal spectral index for an equilateral 
$\omega=1$ and an isosceles $\omega=5/2$ triangle that exited the horizon 
at three different times, namely for $K=(3/2)k_{60}$ (solid curve), 
$K=(3/2)k_{60}/10$ (dashed curve) and $K=(3/2)k_{60}\times10$ (dotted curve). We plot 
the $\omega=5/2$ case only to demonstrate that the results remain qualitatively the same; 
we shall study the effect of different triangle shapes in the next section. The characteristic peaks that $n_K$ exhibits during the 
turning of the fields are inherited from the behaviour of $\f$ at that time and it is a new feature that is absent 
in the time evolution of the power spectrum spectral 
index $n_{\zeta}-1$ (see the right-hand side of figure \ref{fig11}).

In the context of the long-wavelength formalism we are 
restricted to work with the slow-roll approximation at horizon 
exit, so that the slow-roll parameters at that time should be small and vary 
just a little. This should be reflected in the initial value of 
the spectral index, which should be $\mathcal{O}(\e_{k'})$. 
The earlier 
the scale exits, e.g.\  the dashed curve, the smaller are the slow-roll 
parameters evaluated at horizon crossing and hence the 
smaller is the initial $n_K$.  
Indeed, using the definition (\ref{nkt}) with (\ref{fnlin}) for the initial 
value of $f_{\mathrm{NL},in}$, we find for equilateral triangles
\be
n_{K,in}=\frac{2\e_k^2+3\e_k\hpa_k+(\hpe_k)^2-(\hpa_k)^2+\xi^{\parallel}_k}
{\e_k+\hpa_k},\label{ind1}
\ee
which confirms the above statement. 

We notice that the initial, horizon-crossing, differences between the values of $n_K$ for the different horizon-crossing scales
mostly disappear by the end of inflation, after peaking during the turning of the fields. 
The final value of the spectral index is plotted on the right-hand side of figure 
\ref{fig651} and is smaller than its initial value. 
It exhibits a small running of $\mathcal{O}(10\%)$ within the 
range of scales studied, inherited from the initial dispersion of its values at horizon-crossing. 
To verify that $n_K$ describes well the behaviour of $\f$, we have plotted the
approximation (\ref{nkeq}) in figure \ref{fig631} where it can be compared with
the exact result.
We have also verified this for other inflationary models, including the potential (\ref{newpot})
studied in 
\cite{Tzavara:2010ge}, able to produce $\f$ of ${\mathcal O}(1)$. 
The final value of the spectral index in that model is two orders of magnitude smaller than the value for the quadratic model. 
This is related essentially to the fact that for the potential (\ref{newpot}) the turning of the fields, 
and hence the slow-roll breaking, occurs near the end of inflation. This means that at the horizon-crossing times of the scales 
of the triangle, slow-roll parameters change very slowly and as a result the initial variation of $\f$ is much smaller than the one 
for the quadratic potential. 
As a consequence, the final tilt of $\f$ will be smaller.

In appendix \ref{AppD} one can find 
the analytical expression for the final value of the spectral index $n_K$, 
found by differentiating (\ref{fnlfin}). Since the 
result is rather lengthy and does not give any further physical intuition, we moved it to the appendix.

\Section{Changing the shape of the triangle}\label{shape}

\begin{figure}
\begin{center}
\includegraphics[scale=0.8]{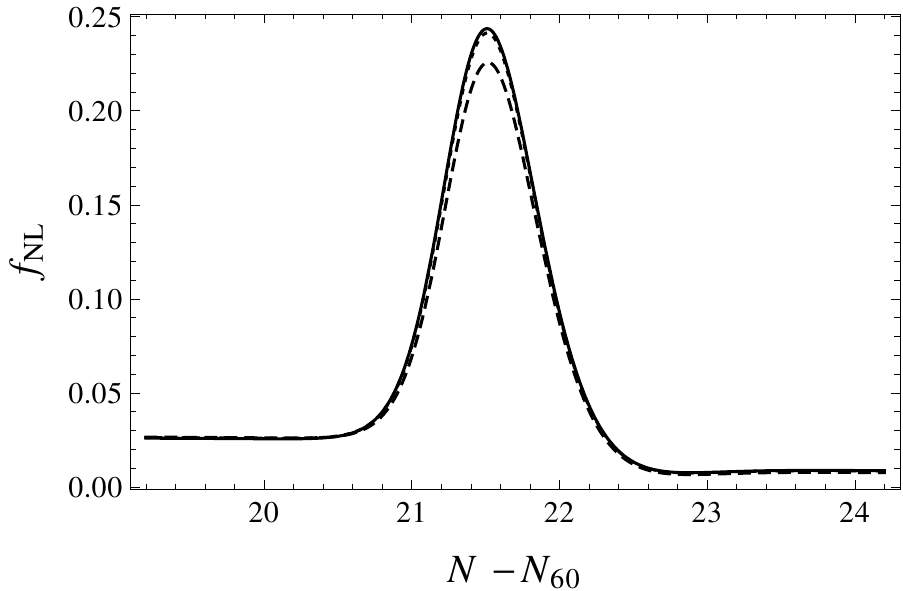}
\end{center}
\caption{The time evolution of $\f$ as a function of the number of e-foldings 
$N-N_{60}$ around the time of the turning of the fields,
for a triangle with $\omega=1$ (solid curve), $\omega=5/2$ (dotted curve) and $\omega=1000$ 
(dashed curve), all with fixed perimeter $K=3/2k_{60}$,
for the model (\ref{qua}) with mass ratio $m_{\phi}/m_{\gs}=9$.}
\label{fig655}
\end{figure}

After studying triangles with the same shape but varying size in the previous
section, we now turn to the scale dependence of $\f$ for triangles of the same perimeter but 
different shape, see the right-hand side of figure~\ref{fig01}. In figure \ref{fig655} we plot the time evolution of $\f$ during 
inflation for an equilateral $\omega=1$ (solid curve), an isosceles  
$\omega=5/2$ (dotted curve) and a squeezed $\omega=1000$ (dashed curve) triangle, 
all of perimeter $K=(3/2)k_{60}$, as a function of the number of e-foldings. The 
profile of the time evolution of $\f$ was discussed in the previous section. Here we 
are interested in the shape dependence of $\f$. 

\begin{figure}
\begin{tabular}{cc}
\includegraphics[scale=0.75]{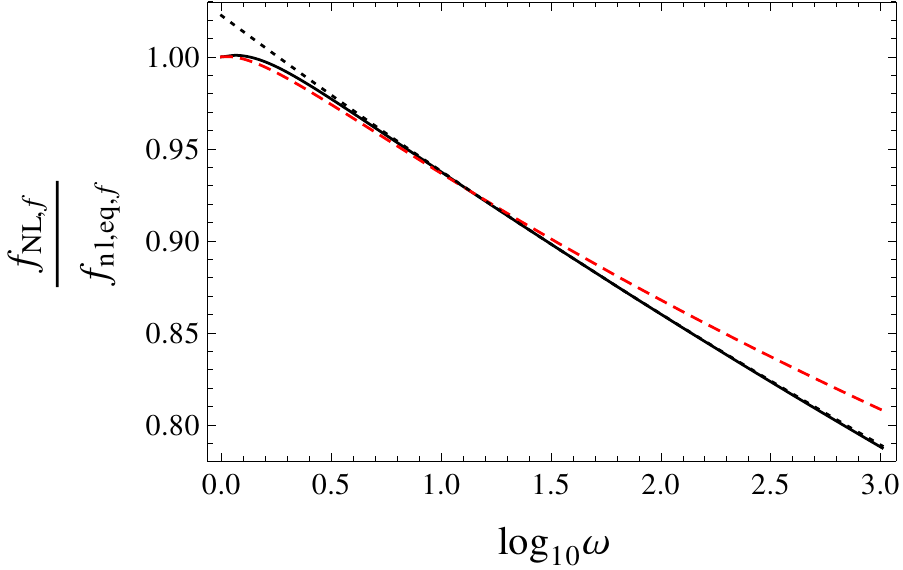}
&\includegraphics[scale=0.75]{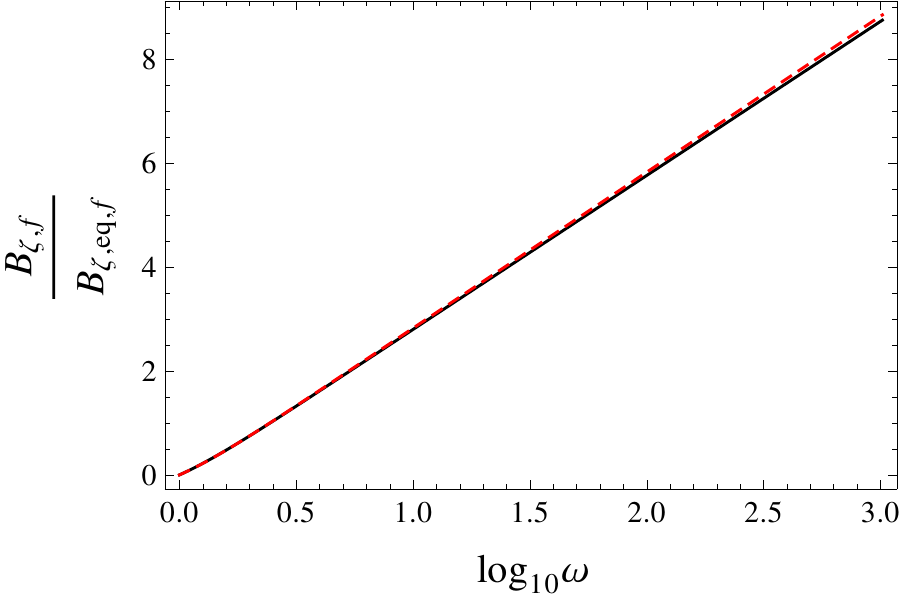}
\end{tabular}
\caption{Left: The final value of $\f$ normalised by the equilateral 
$\f$ as a function of $\omega$ for triangles with $K=
(3/2)k_{60}$ calculated exactly (black curve), using the analytical approximation (\ref{fnlb}) (dashed red curve) 
and using the shape index (\ref{into}) (dotted black curve). 
Right: The logarithm of the final value of the exact local bispectrum 
normalised by the local bispectrum computed on an equilateral triangle as a function of 
$\omega$ for triangles with $K=(3/2)k_{60}$ (black curve) and the same quantity assuming $\f$ scale independent (red dashed curve).  
Both figures are for the quadratic model (\ref{qua}) with mass ratio 
$m_{\phi}/m_{\gs}=9$.}
\label{fig3}
\end{figure}

Although it is during the peak that the variation of $\f$ for different shapes 
is more prominent, its final value is also affected. On the left-hand side of figure 
\ref{fig3} we plot the 
value of $\f$ at the end of inflation for triangles of perimeter $K=(3/2)k_{60}$, 
normalised by its value for the equilateral case ($\omega=1$), as a function
of $\omega$.  
The deviation of the values is small since it is related to 
horizon-exit slow-roll suppressed quantities. 
Within the long-wavelength formalism (or the $\delta N$ formalism) slow roll
at horizon crossing is a requirement. 
Nevertheless, the important conclusion here is that  
$\f$ decreases when the triangle becomes more squeezed. 
This can be attributed 
to the fact that the more squeezed is the triangle, the more the fluctuation $\zeta_k$ is 
frozen and behaves as part of the background when scale $k'$ crosses the horizon.  
As a result 
the correlation between $k$ and $k'$ becomes less and the resulting non-Gaussianity 
is smaller (see also the discussion below equation (\ref{fnlb}).)

An analytical formula can be found when applying the slow-roll approximation 
to expression 
(\ref{fNLgeni}) at the end of inflation, when isocurvature modes have vanished.  
We perform an integration by parts in the integral (see (\ref{gint}) and 
\cite{Tzavara:2010ge}; as before $\tg_{int}=0$). 
More precisely, assuming that we are really 
in the squeezed limit $k\ll k'$, the ratio $\gamma_k^2/\gamma_{k'}^2$ becomes
very large and we can ignore the equilateral terms that depend only on $k'$
and not also on $k$. We also assume 
that the decaying mode has vanished to simplify the expressions for the Green's functions 
(see \cite{Tzavara:2010ge} and the discussion in section \ref{sour}). $G_{12k'k}$ can be set to zero as one can see in figure 
\ref{fig1} (since it is basically equal to $\bv_{12}$ and only involves times
at the very left-hand side of the figure). 
Moreover, the same figure shows that $\bv_{12k}/\bv_{12k'}\approx 1$ (in the
formula these ratios are always multiplied by slow-roll parameters, so that
the deviation from 1 would be like a second-order effect), so that we find 
in the end
\bea
f_{\mathrm{NL},sq,f}
&=&G_{22k'k}f_{\mathrm{NL},eq,f}(k')+\frac{1-G_{22k'k}}{\lh1+\lh\bv_{12k'}\rh^2\rh^2}
\Bigg[\e_{k'}+\hpa_{k'}+\lh\frac{2\hpe_{k'}}{\bv_{12k'}}-\chi_{k'}\rh\lh\bv_{12k'}\rh^2
\Bigg],\label{fnlb}
\eea
where $f_{\mathrm{NL},eq,f}$ is given in equation (\ref{fnlfin}). The only quantity in the above expression that depends on 
the shape of the triangle is $G_{22k'k}$, so it must be $G_{22k'k}$ that is responsible for the decreasing behaviour of $f_{\mathrm{NL},sq,f}$. 
Indeed, increasing $\omega$ for a constant perimeter $K$ 
of the triangle means increasing the interval $t_k-t_{k'}$ and hence decreasing the 
value of $G_{22k'k}$ (see the right-hand side of figure \ref{fig1}, since in the slow-roll regime $\bv_{22}=G_{22}$). 
This means that the interaction of the two modes becomes less important.   
In the complete absence of isocurvature modes $G_{22k'k}=0$ and $f_{\mathrm{NL},sq,f}$ takes its minimal value.  
It is only the isocurvature mode that interacts with itself and the 
greater is the difference between the two momenta the less is the interaction.  Notice that 
the single-field limit of this result would correspond to $G_{22k'k}=0$ and $\bv_{12}=0$. 
\footnote{Inspecting equation (\ref{fnlb}) we notice that we do not 
recover the single-field squeezed limit result of \cite{Maldacena:2002vr}. This is to be expected since 
our $\f$ is the local one produced outside the horizon. As a consequence, we have only used 
the 
first and 
second-order horizon-crossing contributions coming from the redefinitions in the cubic action 
(see \cite{Tzavara:2010ge,Tzavara:2011hn}) 
as initial sources of $\f$ in the context of the 
long-wavelength formalism. 
The result of \cite{Maldacena:2002vr} on the other hand comes from the interaction 
terms in the Langrangian and hence is not the same.} 
 
The decrease of $\f$ for more squeezed triangles seems contradictory 
to the well-known fact that the local bispectrum is maximized 
for squeezed configurations. In order to clarify this subtle point, we stress that 
the left-hand side of figure \ref{fig3} is essentially the ratio of the exact 
bispectrum to the bispectrum 
assuming $\f$ as a constant (\ref{bisp}) and hence the products 
of the power spectrum cancel out. 
We also plot on the right-hand side of figure \ref{fig3} the final value of the 
bispectrum (\ref{bisp}), 
normalised by the value of the bispectrum for equilateral triangles 
with $K=3k_{60}/2$. 
Although $\f$ is maximal 
for equilateral triangles, the bispectrum has the opposite 
behaviour, since it is dominated by the contribution of the products 
of the power spectrum, which leads to an increased bispectrum 
for the more squeezed shape. At the same time though we show 
that there is a small contribution of $\f$ itself, leading to 
smaller values of the bispectrum when compared to a bispectrum 
where $\f$ is assumed to be constant.

\begin{figure}
\begin{tabular}{ll}
\includegraphics[scale=0.75]{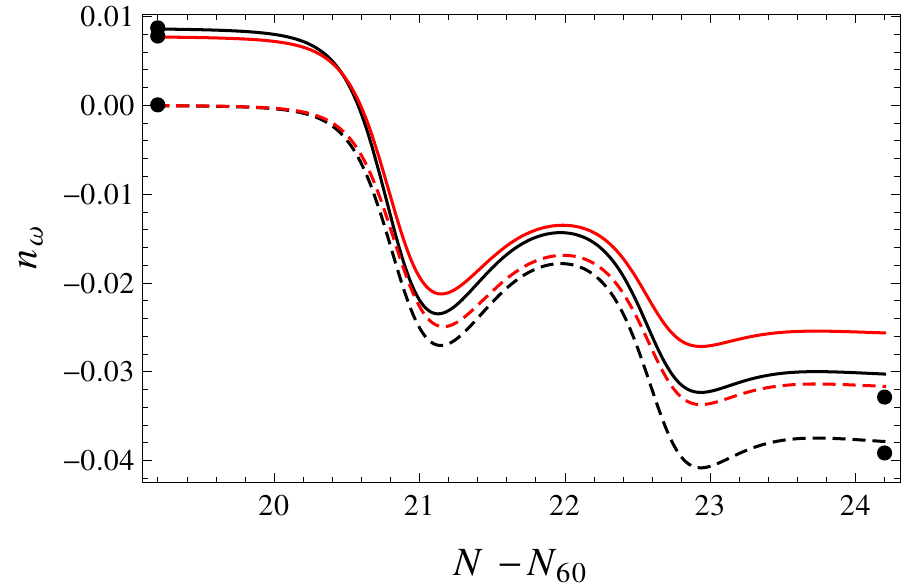}
&\includegraphics[scale=0.75]{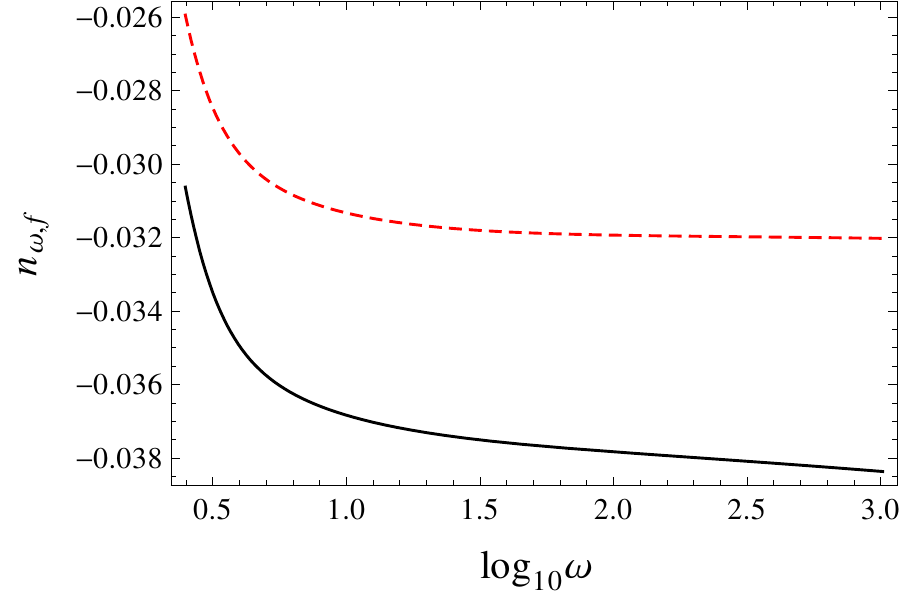}
\end{tabular}
\caption{Left: The time evolution of the shape index 
$n_\omega$ (\ref{intodef}) around the time of the turning of the fields,
for constant $K=(3/2)k_{60}$ (black curves) 
and $K=(3/2)k_{60}/10$ (red curves, above the $K=(3/2)k_{60}$ curves), 
and for the shapes $\omega=5/2$ (solid curve) 
and $\omega=1000$ (dashed curve). 
The points on the left correspond to the analytical values 
of the index as calculated from (\ref{noin}), while the points on the right correspond to the values of the index as calculated from (\ref{A2}) for $\omega\mg 1$.
Right: The final value of the shape index $n_\omega$ 
(\ref{intodef}) for constant $K=(3/2)k_{60}$ (black) and $K=(3/2)k_{60}/10$ (dashed red) 
as a function of $\omega$. Both figures are for the model (\ref{qua}) with mass 
ratio $m_{\phi}/m_{\gs}=9$.}\label{fig41}
\end{figure}

In order to quantify the above results, we examine the shape index $\tn_{\omega}$ (\ref{inddef}), 
assuming $K=\mathrm{const}$ and $\gamma=0$,
\be
 \tn_{\omega}=\frac{\partial\ln{f}}{\partial t_{k'}}\frac{1}{1-\e_{k'}}
 \frac{1}{1+2\omega}
 -\frac{\partial\ln{f}}{\partial t_{k}}\frac{1}{1-\e_{k}}
 \frac{2\omega}{1+2\omega}
.\label{shap}
\ee
In terms of $\tn_{\omega}$, $\f$ takes the form
\be
-\frac{6}{5}\f=\frac{1}{(2\pi^2)^2}\frac{f(k'_0,k'_0)+
2\omega^3f(k'_0,k_0)\lh\frac{\omega}{\omega_0}\rh^{\tn_{\omega}(t_{k'_0},t_{k_0})}}
{\mathcal{P}_\zeta(k'_0)^2\lh\frac{k'}{k'_0}\rh^{2(n_\zeta(t_{k'_0})-1)}
+2\omega^3\mathcal{P}_\zeta(k'_0)\mathcal{P}_\zeta(k_0)\lh\frac{k'}{k'_0}\rh^{n_\zeta(t_{k'_0})-1}\lh\frac{k}{k_0}\rh^{n_\zeta(t_{k_0})-1}},
\label{nbf}
\ee
where $k'/k'_0=(1+\beta)/(1+\beta_0) \propto 2\omega/(\omega+\frac{1}{2})$ and 
$k/k_0=(1-\beta)/(1-\beta_0) \propto 1/(\omega+\frac{1}{2})$. 
This can be further simplified in the squeezed region $\omega\mg1$ to find
\bea
-\frac{6}{5}\f&=&\frac{1}{(2\pi^2)^2}\frac{f(k'_0,k_0)\lh\frac{\omega}{\omega_0}\rh^{\tn_{\omega}(t_{k'_0},t_{k_0})}}
{\mathcal{P}_\zeta(k'_0)\mathcal{P}_\zeta(k_0)
\lh\frac{\omega(\omega_0+\frac{1}{2})}{\omega_0(\omega+\frac{1}{2})}\rh^{n_\zeta(t_{k'_0})-1}
\lh\frac{\omega_0+\frac{1}{2}}{\omega+\frac{1}{2}}\rh^{n_\zeta(t_{k_0})-1}}\nn\\
&\equiv&
-\frac{6}{5}f_{\mathrm{NL},0}
\lh\frac{\omega}{\omega_0}\rh^{\tn_{\omega}(t_{k'_0},t_{k_0})+n_\zeta(t_{k_0})-1}
\equiv
-\frac{6}{5}f_{\mathrm{NL},0}
\lh\frac{\omega}{\omega_0}\rh^{n_{\omega}(t_{k'_0},t_{k_0})},
\label{into}
\eea
with
\be
n_{\omega}\equiv\frac{\d\ln\f}{\d\ln\omega}=\frac{\partial\ln{\f}}{\partial t_{k'}}\frac{1}{1-\e_{k'}}
\frac{1}{1+2\omega}
-\frac{\partial\ln{\f}}{\partial t_{k}}\frac{1}{1-\e_{k}}
\frac{2\omega}{1+2\omega}
.\label{intodef}
\ee
The shape index $n_\omega$ describes the change of $\f$ due to the relative size 
of the two scales, namely due to how squeezed the triangle is, while 
keeping $K$ constant (see the right-hand side of figure 
\ref{fig01}). 

We studied different squeezed  
triangle configurations with constant $K$, varying $\omega$ from $\omega=5/2$ to $\omega=1000$. 
On the left-hand side of figure~\ref{fig41} we plot the time evolution of the shape index. 
The negative values of the index signify the decrease of $\f$ as expected. 
As one can see from the figure, for the more squeezed triangle ($\omega=1000$)
the initial value of $n_\omega$ seems to depend solely on the shape of the 
triangle and not on its magnitude, and even for the less squeezed triangle
($\omega=5/2$) the initial dependence on $K$ is negligible.  
We can find the analytical initial value of $n_\omega$ by differentiating (\ref{fnlin}):
\be
n_{\omega,in}=
\frac{1}{1+2\omega}n_{K,in}+\frac{4\omega}{1+2\omega}G_{22k'k}\frac{(\hpe_{k'})^2}{\e_{k'}+\hpa_{k'}}.\label{noin}
\ee
For $G_{22k'k}=0$, which corresponds to the squeezed limit, $n_{\omega,in}$ is proportional to the initial shape index for equilateral 
triangles times a factor depending on the shape, which also becomes very small in the squeezed limit.

Super-horizon effects, and especially the turning of the fields, 
result in a separation of the curves of $n_\omega$ of the same shape for 
different values of $K$, due to the dependence of the evolution of $\bv_{12k'}$ 
on the scale $k'$. 
The turning of the fields increases the absolute value of $n_\omega$, which is
the opposite of the behaviour of the conformal index $n_K$. 
The shape index depends on the transfer function 
$\bv_{12k'}$ (see appendix \ref{AppD} for an analytical approximation). 
The smaller $K$, the less does the final value of $\bv_{12k'}$ change with respect to its initial value (see figure 
\ref{fig1}) and hence the less the shape index is affected.  
Notice that the slow-roll parameters at horizon-crossing 
have the opposite behaviour: the smaller $K$, the smaller they are. Even though $n_\omega$ also depends on the slow-roll parameters, 
it is $\bv_{12k'}$ that most affects its evolution.

On the right-hand side of figure \ref{fig41} we plot 
the value of the shape index at the end of inflation.  
It exhibits a running of about $20\%$ within the 
range of scales studied, somewhat larger than the conformal index. 
We have analytically computed the shape spectral index for models with $\tg_{int}=0$ and with final $g_{iso,f}=0$ 
and give the result in appendix \ref{AppD}.
 
The dotted curve in the plot on the left-hand side of figure~\ref{fig3} shows
the final value of $\f$ approximated as a simple power law according to 
(\ref{into}).
Within the range of validity of 
our approximation $\omega^3\mg1$ it describes the exact result very well. 
We have also studied the shape spectral index for the potential 
(\ref{newpot}). Similarly to $n_K$, its value is two orders of magnitude smaller than the value for the quadratic potential, 
but the parametrization of 
$\f$ in terms of the shape index is in good agreement with the exact result for a larger range of $\omega\gtrsim3/2$.

%------------------------------ Conclusions ------------------------------------
%\Chapter{Conclusions}\label{Ch7}
\Conclusions\label{Ch7}

%-------------------------------------------------------------------------------------

In this thesis we treated several issues concerning perturbation theory beyond the linear order in two-field inflation models, based on our papers \cite{Tzavara:2010ge,Tzavara:2011hn,Tzavara:2012qq}. In particular we studied in detail gauge invariance at second order along with the cubic action governing the second-order perturbations and refined the long-wavelength formalism to calculate the non-Gaussianity produced during inflation as well as study its scale dependence.

As far as gauge invariance at second order in inflation with more than one fields is concerned, we managed to settle some unresolved issues. Although the gauge-invariant 
curvature perturbation defined through the energy density has been known for many years, 
the energy density is not the quantity that is used in 
calculations of inflationary non-Gaussianity. These use the scalar fields present during inflation instead of their energy. We found this 
gauge-invariant quantity in terms of the fields and discovered that it contains a non-local term unless slow-roll is assumed (see also our paper \cite{Tzavara:2011hn}).

We have also managed to make contact between gauge transformations and the redefinitions of the curvature and 
isocurvature perturbations occurring in the third-order action. 
Since \cite{Maldacena:2002vr} it has been known that the redefinition of the curvature 
perturbation in the action, introduced to remove terms proportional to the first-order equations of motion,  
corresponds to its gauge transformation. 
However, these terms  
appear at first sight to be absent in the flat gauge which would have had as a consequence the absence of 
quadratic contributions of first-order 
curvature perturbations at horizon crossing in this gauge and hence a gauge dependence of the related horizon-crossing 
non-Gaussianity (using Wick's theorem one can calculate the 
three-point correlation function due to these terms, as we did in (\ref{difer}) and our paper  \cite{Tzavara:2010ge}).  
We have extended the calculation for both gauges to second order and proved that in both of them the 
contributions are 
the same. The difference is that, in our perturbative approach, in the uniform energy-density gauge a part of these contributions is due to the first-order corrections 
and the other part to the second-order fields while in the flat gauge they are all due to the second-order fields. 

In addition to the adiabatic one, we also found the gauge-invariant isocurvature perturbation defined in terms of 
the scalar fields by studying the relevant fully non-linear 
spatial gradient defined in \cite{Rigopoulos:2005xx}. Usually isocurvature perturbations are studied in terms of 
the pressure perturbation of the content of the universe. Following \cite{Rigopoulos:2005xx} we found 
a definition using the fields themselves that demonstrates the orthogonality of this quantity to the curvature perturbation. While rewriting the action, these 
isocurvature perturbations appear naturally in the form we have defined them, thus showing that this quantity is the relevant 
one to use during inflation.

The exact cubic action for the perturbations, 
going beyond the 
slow-roll or super-horizon approximations, was computed in section \ref{cub} and in our paper  \cite{Tzavara:2011hn}  (in appendix \ref{AppB} we also give the tensor part of the action). 
This can prove very useful for future calculations. Up to now one had to impose the slow-roll condition at horizon crossing in 
order to calculate the non-Gaussianity. This was because the only two-field action available was that of the fields given in 
\cite{Seery:2005gb}, thus demanding slow-roll at horizon crossing in order to be able to use the long-wavelength 
formalism or the $\delta N$ formalism to find the curvature perturbation bispectrum. 
The action we provide here can be used directly with the in-in formalism \cite{Weinberg:2005vy} in order to 
calculate the exact non-Gaussianity beyond any restrictions, slow-roll or super-horizon. We have also computed the action for models of inflation with non-standard kinetic terms and non-trivial field metrics, which will appear in a paper soon. However, models with non-standard kinetic terms are known to produce non-Gaussianity at horizon crossing, related to the interaction terms of the action ($\f^{(3)}$), so we choose not to present our results in this thesis, since here we are mainly concerned about super-horizon non-Gaussianity. 

At the end of the day, the study of second-order perturbations is used for computing the non-Gaussianity produced by inflation models. The latter has become a
hot topic of research, since the recent observations of Planck allow us to constrain and discriminate 
inflation models based on their non-Gaussian predictions. In this thesis we
investigated the super-horizon bispectral non-Gaussianity produced by 
two-field inflation models. To this end we further worked out the 
long-wavelength formalism developed by Rigopoulos, Shellard, and Van Tent
(RSvT) \cite{Rigopoulos:2004ba,Rigopoulos:2005xx,Rigopoulos:2005ae,
Rigopoulos:2005us}. 

We derived an exact result for the bispectrum parameter $\f$ produced
on super-horizon scales for any two-field inflation model with canonical 
kinetic terms, equation~(\ref{fNLresult}) (see also our paper \cite{Tzavara:2010ge}). The result is expressed in terms 
of the linear perturbation solutions and slow-roll parameters. However, no
slow-roll approximation has been assumed on super-horizon scales,
these parameters should be viewed as short-hand notation and can be
large. In particular this means that the result is valid for models
where the field trajectory makes a sharp turn in field space so that slow
roll is temporarily broken. On the other hand, we need to assume slow roll
to be valid at horizon crossing in order to allow for the decaying mode to vanish and remove any dependence on the
window function. Furthermore, assuming slow roll allows us to use the analytic solutions for the linear mode 
functions. Observations of the scalar spectral index seem to indicate 
that this is a good approximation. 
%Note that the assumption of canonical
%kinetic terms is not a fundamental one: the basic equations of the formalism
%of RSvT are given for more general kinetic terms. We just did not want to
%complicate the notation and expressions in this paper with the covariant
%derivatives and additional curvature terms needed to treat the general case.

The result can be split into the sum of three parts, multiplied by an
overall factor (except for a small slow-roll suppressed term that is
the single-field contribution produced at horizon crossing).  This
overall factor is proportional to the contribution of the isocurvature
mode to the adiabatic mode, which is only non-zero for a truly
multiple-field model where the field trajectory makes a turn in field
space, as parametrized by a non-zero value of the slow-roll parameter
$\hpe$. (Effectively) single-field models do not produce any
non-Gaussianity on super-horizon scales, since the adiabatic
perturbation is conserved in that case. The three parts in the sum
are: 1) a part that only involves slow-roll parameters evaluated at
horizon-crossing and hence is always small; 2) a part proportional to
the pure isocurvature mode; and 3) an integral involving terms
proportional to the pure isocurvature mode. Since the adiabatic mode is not
necessarily constant in the presence of isocurvature modes, we only
consider models where the isocurvature mode has disappeared by the end
of inflation, so that we can directly extrapolate our result at the
end of inflation to recombination and observations of the
CMB. However, this automatically means that the part 2), although
varying wildly during the turn of the field trajectory, cannot give
any persistent non-Gaussianity that can be observed in the CMB. This
means that any large non-Gaussianity on super-horizon scales in models
satisfying this condition will have to come from the integrated effect
in part 3).

The exact equation~(\ref{fNLresult}) is the basis of our numerical studies.
However, to gain further insight we tried to work out the integral 
analytically. For this it turns out that the slow-roll approximation is 
necessary. Even then the integral can only be done explicitly for 
certain specific classes of potentials, among which are product potentials,
$W(\gf,\gs) = U(\gf)V(\gs)$, and generalized sum potentials,
$W(\gf,\gs) = (U(\gf)+V(\gs))^\gn$. We found that, with our assumptions
on the disappearance of the isocurvature mode, no product potential can
give large non-Gaussianity, nor can any simple sum potential with equal powers,
$W(\gf,\gs) = \ga \gf^p + \gb \gs^p$. However, we found conditions
under which the (generalized) sum potential can give large non-Gaussianity
(here defined as $\f$ larger than unity), and we have described an explicit, 
simple model that does. It consists of a heavy field rolling down a quadratic 
potential while a light field sits near the local maximum of a double-well 
potential. When the heavy field reaches zero and starts oscillating,
the light field takes over and rolls down, so that there is a turn of the
field trajectory in field space. We studied this model numerically, using the
exact results, to confirm our analytical predictions.

We included the second-order source term at horizon 
crossing in the long-wavelength formalism of RSvT, a contribution that
had been missing so far. This is the only change of the basic formalism
with respect to the paper \cite{Rigopoulos:2005us} by RSvT.
While this additional term is always small for the models we
consider and hence numerically insignificant, from an analytical point
of view it means we could now compare our results directly to the
$\f^{(4)}$ calculated within the $\gd N$ formalism 
\cite{Choi:2007su,Byrnes:2008wi,Vernizzi:2006ve}. 
Some of the potentials we studied had already been worked out using
that formalism and where available we compared our analytical results
and found perfect agreement. We also compared our exact numerical
results with those obtained using a numerical $\gd N$ treatment for
models where slow roll is broken and the analytic results cannot be
trusted, and again we found excellent agreement.

From our studies it has become clear that the condition on the
disappearance of the isocurvature mode by the end of inflation is a
very strong constraint.  It significantly reduces the possibilities
for a large, observable value of $\f$ produced during inflation. 
Note, however, that we chose to impose this condition only to be able to
neglect the further evolution of the adiabatic mode after inflation; it
is in no way a necessary condition for our formalism during inflation.
In future work we would like to relax this condition, which means that
the adiabatic mode would no longer necessarily be constant after
inflation, and hence will require a much better description and
understanding of the evolution of the perturbations during the
transition at the end of inflation and the subsequent period of
(p)reheating. 
%In conclusion, while a lot of progress has been made
%over the past few years regarding the non-Gaussianity produced in
%multiple-field inflation, more work still remains to be done.

In deriving equation~(\ref{fNLresult}) we assumed that all three scales cross
the horizon at the same moment. However, this is not a necessary assumption,
and we also generalized the result to an arbitrary momentum configuration.
%We find that going to the squeezed limit, where one of the momenta is much
%smaller than the other two, even when remaining within the resolution
%of the Planck satellite ($k' \sim 1000 k$), the result for $\f$ can be 
%reduced by about $10\%$, depending on the model. We stress that we are 
%discussing $\f$ here, so
%this effect is unrelated to the well-known result that the local bispectrum
%peaks on squeezed momentum configurations, which is due to the momentum
%behaviour of the power spectrum, which has been divided out in $\f$.
%However, exactly because of this latter effect, the squeezed limit is very
%relevant for the computation of $\f$.
We then used this result to study the scale dependence of the local non-Gaussianity
parameter $\f$ for 
two-field inflationary models. Multiple-field models with standard kinetic terms
do not exhibit 
the strong scale dependence inherent in models that produce equilateral non-Gaussianity 
at horizon-crossing through quantum mechanical effects. Nevertheless they are not 
scale independent in general and the interesting question is whether we can  
profit from their scale dependence in order to observationally acquire more 
information about inflation. 

We have calculated $\f$ using the long-wavelength formalism. This 
constrains us to assume slow roll at horizon-crossing and hence the relevant 
quantities at that time should not vary much, including the 
scale dependence of $\f$ for any triangle shape. Indeed we confirmed that, 
by introducing the conformal spectral index $n_K$ that measures the tilt of $\f$
for triangles of the same shape but different size ($K$ is a variable 
proportional to the perimeter of the momentum triangle). For the 
quadratic model with mass ratio $m_{\phi}/m_{\gs}=9$ we find $n_K\simeq0.018$, 
pointing to an almost scale-invariant $\f$ (see also our paper \cite{Tzavara:2012qq}).  

We also studied the scale dependence of $\f$ while varying the shape of the 
triangle and keeping its perimeter constant. $\f$ exhibits the opposite behaviour of the full bispectrum, 
i.e.\ it decreases the more squeezed the triangle is (the momentum dependence
of the bispectrum is dominated by that of the products of power spectra, not by
that of $\f$). This variation 
is not related to horizon-crossing quantities, but rather to the fact that 
the more squeezed the isosceles triangle under study, 
the smaller the correlation of its two scales. 
We quantified this effect by introducing the shape spectral index $n_\omega$, which 
for the quadratic model with $m_{\phi}/m_{\gs}=9$ is $n_{\omega}\simeq-0.03$ and has a running of about 
$20\%$ ($\omega$ is defined as the ratio of the
two different sides of an isosceles momentum triangle).

All our calculations have been done numerically in the exact background, 
assuming slow roll only at horizon-crossing, not afterwards. 
Nevertheless, semi-analytical expressions can be easily produced by directly 
differentiating $\f$. If we do assume slow roll we showed that we can even simplify these expressions further 
and find analytical 
formulas for the final value of $\f$ and its spectral indices $n_K$ and 
$n_\omega$, if  
the integral in $\f$ and the isocurvature modes vanish by the end of inflation (but this automatically means that $\f$ is small at the end of inflation), 
which is the case for example for any equal-power sum model. 

We used the two-field quadratic potential in our numerical calculations of momentum dependence.
This potential is easy to examine and allows for simplifications in the relevant 
expressions. Although its final non-Gaussianity is small, $\mathcal{O}(\e_k)$, its 
general behaviour should not be different from other multiple-field 
inflationary models with standard kinetic terms, in the sense 
that the scale dependence of $\f$ 
should always depend on horizon-exit quantities and the evolution of the transfer functions during the turning of the fields.  
Indeed we have checked that for the potential (\ref{newpot}) originally studied in our paper 
\cite{Tzavara:2010ge}, able to produce $\f\sim\mathcal{O}(1)$, the results 
remain qualitatively the same, although the values of the spectral indices are smaller due to the very slow evolution 
of the background at horizon-crossing in that model.

Although 
the effect of the magnitude of the triangle on $\f$ had been considered before, analytical 
and numerical estimates were not available before our paper \cite{Tzavara:2011hn}. In addition, it is the first time that 
the dependence of $\f$ itself (instead of the power spectra in the bispectrum) on the shape of the momentum triangle is studied. Using the long-wavelength 
formalism we have managed to study the two different sources of momentum dependence, i.e.\ the slow-roll parameters at horizon 
crossing and the evolution of the transfer functions, and to understand the role of each for the two different triangle deformations 
that we have studied. In summary, the later a momentum mode exits the horizon, the larger the slow-roll parameters are at that time and 
the larger $\f$ tends to be. In contrast, the final value of $\bv_{12k'}$ and the initial value of $G_{22k'k}$, the 
two transfer functions that are the most important for $\f$, are smaller the later the scale exits, which results 
in decreasing values of $\f$. These two opposite effects manifest themselves in the two different deformations we have studied. When keeping the shape of 
the triangle constant and varying its size, it is the slow-roll parameters at horizon crossing that play the major role in $\f$ and result in an increasing $\f$ for  
larger $K$. When changing the shape of $\f$, it is the correlation between the isocurvature mode at different scales, $G_{22k'k}$, that has the most 
important role, resulting in decreasing values of $\f$ when squeezing the triangle (i.e.\ increasing $\omega$). 

We have verified that the spectral indices of $\f$ ($n_K$ and
$n_\omega$), which we introduced to describe the effect of the two
types of deformations of the momentum triangle, provide a good
approximation over a wide range of values of the relevant scales.  In
the models we studied their values are too small to be detected by
Planck, given that $\f$ itself cannot be big (or it would have been
detected by Planck).  Models that break slow roll at horizon crossing
could in principle have larger spectral indices, but in order to study
such models one would need to go beyond the long-wavelength
formalism. Such models could be studied using the
exact cubic action derived in subsection \ref{cub} and our paper \cite{Tzavara:2011hn}.

We showed that the long-wavelength formalism of RSvT represents a viable
alternative to the $\gd N$ formalism to compute the super-horizon 
non-Gaussianity produced during inflation, allowing us to obtain and
verify results in a different way. Moreover, the long-wavelength formalism
has a number of advantages that can make it preferable in certain situations.
Very importantly, our formalism allows for a simple physical
interpretation of the different parts in terms of adiabatic and
isocurvature modes, providing insight into the behaviour of the different 
transient and persistent contributions to $\f$.
While we did not pursue this in this thesis,
the formalism also provides the solution for the second-order isocurvature
perturbation and hence the isocurvature bispectrum could be computed as
easily as the adiabatic one. Finally it allows us to study the scale dependence of the parameter of non-Gaussianity $\f$ and consider its sources.

Planck has provided us with constraints tighter than ever concerning the primordial non-Gaussianity. In order to employ to its full extent the extraordinary accuracy of the Planck data, it is  necessary to have accurate methods for computing the non-Gaussianity produced during all types of inflation models.  
%Work still needs to be done to that end. 
Regarding our formalism treating two-field models  for example, extending it  beyond the zeroth order in the gradient expansion 
will not only allow us for a more accurate study of the super-horizon regime, but also enable us to study models that break slow roll at horizon crossing. Furthermore, the Planck data allow in principle to test models with strong scale-dependent features like the ones studied in \cite{Achucarro:2012fd}, hence we also need to  better study and understand the momentum dependence of the non-Gaussianity produced in the context of such models.
%
%In the post Planck era, the aim of studying non-Gaussianity has changed.  
%Instead of searching for  models that can produce large, in the sense of measurable by Planck, non-Gaussianity, we want now to focus on models whose predictions are close to the single-field case. Our study of two-field models allows us to claim that these still remain eligible candidates for driving inflation. 

Multiple-field models of inflation can produce  large non-Gaussianity during the turning of the fields, which by the end of inflation tends to vanish. Indeed, models with non-vanishing final non-Gaussianity and vanishing isocurvature modes by the end of inflation seem to be hard to find. 
On the other hand, models with both non-vanishing non-Gaussianity and non-vanishing isocurvature modes by the end of inflation are common. This naturally leads to the need for studying their further evolution during a (p)reheating era. What happens to the isocurvature modes, how they interact with the scalar field responsible for preheating and whether they could produce further non-Gaussianity are a few of the questions one needs to answer. These questions are not only interesting in their own right, but also since on their answer depends the precision of our interpretation of the cosmological data. A study of this highly non-linear and model-dependent era of the universe  demands hard numerical work, but also a better understanding of the theoretical framework, which needs to be further explored. In conclusion, while a lot of progress has been made
over the past few years regarding the non-Gaussianity produced during inflation, the post-Planck era ahead of us raises new, even more demanding questions to be answered.

\appendix

%----------------------------- Appendix A --------------------------------------

{\FirstAppendix{
First and second-order perturbation calculations
}\label{AppA}}
%------------------------------------------------------------------------------------------
%------------------------------------------------------------------------------------------

%In this appendix we present the conventions used throughout this thesis. We start by introducing the system of units we adopt,  called the natural units. In this system we set 
%\be
%\hbar =c=k_B=1,
%\ee
%where $\hbar=h/(2\pi)$ and $h=1.054571596\times10^{-34}\ \mathrm{J\cdot s}$ is the Planck  constant, $c=2.99792458\times 10^8\ \mathrm{m\cdot s^{-1}}$ and $k_B=1.3806503\times 10^{-23}\ \mathrm{J\cdot K^{-1}}$
%
%

In this appendix we present some intermediate steps of the calculation of the perturbed Einstein equations and the gauge transformations used in chapters \ref{Ch2} and \ref{Ch3}.

\Section{First-order Einstein equations}

Perturbing the Einstein equations (\ref{stress}) to first order one can find the linear-order equations  
\bea
&&\mathrm{00:}\quad
%\frac{1}{a^2}
\partial^2\psi_{(1)}=-\frac{N}{Ha^2}\partial^2\alpha_{(1)}+\e\dot{Q}_{(1)}+3\dot{\alpha}_{(1)}+
\lh\e-3\rh\lh N_{(1)}+\e NQ_{(1)}\rh H\label{aee00}\\
&&\mathrm{0i:}\quad N_{(1)}-\frac{\dot{\alpha}_{(1)}}{H}+\e NQ_{(1)}=0\label{aee0i}\\
&&\mathrm{ij:}\quad 
%\frac{N^2}{2a^2}
\frac{N^2}{2}
\partial_i\partial_jB_{(1)}
+\delta_{ij}\Bigg\{-\frac{N^2}{2a^2}\partial^2B-\ddot{\alpha}_{(1)}-HN\lh3-\frac{\dot{N}}{HN^2}\rh\dot{\alpha}_{(1)}+HN\e\dot{Q}_{(1)}+H\dot{N}_{(1)}\nn\\
&&\qquad-H^2N\lh-3+\e+\frac{\dot{N}}{HN^2}\rh N_{(1)}+(HN)^2\e\lh3+\e+\eta\rh Q_{(1)}
\Bigg\}=0\label{aeeij}\\
&&\mathrm{with}\qquad B_{(1)}=
%-\frac{N_{(1)}}{N}
-\frac{N_{(1)}}{a^2N}
%-\frac{H\psi_{(1)}}{N}+\frac{\dot{N}\psi_{(1)}}{N^3}
-\frac{3H\psi_{(1)}}{N}+\frac{\dot{N}\psi_{(1)}}{N^3}
%-\alpha_{(1)}
-\frac{\alpha_{(1)}}{a^2}
-\frac{\dot{\psi}_{(1)}}{N^2},\nn
\eea
where we have defined the shorthand notation 
\be Q_{(r)}\equiv-\frac{H}{\Pi}e_{1A}\varphi^A_{(r)}, \ee
with $r$ the order of the perturbation theory. 
In the single field limit this is simply $Q_{(r)}=-H/\Pi\ \varphi_{(r)}$.

\Section{Second-order $0i$ Einstein equation}

The $0i$ Einstein equation is the only one we are going to need in our study. This takes the form 
%\bea
%&&\frac{1}{N^2}\Big[ \frac{1}{4}\partial^2N_{i\perp(2)}-\frac{N_{(1)}}{N}\partial^2N_{i\perp(1)}-\frac{1}{N}\partial_kN_{(1)}\partial_k\partial_i\psi_{(1)}
%-\frac{1}{N}\partial_kN_{(1)}N_{(i\perp(1),k)}\nn\\
%&&-2\frac{a^2H}{N}\lh N_{k\perp(1)}+\partial_k
%\psi_{(1)}\rh \partial_i\lh N_{k\perp(1)}+\partial_k\psi_{(1)}\rh 
%-\partial^2\alpha_{(1)}\partial_i\psi_{(1)}
%-\partial_i\partial_k\alpha\partial_k\psi\nn\\
%&&-H\partial_iN_{(2)}+\partial_i\alpha_{(2)}
%+6\frac{H}{N}N_{(1)}\partial_iN_{(1)}
%-\frac{2}{N}\dot{\alpha}_{(1)}\partial_iN_{(1)}
%-\frac{4}{N}N_{(1)}\partial_i\dot{\alpha}_{(1)}
%\Big]=\nn\\
%&&\frac{1}{N^2}\Big[
%\e NH\partial_iQ_{(2)}-2\e\lh\dot{\zeta}_{(1)}
%-\dot{\alpha}_{(1)}\rh\partial_i\lh\zeta_{(1)}
%-\alpha_{(1)} \rh-\e(\e+\eta)NH\partial_i\lh\zeta_{(1)}
%-\alpha_{(1)}\rh^2\nn\\
%&&-4\e H N_{(1)} \partial_i\lh\zeta_{(1)}
%-\alpha_{(1)}\rh 
%\Big], 
%\eea
\bea
&&\frac{1}{4}\partial^2N_{i\perp(2)}-\frac{1}{N}\partial_kN_{(1)}\partial_k\partial_i\psi_{(1)}
-2\frac{a^2H}{N}\partial_k
\psi_{(1)}\partial_i\partial_k\psi_{(1)} 
%-\partial^2\alpha_{(1)}\partial_i\psi_{(1)}
-2\partial_i\partial_k\alpha\partial_k\psi
+\partial_k\alpha\partial_i\partial_k\psi
-\partial_i\alpha\partial^2\psi
\nn\\
&&-H\partial_iN_{(2)}+\partial_i\alpha_{(2)}
+6\frac{H}{N}N_{(1)}\partial_iN_{(1)}
-\frac{2}{N}\dot{\alpha}_{(1)}\partial_iN_{(1)}
-\frac{4}{N}N_{(1)}\partial_i\dot{\alpha}_{(1)}
=\nn\\
&&
\e NH\partial_iQ_{(2)}-2\e\lh\dot{\zeta}_{1(1)}
-\dot{\alpha}_{(1)}\rh\partial_i\lh\zeta_{1(1)}
-\alpha_{(1)} \rh-\e(\e+\eta)NH\partial_i\lh\zeta_{1(1)}
-\alpha_{(1)}\rh^2\nn\\
&&-4\e H N_{(1)} \partial_i\lh\zeta_{1(1)}
-\alpha_{(1)}\rh 
-\e\lh\e+\hpa\rh NH\partial_i\zeta_{2(1)}^2
-2\e\dot{\zeta}_{2(1)}\partial_i\zeta_{2(1)}\nn\\
&&
+2\e\hpe NH\lh \zeta_{2(1)}\partial_i(\zeta_{1(1)}
-\alpha_{(1)})
-(\zeta_{1(1)}-\alpha_{(1)})\partial_i\zeta_{2(1)}\rh 
, 
\eea
where we have set $N_{i\perp(1)}=0$ and re-expressed $\phi_{(1)}$ in terms of the first order curvature and adiabatic perturbations. 
Perturbing $NH=\dot{a}/a$ to second order 
\be 
\frac{\dot{\alpha}_{(2)}}{H}-N_{(2)}=
\frac{N}{H}H_{(2)}+\frac{2}{H}N_{(1)}H_{(1)}
\ee
%
%\be 
%\partial_k\alpha\partial_i\partial_k\psi
%-\partial_i\alpha\partial^2\psi
%%-\partial_i\partial_k\alpha\partial_k\psi
%%+\partial^2\alpha\partial_i\psi 
%\ee
and rewriting $H_{(1)}$ in terms of $\rho_{(2)}$ and taking into account that $H_{(1)}=\e H\lh\zeta_{(1)}-\alpha_{(1)}\rh$:
\be
\frac{N}{H}H_{(2)}=-\e N^2H\frac{\rho_{(2)}}{\dot{\rho}}
-\e^2N\lh\zeta_{(1)}-\alpha_{(1)}\rh^2 
\ee
we find
\be
\partial^2N_{i\perp(2)}=0
\ee
and
\bea
&&\!\!\!\!\!\!\!\!\!\!\!\!\!\!\!\!\!\!\!\!\!\!\!\!
-\e NH\Big[
Q_{1(2)}+NH\frac{\rho_{(2)}}{\dot{\rho}}
+(\e-\hpa)\lh\zeta_{1(1)}
-\alpha_{(1)}\rh^2
-(\e+\hpa)\zeta_{2(1)}^2
-2\hpe\zeta_{2(1)}\lh\zeta_{1(1)}
-\alpha_{(1)}\rh 
\nn\\&&\!\!\!\!\!\!
-\frac{2}{NH}\partial^{-2}\partial_i\dot{\zeta}_{2(1)}
\partial_i\zeta_{2(1)}\Big]
+\partial^2\cA
=0,
%-\e\lh\e+\hpa\rh NH\zeta_{2(1)}^2
%+\partial^{-2}\partial_i\Bigg[
%-2\e\dot{\zeta}_{2(1)}\partial_i\zeta_{2(1)}
%+2\e\hpe NH\lh \zeta_{2(1)}\partial_i(\zeta_{1(1)}
%-\alpha_{(1)})
%-\zeta_{1(1)}\partial_i\zeta_{2(1)}\rh \Bigg].
\label{r2}
\eea
where $\partial^2\lambda=\e\lh\dot{\zeta}_{1(1)}-2\hpe NH\zeta_{2(1)}\rh$ and
%\bea
%&&\!\!\!\!\!\!\!\!\!\!\!\!\!\!\!\!\!\!\cA=\partial^{-4}\partial_i\Bigg[
%%2\e\dot{\zeta}_{(1)}\partial_i\lh\zeta_{(1)}
%%-\alpha_{(1)}\rh 
%2\partial^2\lambda\partial_i\lh\zeta_{1(1)}
%-\alpha_{(1)}\rh 
%-\frac{1}{N}\partial_kN_{(1)}\partial_k\partial_i\psi_{(1)}
%-2\frac{a^2H}{N}\partial_k
%\psi_{(1)}\partial_i\partial_k\psi_{(1)}
%-\partial^2\alpha_{(1)}\partial_i\psi_{(1)}
%\nn\\&&\quad\ \ 
%-\partial_i\partial_k\alpha_{(1)}\partial_k\psi
%\Bigg].
%\eea
%plus new terms
%\be 
%\partial_k\alpha\partial_i\partial_k\psi
%-\partial_i\alpha\partial^2\psi
%-\partial_i\partial_k\alpha\partial_k\psi
%+\partial^2\alpha\partial_i\psi 
%\ee
%All together
\bea
&&\!\!\!\!\!\!\!\!\!\!\!\!\!\!\!\!\!\!\cA=\partial^{-4}\partial_i\Bigg[
%2\e\dot{\zeta}_{(1)}\partial_i\lh\zeta_{(1)}
%-\alpha_{(1)}\rh 
2\partial^2\lambda\partial_i\lh\zeta_{1(1)}
-\alpha_{(1)}\rh 
-\frac{1}{N}\partial_kN_{(1)}\partial_k\partial_i\psi_{(1)}
-2\frac{a^2H}{N}\partial_k
\psi_{(1)}\partial_i\partial_k\psi_{(1)}
\nn\\&&\quad\ \ 
-2\partial_i\partial_k\alpha_{(1)}\partial_k\psi
+\partial_k\alpha\partial_i\partial_k\psi
-\partial_i\alpha\partial^2\psi
\Bigg].
\eea

\Section{Gauge transformations}

The expressions of the three terms involved in (\ref{trans}) for the space part of the metric in  an arbitrary gauge take the form

\bea
L_{\beta_{(1)}}^2g_{ij}=2a^2&\Bigg\{&T_1\Big[\frac{\d}{\d t}\lh T_1NH\rh+2\lh NH\rh^2T_1\Big]\delta_{ij} 
+NH\beta_k\partial_kT_1\delta_{ij}+T_1\Big[4NH\partial_{(i}\beta_{j)}+\partial_{(i}\dot{\beta}_{j)}
\Big]\nn\\
&&+\partial_{(j}T_1\dot{\beta}_{i)}-\frac{N^2}{a^2}\partial_iT_1\partial_jT_1
+\beta_k\partial_k\partial_{(i}\beta_{j)}
+\partial_{i}\beta_k\partial_{j}\beta_k
+\partial_{(i}\beta_k\partial_k\beta_{j)}
\Bigg\},
\eea
\be
L_{\beta_{(2)}}g_{ij}=2a^2\Big\{ NHT_2\delta_{ij}+\partial_{(i}\beta_{2j)} \Big\}, 
\ee
\bea
2 L_{\beta_{(1)}}g_{ij(1)}=4a^2&\Bigg\{& T_1\Big[ 2NH\lh \alpha\delta_{ij}+\chi_{ij}\rh +\lh \dot{\alpha}\delta_{ij}+\dot{\chi}_{ij}\rh\Big] 
+\beta_k\partial_k\lh \alpha\delta_{ij}+\chi_{ij}\rh +\partial_{(i}T_1N_{j)}\nn\\
&&+2\lh \alpha\partial_{(i}\beta_{j)}+\partial_{(i}\beta_k\chi_{kj)} \rh\Bigg\}.
\eea
%%The second order $ij$ component of the metric in the uniform energy density gauge can be rewritten as
%\bea
%\tilde{g}_{ij(2)}&=&2a^2\Bigg\{\Big[2\lh \alpha_{(1)}^2+\lh T_1NH\rh^2+2\alpha_{(1)}T_1NH\rh+\tilde{\alpha}_{(2)}\Big]\delta_{ij}+4\lh \alpha_{(1)}+T_1NH\rh\lh\chi_{ij}+\beta_{(i,j)}\rh\nn\\
%&&+\tilde{\chi}_{ij(2)}+2\chi_{ik}\chi_{kj}+2\beta_{(i,k)}\beta_{(k,j)}+2\chi_{ik}\beta_{(k,j)} 
%+2\chi_{kj}\beta_{(k,i)} \Bigg\}
%\eea
%%
Putting everything together we find the second order gauge transformation for the space part of the metric
\bea
&&
\tilde{\alpha}_{(2)}\delta_{ij}+\tilde{\chi}_{ij(2)}
=\alpha_{(2)}\delta_{ij}+\chi_{ij(2)}+NHT_2\delta_{ij}+\beta_{2(i,j)}
+2\partial_{(i}T_1N_{j)}  
+\partial_{(j}T_1\dot{\beta}_{i)}-\frac{N^2}{a^2}\partial_iT_1\partial_jT_1
\nn\\
&&+T_1\Big[\lh\dot{\alpha}+\dot{\zeta}\rh\delta_{ij}
 +2\dot{\chi}_{ij}+D_{ij}\dot{\beta}+\dot{\beta}_{(i\perp,j)}\Big] 
+\beta_k\partial_k\Big[\lh \alpha+\zeta\rh\delta_{ij} 
  +2\chi_{ij}+D_{ij}\beta+\beta_{(i\perp,j)}\Big]\nn\\
&&
+\frac{1}{2}\lh \beta_{k,i}\beta_{k,j}-\beta_{i,k}\beta_{j,k}\rh
%%%+\partial_{(i}\beta_k\partial_{j)}\beta_k
%%%+\partial_{(i}\beta_k\partial_k\beta_{j)}
%%%-2\beta_{(i,k)}\beta_{(k,j)}
+\chi_{kj}\lh\beta_{k,i}-\beta_{i,k}  \rh 
+\chi_{ki}\lh\beta_{k,j}-\beta_{j,k}  \rh. 
%%%+4\partial_{(i}\beta_k\chi_{kj)}
%%%-2\chi_{ik}\beta_{(k,j)} 
%%%-2\chi_{kj}\beta_{(k,i)}
\eea
Setting $\beta_{i(1)}=N_{i\perp(1)}=F=F_i=0$ the above simplifies to
\bea
&&
\tilde{\alpha}_{(2)}\delta_{ij}+\tilde{\gamma}_{ij(2)}
=\alpha_{(2)}\delta_{ij}+\gamma_{ij(2)}+NHT_2\delta_{ij}+\beta_{2(i,j)}
+2\partial_{(i}T_1\partial_{j)}\psi  
-\frac{N^2}{a^2}\partial_iT_1\partial_jT_1
\nn\\
&&+T_1\Big[\lh\dot{\alpha}+\dot{\zeta}\rh\delta_{ij}
 +2\dot{\gamma}_{ij}\Big].  
\eea

{\Appendix{The action}\label{AppB}}
%------------------------------------------------------------------------------------------
%------------------------------------------------------------------------------------------

In this appendix we drop the explicit 
subscript $(1)$ on first-order field and gauge-invariant perturbations to lighten the notation. The single-field limit of any result can be found by setting $\zeta_2=0$ and $\zeta_1=\zeta$ for the perturbations, while $\eta^{(n)\perp}=0$ and $\chi=0$ for the slow-roll parameters.

\Section{Second-order action calculation}
\label{app2}

We start by performing the calculation in the gauge 
\be
e_{1A}\tilde{\varphi}^A=0,\label{ugauge}
\ee
which is the uniform energy-density gauge. 
We first use the energy and momentum constraint (\ref{en}), (\ref{mo}) to find that to first order
\bea
&&\tilde{N}_{(1)}=\frac{\dot{\zeta}_1}{H}, \quad\ \ \qquad\qquad\quad \tilde{N}^i_{\perp}=0,\nn\\
&&\tilde{\psi}_{(1)}=-\frac{N}{Ha^2}\zeta_1+\lambda, \qquad \partial^2\lambda=\e\dot{\zeta}_1-2\e\hpe NH\zeta_2.
\label{n1ni}
\eea
It turns out that we do not need to calculate the shift or the lapse function to higher order, since in the action 
those terms are multiplied by constraint relations and hence vanish.

We start by working out the scalar part of the action. 
Keeping in mind the gauge constraint (\ref{ugauge}) we perturb (\ref{actionexact}) to second order
\bea
\tilde{S}_2=\frac{1}{2}\int\d^4x\Bigg\{&&\!\!\!\!\!\!\!\!\!\!\!\!a^3 e^{3\zeta_1}
\Bigg[
+\frac{1}{N}\lh 1-\frac{\tilde{N}_{(1)}}{N}+\frac{\tilde{N}_{(1)}^2}{N^2}-\frac{\tilde{N}_{(2)}}{2N}\rh 
\Big(-\frac{6}{\kappa^2}(NH+\dot{\zeta_1})^2+\dot{\phi}^2
+2\dot{\phi}^A\dot{\tvarphi}_A+\dot{\tvarphi}^2
\Big)
\nn\\
&&\quad+\lh N+\tilde{N}_{(1)}+\frac{\tilde{N}_{(2)}}{2}\rh \Big(-2W-2W_A\tvarphi^A-W_{AB}\tvarphi^A\tvarphi^B\Big)
\Bigg]\nn\\
&&\!\!\!\!\!\!\!\!\!\!\!\!\!\!\!\!\!-a e^{\zeta_1}\Bigg[\lh 
N+\tilde{N}_{(1)}\rh \frac{2}{\kappa^2}
\lh(\partial\zeta_1)^2+2\partial^2\zeta_1\rh
+N\partial_i\tilde{\varphi}_A
\partial^i\tilde{\varphi}^A\Bigg]\Bigg\},
\eea
where we have omitted a total derivative with respect to $\tilde{\psi}_{(1)}$. 
We then use the background Einstein and field equations to eliminate some terms and find that the term 
proportional to $\tilde{N}_{(2)}$ vanishes.
%Now the second-order action can be written as
%\bea
%\tilde{S}_2=\frac{1}{2}\!\!\int\!\!\d^4x\Bigg\{\!\!\!\!&&\!\!\!\!\!\!\! a^3H\Big[
%\dot{\zeta}_1\Big(\!-4\dot{\phi}^A\dot{\tvarphi}_A
%+\dot{\phi}^2\dot{\zeta}_1\Big)+9\zeta^2_1
%\lh-\frac{6}{\kappa^2}+\dot{\phi}^2\rh-\frac{1}{H^2}W_{AB}\tvarphi^A\tvarphi^B-36\zeta_1\dot{\zeta}_1
%+\dot{\tvarphi}^2\Big]\nn\\
%&&\!\!\!\!\!\!\!\!\!\!\!-aN\Big[2\e(\partial\zeta_1)^2+\partial^i\tilde{\varphi}_A\partial_i\tilde{\varphi}^A\Big]\Bigg\}.
%\label{sin}
%\eea
The terms proportional to $\tilde{\varphi}^A$ can be recast in terms of the curvature perturbations by applying the 
completeness property of the field basis and the expressions 
\bea
&&e_{1A}\dot{\varphi}^A=-(\e+\hpa)N\Pi\lh\zeta_{1(1)}-
\alpha_{(1)}\rh+\hpe N\Pi\zeta_{2(1)}-\frac{\Pi}{H} 
\lh\dot{\zeta}_{1(1)}-
\dot{\alpha}_{(1)}\rh\nn\\
&&e_{2A}\dot{\varphi}^A=-(\e+\hpa)N\Pi\zeta_{2(1)}
-\hpe N\Pi\lh\zeta_{1(1)}-
\alpha_{(1)}\rh-\frac{\Pi}{H} 
\dot{\zeta}_{2(1)},\label{difz}
\eea
found using the time derivatives (\ref{dere}) of the slow-roll parameters and the unit vectors. Using (\ref{difz}) and after integrating by parts and using $\dot{H}=-\e NH^2$ it can be written as
\bea
\tilde{S}_2=\int\d^4x\ \frac{\e}{\kappa^2}\Bigg\{&&\!\!\!\!\!\!\!\!\!\!\!\!\!\frac{a^3}{N}\Big[\dot{\zeta}_1^2+\dot{\zeta}_2^2
-4\hpe NH\dot{\zeta}_1\zeta_2
-3(\chi-\e-\hpa)(NH)^2\zeta_2^2
+2(\e+\hpa)NH\zeta_2\dot{\zeta}_2\nn\\
&&\qquad\!\!\!\!\!\!\!\!\!\!\!\!\!\!\!\!+(\hpe)^2(NH)^2\zeta_2^2
+(\e+\hpa)^2(NH)^2\zeta_2^2\Big]-
aN\Big[ (\partial\zeta_1)^2+(\partial\zeta_2)^2\Big]
\Bigg\}\label{simple}
\eea
or after further integration by parts
\bea
\tilde{S}_2\!=\!\!\int\!\!\d^4x\ \frac{\e}{\kappa^2} \Bigg\{&&\!\!\!\!\!\!\!\!\!\!\!\!\!-aN\Big( (\partial\zeta_1)^2+(\partial\zeta_2)^2\Big)
 +\frac{a^3}{N}\Big(\dot{\zeta}_1^2+\dot{\zeta}_2^2-4\hpe NH \dot{\zeta}_1\zeta_2
+2\chi NH\dot{\zeta}_2\zeta_2\Big)\label{s22}\\
&&\!\!\!\!\!\!\!\!\!\!\!\!\!\!\!\!+a^3NH^2\Big(\!\sqrt{\frac{2\e}{\kappa}}\frac{W_{221}}{3H^2}\!-\!2\e^2\!-\!(\hpa)^2\!+\!3(\hpe)^2\!+\!\frac{2}{3}\hpe\gx^\perp
\!-\!3\e(\hpa\!-\!\chi)\!+\!2\hpa\chi\Big) \zeta_2^2\Bigg\}.\nn
\eea

We can reach the same result while working in the flat gauge $\partial_i{\hat{\alpha}}=0$.  
One can prove that $\hat{N}_1=-\e N\zeta_1,\ \hat{N}^i_{\perp}=0$ and 
$\partial^2\hat{\psi}=\partial^2\lambda=
\e\dot{\zeta}_1-2\e\hpe NH\zeta_2$. The $\hat{\psi}$ terms cancel out and the second-order 
action takes the form
\bea
 \hat{S}_2\!\!\!\!&=&\!\!\!\!\frac{1}{2}\int\d^4x\Bigg\{ a^3\Bigg[\frac{1}{N}\dot{\hat{\varphi}}^2-NW_{AB}\hat{\varphi}^A\hat{\varphi}^B
+\hat{N}_1(-2W_A\hat{\varphi}^A
-2\frac{1}{N^2}\dphi^A\dot{\hat{\varphi}}_A)
+\frac{\hat{N}_1^2}{N^3}(-\frac{6}{\kappa^2}(NH)+\dphi^2)\Bigg]\nn\\
&&\qquad\qquad\!\!-aN\partial^i\hat{\varphi}_A\partial_i
\hat{\varphi}^A\Bigg\}.
\eea
Using the definition of $\zeta_m$, along with (\ref{difz}), this can be rewritten as (\ref{s22}). 

The second-order tensor part of the action in both gauges takes the form 
\be
S_{2\gamma}=\int\d^4xL_{2\gamma}=\frac{1}{2\kappa^2}\int\d^4x\Big\{\frac{a^3}{4N}(\dot{\gamma}_{ij})^2
-\frac{aN}{4}(\partial_k\gamma_{ij})^2\Big\},
\ee 
where $L_{2\gamma}$ is the second-order Lagrangian for the tensor modes. 
We also give the equation of motion of the gravitational waves
\be
\frac{\delta  L_{2\gamma}}{\delta\gamma_{ij}}=-\frac{1}{4\kappa^2}\frac{\d}{\d t}\lh\frac{a^3}{N}\dot{\gamma}_{ij}\rh 
+\frac{aN}{4\kappa^2}\partial^2\gamma_{ij}=0,
\ee
which we are going to use in the next section. In this thesis we will not discuss the evolution and physics of gravitational 
waves, but at linear order this is 
a standard subject in 
the literature, for a discussion see for example \cite{Misner:1974qy}.

\Section{Third-order action calculation}
\label{app3}

In order to compute $S_3$ we follow the same procedure starting from the uniform energy-density gauge. 
Notice that $\tilde{N}_3$ will multiply $(-2W+6H^2-\Pi^2)$ in exact 
analogy with $\tilde{N}_2$ in $S_2$, so it vanishes. Moreover, the overall factor multiplying $\tilde{N}_2$ is 
the first-order energy constraint (\ref{n1ni}), so it can be consistently set to zero as well.

We start by computing the cubic action of the first-order curvature perturbations up to $\tilde{N}_1$ involving only 
scalar quantities
\bea
\tilde{S}_{3(1)}\!=\!\frac{1}{2}\int\!\!\d^4x&&\!\!\!\!\Bigg\{\!a^3e^{3\tilde{\alpha}}\Bigg[ (N+\tilde{N}_1)
           \Big(-2W-2W_A\tvarphi^A-W_{AB}\tvarphi^A\tvarphi^B-\frac{1}{3}W_{ABC}\tvarphi^A\tvarphi^B\tvarphi^C\Big)\nn\\
&&\qquad+\frac{1}{N}\Big[(1-\frac{\tilde{N}_1}{N}+\frac{\tilde{N}_1^2}{N^2}
-\frac{\tilde{N}_1^3}{N^3}
)\Big(-6(NH+\dot{\tilde{\alpha}})^2
+\dot{\phi}^2+2\dot{\phi}^A\dot{\tvarphi}_A
+\dot{\tvarphi}^2\Big)\nn\\
&&\qquad+\Big(\partial^i\partial^j\tilde{\psi}
\partial_i\partial_j\tilde{\psi}
-(\partial^2\tilde{\psi})^2\Big)(1-\frac{\tilde{N}_1}{N})-4\partial^i\tilde{\psi}\partial_i\zeta_1\partial^2\tilde{\psi}
-2\dot{\tilde{\varphi}}_A\partial^i\tilde{\psi}\partial_i\tilde{\varphi}^A\Big]\Bigg]\nn\\
&&\!\!\!\!\!-ae^{\tilde{\alpha}}\Big[(N+\tilde{N}_1)\Big(\partial^i\tilde{\varphi}_A\partial_i\tilde{\varphi}^A
+4\partial^2\zeta_1+2(\partial\zeta_1)^2\Big)
  \Big]\Bigg\}.\label{s311}
\eea
After using the background equations and the definitions of the perturbations, eq. (\ref{s311}) takes the form
\bea
\tilde{S}_{3(1)}=\!\!\int\!\!\d^4x\Bigg\{&&\!\!\!\!\!\!\!\!\!\!\!\!\!
\frac{a^3e^{3\zeta}}{N}\Bigg[\e\lh1-
\frac{\dot{\zeta}_1}{NH}\rh 
\Big(\!\dot{\zeta}_1^2+\dot{\zeta}_2^2
+2NH(\e+\hpa)\dot{\zeta}_2\zeta_2
+(NH)^2\!\lh\!(\hpe)^2\!+\!(\e+\hpa)^2\rh\zeta_2^2\nn\\
&&\ \ \qquad\qquad\qquad\quad-2NH\hpe \zeta_2\dot{\zeta}_1\!\Big)-2NH\e\hpe \zeta_2\dot{\zeta}_1+\e (NH)^2\sqrt{\frac{2\e}{\kappa}}\frac{W_{222}}{3H^2}\zeta_2^3\nn\\
&&\ \ 
-3(NH)^2\lh 1+\frac{\dot{\zeta}_1}{NH}\rh 
\e(\chi-\e-\hpa)\zeta_2^2
-2\partial^i\tilde{\psi}\partial_i\zeta_1\partial^2
\tilde{\psi}\nn\\
&&\quad+\frac{1}{2}\Big(\partial^i\partial^j\tilde{\psi}\partial_i\partial_j\tilde{\psi}
-(\partial^2\tilde{\psi})^2\Big)\lh1-\frac{\dot{\zeta}_1}{NH}\rh\nn\\&&\quad
-2\e\partial^i\tilde{\psi}\Big((\e+\hpa)\zeta_2
\partial_i\zeta_2
+\frac{1}{NH}\dot{\zeta}_2\partial_i\zeta_2\Big)\Bigg]\nn\\
&&\!\!\!\!\!\!\!\!\!\!\!
-aN\lh\zeta_1+\frac{\dot{\zeta}_1}{NH}\rh\Big[2\partial^2\zeta+(\partial\zeta_1)^2+\e(\partial\zeta_2)^2\Big]\Bigg\}.\label{s3111}
\eea
By performing integrations by parts in (\ref{s3111}) we find 
\bea
\tilde{S}_{3(1)}\!=\!\!\int\!\!\d^4x\!\!&\Bigg\{&\!\!\!\!\frac{a^3\e}{N}
\Bigg[\e\zeta_1(\dot{\zeta}_1^2+\dot{\zeta}_2^2)
-2\dot{\zeta}_1\partial^i\lambda\partial_i\zeta_1-2\dot{\zeta}_2\partial^i\lambda\partial_i\zeta_2
-2NH\Big(\e\hpa\!+\!(\hpa)^2\!+\!(\hpe)^2\!\Big)\zeta_1\zeta_2\dot{\zeta}_2\nn\\
&&
\quad\ \ +NH(3\e\hpe+\gx^\perp)\zeta_1^2\dot{\zeta}_2
+2NH(\e\hpe+\gx^\perp)\zeta_1\zeta_2\dot{\zeta}_1\nn\\
&&\quad\ \ 
-NH\Big(\e^2+2\e\hpa+(\hpa)^2+(\hpe)^2\Big)
\zeta_2^2\dot{\zeta}_1 \nn\\
&&\quad\ \ +NH\Big(2\e^2+3\e\hpa-\!(\hpa)^2-\!(\hpe)^2+\gx^\parallel\Big)\zeta_1^2\dot{\zeta}_1
+(NH)^2\sqrt{\frac{2\e}{\kappa}}\frac{W_{222}}{3H^2}\zeta_2^3 \nn\\
&&\quad\ \ +(NH)^2\Bigg(2(\e+\hpa)\gx^\perp+\hpe\Big(2\e(3+\e)
+6\hpa-\gx^\parallel-3\chi\Big)\Bigg)
\zeta_1^2\zeta_2
\nn\\
&&\quad\ \ +\Bigg(\!\!-\e\Big(4\e^2+6\e+12\e\hpa+\hpa(9+8\hpa)
+8(\hpe)^2+2\gx^\parallel-3\chi\Big)
\!+\!\sqrt{\frac{2\e}{\kappa}}\frac{W_{221}}{H^2}\nn\\
&&\qquad\ \ -3(\hpa)^2-3(\hpe)^2-2\hpa\gx^\parallel
-2\hpe\gx^\perp\Bigg)(NH)^2\zeta_2^2\zeta_1
-2(\e+\hpa)\zeta_2\partial^i\lambda\partial_i\zeta_2\nn\\
&&\quad\ \ +4\hpe\zeta_2\partial^i\lambda\partial_i\zeta_1
+\frac{1}{2}\zeta_1(\partial^i\partial^j\lambda\partial_i\partial_j\lambda-(\partial^2\lambda)^2)\Bigg]
+a\e^2N\zeta_1\Big[(\partial\zeta_1)^2\!+\!(\partial\zeta_2)^2
\Big]\nn\\
&&\!\!\!\!\!\!\!\!\!\!-\frac{\delta L_2}{\delta\zeta_1}\Big(\frac{\e+\hpa}{2}\zeta_1^2-\hpe\zeta_1\zeta_2
+\frac{1}{NH}\dot{\zeta}_1\zeta_1-\frac{N^2}{4a^2}(\partial\zeta_1)^2
+\frac{N^2}{4a^2}\partial^{-2}\partial^i\partial^j(\partial_i\zeta_1\partial_j\zeta_1)\nn\\
&&\quad+\frac{1}{2}\partial^i\zeta_1\partial_i\lambda
-\frac{1}{2}\partial^{-2}\partial^i\partial^j
(\partial_i\lambda\partial_j\zeta_1)\Big)\nn\\
&&\!\!\!\!\!\!\!\!\!\!
-\frac{\delta L_2}{\delta\zeta_2}
\lh\!(\e+\hpa)\zeta_1\zeta_2+\frac{1}{NH}\dot{\zeta}_2\zeta_1+\frac{\hpe}{2}\zeta_1^2\rh\!\!\Bigg\},
\eea
where $\delta L_2/\delta\zeta_m$ are the first-order equations of motion. 
We can further integrate by parts the rest of the action to simplify it and prove that it takes the form of the 
flat gauge action (\ref{sflatap}), as expected since the action should be gauge-invariant.
The terms involving $\lambda$ along with the terms with 
space gradients vanish outside the horizon in the long-wavelength approximation, since $\lambda$ is equal to the 
first-order super-horizon energy constraint (\ref{mc}).

Finally we include the second-order fields. The extra terms in the action are
\bea
\tilde{S}_{3(2)}\!=\!\frac{1}{2}\int\!\!\d^4x\!&\Bigg\{&\!\!\!\!a^3e^{3\zeta_1}\!\Big[(N+\tilde{N}_1)
(-W_A\tvarphi^A_{(2)}\!-\!W_{AB}
\tvarphi^A_{(2)}\tvarphi^B)\!+\!\frac{1}{N}\lh 1-\frac{\tilde{N}_1}{N}\rh (\dphi_A\tdvphi^A_{(2)}+\tdvphi_A\tdvphi^A_{(2)})\nn\\
&&
\qquad-\frac{1}{N}\dot{\phi}^A\partial^i\tilde{\psi}\partial_i
\tilde{\varphi}_{(2)}+\frac{2}{N}\dot{\zeta}_{1(2)}\partial^2\tilde{\psi}\Big]\nn\\
&&
\!\!\!\!\!\!\!\!-aN \Big[-\partial^i\zeta_1\partial_i\zeta_{1(2)}+
\frac{1}{NH}\dot{\zeta}_1\partial^2\zeta_{1(2)}
+\partial^i\tilde{\varphi}^A\partial_i\tilde{\varphi}_{A(2)}\Big]\Bigg\},\label{s32}
\eea
where $\varphi^A$ without a subscript always denotes the first-order perturbation.  
After performing integrations by parts we find
\bea
\tilde{S}_{3(2)}=\int\d^4x\Bigg\{\frac{\delta L_2}{\delta\zeta_1}\lh\frac{\zeta_{1(2)}}{2}+\frac{\tilde{Q}_{1(2)}}{2}\rh
+\frac{\delta L_2}{\delta\zeta_2}\frac{\tilde{Q}_{2(2)}}{2}\Bigg\}.\label{s32s}
\eea

Next, we perform the same calculation for the flat gauge, starting from
\bea
\hat{S}_{3(1)}=\frac{1}{2}\int\d^4x\!\!\!\!&&\!\!\!\!\Bigg\{a^3\Bigg[ (N+\hat{N}_1)
           \Big(-2W-2W_A\hat{\varphi}^A-W_{AB}\hat{\varphi}^A\hat{\varphi}^B-\frac{1}{3}W_{ABC}\hat{\varphi}^A\hat{\varphi}^B
\hat{\varphi}^C\Big)\\
 &&+\frac{1}{N}\lh1-\frac{\hat{N}_1}{N}+\frac{\hat{N}_1^2}{N^2}
-\frac{\hat{N}_1^3}{N^3}\rh \Big(-6+\dot{\phi}^2+2\dot{\phi}^A\dot{\hat{\varphi}}_A+\dot{\hat{\varphi}}^2
+4\partial^2\hat{\psi}\nn\\
&&+\partial^i\partial^j\hat{\psi}\partial_i\partial_j\hat{\psi}-(\partial^2\hat{\psi})^2
-2\partial^i\hat{\psi}\dot{\phi}^A\partial_i\hat{\varphi}_A
-2\partial^i\hat{\psi}\dot{\hat{\varphi}}^A
\partial_i\hat{\varphi}_A\Big)\Bigg]
-a\hat{N}_1\partial^i\hat{\varphi}^A\partial_i\hat{\varphi}_A\Bigg\},\nn
\eea
again taking into account that $\hat{N}_2$ multiplies the first-order energy constraint and thus we set it to zero. 
We find using the definition of $\zeta_m$, along with (\ref{eq2}), (\ref{srpara2}) and (\ref{difz})
\bea
\hat{S}_{3(1)}&=&\!\!\int\!\!\d^4x
\Bigg\{\!\frac{a^3\e}{N} H\Bigg[\e\zeta_1(\dot{\zeta}_1^2+\dot{\zeta}_2^2)
-2\dot{\zeta}_2\partial^i\lambda\partial_i\zeta_2-2\dot{\zeta}_1\partial^i\lambda\partial_i\zeta_1
\nn\\
&&\qquad\qquad\quad+2NH\e(\e+\hpa)\zeta_1\zeta_2
\dot{\zeta}_2
+2NH\e\hpe \zeta_1^2\dot{\zeta}_2 
+NH\e(3\hpa+2\e)\zeta_1^2\dot{\zeta}_1 \nn\\
&&\qquad\qquad\quad+(NH)^2\Bigg(\!\!\sqrt{\frac{2\e}{\kappa}}\frac{W_{211}}{H^2}-2\e(\e\hpe+\hpa\hpe+\gx^\perp+3\hpe)\Bigg)
\zeta_1^2\zeta_2\nn\\
&&\qquad\qquad\quad
+(NH)^2\sqrt{\frac{2\e}{\kappa}}\frac{W_{222}}{3H^2}\zeta_2^3\nn\\
&&\qquad\qquad\quad+(NH)^2\Bigg(\!\!\sqrt{\frac{2\e}{\kappa}}\frac{W_{221}}{H^2}+\e\lh-3(\hpe)^2+
(\e+\hpa)^2\rh+3\e(\chi-\e-\hpa)\Bigg)
\zeta_2^2\zeta_1\nn\\
&&\qquad\qquad\quad+(NH)^2\lh\!\!\sqrt{\frac{2\e}{\kappa}}\frac{W_{111}}{3H^2}-\e\lh\gx^\parallel+3\hpa-(\hpe)^2-(\hpa)^2\rh
\rh \zeta_1^3
\nn\\
&&\qquad\qquad\quad-2(\e+\hpa)\zeta_2\partial^i\lambda\partial_i\zeta_2+4\hpe\zeta_2\partial^i\lambda\partial_i\zeta_1
+\frac{1}{2}\zeta_1(\partial^i\partial^j\lambda\partial_i\partial_j\lambda-(\partial^2\lambda)^2)\Bigg]
\nn\\
&&\qquad\!+a\e^2N
\zeta_1\Big((\partial\zeta_1)^2+(\partial\zeta_2)^2\Big)\Bigg\}.\label{sflatap}
\eea

Finally we include the second-order fields. The surviving terms in the action are
\bea
\hat{S}_{3(2)}\!=\!\frac{1}{2}\int\!\!\d^4x\!&\Bigg\{&\!\!\! a^3\Big[(N+\hat{N}_1)(-W_A\hat{\varphi}^A_{(2)}-W_{AB}
\hat{\varphi}^A_{(2)}\hat{\varphi}^B)
+\frac{1}{N}\lh 1-\frac{\hat{N}_1}{N}\rh 
(\dphi_A\dot{\hat{\varphi}}^A_{(2)}+\dot{\hat{\varphi}}_A\dot{\hat{\varphi}}^A_{(2)})\nn\\
&&\ \ -\frac{1}{N}\dot{\phi}^A\partial^i\hat{\psi}\partial_i\hat{\varphi}_{A(2)}\Big]
-aN\partial^i\hat{\varphi}^A\partial_i\hat{\varphi}_{A(2)}\Bigg\}
\eea
and they can be rewritten as 
\bea
\hat{S}_{3(2)}=\int\d^4x\Bigg\{\frac{\delta L_2}{\delta\zeta_1}\frac{\hat{Q}_{1(2)}}{2}
+\frac{\delta L_2}{\delta\zeta_2}\frac{\hat{Q}_{2(2)}}{2}\Bigg\}.
\eea

In the last part of this appendix we consider the tensor scalar part of the action. There will be no contributions from the 
second-order fields, since these 
cancel due to $\gamma_{ij}$ being transverse. We start from the action for two scalar and one tensor modes in the 
uniform energy-density gauge
\bea
\tilde{S}_{\zeta\zeta\gamma}=
\int\d^4x&\Big\{&aN\Big[-\frac{2}{NH}\gamma_{ij}\partial^i\dot{\zeta}_1\partial^j\zeta_1
-\gamma_{ij}\partial^i\zeta_1\partial^j\zeta_1+\e\gamma_{ij}\partial^i\zeta_2\partial^j\zeta_2\Big]\nn\\
&&\!\!\!\!\!\!\!+\frac{1}{2}\frac{a^3}{N}\Big[-\lh 3\zeta_1-\frac{\dot{\zeta}_1}{NH}\rh \dot{\gamma}_{ij}\partial^i\partial^j\tilde{\psi}
+\partial_k\gamma_{ij}\partial^i\partial^j\tilde{\psi}\partial^k\tilde{\psi}\Big]\Big\},
\eea
which after integrations by parts becomes
\bea
\tilde{S}_{\zeta\zeta\gamma}\!\!=\!\!\!\int\!\!\d^4x\!&\Big\{&\!\!\!\frac{a^3}{N}\Big[
\frac{\e}{2}\dot{\gamma}_{ij}\partial^i\zeta_1\partial^j\lambda
+\frac{1}{4}\partial^2\gamma_{ij}\partial^i\lambda\partial^j\lambda\Big]
+aN
\e\gamma_{ij}\Big[\partial^i\zeta_1\partial^j\zeta_1+
\partial^i\zeta_2\partial^j\zeta_2\Big]\label{szzg}\\
&&\!\!\!\!\!\!+\frac{\delta L_{2\gamma}}{\delta\gamma_{ij}}\Big(\frac{N^2}{a^2}\partial_i\zeta_1\partial_j\zeta_1
-(\partial_i\zeta_1\partial_j\lambda+\partial_j\zeta_1\partial_i\lambda)\Big)
+\frac{\delta L_2}{\delta\zeta_1}\frac{1}{4}\partial^{-2}(\dot{\gamma}_{ij}\partial^i\partial^j\zeta_1)\Big\}.\nn
\eea
In the flat gauge one can find directly after substitution in (\ref{actionexact}) the first line of (\ref{szzg}), so that 
there are no redefinitions.

Finally we calculate the part of the action consisting of one scalar and two tensor modes, starting from the uniform 
energy-density gauge
\bea
\tilde{S}_{\zeta\gamma\gamma}=\frac{1}{2}\int\d^4x\Big\{\frac{a^3}{N}\Big[\frac{1}{4}\lh 3\zeta_1-\frac{\dot{\zeta}_1}{NH}\rh 
(\dot{\gamma}_{ij})^2-\frac{1}{2}\dot{\gamma}_{ij}
\partial_k\gamma^{ij}\partial^k\tilde{\psi}\Big]
-\frac{aN}{4}\lh\zeta_1+\frac{\dot{\zeta}_1}{NH}\rh 
(\partial_k\gamma_{ij})^2\Big\}
\eea
or equivalently
\bea
\tilde{S}_{\zeta\gamma\gamma}&=&\int\d^4x\Big\{-\zeta_1\dot{\gamma}_{ij}\frac{\delta L_{2\gamma}}{\delta\gamma_{ij}}
+\frac{a^3}{N}\Big[\frac{\e}{8}\zeta_1 (\dot{\gamma}_{ij})^2-\frac{1}{4}\dot{\gamma}_{ij}\partial_k\gamma^{ij}\partial^k\lambda\Big]
+\frac{aN}{8}\e\zeta_1(\partial_k\gamma_{ij})^2\Big\},
\eea
while in the flat gauge we find directly
\bea
\hat{S}_{\zeta\gamma\gamma}=\frac{1}{2}\int\d^4x\Big\{\frac{a^3}{N}\Big[\frac{\e}{4}\zeta_1(\dot{\gamma}_{ij})^2-\frac{1}{2}\dot{\gamma}_{ij}
\partial_k\gamma^{ij}\partial^k\lambda\Big]+\frac{aN}{4}\e\zeta_1(\partial_k\gamma_{ij})^2\Big\}.
\eea

The three tensor modes action does not contain any redefinitions. For details the reader may look in \cite{Maldacena:2002vr}.

%----------------------------- Appendix D --------------------------------------

{\Appendix{Long-wavelength calculations}\label{AppC}}

\Section{Computation of the second-order source term}
\label{appSecondSource}

To compute the second-order source term $b_{ia}^{(2)}$ of
(\ref{secondsource}) we will consider the consequences of the perturbation redefinitions in the cubic action (\ref{finu}). Immediately after horizon-crossing these become
\bea
\zeta_1+\frac{\zeta_1^{(2)}}{2}=\zeta_{1c}+\frac{\dot{\zeta}_1\zeta_1}{NH}
-\frac{\dot{\zeta}_2\zeta_2}{NH}
+\frac{\e+\hpa}{2}\lh\zeta_1^2-\zeta_2^2\rh-\hpe\zeta_1\zeta_2
+\partial^{-2}\partial^i\lh\frac{\zeta_2}{NH}\partial_i\dot{\zeta}_2\rh,
\eea
\bea
\zeta_2+\frac{\zeta_2^{(2)}}{2}=\zeta_{2c}+\frac{\zeta_2\dot{\zeta}_1}{NH}
+\frac{\zeta_1\dot{\zeta}_2}{NH}+\frac{\hpe}{2}\lh\zeta_1^2-\zeta_2^2\rh
+(\e+\hpa)\zeta_1\zeta_2
-\partial^{-2}\partial^i\lh\frac{\zeta_2}{NH}\partial_i\dot{\zeta}_1\rh,
\eea
where we have dropped the $(1)$ subscript for the first-order perturbations.  
Inspecting (\ref{zai}) and (\ref{z2u}) we see that in the uniform energy density gauge $\tilde{\zeta}_{mi}=\partial_i\zeta_{m}$ both for the 
adiabatic and the isocurvature component. Since we perform our main 
calculation in the flat gauge and we use the variable 
$\zeta_{mi}$ rather than $\zeta_m$, we want to transform the 
above redefinitions to this gauge by the simple gauge transformation
(\ref{gat}):
\be
\zeta_{mi(2)}=\tilde{\zeta}_{mi(2)}
-\frac{1}{NH}\zeta_{1}\dot{\zeta}_{mi}.
\ee
We find
\be
\zeta_{1i}+\frac{\zeta_{1i}^{(2)}}{2}=\partial_i\Big[\zeta_{1c}
+\dot{\zeta}_1\zeta_1
-\dot{\zeta}_2\zeta_2
+\frac{\e+\hpa}{2}\lh\zeta_1^2-\zeta_2^2\rh-\hpe\zeta_1\zeta_2
+\partial^{-2}\partial^j\lh\zeta_2\partial_j\dot{\zeta}_2\rh\Bigg]
-\zeta_1\partial_i\dot{\zeta}_1,
\ee
\be
\zeta_{2i}+\frac{\zeta_{2i}^{(2)}}{2}=\partial_i\Big[\zeta_{2c}
+\zeta_2\dot{\zeta}_1
+\zeta_1\dot{\zeta}_2+\frac{\hpe}{2}\lh\zeta_1^2-\zeta_2^2\rh
+(\e+\hpa)\zeta_1\zeta_2
-\partial^{-2}\partial^j\lh\zeta_2\partial_j\dot{\zeta}_1\rh
\Bigg]-\zeta_1\partial_i\dot{\zeta}_2,
\ee
where we have made the time coordinate choice $NH=1$ (the natural choice when working in the flat gauge choice $\bN\bH=1$). 
Note that after horizon exit $\dot{\zeta}_1=2\hpe\zeta_2$ and 
$\dot{\zeta}_2=-\chi\zeta_2$ (the latter is valid under the slow-roll 
approximation only, see subsection \ref{tem}), so that the expressions simplify to
\bea
&&\zeta_{1i}+\frac{\zeta_{1i}^{(2)}}{2}
=
\partial_i\Big[\zeta_{1c}
+\frac{\e+\hpa}{2}\zeta_1^2
+\hpe\zeta_1\zeta_2
-\frac{\e+\hpa-\chi}{2}\zeta_2^2\Bigg]
-2\hpe\zeta_1\partial_i\zeta_2
\label{z1fgsr}\\
&&\zeta_{2i}+\frac{\zeta_{2i}^{(2)}}{2}=\partial_i\Big[\zeta_{2c}
+(\e+\hpa-\chi)\zeta_1\zeta_2
+\frac{\hpe}{2}\lh\zeta_1^2+\zeta_2^2\rh
\Bigg]
+\chi\zeta_1\partial_i\zeta_2,
\label{z2fsr}
\eea

For a field redefinition of the form $\zeta=\zeta_c+A\zeta_c^2$ 
(note that in the equations above we have not added the subscript $c$
explicitly in the quadratic terms, since to second order it makes no 
difference) the three-point correlation function can be written as
\be
\langle\zeta_\mathbf{\vc{k}_1}\zeta_\mathbf{\vc{k}_2}
\zeta_\mathbf{\vc{k}_3}
\rangle
=\langle\zeta_{c\mathbf{\vc{k}_1}}
\zeta_{c\mathbf{\vc{k}_2}}
\zeta_{c\mathbf{\vc{k}_3}}\rangle
+2A[\langle\zeta_{c\mathbf{\vc{k}_1}}
\zeta_{c\mathbf{\vc{k}_2}}\rangle
\langle\zeta_{c\mathbf{\vc{k}_1}}
\zeta_{c\mathbf{\vc{k}_3}}\rangle
+\mathrm{cyclic}].
\ee
In terms of our key quantities $\zeta_{mi}$ the perturbations are rewritten as $\zeta_m=\partial^2\partial_i\zeta_{mi}$. Hence the elements of $L_{1ab}$ and $N_{1ab}$ are just the 
coefficients of the various products of $\zeta_1$ and $\zeta_2$ in the 
redefinition of $\zeta_1$ multiplied by $2$, and similarly for 
$L_{2ab}$ and $N_{2ab}$. Note that the local terms $L_{abc}$ correspond
to the terms between the square brackets, and the non-local terms $N_{abc}$ 
to the terms outside. 
As an example we consider the contribution  of the first local term of the redefinition (\ref{z1fgsr}), $(\e+\hpa)\zeta_1^2/2$, in the bispectrum of the adiabatic perturbations
\bea
&&\langle\zeta_{1\mathbf{\vc{k}_1}}
\zeta_{1\mathbf{\vc{k}_2}}
\zeta_{1\mathbf{\vc{k}_3}}
\rangle
\supseteq
\frac{\e_{k_1}+\hpa_{k_1}}{2}\langle
\int\frac{\d^3\vc{k}}{(2\pi)^{3/2}}
\zeta_{1\mathbf{\vc{k}}}\zeta_{1\mathbf{\vc{k}_1}-\vc{k}}
\zeta_{1\mathbf{\vc{k}_2}}
\zeta_{1\mathbf{\vc{k}_3}}\rangle
+\mathrm{cyclic}\nn\\
&&
=\frac{\e_{k_1}+\hpa_{k_1}}{2}\Bigg[\langle\zeta_{1\mathbf{\vc{k}}}
\zeta_{1\mathbf{\vc{k}_2}}\rangle
\langle
\zeta_{1\mathbf{\vc{k}_1}-\vc{k}}\zeta_{1\mathbf{\vc{k}_3}}
\rangle
+
\langle\zeta_{1\mathbf{\vc{k}}}
\zeta_{1\mathbf{\vc{k}_3}}\rangle
\langle
\zeta_{1\mathbf{\vc{k}_1}-\vc{k}}\zeta_{1\mathbf{\vc{k}_2}}
\rangle\Bigg]+\mathrm{cyclic}
\eea
This leads to $L_{111*}=\e_{k_1}+\hpa_{k_1}$. Working likewise one can find  the explicit slow-roll expressions given in (\ref{nl}). Their contribution to $\f$ in the equal-momenta case turns out to be 
\be
-\frac{6}{5}\f=\frac{\es+\hpas+\hpes\bv_{12}}{1+(\bv_{12})^2},\label{difer}
\ee
where the index $*$ indicates the time when the scale exits the horizon (this term is part of $g_{sr}$ in (\ref{gisosrint})). 
Directly after horizon crossing or equivalently in the 
single-field limit, when $\bv_{12}=0$, this reduces to the well-known result by Maldacena $-6/5\f=\es+\hpas$.

\Section{Gradients and locality}
\section*{}\label{appGauge}

As a consistency check we want to verify that $\tilde{\zeta}_{1i(2)}$ is
indeed a total gradient, as it should be according to  (\ref{zai}). Taking expression
(\ref{u2hor}) (corrected by the gauge transformation),
\bea
 \tilde{\zeta}_{1i(2)}\!\!&=&\!\!-(\partial_iv_{e*})v_{f*}
\Bigg\{\!\!-2\hpe G_{1f}(t,t_*)G_{2e}(t,t_*)
+\int_{t_*}^t\!\!\mathrm{d} t' G_{1a}(t,t')
\bar{A}_{abc}G_{be}(t',t_*)G_{cf}(t',t_*)\Bigg\}\nn\\
 &&\!\!+G_{1a}(t,t_*)L_{aef*}\partial_i(v_{e*}v_{f*})+G_{1a}(t,t_*)N_{aef*}(\partial_iv_{e*})v_{f*},
\eea
and rewriting it using (\ref{timederA}) we find
\bea
 \tilde{\zeta}_{1i(2)}\!\!&=&\!\!-(\partial_iv_{e*})v_{f*}
\Bigg\{\!\!A_{ab}(t_*)G_{1a}(t,t_*)\delta_{be}\delta_{1f}
+\int_{t_*}^t\!\!\mathrm{d} t' G_{1a}(t,t')
\bar{A}_{ab\bar{c}}G_{be}(t',t_*)G_{\bar{c}f}(t',t_*)\nn\\
 &&\qquad\qquad\ \ \  -\int_{t_*}^t\!\!\mathrm{d}t'A_{ab}A_{1c}G_{1a}(t,t')G_{be}(t',t_*)G_{cf}(t',t_*)-G_{1a}(t,t_*)N_{aef*}\Bigg\}\nn\\
 &&\!\!+G_{1a}(t,t_*)L_{aef*}\partial_i(v_{e*}v_{f*}).\label{zu}
\eea
This expression should be symmetrical under the interchange of the indices 
$e$ and $f$. Notice that the last term is automatically symmetrical.

The anti-symmetrical
part of the two integrands turns out to be proportional to 
\be
T_a=G_{23}(t',t_*)G_{32}(t',t_*)-G_{22}(t',t_*)G_{33}(t',t_*).
\label{sym}
\ee
We explicitly check the exact numerical value of this quantity and find
it to be zero. For this it is crucial that we have defined $t_*$ as the time
a few (about 3) e-folds after horizon crossing.
The reason is that the long-wavelength 
approximation we use in all our derivations is only valid once the rapidly
decaying mode can be neglected, which takes a few e-folds. If the above
quantity were to be evaluated before that time, it would not yet be zero.
Note that in the slow-roll case $T_a$ is identically zero according to 
(\ref{GsolSR}), since within the slow-roll approximation the decaying mode 
is neglected by construction.
The above means that $\tilde{\zeta}_{1i}$ is well defined only after it is 
well outside the horizon, where we can neglect the decaying mode. 
The case where the decaying mode can remain important is treated in the 
one-field case in \cite{Takamizu:2010xy}.

The remaining non-integral terms between the braces can be explicitly checked 
to cancel when taking the slow-roll limit at horizon crossing: the first term 
of the first line of (\ref{zu}) gives
\be
-(\partial_iv_{e*})v_{f*}
A_{ab}(t_*)G_{1a}(t,t_*)\delta_{be}\delta_{1f}=\lh2\hpe
-\chi G_{12}(t,t_*)\rh\lh v_{1*}\partial_iv_{2*}-v_{2*}\partial_iv_{1*}\rh,
\ee
while the terms arising from the non-local contribution $N_{aef*}$ gives 
exactly the same but with opposite sign. Hence we see that within the
conditions of the long-wavelength approximation $\tilde{\zeta}^{(2)1}_i$
is indeed a total gradient, as it should be.

\Section{Detailed calculations}
\section*{}

\subsection{Relation between space and time derivatives}
\label{appTimeder}

We begin by proving that
	\be \label{timederA}
	\bA_{ab1} = - \frac{1}{NH} \der_t A_{ab}
	\ee 
for all gauges with $\der_i \alpha = 0$. Actually the statement is more general:
the prefactor of $\gz_i^1$ in the expression for $\der_i f$, where $f$ is any
function of $H,\Bgf,\BgP$, is equal to $-(\der_t f)/(NH)$ for gauges satisfying
$\der_i \alpha = 0$.

We start by computing the time derivative of $f$:
	\bea
	\frac{\der_t f(H,\Bgf,\BgP)}{NH} & = & \frac{1}{NH} 
	\lh \der_H f \, \der_t H + \Bnabla_\Bgf f \cdot
	\der_t \Bgf + \Bnabla_\BgP f \cdot \der_t \BgP \rh \non\\
	& = & -H \ge \, \der_H f 
	+ \frac{\sqrt{2\ge}}{\gk} \, \vc{e}_1 \cdot \Bnabla_\Bgf f
	+ \frac{\sqrt{2\ge}}{\gk} H \Bget \cdot \Bnabla_\BgP f,
	\eea
where we used the definitions of the slow-roll parameters $\ge$ and
$\Bget$, and of the basis vector $\vc{e}_1$.
On the other hand, the spatial derivative of $f$, in a $\der_i \alpha = 0$ gauge,
is given by
	\bea
	\der_i f(H,\Bgf,\BgP) & = &  
	\der_H f \, \der_i H + \Bnabla_\Bgf f \cdot
	\der_i \Bgf + \Bnabla_\BgP f \cdot \der_i \BgP \non\\
	& = & H \ge \, \der_H f \, \vc{e}_1 \cdot \Bgz_i
	- \frac{\sqrt{2\ge}}{\gk} \, \Bgz_i \cdot \Bnabla_\Bgf f\\
	&&- \frac{\sqrt{2\ge}}{\gk} H \lh {\textstyle\frac{1}{NH}} \, \Bgth_i
	+(\ge+\get^\parallel)\Bgz_i -\ge (\vc{e}_1 \cdot \Bgz_i)
	\vc{e}_1 \rh \cdot \Bnabla_\BgP f, \non
	\eea
using the constraint relations for $\der_i H$, $\der_i \Bgf$, and $\der_i \BgP$
given in \cite{Rigopoulos:2005xx,Rigopoulos:2005us}. (Note that some time
and space derivatives have to be replaced with their covariant (in field space)
version to make contact with the more general expressions given in those papers
that take into account a non-trivial field metric.)
Taking the components of $\Bgz$ and $\Bgth$ in the field basis, not forgetting the
relation $\vc{e}_m \cdot \Bgth_i = \gth_i^m + NH Z_{mn} \gz_i^n$, with
$Z_{21}=\get^\perp$ \cite{Rigopoulos:2005xx,Rigopoulos:2005us}, we prove the
stated relation, of which (\ref{timederA}) is a special case.

\subsection{Derivation of equation (\ref{fNLresult})}
\label{derivation}

In this appendix we work out the last term of (\ref{fNLint}), which has to
be added to the result for the second term ($I$) given in (\ref{Ires}), to
derive the final expression (\ref{fNLresult}) for $\f$. We call the sum of
these two terms $J$: 
\be 
\frac{J}{\gamma_*^2} \equiv \frac{I}{\gamma_*^2} 
+ \int_{t_*}^t \d t' G_{1a} \bA_{ab\bc} \bv_{bm} \bv_{\bc n},
\ee 
which is
        \bea
	\frac{J}{\gamma_*^2} = & \gd_{m2} \gd_{n1} 
	\Big( -2 \get^\perp_* + & \!\!\!\!\gc_* G_{12}(t,t_*)+ A_{32*} G_{13}(t,t_*)
	- \gc_* A_{33*} G_{13}(t,t_*) \Big) \nn\\
	&- \gd_{m2} \gd_{n2} \, 2 \get^\perp_* \gc_*
	& \!\!\!\!\!\!G_{13}(t,t_*) \\
 & + \gd_{m2} \gd_{n2} \int_{t_*}^t \d t' \Big [ & 
\!\!\!\!\lh \bA_{122} - 4(\get^\perp)^2 + \bA_{322} G_{13} \rh (\bv_{22})^2
+ \lh \bA_{333} + 2\get^\perp \rh G_{13} (\bv_{32})^2 \nn\\
	&&  \!\!\!\!+ \lh \bA_{123} - 4\get^\perp G_{12} 
	+ (\bA_{323} + \bA_{332}
	+ 2\get^\perp A_{33} + 2 \dot{\get}^\perp) G_{13} \rh 
	\bv_{22} \bv_{32} 
	\Big] . \non
	\eea
We remind the reader that the bar on top of an index ($\bc$) means that it 
does not take the value 1 and that a subscript $*$ means that a quantity is
evaluated at $t_*$.
The explicit form of the matrix $\mx{\bA}$ is given in \cite{Rigopoulos:2005us}:
\bea
\bA_{121} & = & 2\ge\get^\perp
-4\get^\parallel\get^\perp + 2\gx^\perp, \non\\
\bA_{122} & = & -6\gc - 2\ge\get^\parallel - 2(\get^\parallel)^2 
- 2(\get^\perp)^2, \non\\
\bA_{123} & = &	-6 - 2\get^\parallel, \non\\
\bA_{321}&=&-12\ge\get^\parallel - 12(\get^\perp)^2 
- 6\ge\gc - 8\ge^3 - 20\ge^2 \get^\parallel 
- 4\ge(\get^\parallel)^2 - 12\ge(\get^\perp)^2\nn\\
&&+ 16\get^\parallel(\get^\perp)^2 - 6\ge\gx^\parallel
- 12\get^\perp\gx^\perp + 3 (\tilde{W}_{111}-\tilde{W}_{221}),\nn\\
\bA_{322}&=&-24\ge\get^\perp - 12\get^\parallel\get^\perp 
+ 24\get^\perp\gc - 12\ge^2\get^\perp + 8(\get^\parallel)^2\get^\perp 
+ 8(\get^\perp)^3\nn\\
&& - 8\ge\gx^\perp - 4\get^\parallel\gx^\perp
+ 3 (\tilde{W}_{211}-\tilde{W}_{222}),\nn\\
\bA_{323}&=& 12\get^\perp - 4\ge\get^\perp + 8\get^\parallel\get^\perp
- 4\gx^\perp, \non\\
\bA_{331} & = & -2\ge^2 - 4\ge\get^\parallel +
2(\get^\parallel)^2 - 2(\get^\perp)^2 - 2\gx^\parallel, \non\\
\bA_{332} & = &	-4\ge\get^\perp - 2\gx^\perp, \non\\ 
\bA_{333} & = & -2\get^\perp,
\label{Abar}
\eea
while the rest of the matrix elements are zero.
Using these expressions we have
	\bea
	\bA_{333} + 2 \get^\perp = 0,
	\non\\
	\bA_{323} + \bA_{332} + 2 \get^\perp A_{33} + 2 \dot{\get}^\perp
	= 18 \get^\perp - 4 \dot{\get}^\perp,
	\non\\
	\bA_{123} = -2 A_{33} + 2 \ge + 2 \get^\parallel,
	\non\\
	\bA_{122} - 4 (\get^\perp)^2 
	= -2 A_{32} + 2 \dot{\ge} + 2 \dot{\get}^\parallel,
	\eea
so that we can write
	\bea
	\frac{J}{\gamma_*^2} = & \gd_{m2} \gd_{n1} 
	\Big( -2 \get^\perp_* + &\!\!\!\! \gc_* G_{12}(t,t_*)+ A_{32*} G_{13}(t,t_*)
	- \gc_* A_{33*} G_{13}(t,t_*) \Big) \nn\\
	&- \gd_{m2} \gd_{n2} \, 2 \get^\perp_* \gc_*
	&\!\!\!\!\!\! G_{13}(t,t_*) \non\\
 & + \gd_{m2} \gd_{n2} \int_{t_*}^t \d t' \Big [ & 
	\!\!\!\!2 (\bv_{22})^2  \frac{\d}{\d t'}(\ge+\get^\parallel)
	+ 2 \bv_{22} \frac{\d}{\d t'} \bv_{32}
	- 4 \lh \get^\perp G_{12} + \dot{\get}^\perp G_{13} \rh
	\frac{1}{2} \frac{\d}{\d t'} (\bv_{22})^2 \non\\
 &&\!\!\!\! + 2 (\ge+\get^\parallel) \bv_{22} \bv_{32}
	+ \bA_{322} G_{13} (\bv_{22})^2
	+ 18\get^\perp G_{13} \bv_{22} \bv_{32} 
	\Big ].
	\eea
Doing integrations by parts on the three terms in the third line we obtain
	\bea
	\frac{J}{\gamma_*^2} = & \gd_{m2} \gd_{n1} 
	\Big( -2 & \!\!\!\!\get^\perp_* + \gc_* G_{12}(t,t_*)+ A_{32*} G_{13}(t,t_*)
	- \gc_* A_{33*} G_{13}(t,t_*) \Big) \nn\\
	& + 2 \gd_{m2} \gd_{n2} \Biggl (
	&\!\!\!\! - \get^\perp_* \gc_* G_{13}(t,t_*)
	- (\ge_* + \get^\parallel_*) + \gc_*
	+ \get^\perp_* G_{12}(t,t_*)\nn\\
 	&&\!\!\!\!
	+ \dot{\get}^\perp_* G_{13}(t,t_*)
	+ (\ge + \get^\parallel) (\bv_{22})^2
	+ \bv_{22} \bv_{32} \Biggr ) \nn\\
	& + 2 \gd_{m2} \gd_{n2} &\!\!\!\! \int_{t_*}^t \d t' \Big [ 
	- 2 (\get^\perp)^2 (\bv_{22})^2
	- (\ge+\get^\parallel) \bv_{22} \bv_{32} 
	- (\bv_{32})^2
	+ 9 \get^\perp G_{13} \bv_{22} \bv_{32}\nn\\
	&& \qquad\quad
	+ \frac{1}{2} \lh \bA_{322} + 2 \ddot{\get}^\perp 
	+ 2 \dot{\get}^\perp A_{33} + 2 \get^\perp A_{32} 
	\rh G_{13} (\bv_{22})^2 \Big ] .
	\eea

The following relation (derived by taking two time derivatives of the 
field equation) can be used to remove higher-order slow-roll parameters:
	\be \label{Wm11rel}
	\tilde{W}_{m11}
	= -\frac{\get^{(4)}_m}{3} - \lh 1-\frac{\get^\parallel}{3} \rh \gx_m
	+ (2\ge+\get^\parallel) \get_m + \ge \get^\parallel \gd_{m1}
	- \get^\perp \tilde{W}_{m2}.
	\ee 
Explicitly, for $m=1$ and $m=2$ in the case of two fields, this becomes
	\bea\label{W111W211rel}
	\tilde{W}_{111}
	& = & -\frac{1}{3} \get^{(4)\,\parallel} 
	- \lh 1-\frac{1}{3}\get^\parallel \rh \gx^\parallel
	+ 3\ge\get^\parallel + (\get^\parallel)^2 + (\get^\perp)^2
	+ \frac{1}{3} \get^\perp \gx^\perp,
	\non\\
	\tilde{W}_{211}
	& = & -\frac{1}{3} \get^{(4)\,\perp} 
	- \lh 1-\frac{1}{3}\get^\parallel \rh \gx^\perp
	+ 3\ge\get^\perp + 2 \get^\parallel \get^\perp
	- \get^\perp \gc.
	\eea
Using the second of these relations, as well as the explicit expression for 
$\bA_{322}$, we find that
\bea
	&&\bA_{322} + 2 \ddot{\get}^\perp + 2 \dot{\get}^\perp A_{33}
	+ 2 \get^\perp A_{32}\\
&&\qquad\qquad	
        \!\!=24 \get^\perp \gc - 12 \get^\parallel \get^\perp
	+ 12 (\get^\parallel)^2 \get^\perp
        + 12 (\get^\perp)^3
	- 4 \get^\perp \gx^\parallel - 4 \get^\parallel \gx^\perp
	- 3(\tilde{W}_{211} + \tilde{W}_{222}). \non
	\eea
We now drop boundary terms that are second order in the slow-roll parameters 
{\em at horizon crossing}, since it would be inconsistent to include them given
that the linear solutions used at horizon crossing are only given up to first 
order. Then the result is
	\bea
	\frac{J}{\gamma_*^2} = & \gd_{m2} \gd_{n1} 
	 ( -2 \get^\perp_* + \gc_* & \bv_{12} ) 
	+ 2 \gd_{m2} \gd_{n2} \lh
	- \ge_*\! - \get^\parallel_*\! + \gc_*
	+ \get^\perp_* \bv_{12}
	+ (\ge + \get^\parallel) (\bv_{22})^2
	+ \bv_{22} \bv_{32} \rh \nn\\
 & + 2 \gd_{m2} \gd_{n2} \int_{t_*}^t \d t' \Bigg [ 
	& \!\!\!\!\!\!\!\!- 2 (\get^\perp)^2 (\bv_{22})^2
	- (\ge+\get^\parallel) \bv_{22} \bv_{32}
	- (\bv_{32})^2 + 9 \get^\perp G_{13} \bv_{22} \bv_{32} \non\\
 &&  \!\!\!\!\!\!\!\!
	+ \Bigg( 12 \get^\perp \gc - 6 \get^\parallel \get^\perp
	+ 6 (\get^\parallel)^2 \get^\perp  + 6 (\get^\perp)^3
	- 2 \get^\perp \gx^\parallel - 2 \get^\parallel \gx^\perp \non\\
 && \qquad\!\!\!\!\!\!\!\!
	- \frac{3}{2}( \tilde{W}_{211} + \tilde{W}_{222})
	\Bigg) G_{13} (\bv_{22})^2 \Bigg ] .
	\eea
Inserting this into (\ref{fNL}) gives the final result for $\f$ in
(\ref{fNLresult}).

%----------------------------- Appendix E --------------------------------------

{\Appendix{Analytical expressions for the spectral indices}\label{AppD}}

By differentiating (\ref{fnlfin}) and using (\ref{con}) we can find the final value of $n_K$ for equilateral triangles in the slow-roll approximation, 
assuming that isocurvature modes have vanished 
for an equal-power sum potential (for which the $\tg_{int}$ contribution is zero, see (\ref{gint}) and our paper \cite{Tzavara:2010ge}):
\bea
&&\!\!\!\!\!\!\!\!
n_{K,eq,f}=\label{A1}\\
&&\!\!\!\!\!\!\!\!
-4\frac{\bv_{12k}(\bv_{12k}\chi_k-2 \hpe_k)}{1+(\bv_{12k})^2}
-\frac{1}{f_{\mathrm{NL},eq,f}(1+(\bv_{12k})^2)^2} \Bigg[-2\e_k^2-3\e_k\hpa_k+(\hpa_k)^2+5(\hpe_k)^2-\xi^\parallel_k\nn\\
&&\!\!\!\!\!\!\!\!
+3\bv_{12k}\Bigg(\hpe_k(3\e_k+6\hpa_k-5\chi_k)-\xi^\perp_k\Bigg)+3(\bv_{12k})^2\Big(\tilde{W}_{221k}+4(\hpe_k)^2
-2(\e_k+\hpa_k-\chi_k)(\e_k+2\chi_k)\Big)\nn\\
&&\!\!\!\!\!\!\!\!
+\frac{(\bv_{12k})^3}{\hpe_k} 
\Bigg(\chi_k\Big(3\e_k^2-2\tilde{W}_{221k}+4 \e_k\hpa_k +3 (\hpa_k)^2 - 8(\hpe_k)^2 +\hpa\chi_k -3\chi_k^2\Big)
+\xi^\parallel_k(\e_{k}+\hpa_{k}-\chi_{k})\nn\\
&&\!\!\!\!\!\!\!\!
+\hpa_k(\e_k^2 - (\hpa_k)^2)+\tilde{W}_{221k}(\e_k+\hpa_k)+(\hpe_k)^2(2\e_k+5\hpa_k)-\xi^\perp_k\lh\hpe_{k}+\frac{(\e_{k}+\hpa_{k}+\chi_{k})\chi_{k}}{\hpe_{k}}\rh
\Bigg)
\Bigg],\nn
\eea
where $\tilde{W}_{221}=(\sqrt{2\e}/\kappa)W_{221}/(3H^2)$. 
We have checked this approximation and we find good agreement with the exact conformal index for equilateral triangles.

We repeat the calculation for the shape index $n_\omega$ (\ref{intodef}), differentiating the squeezed $\f$ (\ref{fnlb}). Where needed we use the slow-roll 
approximation $G_{32k'k}=-\chi_{k'}G_{22k'k}$ and $G_{23k'k}=G_{22k'k}/3$.
The result is:
\bea
 &&\!\!\!\!\!\!\!\!
 n_{\omega,sq,f}=\label{A2}\\
 &&\!\!\!\!\!\!\!\!
 \frac{1}{f_{\mathrm{NL},sq,f}(1+(\bv_{12})^2)^2}\frac{1}{1+2\omega}\Bigg\{
 \frac{2G_{22k'k}\bv_{12k'}}{1+(\bv_{12})^2}\lh4-\lh\omega
 +(2+\omega)\bv_{12k'}\rh\frac{\chi_{k'}}{\hpe_{k'}}\rh\nn\\
 &&\!\!\!\!\!\!\!\!
 \times\Bigg[\hpe_{k'}+\bv_{12k'}\lh3(\e_{k'}+\hpa_{k'})-2\chi_{k'}\rh+(\bv_{12k'})^2\lh \hpe_{k'}-\frac{(\e_{k'}+\hpa_{k'}-\chi_{k'})
 \chi_{k'}}{\hpe_{k'}}\rh\Bigg]
 \nn\\
 &&\!\!\!\!\!\!\!\!
 -G_{22k'k}\Bigg[2(\hpe_{k'})^2+\bv_{12k'}\lh-\xi^\perp_{k'}+\hpe_{k'}(11\e_{k'}+14\hpa_{k'}-8\chi_{k'})\rh
 +\frac{(\bv_{12k'})^3}{\hpe_{k'}}\Bigg(
 \chi_{k'}\Big(3\e_{k'}^2-2\tilde{W}_{221k'}\nn\\
 &&\!\!\!\!\!\!\!\!
 +4\e_{k'}\hpa_{k'}+3(\hpa_{k'})^2-7(\hpe_{k'})^2
 -2\chi_{k'}^2-\e_{k'}\chi_{k'}\Big)+\xi^\parallel(\e_{k'}+\hpa_{k'}-\chi_{k'})+\hpa_{k'}(\e_{k'}^2-(\hpa_{k'})^2)\nn\\
&&\!\!\!\!\!\!\!\!
+\tilde{W}_{221k'}(\e_{k'}+\hpa_{k'})
 +(\hpe_{k'})^2(2\e_{k'}+5\hpa_{k'})
 -\xi^\perp_{k'}\lh\hpe_{k'}+\frac{(\e_{k'}+\hpa_{k'}+\chi_{k'})\chi_{k'}}{\hpe_{k'}}\rh \Bigg)\nn\\
 &&\!\!\!\!\!\!\!\!
 +(\bv_{12k'})^2\lh2\tilde{W}_{221k'}-6\e_{k'}^2+(\hpa_{k'})^2+9(\hpe_{k'})^2-\xi_{k'}^\parallel-9\hpa_{k'}\chi_{k'}+8\chi_{k'}^2-\e_{k'}(7\hpa_{k'}+5\chi_{k'})\rh
 \Bigg]\nn\\
 &&\!\!\!\!\!\!\!\!
 -\frac{1}{1+(\bv_{12k'})^2}\Bigg[(\hpa_{k'})^2+3(\hpe_{k'})^2-2\e_{k'}^2-3\e_{k'}\hpa_{k'}-\xi_{k'}^\parallel
 -2(\bv_{12k'})^3\lh\xi_{k'}^\perp+\hpe_{k'}(\e_{k'}-2\hpa_{k'}-5\chi_{k'})\rh\nn\\
 &&\!\!\!\!\!\!\!\!
 +(\bv_{12k'})^4\lh \tilde{W}_{221k'} +(\e_{k'}-\hpa_{k'})\hpa_{k'}+3(\hpe_{k'})^2+\xi_{k'}^\parallel+2\chi_{k'}(\e_{k'}+\chi_{k'})\rh
 -2\bv_{12k'}\big( \xi_{k'}^\perp+\hpe_{k'}(5\e_{k'}\nn\\
 &&\!\!\!\!\!\!\!\!
 +2\hpa_{k'}+3\chi_{k'})\big)
 +(\bv_{12k'})^2\lh \tilde{W}_{221k'} +2\lh-\e_{k'}(\e_{k'}+\hpa_{k'})-5(\hpe_{k'})^2+(3\e_{k'}+3\hpa_{k'}+\chi_{k'})\chi_{k'}\rh\rh
 \Bigg]\Bigg\}\nn
 \eea
We have checked this approximation and we find good agreement with the exact shape index for $\omega\gtrsim3$.

\renewcommand\listfigurename{List of figures}

\fancyhead[LE]{\thepage~~~~List of figures}
\fancyhead[RO]{List of figures~~~~~\thepage}

\listoffigures
%\addcontentsline[lof]{lof}{entry}
\newpage~\thispagestyle{empty}

% Change title of listoftables
\renewcommand\listtablename{List of tables}
%Define non-default headers
\fancyhead[LE]{\thepage~~~~List of tables}
\fancyhead[RO]{List of tables~~~~~\thepage}
%I make it to appear in Contents
%\addcontentsline{toc}{chapter}{List of tables}
\listoftables
%\addcontentsline[lot]{lot}{entry}
\newpage~\thispagestyle{empty}

%-------------------------------------------------------------------------------
%-------------------------------------------------------------------------------

\fancyhead[LE]{\thepage~~~~Bibliography}
\fancyhead[RO]{Bibliography~~~~~\thepage}

\bibliography{bib_new}{}

\bibliographystyle{utphys.bst}

\end{document}